\newcommand*{\ATLASLATEXPATH}{latex/}
\documentclass[cernpreprint,UKenglish,texlive=2011,txfonts,UKenglish]{\ATLASLATEXPATH atlasdoc}
\pdfoutput=1
\usepackage[block=space]{\ATLASLATEXPATH atlaspackage}
\usepackage{\ATLASLATEXPATH atlasbiblatex}
\usepackage{\ATLASLATEXPATH atlascontribute}
\usepackage[process]{\ATLASLATEXPATH atlasphysics}
\bibliography{ZPTPhiStarPaperReferences.bib}

\graphicspath{{logos/}{figures/}}

\usepackage{\ATLASLATEXPATH atlascover}
\usepackage{multirow}

\AtlasTitle{Measurement of the transverse momentum and $\phi^*_{\eta}$ distributions of Drell--Yan lepton pairs in proton--proton collisions at $\sqrt{s}=8$~TeV with the ATLAS detector}

\author{The ATLAS Collaboration}

\AtlasRefCode{STDM-2014-12}
\PreprintIdNumber{CERN-PH-EP-2015-275}

 \AtlasJournal{Eur. Phys. J. C }

\AtlasAbstract{%
Distributions of transverse momentum $p_T^{\ell\ell}$ and the related angular variable $\phi^*_\eta$ of Drell--Yan lepton pairs are
measured in 20.3~fb$^{-1}$ of proton--proton collisions at $\sqrt{s}=8$~TeV with the ATLAS detector at the
LHC.  Measurements in electron-pair and muon-pair final states are corrected for detector effects and
combined.  Compared to previous measurements in proton--proton collisions at $\sqrt{s}=7$~TeV, these
new measurements benefit from a larger data sample and improved control of systematic uncertainties.
Measurements are performed in bins of lepton-pair
mass above, around and below the $Z$-boson mass peak.  The data are compared to predictions from perturbative and
resummed QCD calculations.  For values of $\phi^*_\eta < 1$ the predictions from the Monte
Carlo generator \textsc{ResBos} are generally consistent with the data within the theoretical uncertainties.
However, at larger values of $\phi^*_\eta$ this is not the case.  Monte Carlo generators based on the
parton-shower approach are unable to describe the data over the full range of $p_T^{\ell\ell}$ while the
fixed-order prediction of \textsc{Dynnlo} falls below the data at high values of $p_T^{\ell\ell}${}.
\textsc{ResBos} and the parton-shower Monte Carlo generators provide a much better description of the evolution of the $\phi^*_\eta$ and
$p_T^{\ell\ell}$ distributions as a function of lepton-pair mass and rapidity than the basic shape of the data.
}

 \AtlasCoverSupportingNote{Support documentation}{https://cds.cern.ch/record/1694674}
\AtlasCoverCommentsDeadline{02 December 2015}

 \AtlasCoverAnalysisTeam{Mikhail Karnevskiy, Tai-Hua Lin, Mykhailo Lisovyi, Matthias Schott, Lee Tomlinson, Samuel Webb, Terry Wyatt, Christoph Zimmermann}

 \AtlasCoverEdBoardMember{Nathalie Besson, Daniel Fournier (chair), Andreas Warburton, Tony Weidberg}

   \AtlasCoverEgroupEditors{atlas-stdm-2014-12-editors@cern.ch} 
   
 \AtlasCoverEgroupEdBoard{atlas-stdm-2014-12-editorial-board@cern.ch}

\hypersetup{pdftitle={ATLAS paper},pdfauthor={The ATLAS Collaboration}}

\usepackage{booktabs}

\newcommand{\roots}{\ensuremath{\sqrt{s}}}

\newcommand{\Acermc}{\textsc{AcerMC}}
\newcommand{\Pythia}{\textsc{Pythia}}

\newcommand{\Powheg}{\textsc{Powheg}}
\newcommand{\Mcatnlo}{\textsc{MC@NLO}}
\newcommand{\Herwig}{\textsc{Herwig}}
\newcommand{\Photos}{\textsc{Photos}}
\newcommand{\Jimmy}{\textsc{Jimmy}}

\newcommand{\Alpgen}{\textsc{Alpgen}}
\newcommand{\Sherpa}{\textsc{Sherpa}}
\newcommand{\Resbos}{\textsc{ResBos}}
\newcommand{\Dynnlo}{\textsc{Dynnlo}}
\newcommand{\Geant}{\textsc{Geant}}

\newcommand{\WZpt}{\ensuremath{p_\mathrm{T}^{(W,Z)}}}
\newcommand{\Zpt}{\ensuremath{p_\mathrm{T}^{\ell\ell}}}
\newcommand{\dR}{\ensuremath{\Delta R}}
\newcommand{\mody}{\ensuremath{|y_{\ell\ell}|}}
\newcommand{\modeta}{\ensuremath{|\eta|}}
\newcommand{\Mll}{\ensuremath{m_{\ell\ell}}}
\newcommand{\phistar}{\ensuremath{\phi^*_{\eta}}}
\newcommand{\PhiStar}{\ensuremath{\phi^*_{\eta}}}

\newcommand{\Imu}{{\ensuremath{I_\mu}}}
\newcommand{\Ie}{{\ensuremath{I_e}}}
\newcommand{\Iemin}{{\ensuremath{I_e^{\textrm{min}}}}}
\newcommand{\Iestar}{{\ensuremath{I_e^*}}}

\newcommand{\Vast}{\Bigg@{5}}

\begin{document}
\maketitle

\newpage
\tableofcontents

\newpage
\clearpage

\section{Introduction}

In high-energy hadron--hadron collisions the vector bosons $W$ and $Z/\gamma^*$ are produced via
quark--antiquark annihilation, and may be observed with very small backgrounds in their leptonic decay
modes.
The vector bosons may have non-zero momentum transverse to the beam direction \WZpt\ due to the
emission of quarks and gluons from the initial-state partons as well as to the intrinsic
transverse momentum of the initial-state partons in the proton.
Phenomenologically, the spectrum at low \WZpt{} can be 
described using soft-gluon resummation~\cite{Collins:1984kg} together with a non-perturbative contribution
from the parton intrinsic transverse momentum.
At high \WZpt\ the spectrum may be described by fixed-order perturbative QCD
predictions~\cite{Melnikov:2006kv,Li:2012wna,Catani:2009sm}.
Parton-shower models~\cite{Corcella:2000bw,Sjostrand:2006za} may be used to compensate for missing higher-order corrections in the
fixed-order QCD predictions.

Measurements of \WZpt{} thus test several aspects of QCD.
The correct modelling of \WZpt{} is also
important in many physics analyses at the LHC for which the production of $W$ and/or $Z$ bosons constitutes a
background.
Moreover, it is a crucial ingredient for a precise measurement of the $W$-boson
mass, at both the LHC and the Tevatron.
Measurements of the dependence of \WZpt{}  on the boson rapidity~\cite{Coordinates} are sensitive to the gluon distribution function of the proton~\cite{Brandt:2013hoa}.
High-precision measurements at large values of \WZpt\ could be sensitive to electroweak (EW) corrections~\cite{Dittmaier:2014qza}.

Drell--Yan events with final states including $e^+ e^-$ or $\mu^+ \mu^-$ (`Drell--Yan lepton pairs') allow the transverse momentum  \Zpt\ of $Z/\gamma^*$
bosons to be measured with greater precision than is possible in the case of $W$ bosons, because of the
unobserved neutrino produced in $W$ leptonic decays.
Measurements of \Zpt{} for lepton-pair masses, \Mll, around the $Z$-boson mass peak have been made by the
CDF Collaboration~\cite{Affolder_1999} and the D0 Collaboration~\cite{Abbott:1999yd,Abazov_2007,Abazov:2010kn} at the Tevatron, and the ATLAS Collaboration~\cite{Aad:2011gj,Aad:2014xaa},
the CMS Collaboration~\cite{Chatrchyan:2011wt,Khachatryan:2015oaa} and the LHCb Collaboration~\cite{Aaij:2012mda,Aaij:2015gna,Aaij:2015vua} at the LHC.
Measurements of \Zpt{} require a precise understanding of the transverse momentum \pt{} calibration and resolution of the
final-state leptons.
Associated systematic uncertainties affect the resolution in \Zpt\ and limit the ultimate precision of
the measurements, particularly in the low-\Zpt\ domain.
To minimise the impact of these uncertainties, the $\phi_\eta^*$ observable was introduced~\cite{Banfi:2010cf} as an alternative
probe of \Zpt{}.
It is defined as 

\begin{equation}
\phi_\eta^* = \tan\left(\frac{\pi-\Delta\phi}{2}\right) \cdot \sin(\theta^*_\eta)\,,
\end{equation}

where $\Delta \phi$ is the azimuthal angle in radians between the two leptons.
The angle $\theta_\eta^*$ is a measure of the scattering angle of the leptons with respect to the proton
beam direction in the rest frame of the dilepton system and is defined by
$\cos(\theta_\eta^*) = \tanh[(\eta^- - \eta^+)/2]$,
where $\eta^-$ and $\eta^+$ are the pseudorapidities of the negatively and positively charged lepton,
respectively~\cite{Banfi:2010cf}.
Therefore, \phistar\ depends exclusively on the directions of the two leptons, which are more precisely measured than their momenta. 
Measurements of \phistar\ for \Mll\ around the $Z$-boson mass peak were first made by the D0 Collaboration~\cite{Abazov:2010mk} at the
Tevatron and subsequently by the ATLAS Collaboration~\cite{Aad:2012wfa} for $\roots=7 \TeV$  and the LHCb Collaboration for $\roots=7 \TeV$~\cite{Aaij:2012mda,Aaij:2015gna} and~$8 \TeV$~\cite{Aaij:2015vua} at the LHC.
First measurements of $\phi^*_\eta$ for ranges of \Mll{} above and below the
$Z$-boson mass peak were recently presented by the D0 Collaboration~\cite{Abazov:2014mza}.

Measurements are presented here of \phistar\ and \Zpt\ for Drell--Yan lepton-pair events using the complete $\roots=8 \TeV$
data set of the ATLAS experiment at the LHC, corresponding to an integrated luminosity of 20.3~\ifb.
The data are corrected for detector effects.
The measurements are presented for $e^+ e^-$ and $\mu^+ \mu^-$  final states, in bins of \Mll, above and
below, as well as at the
$Z$-boson mass peak, and in bins of the $Z/\gamma^*$-boson rapidity \mody{}.
In addition, integrated fiducial cross sections are provided for six regions of \Mll{}.

The ATLAS experiment is
briefly described in Section~\ref{sec:ATLAS}.
A general overview of the measurement methods is given in Section~\ref{sec:Measurement}, which has specific sections
on the event simulation, event reconstruction, event
selection, background estimation, corrections for detector effects, and the evaluation of the systematic uncertainties.
The combination of the measurements in the  $e^+ e^-$ and $\mu^+ \mu^-$  final states is described in Section~\ref{sec:Results}.
The corrected differential cross sections are compared to various theoretical predictions in Section~\ref{sec:Comparison}.
A short summary and conclusion are given in Section~\ref{sec:Conclusion}.
The values of the normalised differential cross sections $(1/\sigma)\, \mathrm{d}\sigma / \mathrm{d}\phistar$ and  $(1/\sigma)\, \mathrm{d}\sigma / \mathrm{d}\Zpt$ are given in tables in the
appendix for each region of \Mll\ and \mody\ considered.

\section{\label{sec:ATLAS}The ATLAS detector}

\newcommand{\AtlasCoordFootnote}{%
ATLAS uses a right-handed coordinate system with its origin at the nominal interaction point (IP)
in the centre of the detector and the $z$-axis along the beam pipe.
The $x$-axis points from the IP to the centre of the LHC ring,
and the $y$-axis points upwards.
Cylindrical coordinates $(r,\phi)$ are used in the transverse plane, 
$\phi$ being the azimuthal angle around the $z$-axis.
The pseudorapidity is defined in terms of the polar angle $\theta$ as $\eta = -\ln \tan(\theta/2)$.
Angular distance is measured in units of $\Delta R \equiv \sqrt{(\Delta\eta)^{2} + (\Delta\phi)^{2}}$.}

The ATLAS detector~\cite{TheATLASExperiment} at the LHC covers nearly the entire solid angle around the collision point.
It consists of an inner tracking detector (ID) surrounded by a thin superconducting solenoid, electromagnetic and hadronic calorimeters,
and a muon spectrometer (MS) incorporating three large superconducting toroid magnets.
The ID is immersed in a \SI{2}{\tesla} axial magnetic field 
and provides charged-particle tracking in the range $|\eta| < 2.5$.
A high-granularity silicon pixel detector typically provides three measurements per track, and
is followed by a silicon microstrip tracker, which usually provides four three-dimensional measurement points per track.
These silicon detectors are complemented by a transition radiation tracker,
which enables radially extended track reconstruction up to $|\eta| = 2.0$. 
The transition radiation tracker also provides electron identification information 
based on the fraction of hits (typically 30 in total) above a higher energy-deposit threshold corresponding to transition radiation.

The calorimeter system covers the pseudorapidity range $|\eta| < 4.9$.
Within the region $|\eta|< 3.2$, electromagnetic calorimetry is provided by barrel and 
endcap high-granularity lead/liquid-argon (LAr) electromagnetic calorimeters,
with an additional thin LAr presampler covering $|\eta| < 1.8$,
to correct for energy loss in material upstream of the calorimeters.
Hadronic calorimetry is provided by the steel/scintillator-tile calorimeter,
segmented into three barrel structures within $|\eta| < 1.7$, and two copper/LAr hadronic endcap calorimeters.
The solid angle coverage is completed with forward copper/LAr and tungsten/LAr calorimeter modules
optimised for electromagnetic and hadronic measurements, respectively.

The MS comprises separate trigger and
precision tracking chambers measuring the deflection of muons in a magnetic field generated by superconducting air-core toroids.
The precision chamber system covers the region $|\eta| < 2.7$ with three layers of monitored drift tubes,
complemented by cathode-strip chambers in the forward region, where the background is highest.
The muon trigger system covers the range $|\eta| < 2.4$ with resistive-plate chambers in the barrel, and thin-gap chambers in the endcap regions.

A three-level trigger system is used to select interesting events~\cite{Aad:2012xs}.
The Level-1 trigger is implemented in hardware and uses a subset of detector information
to reduce the event rate to a design value of at most \SI{75}{\kHz}.
This is followed by two software-based trigger levels which together reduce the event rate to about \SI{400}{\Hz}.

\section{\label{sec:Measurement}Analysis methods}

This section describes the particle-level measurements presented in this paper (Section~\ref{sec:measurementdefinition}), the simulation of signal and background Monte Carlo (MC) samples (Section~\ref{sec:eventsimulation}),
the event reconstruction and selection criteria (Section~\ref{sec:eventreconstruction}),
the estimation of backgrounds (Section~\ref{sec:background}),
corrections to the distributions of \phistar\ and \Zpt\ for detector effects and final-state radiation (Section~\ref{sec:corrections}),
and the estimation of systematic uncertainties (Section~\ref{sec:systematics}).

\subsection{Description of the particle-level measurements}
\label{sec:measurementdefinition}

Drell--Yan signal MC simulation is used to correct the background-subtracted data for detector resolution
and inefficiency. Three different `particle-level' definitions are employed,
which differ in their treatment of final-state photon radiation (FSR).
The Born and bare levels are defined from the lepton kinematics before and after FSR, respectively.
The dressed level is defined by combining the bare four-momentum of each lepton with that of
photons radiated within a cone defined by $\Delta R = 0.1$~\cite{Coordinates} around the lepton.
The muon-pair data are corrected to the bare, dressed, and Born levels.
The electron-pair data are corrected to the dressed and Born levels.
The two lepton-pair channels are combined at the Born level.
The bare and dressed particle-level definitions reduce the dependence on the MC FSR model used to correct the data,
which results (particularly for events with \Mll\ below the $Z$-boson mass peak) in a lower systematic uncertainty.
Corrections to a common particle-level definition (Born level) for the combination of the two channels
allow comparisons to calculations that do not account for the effects of FSR, albeit at the cost of an
increased systematic uncertainty on the corrected data.

The data are corrected to the particle level within fiducial regions in lepton $\pt$ and
\modeta{}, and in lepton-pair \Mll\ and \mody\ that correspond closely to the selection criteria applied to the data. 
The fiducial regions common to the measurements of \phistar{} and \Zpt{} are described first.
The two leptons are required to have $\pt{} > 20 \GeV$ and $|\eta|<2.4$.
Measurements of the normalised differential cross sections $(1/\sigma)\, \mathrm{d}\sigma / \mathrm{d}\phistar$ and $(1/\sigma)\, \mathrm{d}\sigma / \mathrm{d}\Zpt$,
and of the absolute differential cross section $\mathrm{d}\sigma / \mathrm{d}\Zpt$, are made in
three \Mll\ regions within $46 \GeV < \Mll < 150 \GeV$ for $\mody<2.4$.
In the mass region $66 \GeV < \Mll < 116 \GeV$, measurements are
made in six equally sized regions of \mody.
The distributions of $(1/\sigma)\, \mathrm{d}\sigma / \mathrm{d}\phistar$ and $(1/\sigma)\, \mathrm{d}\sigma / \mathrm{d}\Zpt$ are
individually normalised in each region of \mody.
Measurements of  $(1/\sigma)\, \mathrm{d}\sigma / \mathrm{d}\phistar$ in the regions of \Mll\ above and below the $Z$-boson
mass peak, $46 \GeV < \Mll < 66 \GeV$ and
$116 \GeV < \Mll < 150 \GeV$, are made in three equally-sized regions of $\mody$.
For $\Zpt>45 \GeV$, measurements of \Zpt{} are made in three additional mass regions below $46 \GeV$.

A synopsis of the \phistar{} and \Zpt{} measurements, and of the fiducial-region definitions used is given in Table~\ref{tab:FiducialDefinition}.

\begin{table*}
\centering
\caption{Synopsis of the \phistar{} and \Zpt{} measurements, and of the fiducial region definitions used. Full details including the definition of the Born, bare and dressed particle levels are provided in the text. Unless otherwise stated criteria apply to both \phistar{} and \Zpt{} measurements.}
\begin{tabular}{l l}
\toprule

\multicolumn{2}{l}{\underline{Particle-level definitions (Treatment of final-state photon radiation)}} \\
electron pairs        & \multicolumn{1}{l}{dressed;\, Born}                                                                             \\
muon pairs            & \multicolumn{1}{l}{bare;\, dressed;\, Born}                                                                       \\
combined              & \multicolumn{1}{l}{Born}                                                                                      \\
\midrule

\multicolumn{2}{l}{\underline{Fiducial region}}                                                                   \\
Leptons               & \multicolumn{1}{l}{$p_T > 20 \GeV$ and $\modeta < 2.4$}                                                           \\
Lepton pairs          & \multicolumn{1}{l}{$\mody < 2.4$}                                                                             \\
\midrule

\multicolumn{2}{l}{\underline{Mass and rapidity regions}} \\
$46 \GeV<\Mll<66 \GeV$      & $\mody<0.8$;\, \ $0.8<\mody<1.6$;\, \ $1.6<\mody<2.4$   \\
 & (\phistar\ measurements only) \\
\cmidrule{2-2}
                            & $\mody<2.4$            \\
\cmidrule{1-2}
$66 \GeV<\Mll<116 \GeV$     & $\mody<0.4$;\, \ $0.4<\mody<0.8$;\, \ $0.8<\mody<1.2$;\, \\
 & $1.2<\mody<1.6$;\, \ $1.6<\mody<2.0$;\, \  $2.0<\mody<2.4$;\, \\
 & $\mody<2.4$          \\
\cmidrule{1-2}

$116 \GeV<\Mll<150 \GeV$      & $\mody<0.8$;\, \ $0.8<\mody<1.6$;\, \ $1.6<\mody<2.4$   \\
 & (\phistar\ measurements only) \\
\cmidrule{2-2}
      & $\mody<2.4$            \\

\midrule
\midrule

\multicolumn{2}{l}{\underline{Very-low mass regions}} \\

$12 \GeV<\Mll<20 \GeV$      & \multirow{3}{*}{ \Bigg \} \,  $\mody<2.4$, $\Zpt>45 \GeV$, \Zpt{} measurements only}      \\
$20 \GeV<\Mll<30 \GeV$      & \\
$30 \GeV<\Mll<46 \GeV$      & \\

\bottomrule

  \end{tabular}

  \label{tab:FiducialDefinition}  
\end{table*}

\subsection{Event simulation}
\label{sec:eventsimulation}

MC simulation is used to estimate backgrounds and to correct the data for detector
resolution and inefficiencies, as well as for the effects of FSR.

Three generators are used to produce samples of Drell--Yan lepton-pair signal events.
The first is \Powheg{}~\cite{Alioli:2008gx,Alioli:2010xd} which uses the
CT10 set of parton distribution functions (PDFs)~\cite{Gao:2013xoa} and is interfaced to
\Pythia{} 8.170~\cite{Sjostrand:2006za,Sjostrand:2007gs} with the AU2 set of tuned parameters (tune)~\cite{ATLAS:2012uec}
to simulate the parton shower, hadronisation and underlying event, and to \Photos{}~\cite{Golonka:2005pn} to simulate FSR.
This is referred to as \Powheg{}+\Pythia{} in the text.
The second is \Powheg{} interfaced to \Herwig{} 6.520.2~\cite{Corcella:2000bw} for the parton shower and hadronisation, \Jimmy{}~\cite{Butterworth:1996zw} for the underlying event, and \Photos{} for FSR (referred to as \Powheg{}+\Herwig{}).
The \Sherpa{} 1.4.1~\cite{Gleisberg:2008ta} generator is also used, which has its own implementation of the parton
shower, hadronisation, underlying event and FSR, and which again uses the
CT10 PDF set.
Differences between the results obtained using these three generators are used to estimate systematic uncertainties related to the choice of generator.

Background events from the process $Z\rightarrow\tau\tau$ are produced using \Alpgen{}~\cite{Mangano:2002ea}
interfaced to \Herwig{} to simulate the parton shower and \Jimmy{} to simulate the underlying event.
Single $W$-boson decays to electrons, muons and $\tau$ leptons are produced with \Sherpa, and the diboson processes $WW$, $WZ$ and $ZZ$ are produced with \Herwig.
The $t\bar{t}$ process is simulated with \Mcatnlo{}~\cite{Frixione:2002ik} interfaced to \Jimmy{},
as is the single-top process in the \textit{s}-channel and \textit{Wt}-channel.
The \textit{t}-channel is generated with \Acermc{}~\cite{Kersevan:2004yg} interfaced to \Pythia{}.
Exclusive $\gamma\gamma\rightarrow\ell\ell$ production is generated using the \Herwig{}++ 2.6.3
generator~\cite{Bahr:2008pv}. Photon-induced single-dissociative dilepton production,
is simulated using Lpair 4.0~\cite{Vermaseren:1982cz} with the Brasse~\cite{Brasse:1976bf} and Suri--Yennie~\cite{Suri:1971yx}
structure functions for proton dissociation. For
double-dissociative $\gamma\gamma\rightarrow \ell\ell$ reactions, \Pythia{} 8.175~\cite{Sjostrand:2007gs} is used with the
MRST2004QED~\cite{Martin:2004dh} PDFs.

The effect of multiple interactions per bunch crossing (pile-up) is simulated by overlaying MC-generated minimum bias events~\cite{Aad:2010ah}.
The simulated event samples are reweighted to describe the distribution of the number of pile-up
events in the data.
The \Geant4~\cite{Agostinelli:2002hh} program is used to simulate the passage of particles through the
ATLAS detector.
Differences in reconstruction, trigger, identification and isolation
efficiencies between MC simulation and data are evaluated using a tag-and-probe method~\cite{ElectronEfficiency_2014,Aad:2014rra} and are corrected for by
reweighting the MC simulated events.
Corrections are also applied to MC events for the description of the lepton energy and momentum scales and resolution,
which are determined from fits to the observed \Zboson{}-boson line shapes in data and MC simulation~\cite{Aad:2014nim,Aad:2014rra}.
The MC simulation is also reweighted to better describe the distribution of the longitudinal position of the primary $pp$ collision vertex~\cite{ATLAS:2010lca} in data.

Three additional samples of Drell--Yan lepton-pair signal events are produced without detector simulation, for the purpose of comparison with the corrected data in Section~\ref{sec:Comparison}.
The MC generators used are \Resbos{}, \Dynnlo{}, and \Powheg{}+\Pythia{} (AZNLO tune).

\Resbos{}~\cite{Balazs:1995nz} simulates vector-boson production and decay, but does not include a
description of the hadronic activity in the event nor of FSR.
 Initial-state QCD corrections to $Z$-boson production are simulated at approximately next-to-next-to-leading-order (NNLO) accuracy using 
approximate NNLO (i.e. $\mathcal{O}(\alpha_s^2)$) Wilson coefficient functions~\cite{Guzzi:2013aja}.
The contributions from $\gamma^{*}$ and from $Z/\gamma^{*}$ interference are 
simulated at next-to-leading-order (NLO) accuracy (i.e. $\mathcal{O}(\alpha_s)$). 
\Resbos{} uses a resummed treatment of soft-gluon emissions at next-to-next-to-leading-logarithm (NNLL) accuracy.
It uses the GNW parameterisation~\cite{Guzzi:2013aja,Guzzi:2012jc} of non-perturbative
effects at small \Zpt{}, as optimised using the
D0 \phistar\ measurements in Ref.~\cite{Abazov:2010mk}.
The CT14 NNLO PDF sets~\cite{Dulat:2015mca} are used and the corresponding 90\% confidence-level PDF uncertainties are
evaluated and rescaled to 68\% confidence level.
The choices~\cite{ScaleVariations} of central values and
range of systematic uncertainty variations
for QCD scales and the non-perturbative parameter $a_Z$ are made following Ref.~\cite{Guzzi:2013aja}. These differ from the choices made for the ATLAS $7 \TeV$ \Zpt{} and \phistar{} papers~\cite{Aad:2014xaa,Aad:2012wfa}.

\textsc{Dynnlo1.3}~\cite{Catani:2009sm} simulates initial-state QCD corrections to NNLO accuracy.
The CT10 NNLO PDF sets are used.
The \Dynnlo{} calculation is performed in the $G_{\mu}$ electroweak parameter scheme~\cite{Dittmaier:2009cr}.
Additional NLO electroweak virtual corrections~\cite{DYNNLOTechnical} are provided by the authors of Ref.~\cite{Denner:2011vu}. 
\Dynnlo\ does not account for the effects of multiple soft-gluon emission and therefore is not
able to make accurate predictions at low \phistar\ and \Zpt{}.

An additional \Powheg{}+\Pythia{} sample is produced which uses the AZNLO tune~\cite{Aad:2014xaa}. This tune includes the ATLAS $7 \TeV$ \phistar{} and \Zpt{} results in a mass region around the \Zboson{} peak. The sample uses \Pythia{} version 8.175 and the CTEQ6L1 PDF set~\cite{Pumplin:2002vw} for
the parton shower, while CT10 is used for the \Powheg{} calculation.

\subsection{Event reconstruction and selection}
\label{sec:eventreconstruction}
The measurements are performed using proton--proton collision data recorded at $\sqrt{s} = 8\,\TeV$. The data were collected between April and December 2012 and correspond to an integrated luminosity of $20.3\,\ifb$. Selected events are required to be in a data-taking period in which there were stable beams and the detector was fully operational.

For measurements of \phistar{}, candidate electron-pair events were obtained using a dielectron trigger,
whilst for measurements of \Zpt{}, a combination of a single-electron trigger
(to select events with the leading reconstructed electron $\pt > 60 \GeV$ and the sub-leading electron $\pt > 25 \GeV$) and a dielectron trigger (to select all other events) was used.
The motivation for using a slightly different trigger selection for measurements of the \Zpt\ observable
is to obtain a higher efficiency for electron pairs with $\dR < 0.35$, which is relevant to maintain a high acceptance for $\Mll<46 \GeV$.
Electron candidates are reconstructed from clusters of energy in the electromagnetic calorimeter matched to ID tracks~\cite{Aad:2014fxa}.
They are required to have $\pt > 20 \GeV$ and $|\eta| < 2.4$, but excluding the transition regions between the barrel and the endcap electromagnetic calorimeters,
$1.37 < |\eta| < 1.52$. The electron candidates must satisfy a set of `medium' selection criteria~\cite{Aad:2014fxa} that have been reoptimised
for the larger number of proton--proton collisions per beam crossing observed in the 2012 data.
Events are required to contain exactly two electron candidates.
Except for the \Mll\ region around the $Z$-boson mass peak, the electron candidates are required to be
isolated, satisfying \Ie~$<$~0.2, where \Ie\ is
the scalar sum of the $\pt$ of tracks with $\Delta R < 0.4$ around the electron track divided by the \pt{} of the electron.
For measurements of \Zpt{}, this requirement is not applied when the two electrons are separated by $\Delta R < 0.5$.
For measurements of \Zpt{} the two electron candidates must satisfy $\Delta R > 0.15$.

Candidate muon-pair events  are retained for further analysis using a combination of a
single-muon trigger (for $\pt > 25$~GeV) and a dimuon trigger (for $20 < \pt < 25$~GeV). Muon
candidates are reconstructed by combining tracks reconstructed in both the inner detector and
the MS~\cite{Aad:2014rra}. They are required to have $\pt > 20 \GeV$ and
$|\eta| < 2.4$. In order to suppress backgrounds, track-quality requirements are imposed for muon
identification, and longitudinal and transverse impact-parameter requirements ensure that the muon candidates originate from a common
primary proton--proton interaction vertex. The muon candidates are also required to be isolated, satisfying \Imu~$<$~0.1, where
\Imu\ is the scalar sum of the \pt{} of
tracks within a cone of size $\Delta R = 0.2$ around the muon divided by the \pt{} of the
muon.  Events are required to contain exactly two muon candidates of opposite charge satisfying the above criteria.

Precise knowledge of the lepton directions is particularly important for the \phistar{} measurements. These are determined for electron candidates by the track direction in the ID, and for muon candidates from a combination of the track direction in the ID and in the MS.

Tables~\ref{tab:DataZ_electron} and~\ref{tab:DataZ_muon} show the number of events satisfying the
above selection criteria in the electron-pair and muon-pair channels, respectively, for six regions of \Mll. Also
given is the estimated contribution to the data from the various background sources considered (described in Section~\ref{sec:background}).

\begin{table*}
\centering
\caption{The number of events in data satisfying the selection criteria in the electron-pair channel for six different regions of \Mll\ and the estimated contribution to this value from the various background sources considered. The uncertainties quoted on the background samples include contributions from statistical and systematic sources. }
\resizebox{\textwidth}{!}{
\centering
\begin{tabular}{crr@{$\pm$}lr@{$\pm$}lr@{$\pm$}lr@{$\pm$}lr@{$\pm$}lr@{$\pm$}lr@{$\pm$}l} \toprule
$m_{\ell\ell}$ [GeV] & \multicolumn{1}{c}{Data} & \multicolumn{2}{c}{Total Bkg} & \multicolumn{2}{c}{Multi-jet} & \multicolumn{2}{c}{$t\bar t$, single top} & \multicolumn{2}{c}{$Z\to\tau\tau$} & \multicolumn{2}{c}{$W\to \ell\nu$} & \multicolumn{2}{c}{$WW/WZ/ZZ$} & \multicolumn{2}{c}{$\gamma\gamma\rightarrow \ell\ell$} \\
 \cmidrule(r){1-1}
 \cmidrule(r){2-4}
   \cmidrule(r){5-16}

12--20&       17\,729  &       2\,220  &      470  &        1\,370  &      460  &         509  &       27  &           7  &        1  &         215  &       44  &          81  &        7  &          41  &       16  \\
20--30&       13\,322  &       1\,860  &      210  &         600  &      200  &         873  &       46  &          33  &        3  &         144  &       36  &         158  &       11  &          54  &       21  \\
30--46&       14\,798  &       3\,290  &      260  &         570  &      230  &        1\,920  &      100  &         228  &       23  &         192  &       48  &         314  &       25  &          75  &       30  \\
46--66&      201\,613  &      25\,600  &     3\,900  &        6\,200  &     3\,400  &        3\,990  &      210  &        9\,360  &      940  &         670  &      170  &        1\,060  &       88  &        4\,300  &     1\,700  \\
66--116&     6\,671\,873  &   59\,400  &     9\,500  &       23\,500  &     9\,200  &       13\,040  &      680  &        3\,560  &      360  &        3\,860  &      930  &       10\,450  &      320  &        5\,000  &     2\,000  \\
116--150&       77\,919  &     8\,280  &      170  &         910  &      170  &        4\,590  &      240  &          82  &        8  &         530  &      130  &        1\,097  &       90  &        1\,070  &      430  \\

\bottomrule 
\end{tabular}
}
\label{tab:DataZ_electron}
\end{table*}

\begin{table*}
\centering
\caption{The number of events in data satisfying the selection criteria in the muon-pair channel for six different regions of \Mll\ and the estimated contribution to this value from the various background sources considered. The uncertainties quoted on the background samples include contributions from statistical and systematic sources.}
\resizebox{\textwidth}{!}{
\centering
\begin{tabular}{cr r@{$\pm$}l r@{$\pm$}l r@{$\pm$}l r@{$\pm$}l r@{$\pm$}l r@{$\pm$}l r@{$\pm$}l} \toprule
$m_{\ell\ell}$ [GeV] & \multicolumn{1}{c}{Data} & \multicolumn{2}{c}{Total Bkg} & \multicolumn{2}{c}{Multi-jet} & \multicolumn{2}{c}{$t\bar t$, single top} & \multicolumn{2}{c}{$Z\to\tau\tau$} & \multicolumn{2}{c}{$W\to \ell\nu$} & \multicolumn{2}{c}{$WW/WZ/ZZ$} & \multicolumn{2}{c}{$\gamma\gamma\rightarrow \ell\ell$} \\
 \cmidrule(r){1-1}
 \cmidrule(r){2-4}
   \cmidrule(r){5-16}

12--20&    25\,297  &        1\,220  &       180  &         440  &         170  &         605  &          32  &           1  &           0  &           9 &         2  &         107  & 10  &          64  &          26  \\
20--30&       19\,485  &         2\,100  &         250  &         590  &         240  &        1\,156  &          61  &          20  &           2  &           8  &           2  &         241  &          19  &          84  &          33 \\ 
30--46&       20\,731  &         3\,980  &         330  &         730  &         290  &        2\,540  &         130  &         156  &          16  &          12  &           3  &         429  &          36  &         114  &          45 \\ 
46--66&      318\,117  &         30\,900  &        4\,100  &        7\,400  &        3\,000  &        5\,370  &         280  &        9\,940  &         990  &         174  &          35  &        1\,460  &         120  &        6\,600  &        2\,600 \\ 
66--116&     9\,084\,639  &      46\,500  &        4\,200  &        7\,400  &        3\,000  &       13\,730  &         720  &        4\,150  &         420  &         870  &         170  &       13\,640  &         420  &        6\,700  &        2\,700 \\ 
 116--150&      100\,697  &      9\,960  &         520  &        1\,270  &         520  &        5\,790  &         300  &          58  &           6  &         153  &          38  &        1\,310  &         110  &        1\,380     &      550 \\

\bottomrule                     
\end{tabular}
}
\label{tab:DataZ_muon}
\end{table*}

Figure~\ref{fig:ZeeControl_WideMass} shows the distributions of \Mll{} and
$\eta$  for electron-pair events passing the selection requirements described above.
Figure~\ref{fig:ZmmControl_WideMass} shows the equivalent distributions for the dimuon channel.
The MC signal sample is simulated using \Powheg{}+\Pythia{}.
The predictions from the model are in qualitative agreement with the data.

\begin{figure}[ht]
  \centering
  \includegraphics[width=0.45\textwidth]{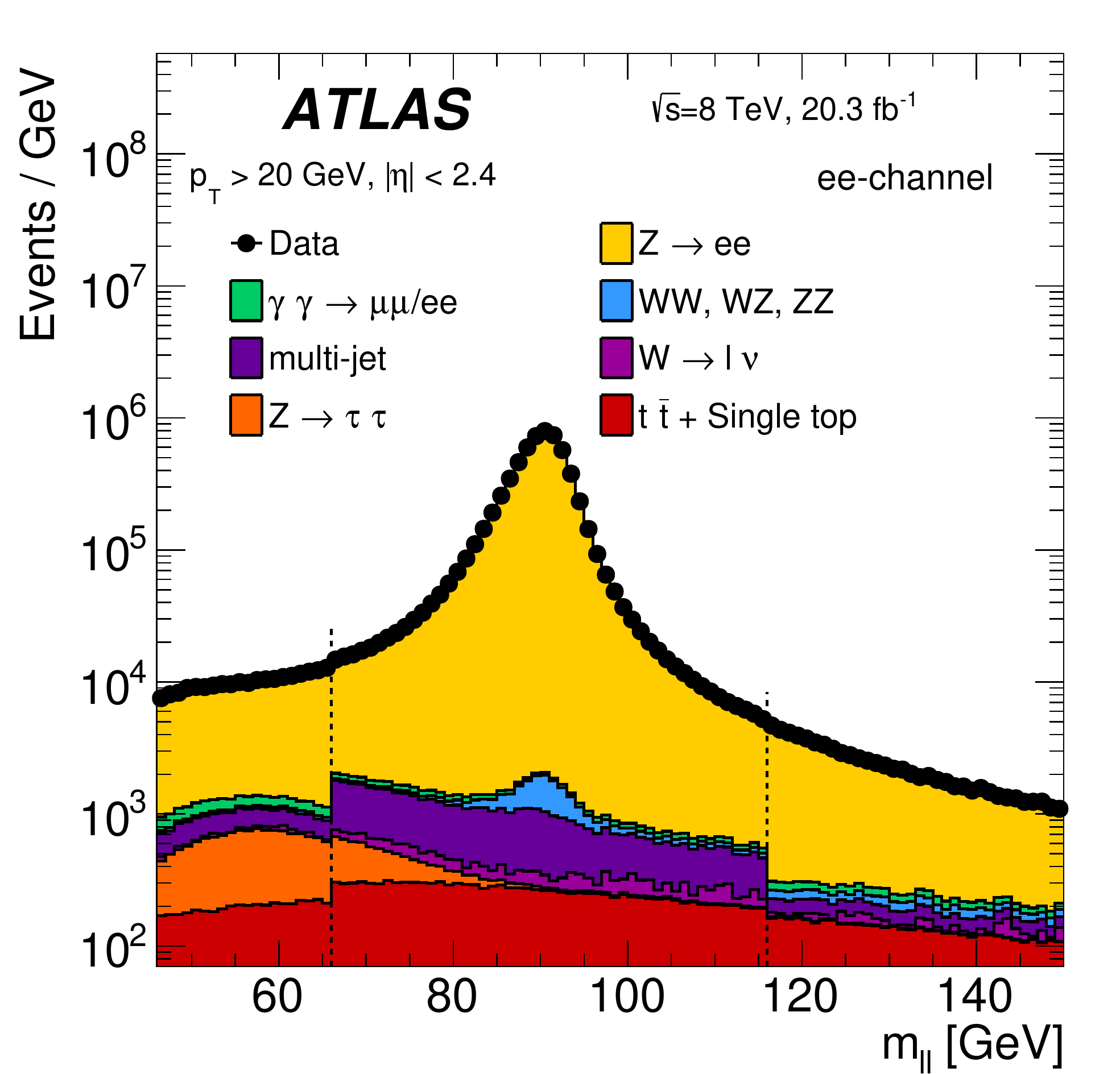}
  \includegraphics[width=0.45\textwidth]{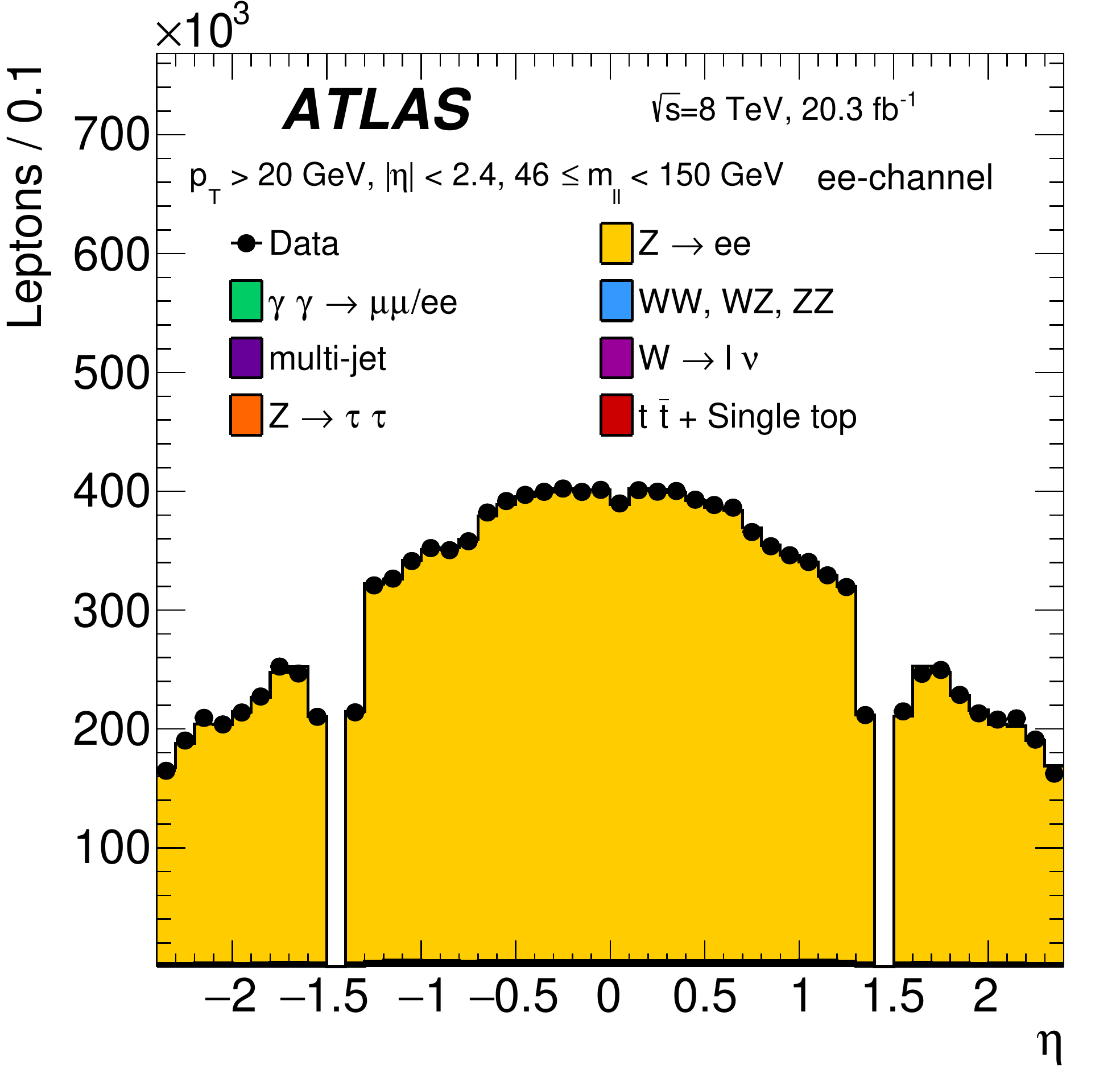}\\
  \caption{The distribution of events passing the selection requirements in the electron-pair channel as
    a function of dilepton invariant mass \Mll\ (left) and electron pseudorapidity $\eta$ (right).
    Events are shown for the \Mll\ range $46 \GeV$ to $150 \GeV$.
    The MC signal sample (yellow) is simulated using \Powheg{}+\Pythia{}.
    The statistical uncertainties on the data points are smaller than the size of the markers and the
    systematic uncertainties are not plotted.
    The prediction is normalised to the integral of the data.
    The vertical dashed lines on the left-hand plot  at \Mll{} values of $66 \GeV$ and
    $116 \GeV$ indicate the boundaries between the three principal
     \Mll{} regions employed in the analysis.
    The small discontinuities in the \Mll{} distribution at $66 \GeV$ and $116 \GeV$ are due to the absence of the isolation requirement around the \Zboson{}-boson mass peak.}
  \label{fig:ZeeControl_WideMass}
\end{figure}

\begin{figure}[htb]
  \centering
  \includegraphics[width=0.45\textwidth]{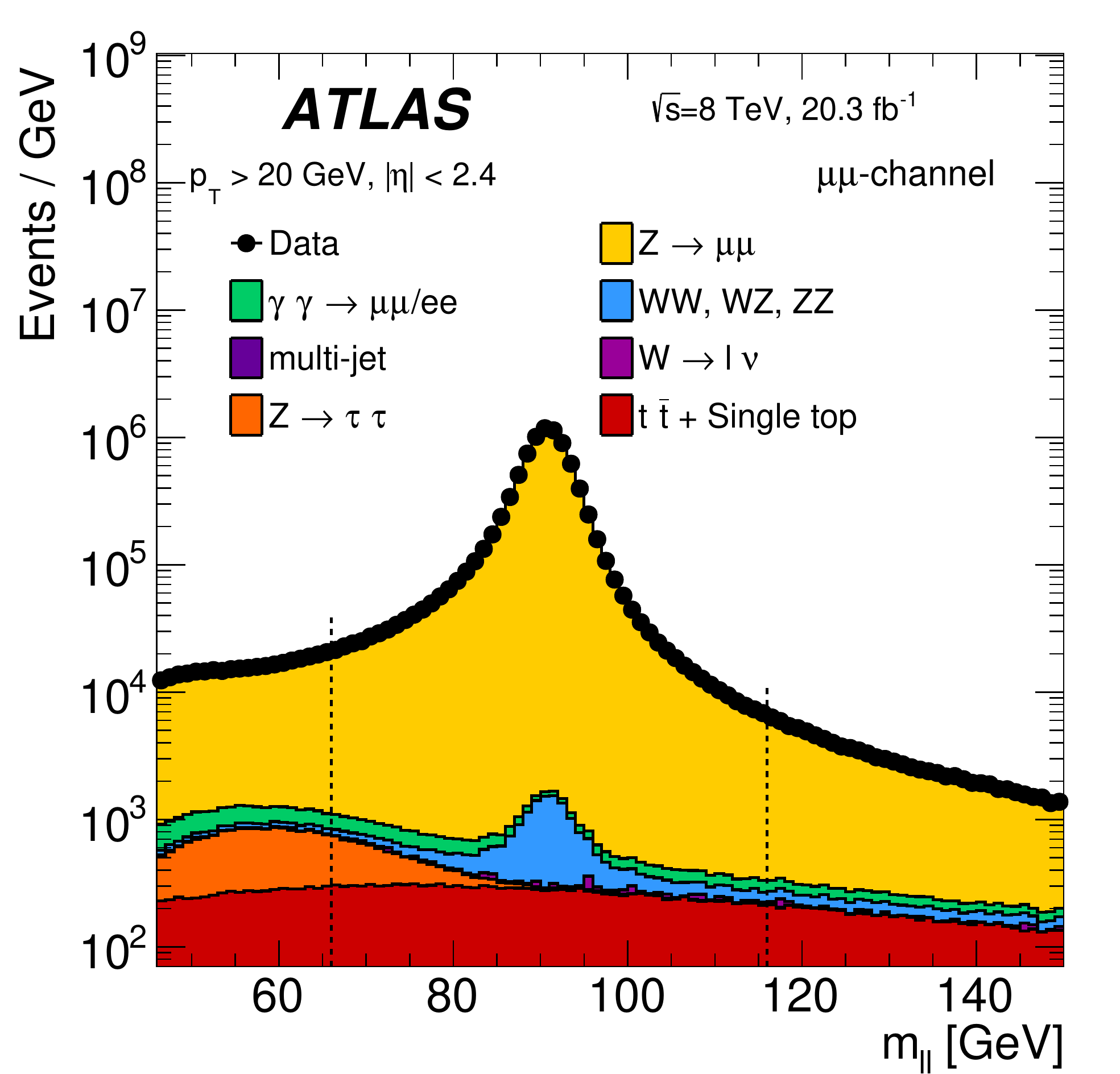}
  \includegraphics[width=0.45\textwidth]{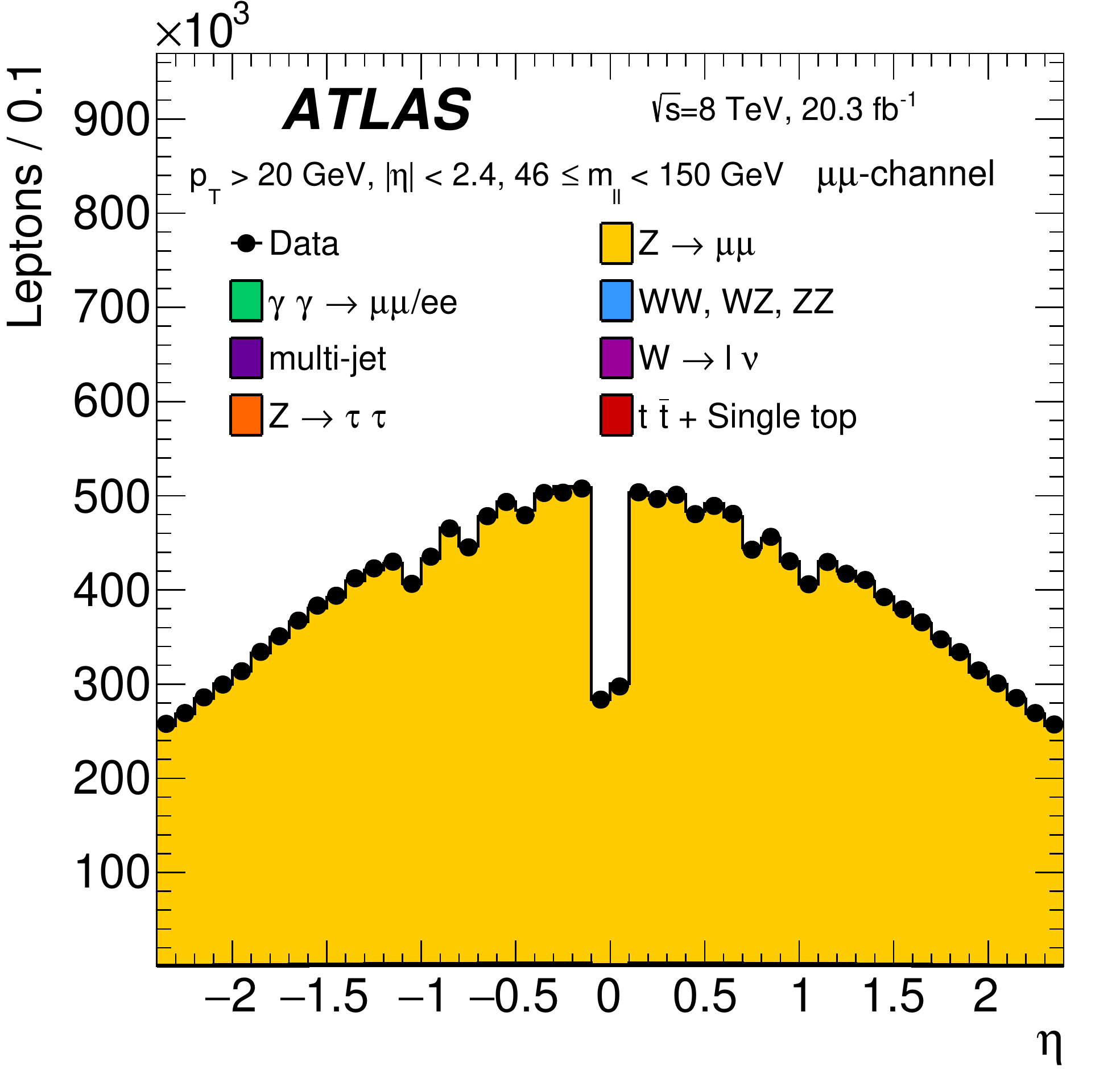}\\
  \caption{The distribution of events passing the selection requirements in the muon-pair channel as a
    function of dilepton invariant mass \Mll\ (left) and muon pseudorapidity $\eta$ (right). Events are shown for the \Mll\ range $46 \GeV$ to $150 \GeV$. The MC signal sample (yellow) is simulated using \Powheg{}+\Pythia{}. The statistical uncertainties on the data points are smaller than the size of the markers and the systematic uncertainties are not plotted. The prediction is normalised to the integral of the data.    The vertical dashed lines on the left hand plot at \Mll{} values of $66 \GeV$ and
    $116 \GeV$ indicate the boundaries between the three principal
     \Mll{} regions employed in the analysis.}
  \label{fig:ZmmControl_WideMass}
\end{figure}

\subsection{Estimation of backgrounds}
\label{sec:background}

The number and properties of the background events arising from multi-jet processes are estimated using a
data-driven technique. A background-dominated sample is selected using a modified version of the signal-selection criteria.
In the electron-pair channel, both electrons are required to satisfy the `loose' identification
criteria~\cite{Aad:2014fxa}, but not the `medium' criteria, and are also required to have the same
charge.
For the muon-pair channel, two samples of lepton pairs are used: the light-flavour
background is estimated by requiring a pair of muons with the same charge, whilst the heavy-flavour
background is estimated by requiring one electron and one muon with opposite charge. The electron is required to be identified as `loose' and the electron isolation cut is inverted.
It is assumed that in all other variables the shape of the distribution of the multi-jet events is the same in both the signal- and background-dominated samples.

The normalisation of the multi-jet background is determined by performing a $\chi^2$ minimisation in a
variable that discriminates between the signal and multi-jet background. The contribution from all
sources other than the multi-jet background is taken from MC simulation.
Two independent fits are performed, using lepton isolation and \Mll\ as discriminating variables.
The signal event-selection criteria are applied, except that the selection
criteria on the isolation variables are removed for the fit that uses lepton isolation.
In the muon-pair final state, the fit using isolation is performed using the values of \Imu.
In the electron-pair final state, the isolation variable \Iestar{} is defined as the scalar sum of the $E_\mathrm{T}$
of energy deposits in the calorimeter within a cone of size $\Delta R = 0.2$ around the electron cluster divided by the \pt{} of the electron.
The $E_\mathrm{T}$ sum excludes cells assigned to the electron cluster and can be negative due to cell noise and negative signal contribution
from pile-up in neighbouring bunches~\cite{ATLAS:2014wka}.
The fit is performed using the quantity \Iemin{}, where \Iemin{} is the smaller of the \Iestar{} values of the two electrons in an event.
Example results of fits to the isolation variables for the electron- and muon-pair channels are shown in
Figure~\ref{fig:ZeeBkg} for the \Mll\ region around the \Zboson{}-boson mass peak.
The difference in the results of the fits to isolation and \Mll\ is taken as the systematic uncertainty on the normalisation
of the multi-jet background. As a cross-check the procedure is repeated in bins of \mody{} and gives results consistent with the fit
performed inclusively in \mody{}.

\begin{figure}[htb]
  \centering
\includegraphics[width=0.45\textwidth]{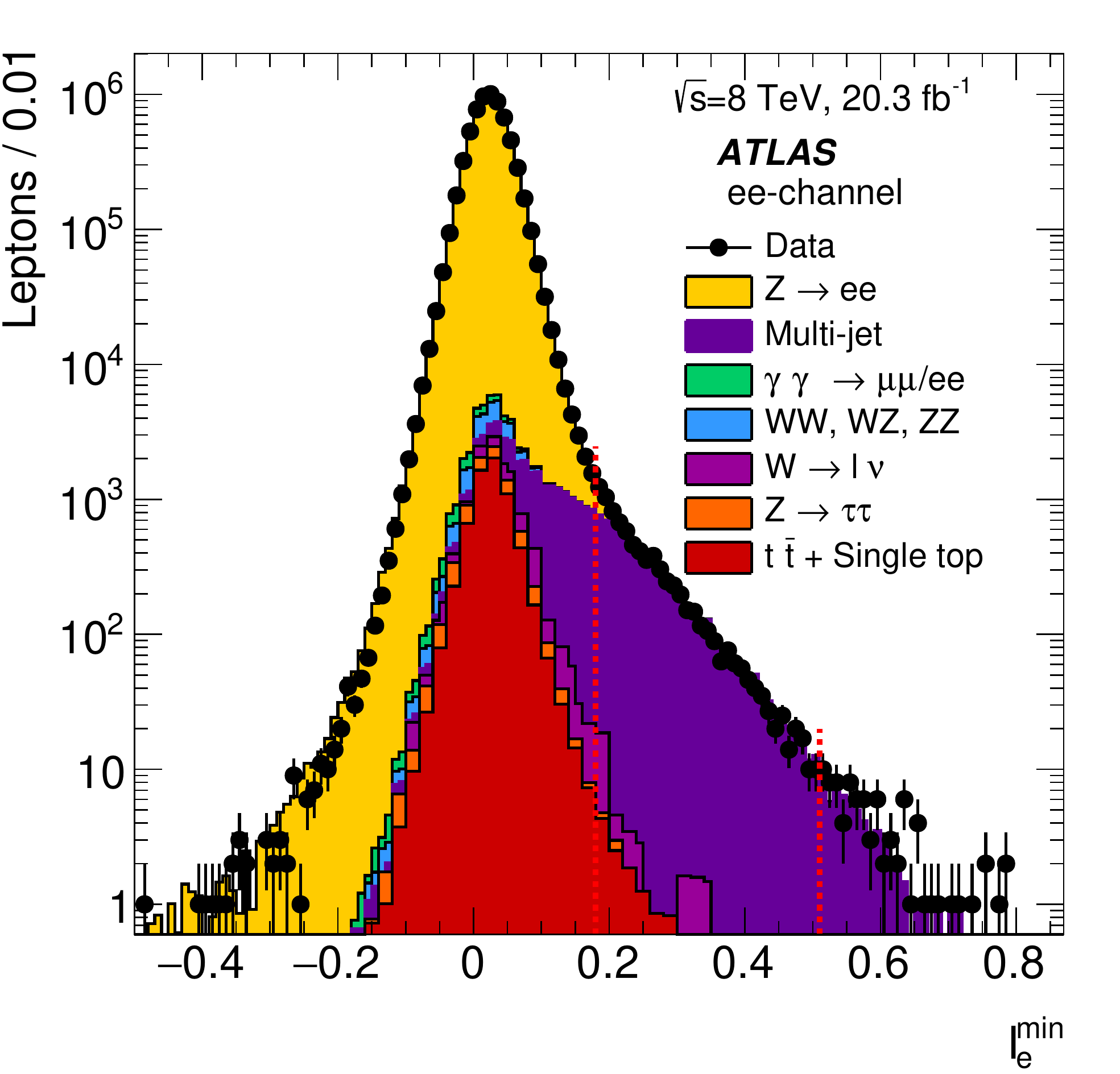}
\includegraphics[width=0.45\textwidth]{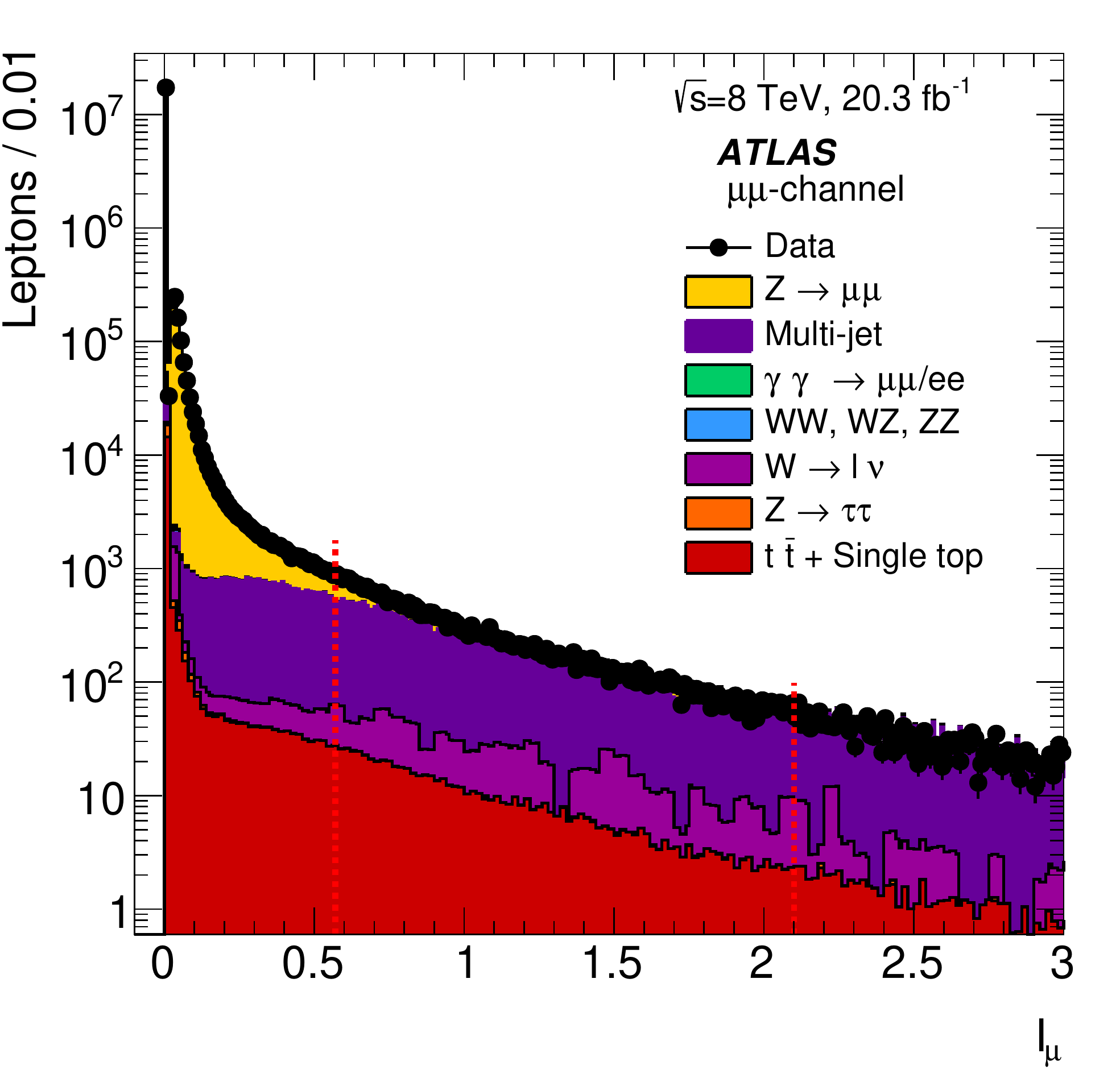}
\caption{Left: the distribution of the smallest of the
  isolation variables of the two electrons \Iemin{}.  Right: the distribution of the muon isolation variable \Imu. The data for $66 \GeV < \Mll < 116 \GeV$ are compared to the
   sum of the estimated multi-jet background and all other processes, which are estimated from MC simulation. The red dashed lines indicate the range over which the fit is performed.}
  \label{fig:ZeeBkg}
\end{figure}

The backgrounds from all
sources other than multi-jet processes are estimated using the MC samples detailed in
Section~\ref{sec:eventsimulation}. These estimates are cross-checked by comparing MC simulation to data in control
regions, selected using criteria that increase the fraction of background. The \Ztautau{} and \ttbar{}
backgrounds are enhanced by requiring exactly one electron and one muon candidate per event according to
the criteria described in Section~\ref{sec:eventreconstruction}. The MC simulation is found to be consistent with
the data within the assigned uncertainties on the cross sections (see Section~\ref{sec:systematics}). In addition,
a subset of these events is studied in which two jets with $\pt{} > 25 \GeV$
are identified, which significantly enhances the contribution from the \ttbar{} background. Again, the MC simulation is consistent with
the data within the assigned uncertainties.

Around the \Zboson{}-boson mass peak and at low values of \phistar{} and \Zpt{}, the background is dominated
by multi-jet and $\gamma\gamma\rightarrow \ell\ell$ processes which together amount to less than 1\% of the selected electron-pair or muon-pair event sample. At high \phistar{} and \Zpt{}, 
\ttbar{} and diboson processes dominate and constitute a few percent of the selected data.
In the regions of \Mll\ below the $Z$-boson mass peak, \ttbar{} continues to be a dominant background at larger values of \phistar{} and \Zpt{} (forming up to 20\% of the selected data), whilst at lower values of \phistar{} and \Zpt{} the dominant contribution is from $\gamma\gamma\rightarrow \ell\ell$ processes with other contributions from \Ztautau and multi-jet processes (totalling between 10\% and 20\% of the selected data). The fraction of \ttbar{} background in the \Mll{} regions below $46 \GeV$ is enhanced by the requirement that \Zpt{} be greater than $45 \GeV$.
In the region of \Mll\ above the $Z$-boson mass peak, the \ttbar{} background forms more than 30\% of the selected data at higher values of \phistar{} and \Zpt{}. The total background is smaller at low values (approximately 10\% of the selected data) with the dominant contribution again coming from $\gamma\gamma\rightarrow \ell\ell$ processes.

\subsection{Corrections for detector effects and FSR}
\label{sec:corrections}

After the estimated total background is subtracted from the data, Drell--Yan signal
MC simulation is used to correct to the particle level, accounting for detector resolution and inefficiencies
and the effects of FSR.

Since the experimental resolution in \phistar\ is smaller
than the chosen bin widths, the fractions of accepted
events that fall within the same bin in \phistar\ at the particle
level and reconstructed detector level in the MC simulation
are high, having typical values of around 90\%.
Therefore, simple bin-by-bin corrections of the
\phistar\ distributions are sufficient.
A single iteration is performed by reweighting the signal MC events at particle level to the corrected data and
rederiving the correction factors.
The correction factors are estimated using an average over all available signal MC samples (as described in Section~\ref{sec:eventsimulation}).

The detector resolution has a larger effect in the measurement of \Zpt.
An iterative Bayesian unfolding method~\cite{D'Agostini:1994zf,D'Agostini:2010,Adye:2011gm} with seven iterations is used to correct the \Zpt{} distribution to particle level.
The response matrix, which connects the \Zpt{} distribution at reconstruction and particle levels is
estimated using the \Powheg{}+\Pythia{} signal MC sample.

\subsection{Systematic uncertainties}
\label{sec:systematics}

In this section the principal sources of uncertainty on the measurements are discussed, as well as the degree to which these uncertainties
are correlated (between bins in \phistar\ or \Zpt, or between the electron-pair and muon-pair channels)
when combining the electron-pair and muon-pair results and in quoting the final results.
Figure~\ref{fig:Systematics} provides a summary of the uncertainties arising from data statistics, mis-modelling of the detector, background
processes, and of the MC signal samples used to correct the data. These are given for both the electron (dressed level) and muon (bare level) channels as a function of \phistar{} and \Zpt{} for events with $66 \GeV < \Mll < 116 \GeV$ and $\mody<2.4$.

\begin{figure}
  \centering
  \includegraphics[width=0.45\textwidth]{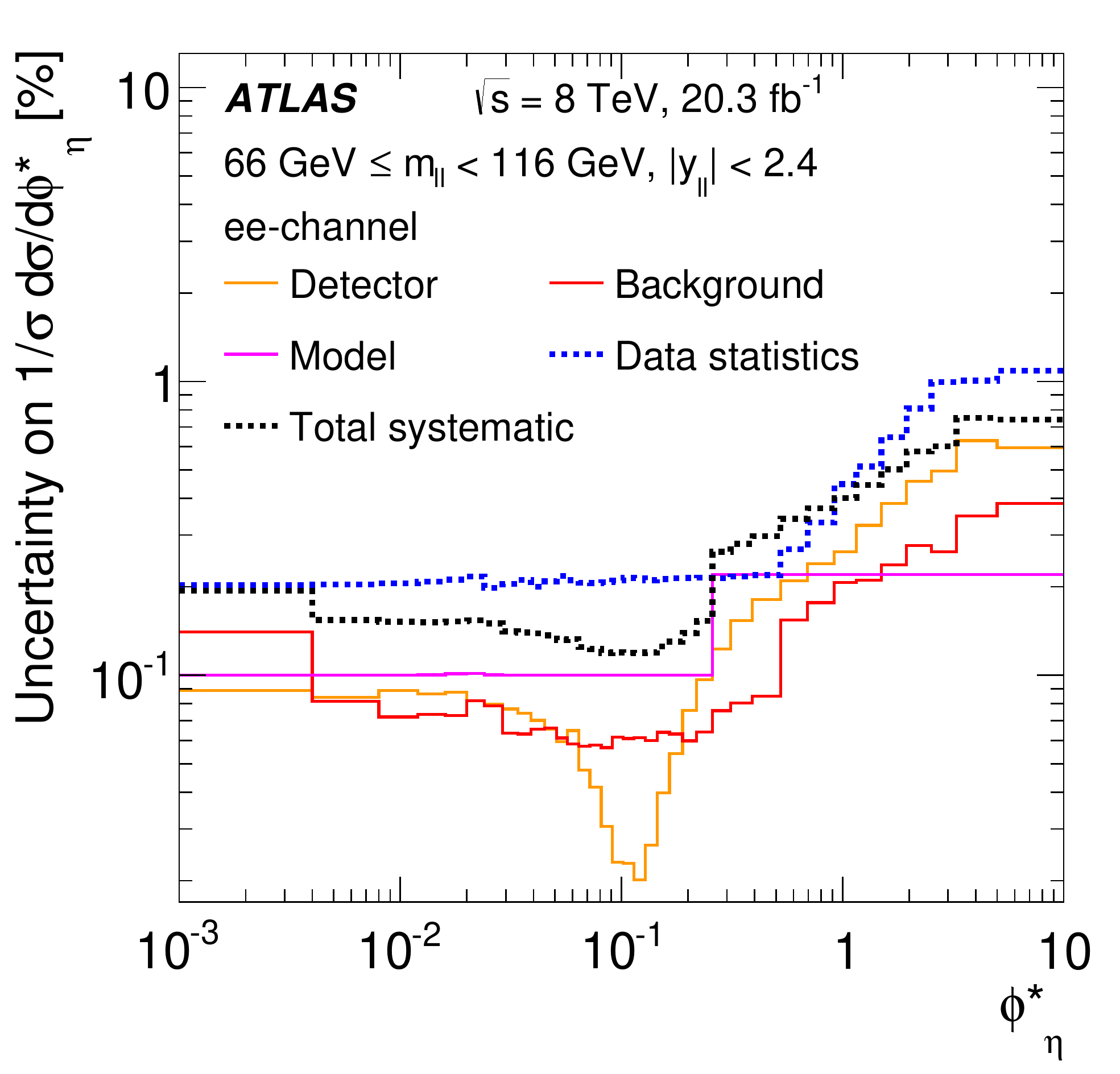}
  \includegraphics[width=0.45\textwidth]{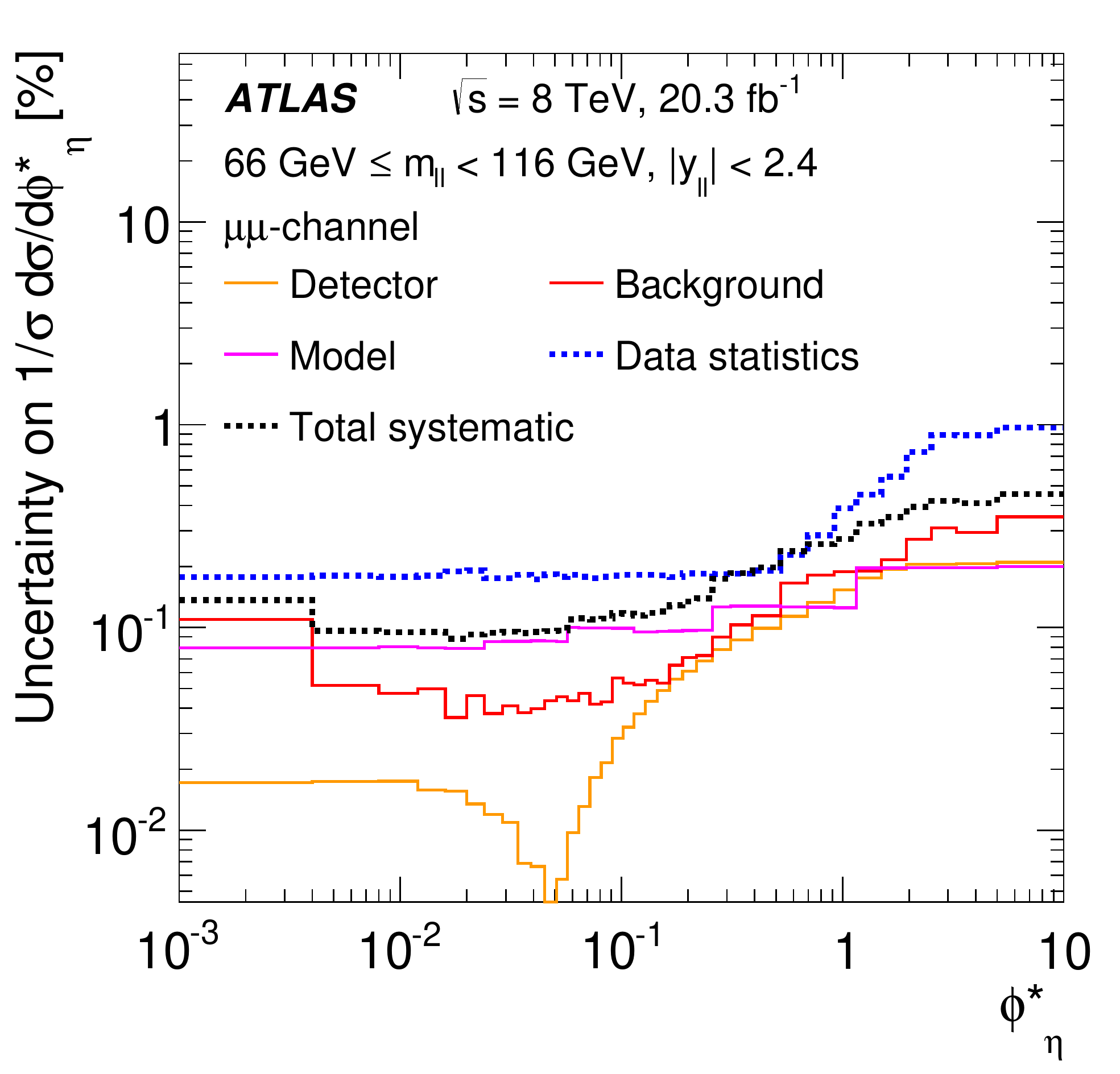}
  \includegraphics[width=0.43\textwidth]{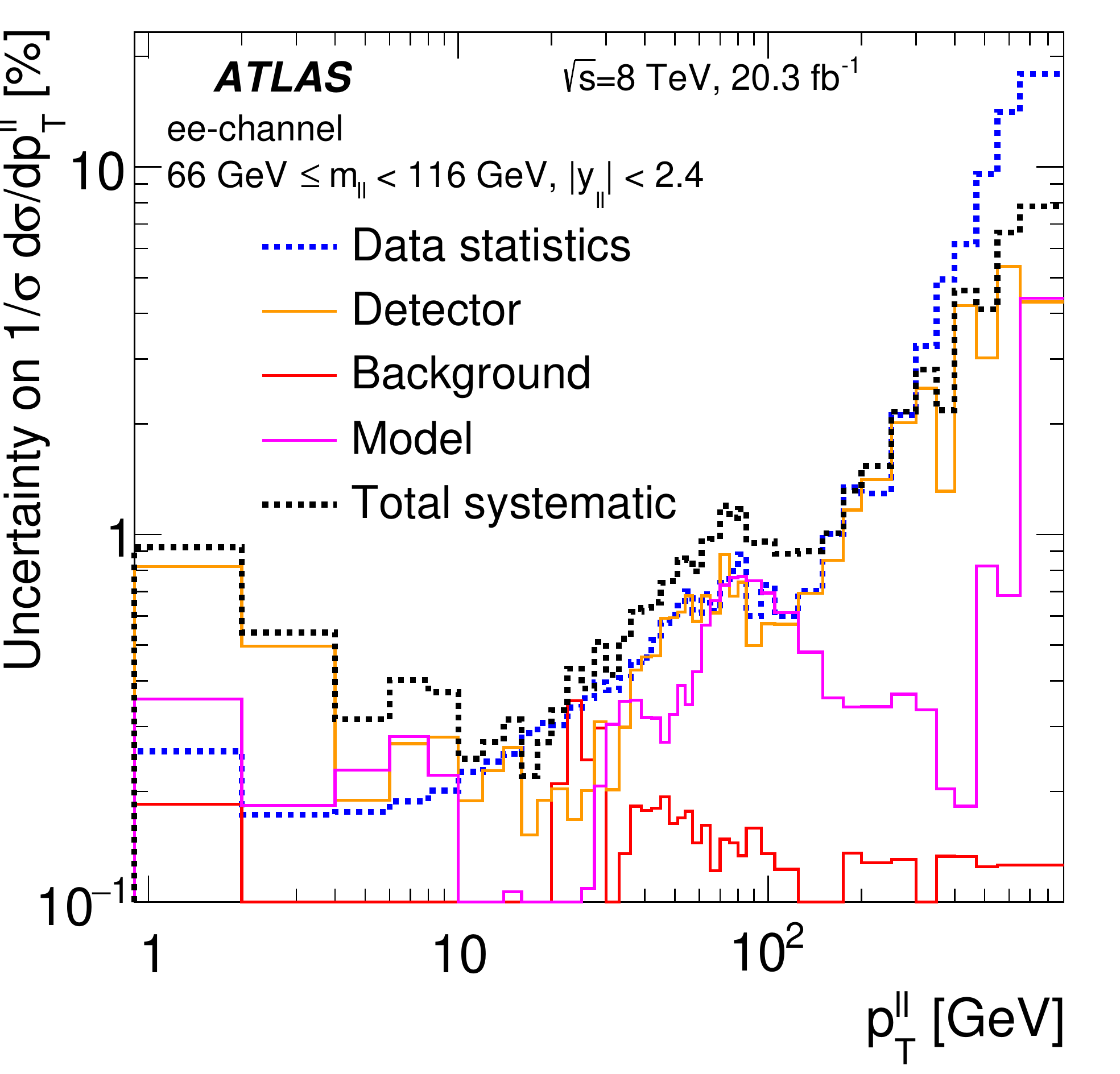}\quad
  \includegraphics[width=0.43\textwidth]{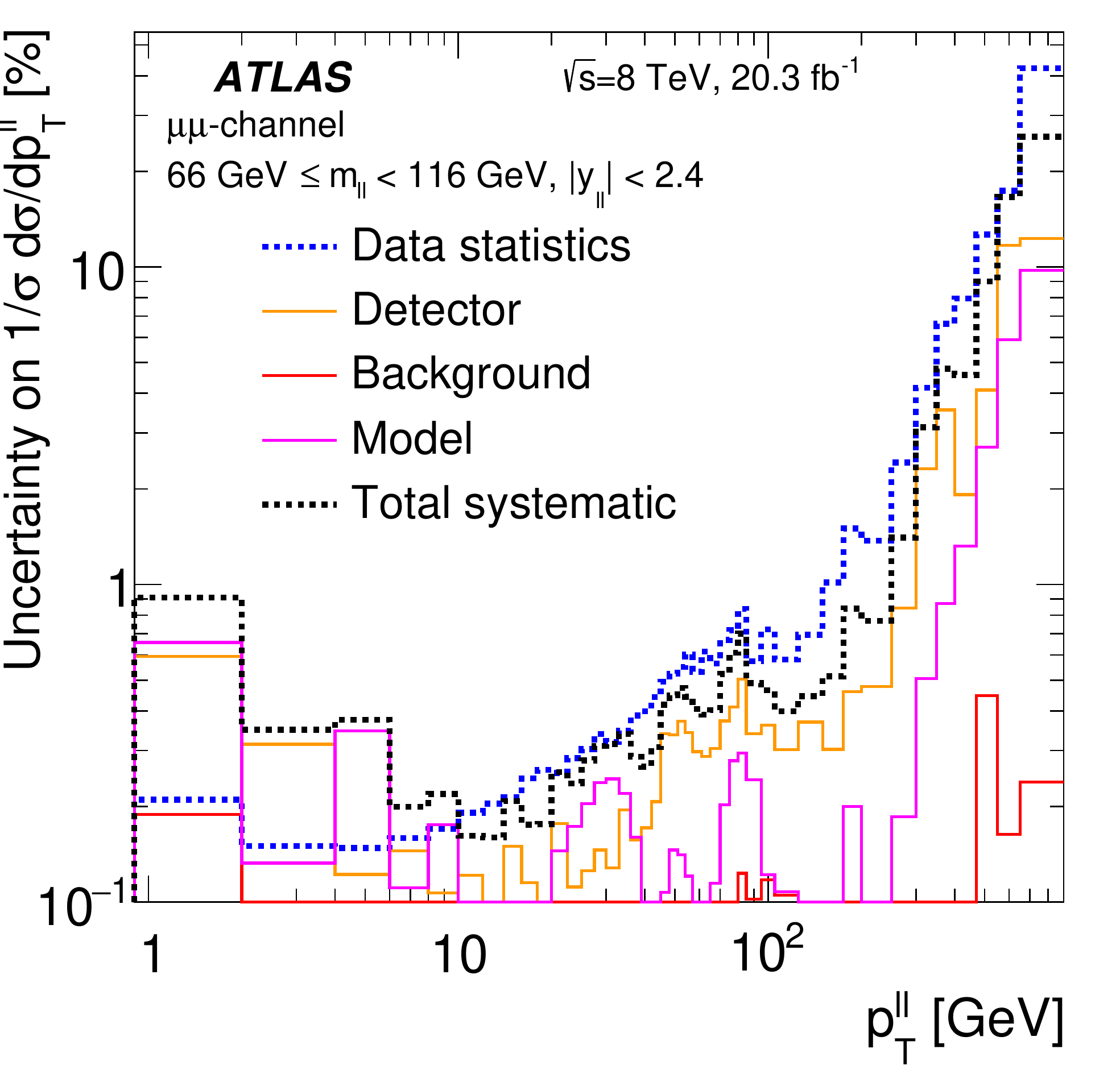}
    \caption{Uncertainty from various sources on $(1/\sigma)\, \mathrm{d}\sigma / \mathrm{d}\phistar$ (top) and $(1/\sigma)\, \mathrm{d}\sigma / \mathrm{d}\Zpt$  (bottom) for events with $66 \GeV < \Mll < 116 \GeV$ and $\mody<2.4$. Left: electron-pair channel at dressed level. Right: muon-pair channel at bare level.}
  \label{fig:Systematics}
\end{figure}

The statistical uncertainties on the data, and on the MC samples used to correct the data, are considered as uncorrelated between bins 
and between channels.
In most kinematic regions the statistical uncertainty on the data is larger than the total systematic uncertainty in both \phistar{} and \Zpt{} 
(for the normalised measurements) and is always a large contribution to the total uncertainty.

Most sources of systematic uncertainty from the modelling of the detector and beam conditions are treated as fully correlated between bins.
These comprise possible mis-modelling of the lepton energy (electron) and momentum (muon) scales and their resolution
as well as mis-modelling of the lepton reconstruction, identification, trigger and isolation efficiencies~\cite{ElectronEfficiency_2014,Aad:2014nim,Aad:2014rra}.
Some of the detector uncertainties have a statistical component, which for the \Zpt{} and integrated cross-section measurements
is non-negligible and is propagated to the final measurements using a toy MC method.
The above uncertainties are treated as uncorrelated between the two channels and are generally a small fraction of the total systematic uncertainty in the individual channels and on the combined result.
The exceptions are the energy and momentum scale uncertainties, which become significant for the \Zpt{}
measurements at high values of \Zpt{}.
Also considered are uncertainties due to mis-modelling of the pile-up distribution and of the
distribution of the longitudinal position of the primary vertex, which are
estimated by varying the associated MC scaling factor and are treated as correlated between channels.
The pile-up uncertainty is a small, but non-negligible contribution to the total systematic uncertainty in most kinematic regions and the vertex uncertainty is generally even smaller.
An uncertainty is estimated for the possible mis-modelling of the lepton angular resolution.
This uncertainty is relevant only for the measurements of \phistar{} and its size is found to be of an order similar to that of the pile-up uncertainty.

Important contributions to the total systematic uncertainty on both \phistar{} and \Zpt{} arise from the modelling of the background processes.
The uncertainty arising from varying the normalisation of each MC background within its theoretical cross-section uncertainty is
treated as correlated between channels.
This source makes a small contribution to the total systematic uncertainty in the \Mll\ region around the
\Zboson{}-boson mass peak (where the total background is small),
but becomes more significant in regions away from the peak.
The dominant uncertainty on the multi-jet background arises from the difference in normalisation obtained from template fits
performed in the distribution of the isolation variable or in \Mll{}.
This is treated as fully correlated between bins and is generally a small contribution to the total uncertainty,
becoming more important for the \Mll\ regions below the \Zboson{} peak.
The statistical uncertainty on the multi-jet background is considered as uncorrelated between bins and
channels, and is small.

Several sources of systematic uncertainty are considered, arising from mis-modelling of the underlying physics distributions
by the Drell--Yan signal MC generator.

The effect of any mis-modelling of the underlying \phistar{} and \Zpt{} distributions is evaluated as follows.
For \phistar{} a second iteration of the bin-by-bin correction procedure (see
Section~\ref{sec:corrections}) is made and any difference with respect 
to the first iteration is treated as a systematic uncertainty.
This is found to be negligible in all kinematic regions, due to the very small bin-to-bin migration in \phistar.
For \Zpt{} the MC simulation is reweighted at particle level to the unfolded data and the unfolding is repeated.
Any change is treated as a systematic uncertainty, which is always found to be a small fraction of the total uncertainty.

The systematic uncertainty due to the choice of signal MC generator used to correct the data
is evaluated as follows.
For \phistar{} an uncertainty envelope is chosen that encompasses the difference in the bin-by-bin
correction factors obtained using any individual signal MC sample compared to the central values.
(As described in Section~\ref{sec:corrections}, the central values are obtained from an average over all
available signal MC samples.)
For \Zpt{} the uncertainty is quoted as the difference in the results obtained when  unfolding the data
with \Sherpa{}, as compared to \Powheg{}+\Pythia{}, which is used for the central values.
This source results in a significant contribution to the systematic uncertainty in both  \phistar{} and \Zpt{}
for the \Mll\ region around the \Zboson{}-boson mass peak.
The systematic uncertainty on the Born-level measurements below the \Zboson{}-boson mass peak receives a  significant
contribution due to the differences in FSR modelling between \Photos{} and \Sherpa{}.

Potential uncertainties on the final \phistar{} and \Zpt{} distributions could arise from the modelling of the PDFs in the MC generators used
to correct data to particle level. These are estimated using the CT10 error sets~\cite{Gao:2013xoa} using the LHAPDF
interface~\cite{Whalley:2005nh}, and are found to be negligible.
A correction is applied to the \Powheg{}+\Pythia{} sample, which implements a running coupling for the photon
exchange and a running width in the \Zboson{}-boson propagator.
This correction is found to have a negligible effect on the final results.

\Powheg{}+\Pythia{} provides a poor description of the data for the samples with very low mass, $\Mll < 46 \GeV$ and $\Zpt > 45 \GeV$.
The prediction from \Powheg{}+\Pythia{} is reweighted to that from \Sherpa{} in order to evaluate an uncertainty due to this effect,
which is found to be a small fraction of the total systematic uncertainty.

The Bayesian unfolding procedure used to correct the \Zpt\ distributions for the effects of detector resolution and FSR has associated uncertainties.
A statistical component is estimated using the bootstrap method~\cite{efron1979} and the difference in the unfolded result between using six and seven iterations is treated as a systematic uncertainty,
which is assumed fully correlated between bins of \Zpt{} and found to be a small fraction of the total systematic uncertainty.

The uncertainty on the integrated luminosity is 2.8\%, which is determined following the
methodology described in Ref.~\cite{Luminosity_2011}.
This has a negligible impact on the uncertainty in the normalised differential distributions
 $(1/\sigma)\, \mathrm{d}\sigma / \mathrm{d}\phistar$ and $(1/\sigma)\, \mathrm{d}\sigma / \mathrm{d}\Zpt$. 

The total systematic uncertainties are generally smaller than the statistical uncertainties on the data.
In \phistar\ the total systematic uncertainties at the $Z$-boson mass peak are at the level of around 1\textperthousand\ at low
\phistar, rising to around 0.5\% for high \phistar.
In \Zpt\  the total systematic uncertainties at the $Z$-boson mass peak are at the level of around 0.5\% at low
\Zpt, rising to around 10\% for high \Zpt.

The full results for $(1/\sigma)\, \mathrm{d}\sigma / \mathrm{d}\phistar$ and $(1/\sigma)\, \mathrm{d}\sigma / \mathrm{d}\Zpt$ are presented in the appendix in bins of \mody, for which the size of the data statistical
uncertainties relative to the systematic uncertainties are larger still.

\section{Results}
\label{sec:Results}
       
\subsection{Combination procedure}

The differential and integrated cross-section measurements in the electron-pair and muon-pair channels are combined at Born level using the HERA averager tool, 
which performs a $\chi^2$ minimisation in which correlations between bins and between the two channels are taken into account~\cite{Glazov_2005}.
The combinations for the \Zpt{} and \phistar{} measurements are performed separately in each region of \Mll{} and \mody{}.

\subsection{Differential cross-section measurements}

Figure~\ref{fig:ZphiMComb} shows the combined Born-level distributions of $(1/\sigma)\, \mathrm{d}\sigma / \mathrm{d}\phistar$, in three \Mll\ regions from $46 \GeV$ to $150 \GeV$
for $\mody<2.4$.
The central panel of each plots in Figure~\ref{fig:ZphiMComb} shows the ratios of the
values from the individual channels to the combined values and the lower
panel of each plot shows the difference between the electron-pair and muon-pair values divided by the uncertainty on that difference (pull).
The $\chi^2$ per degree of freedom is given.
The level of agreement between the electron-pair and muon-pair distributions is good.
Figure~\ref{fig:ZptMComb} shows the equivalent set of plots for the distributions of $(1/\sigma)\, \mathrm{d}\sigma / \mathrm{d}\Zpt$
for the six regions of \Mll{} from $12 \GeV$ to $150 \GeV$. Again the level of agreement between the two channels is good.

\begin{figure}[htb]
  \centering
\includegraphics[width=0.43\textwidth]{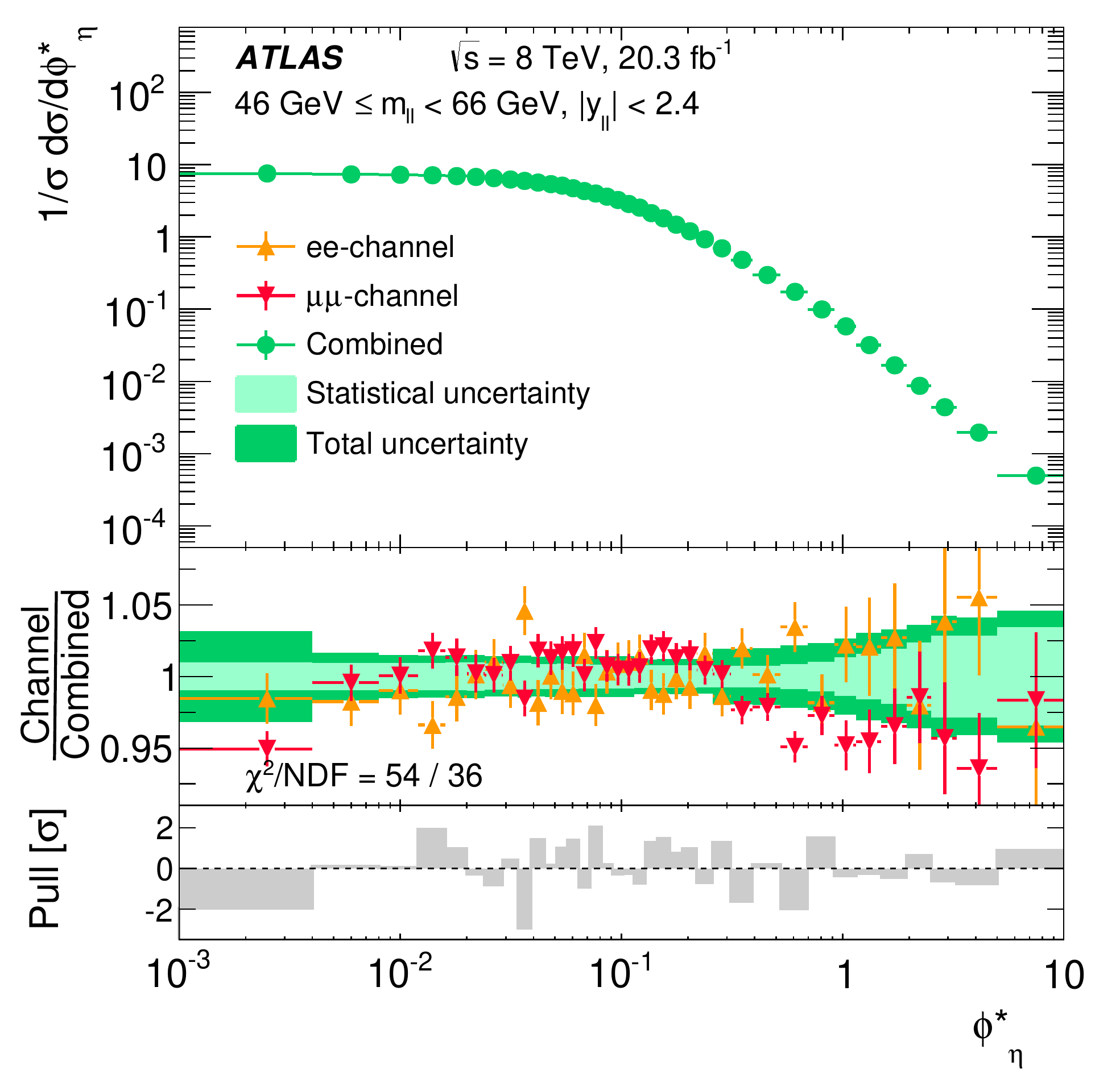}

\includegraphics[width=0.43\textwidth]{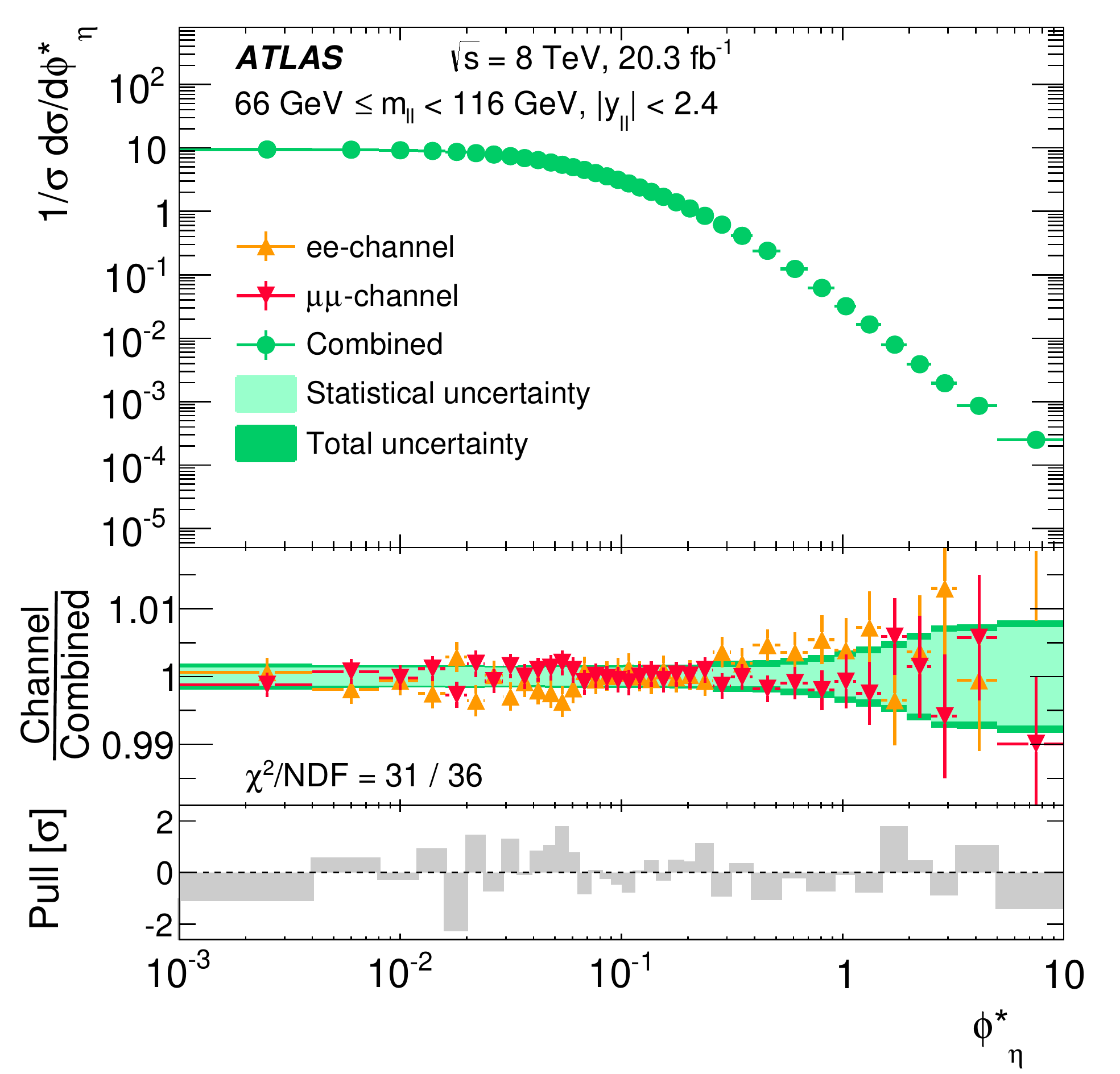}

\includegraphics[width=0.43\textwidth]{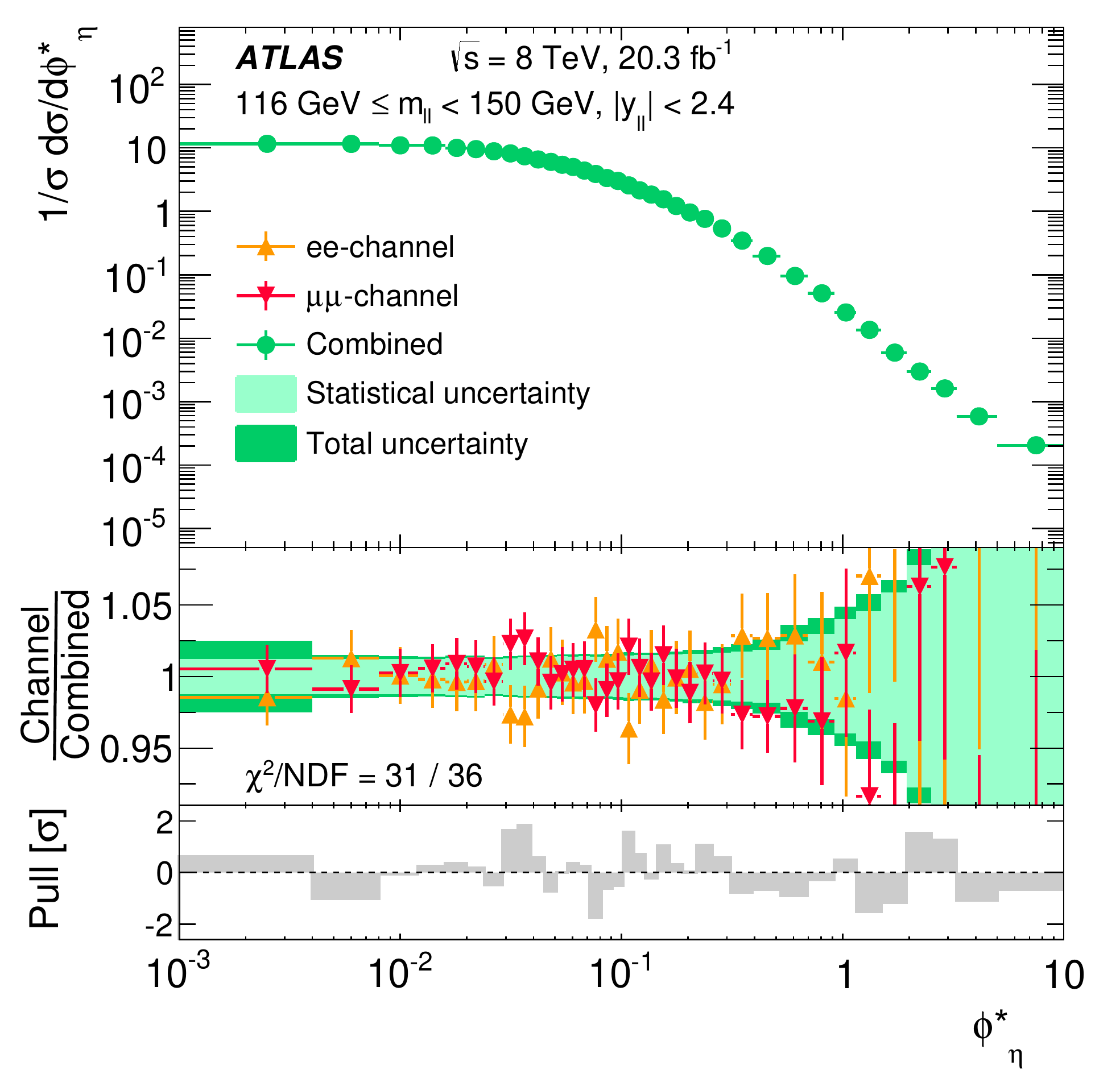}

\caption{The Born-level distributions of $(1/\sigma)\, \mathrm{d}\sigma / \mathrm{d}\phistar$ for the combination of the electron-pair 
and muon-pair channels, shown in three \Mll\ regions from $46 \GeV$ to $150 \GeV$ for $\mody<2.4$. The central panel of each plot shows the ratios of the
values from the individual channels to the combined values, where the error
bars on the individual-channel measurements represent the total uncertainty uncorrelated between
bins. The light-green band represents the data statistical uncertainty on the combined value and the
dark-green band represents the total uncertainty (statistical and systematic).
The $\chi^2$ per degree of freedom is given.
The lower panel of each plot shows the pull, defined as the difference between the electron-pair and muon-pair values divided by the uncertainty on that difference.}
\label{fig:ZphiMComb}
\end{figure}

\begin{figure}[htb]
  \centering
\includegraphics[width=0.42\textwidth]{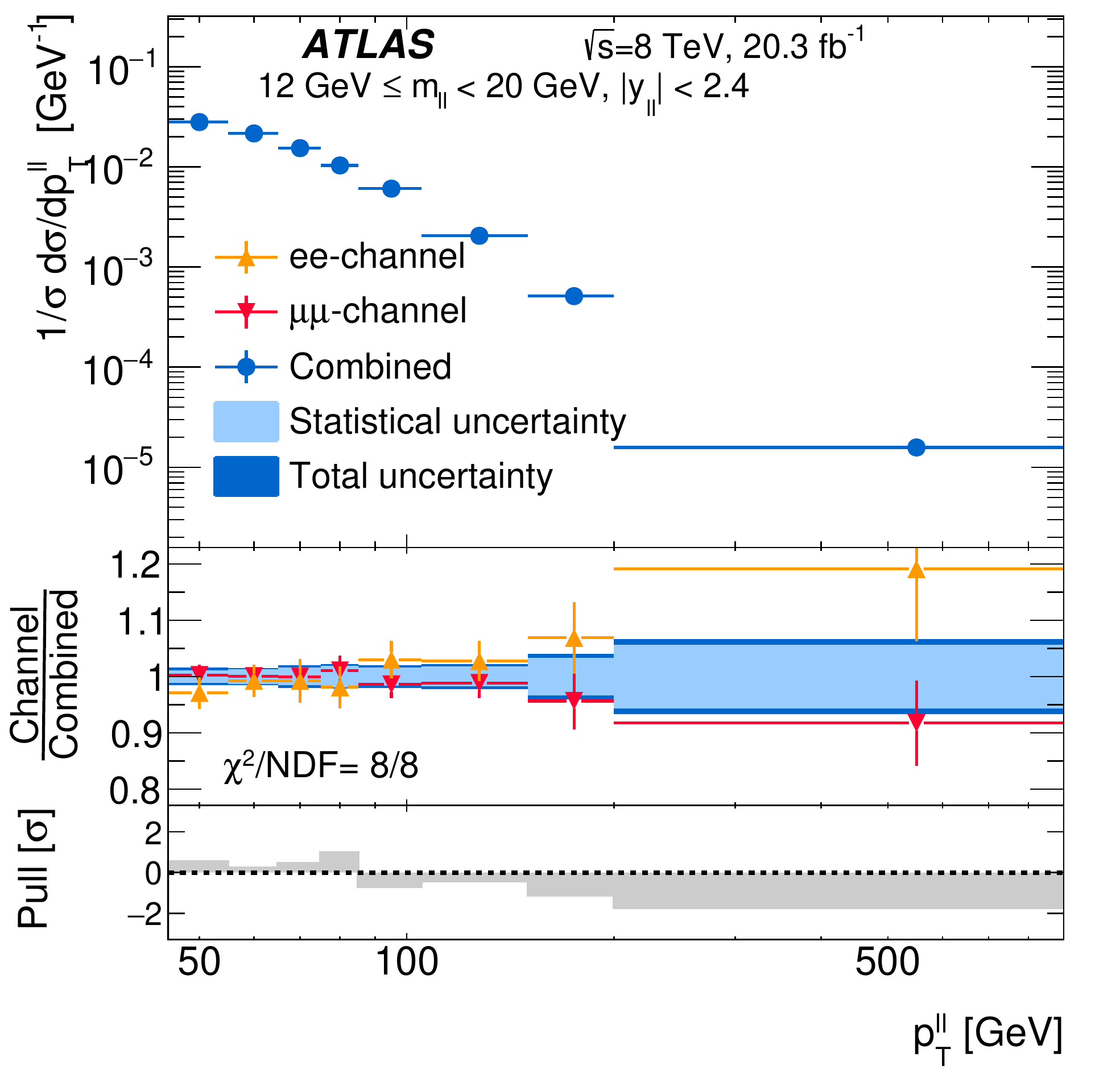}
\includegraphics[width=0.42\textwidth]{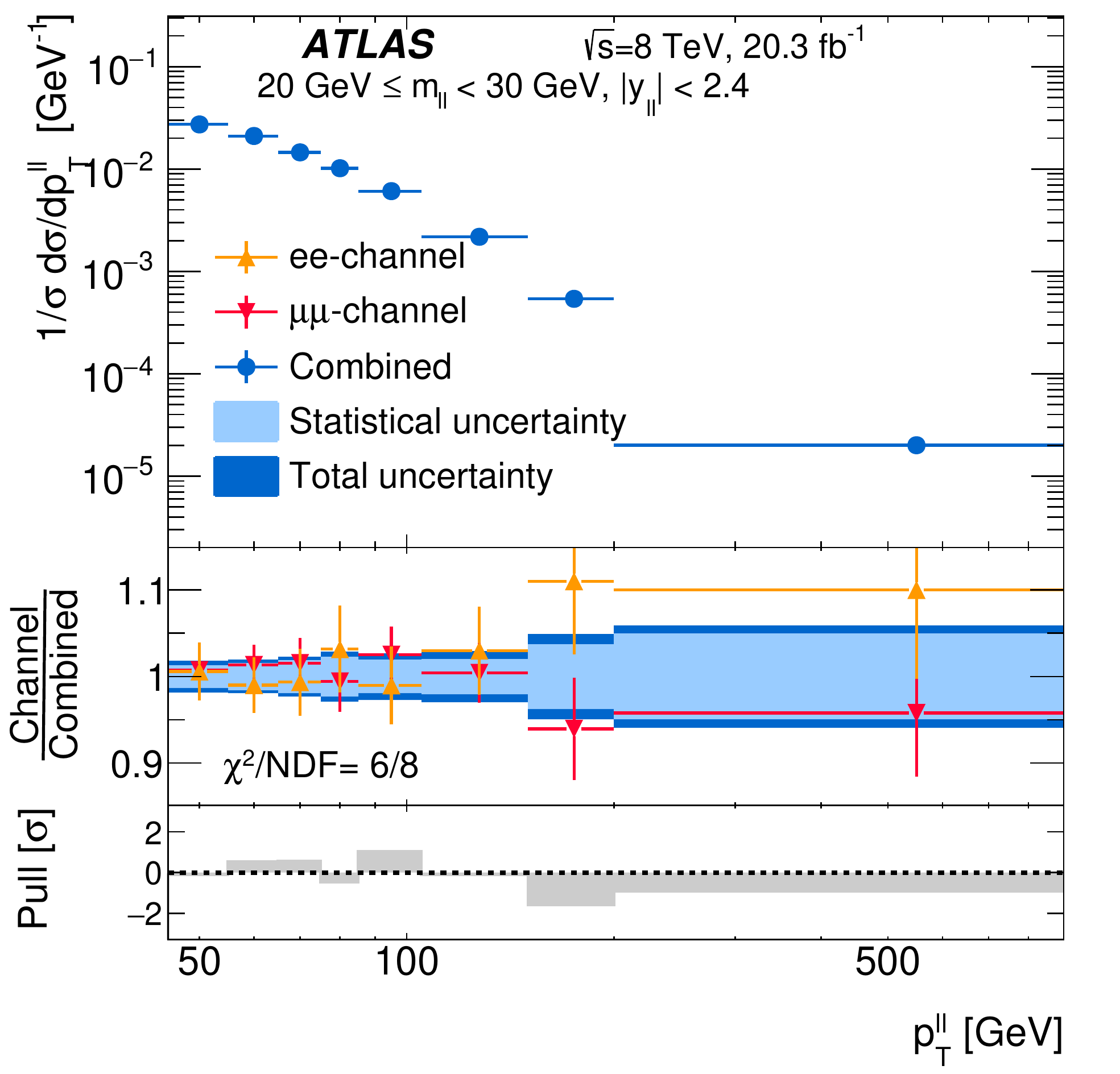}

\includegraphics[width=0.42\textwidth]{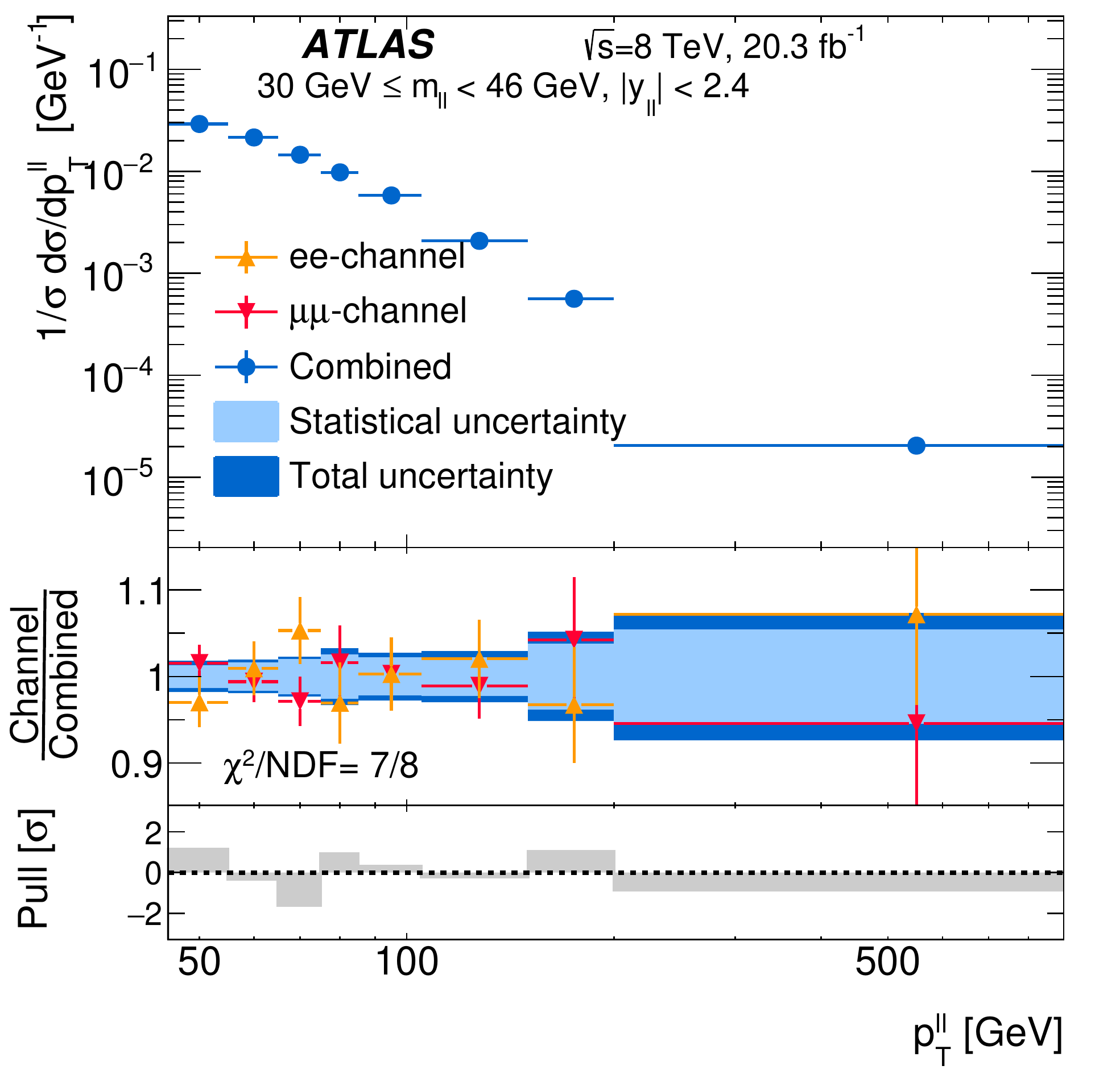}
\includegraphics[width=0.42\textwidth]{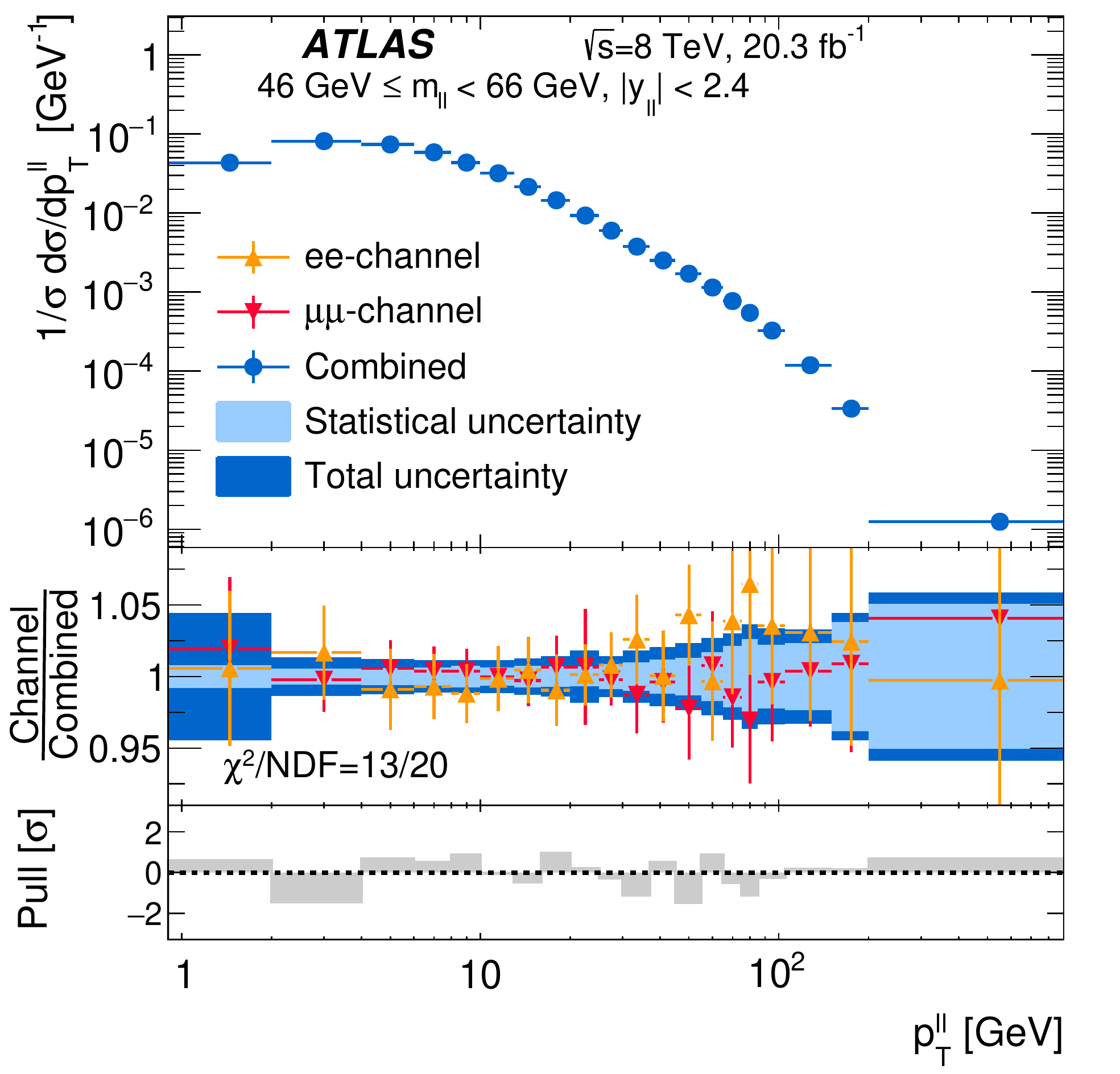}

\includegraphics[width=0.42\textwidth]{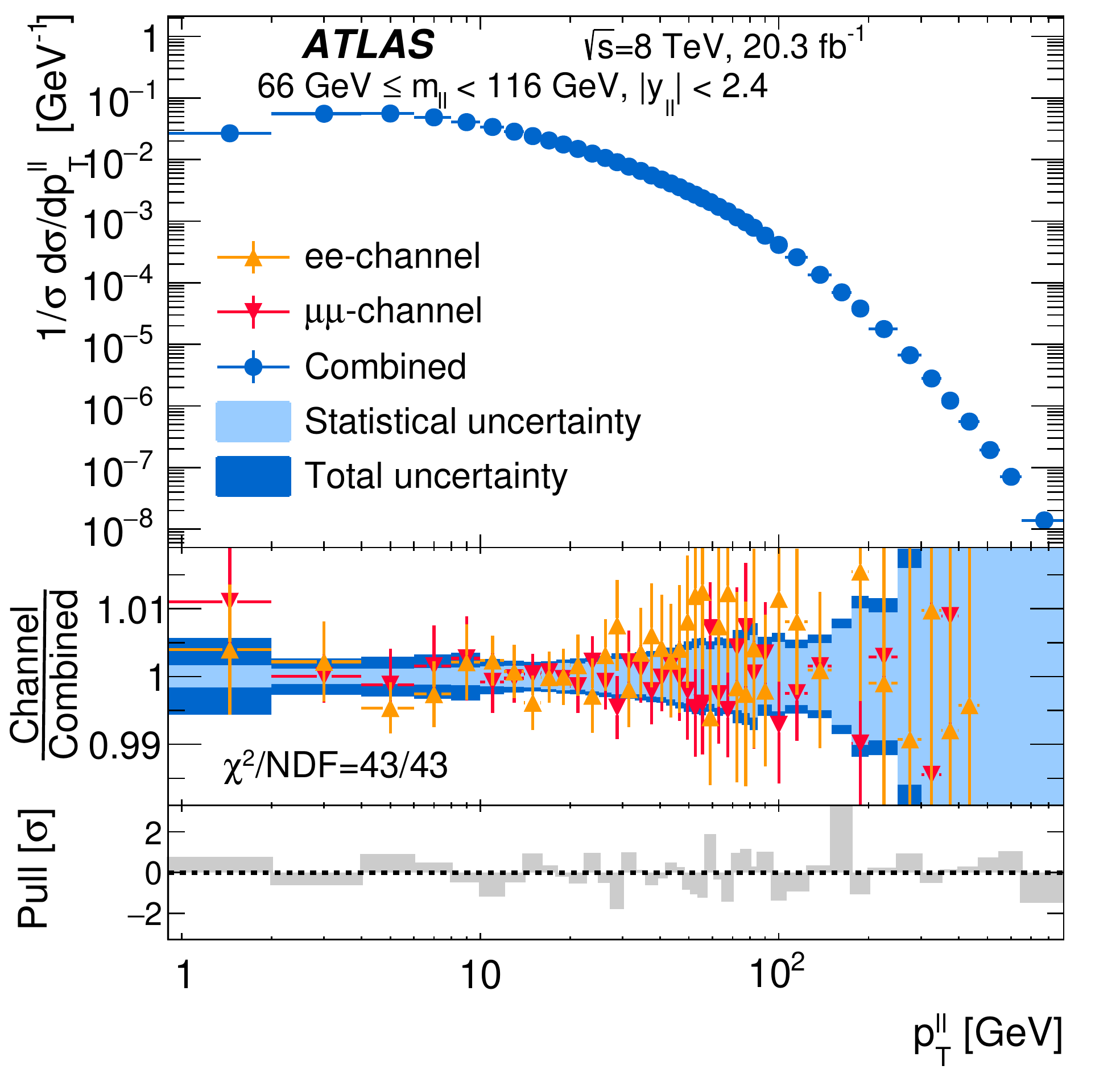}
\includegraphics[width=0.42\textwidth]{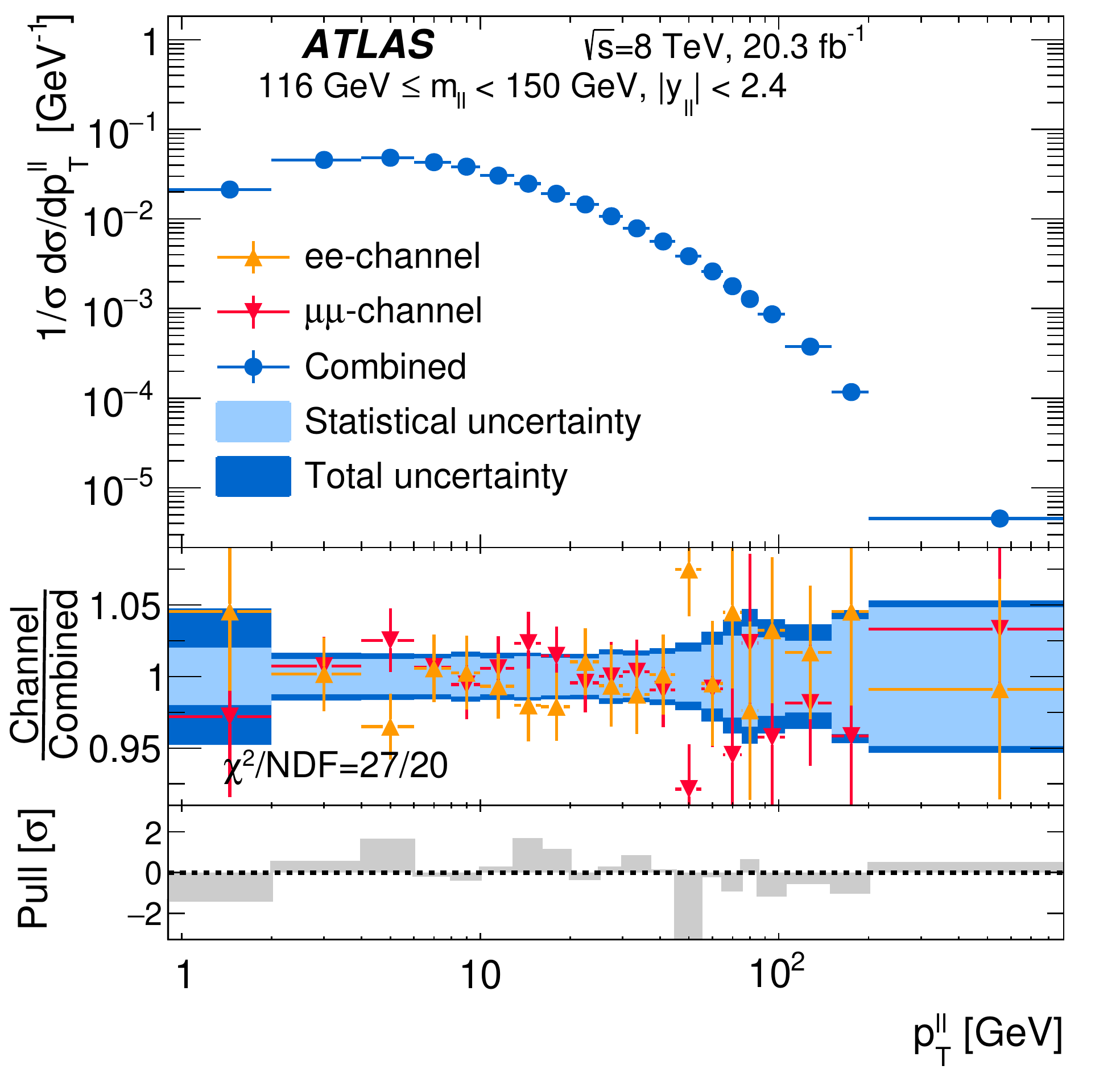}
\caption{The Born-level distributions of $(1/\sigma)\, \mathrm{d}\sigma / \mathrm{d}\Zpt$ for the combination of the electron-pair
  and muon-pair channels, shown in six \Mll\ regions for  $\mody<2.4$. The central panel of each plot shows the ratios of the
values from the individual channels to the combined values, where the error
bars on the individual-channel measurements represent the total uncertainty uncorrelated between
bins. The light-blue band represents the data statistical uncertainty on the combined value and the
dark-blue band represents the total uncertainty (statistical and systematic).
The $\chi^2$ per degree of freedom is given.
The lower panel of each plot shows the pull, defined as the difference between the electron-pair and muon-pair values divided by the uncertainty on that difference.}
\label{fig:ZptMComb}
\end{figure}

The values of $(1/\sigma)\, \mathrm{d}\sigma / \mathrm{d}\phistar$ and  $(1/\sigma)\, \mathrm{d}\sigma / \mathrm{d}\Zpt$  are given in tables in the
appendix for each region of \Mll\ and \mody\ considered. The electron-pair results are given at the
dressed and Born levels, and the muon-pair results at the bare, dressed and Born levels.
The Born-level combined results are also given. The associated statistical and systematic uncertainties
(both uncorrelated and correlated between bins in \phistar\ or \Zpt) are provided in percentage form.

\subsection{Integrated cross-section measurements}

In addition to detailed differential studies in \phistar{} and \Zpt{}, integrated fiducial cross sections are provided for six regions in \Mll{} from $12 \GeV$ to $150 \GeV$.
The fiducial phase space is the same as for the \Zpt{} measurements defined in Table~\ref{tab:FiducialDefinition}. 
The Born-level fiducial cross sections are provided in Table~\ref{tab:FidCrossSections}
for the electron-pair and muon-pair channels separately, as well as for their combination. Uncertainties arising from data statistics, mis-modelling of the detector, background processes and of the MC signal samples used to correct the data are provided as a percentage of the cross section. The individual uncertainty sources after the combination are not necessarily orthogonal and also do not include uncertainties uncorrelated between bins of \Mll{}. Therefore their quadratic sum may not give the total systematic uncertainty.

These results are displayed in Figure~\ref{fig:ZcombMFullBorn}. 
In the channel combination the $\chi^2$ per degree of freedom is 8/6, showing that the electron-pair and muon-pair measurements are
consistent. A total uncertainty of 0.6\%, not including the uncertainty of 2.8\% on the integrated
luminosity, is reached in the region of the \Zboson{}-boson mass peak.
The fact that in some individual \Mll\ bins the combined cross section
does not lie at the naive weighted average of the individual channel values
is due to the effect of systematic uncertainties that are correlated among \Mll\ bins, but
uncorrelated between channels
(see, for example, Refs.~\cite{Cowan:1998} and~\cite{Lyons:1988rp}).

\begin{table*}
\centering
\centering 
\caption{Fiducial cross sections at Born level in the electron- and muon-pair channels as well as the combined value. The statistical and systematic uncertainties are given as a percentage of the cross section. An additional uncertainty of 2.8\% on the integrated luminosity, which is fully correlated between channels and among all \Mll\ bins, pertains to these measurements. The individual uncertainty sources after the combination are not necessarily orthogonal and also do not include uncertainties uncorrelated between bins of \Mll{}. Therefore their quadratic sum may not give the total systematic uncertainty.
} 
\begin{tabular}{ c  cccccc }  
\toprule  
$m_{\ell\ell}$ [GeV]	&	12--20	&	20--30	&	30--46	&	46--66	&	66--116	&	116--150	\\ 
\midrule 
$\sigma(Z/\gamma^*\rightarrow e^+e^-)$	[pb]   &       1.42  &       1.04  &       1.01  &      15.16  &     537.64  &       5.72 \\ 
Statistical uncertainty [\%]  &       0.91  &       1.05  &       1.13  &       0.28  &       0.04  &       0.41 \\ 
Detector uncertainty [\%]  &       2.28  &       2.12  &       1.79  &       3.47  &       0.83  &       0.87 \\ 
Background uncertainty [\%]  &       3.16  &       1.97  &       2.36  &       2.77  &       0.14  &       0.83 \\ 
Model uncertainty [\%]  &       5.11  &       4.38  &       3.59  &       1.59  &       0.16  &       0.74 \\ 
Total systematic uncertainty [\%]  &       6.43  &       5.25  &       4.66  &       4.72  &       0.86  &       1.41 \\ 
\midrule 
$\sigma(Z/\gamma^*\rightarrow \mu^+\mu^-)$	[pb]  &       1.45  &       1.04  &       0.97  &      14.97  &     535.25  &       5.48 \\ 
Statistical uncertainty [\%]  &       0.69  &       0.82  &       0.91  &       0.21  &       0.03  &       0.37 \\ 
Detector uncertainty [\%]  &       1.07  &       1.08  &       1.01  &       1.10  &       0.71  &       0.84 \\ 
Background uncertainty [\%]  &       0.75  &       2.19  &       2.00  &       1.48  &       0.04  &       0.97 \\ 
Model uncertainty [\%]  &       2.59  &       1.81  &       2.36  &       0.75  &       0.31  &       0.31 \\ 
Total systematic uncertainty [\%]  &       2.90  &       3.04  &       3.25  &       2.00  &       0.78  &       1.32 \\ 
\midrule 
$\sigma(Z/\gamma^*\rightarrow \ell^+\ell^-)$ [pb]  &       1.45  &       1.03  &       0.97  &      14.96  &     537.10  &       5.59 \\ 
Statistical uncertainty [\%]  &       0.63  &       0.75  &       0.83  &       0.17  &       0.03  &       0.31 \\ 
Detector uncertainty [\%]  &       0.84  &       0.99  &       0.87  &       1.05  &       0.40  &       0.56 \\ 
Background uncertainty [\%]  &       0.18  &       0.85  &       1.42  &       1.28  &       0.06  &       0.77 \\ 
Model uncertainty [\%]  &       1.84  &       2.24  &       2.27  &       0.89  &       0.19  &       0.50 \\ 
Total systematic uncertainty [\%]  &       2.06  &       2.44  &       2.38  &       1.82  &       0.45  &       1.03 \\ 
\bottomrule 
\end{tabular}
\label{tab:FidCrossSections}
\end{table*}

\begin{figure}[htb]
  \centering
  \includegraphics[width=0.6\textwidth]{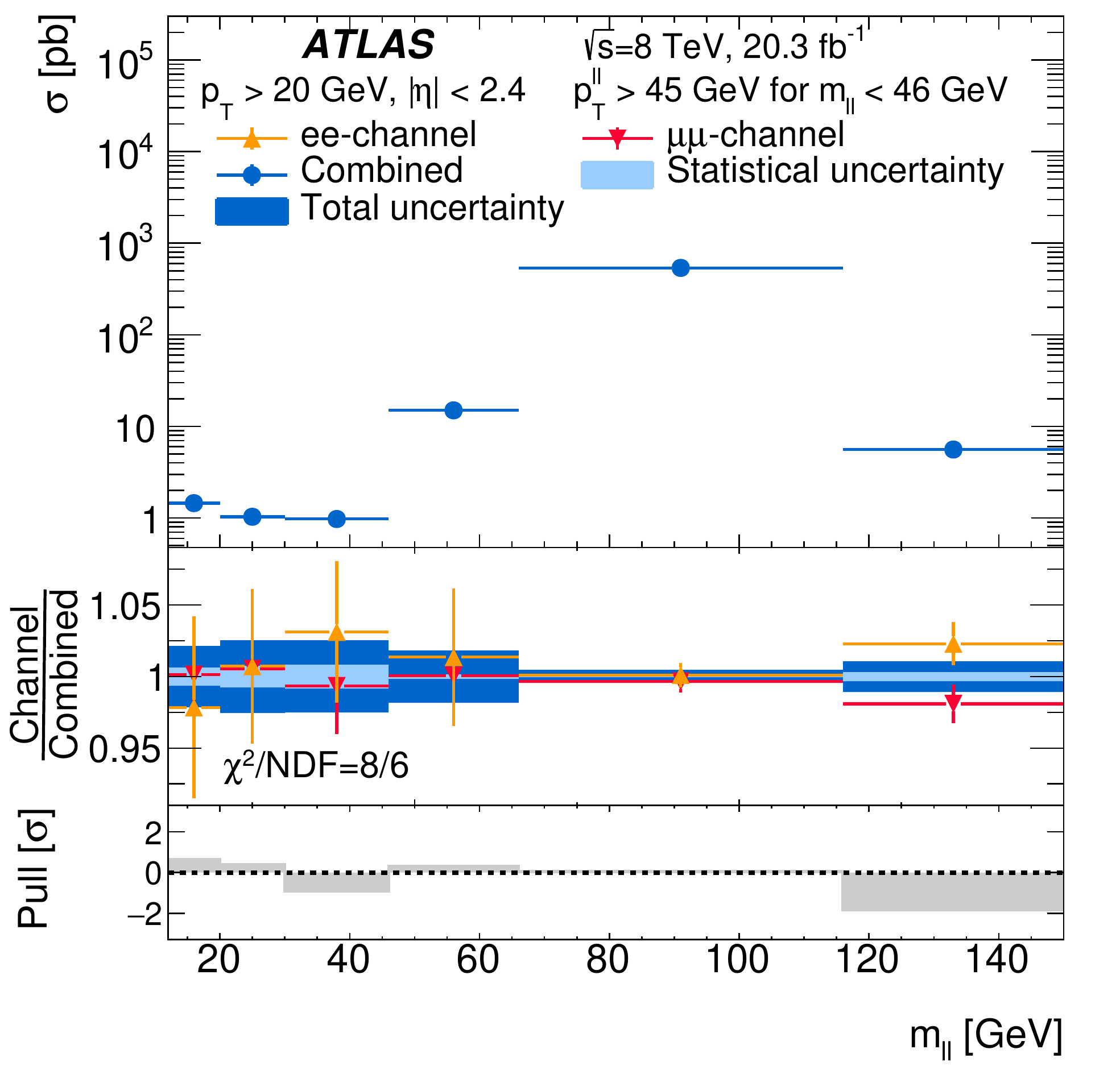}
\caption{Born-level fiducial cross sections in bins of \Mll\
  for the combination of the electron-pair and muon-pair channels.
  The middle plot shows the ratios of the
values from the individual channels to the combined values, where the error
bars on the individual-channel measurements represent the total uncertainty uncorrelated between
bins. The light-blue band represents the data statistical uncertainty on the combined value.
The dark-blue band represents the total uncertainty (statistical and systematic), except for
the uncertainty of 2.8\% on the integrated luminosity, which is fully correlated between channels and among all \Mll\ bins.
The $\chi^2$ per degree of freedom is given.
The lower plot shows the pull, defined as the difference between the electron-pair and muon-pair values
divided by the uncertainty on that difference.
The fiducial regions to which these cross sections correspond are specified in Table~\protect\ref{tab:FiducialDefinition}. 
Note that \Zpt{} is required to be greater than $45 \GeV$ for $\Mll{}<46 \GeV$.
}
\label{fig:ZcombMFullBorn}
\end{figure}

\section{\label{sec:Comparison}Comparison to QCD predictions}
\subsection{Overview}

The combined Born-level measurements of \phistar\ and \Zpt\ presented in Section~\ref{sec:Results} are
compared in this section to a series of theoretical predictions.

A first general comparison is provided by  Figure~\ref{fig:Theory_Phistar_Zpt_Comparison}.
This shows the ratio of the predictions of \Resbos\ for the $Z$-boson mass peak and for \mody~$<$~2.4 to the
combined Born-level data for  $(1/\sigma)\,\mathrm{d}\sigma / \mathrm{d}\phistar$ and  $(1/\sigma)\,\mathrm{d}\sigma / \mathrm{d}\Zpt$.
In order to allow the features of these two distributions to be compared easily, the scales on the
abscissae in Figure~\ref{fig:Theory_Phistar_Zpt_Comparison} are aligned according to the approximate relationship~\cite{phistarToZpT,Banfi:2010cf}
\mbox{$\sqrt{2} m_Z \phistar \approx \Zpt$}.
The general features of the two distributions in Figure~\ref{fig:Theory_Phistar_Zpt_Comparison} are 
similar.
At low values of \phistar\ and \Zpt, in which non-perturbative effects and soft-gluon resummation are most important, the predictions from
\Resbos\ are consistent with the data within the assigned theoretical uncertainties.
However, at high values of \phistar\ and \Zpt, which are more sensitive to the emission of hard partons, the predictions from
\Resbos\ are not consistent with the data within theoretical uncertainties.
Figure~\ref{fig:Theory_Phistar_Zpt_Comparison} illustrates the particular power of \phistar\ to probe the
region of low \Zpt.
Finer binning is possible in \phistar\ than in \Zpt\ whilst maintaining smaller systematic uncertainties
from experimental resolution.

\begin{figure}[htb]
  \centering
  \includegraphics[width=0.65\textwidth]{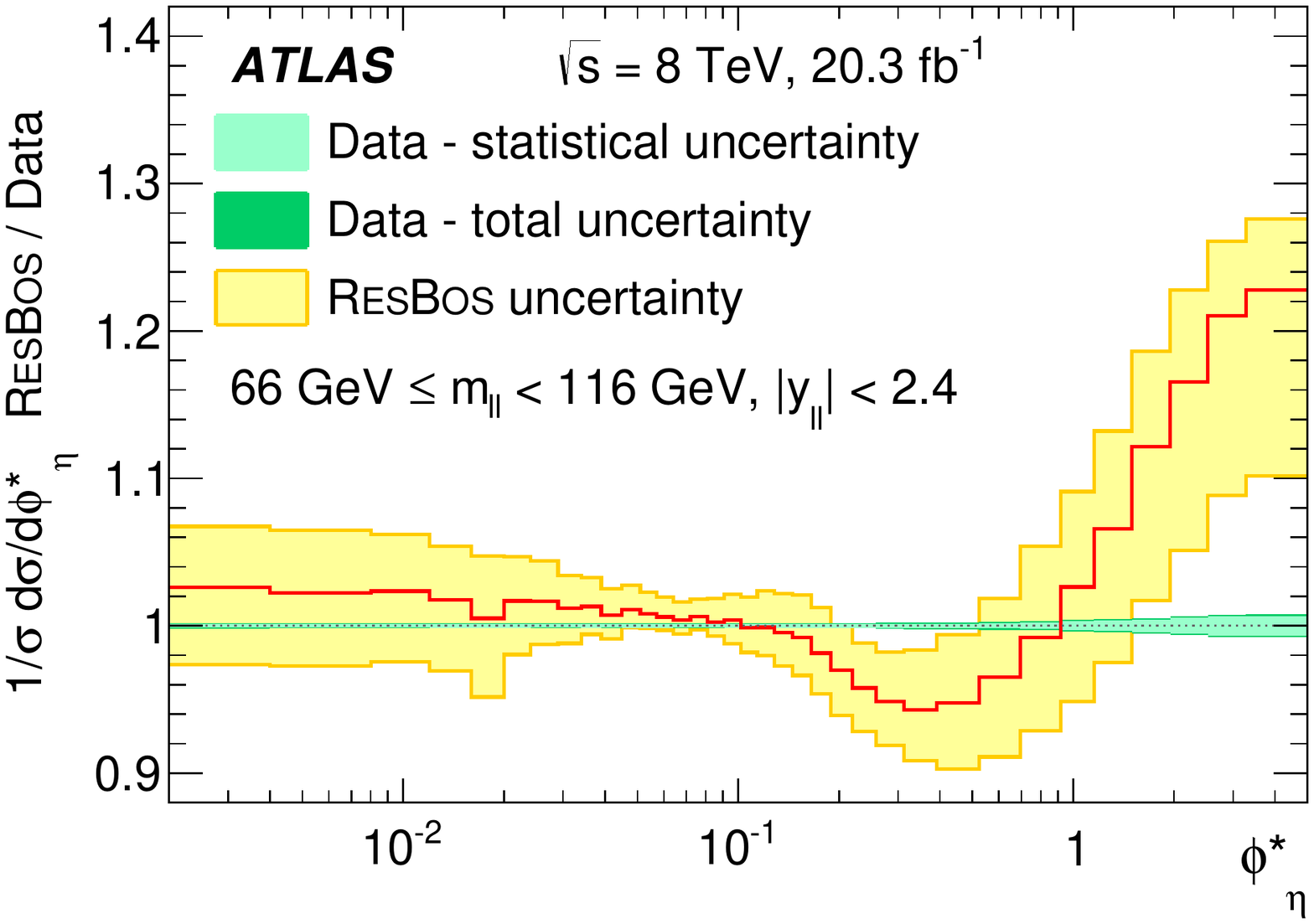}\\
  \includegraphics[width=0.65\textwidth]{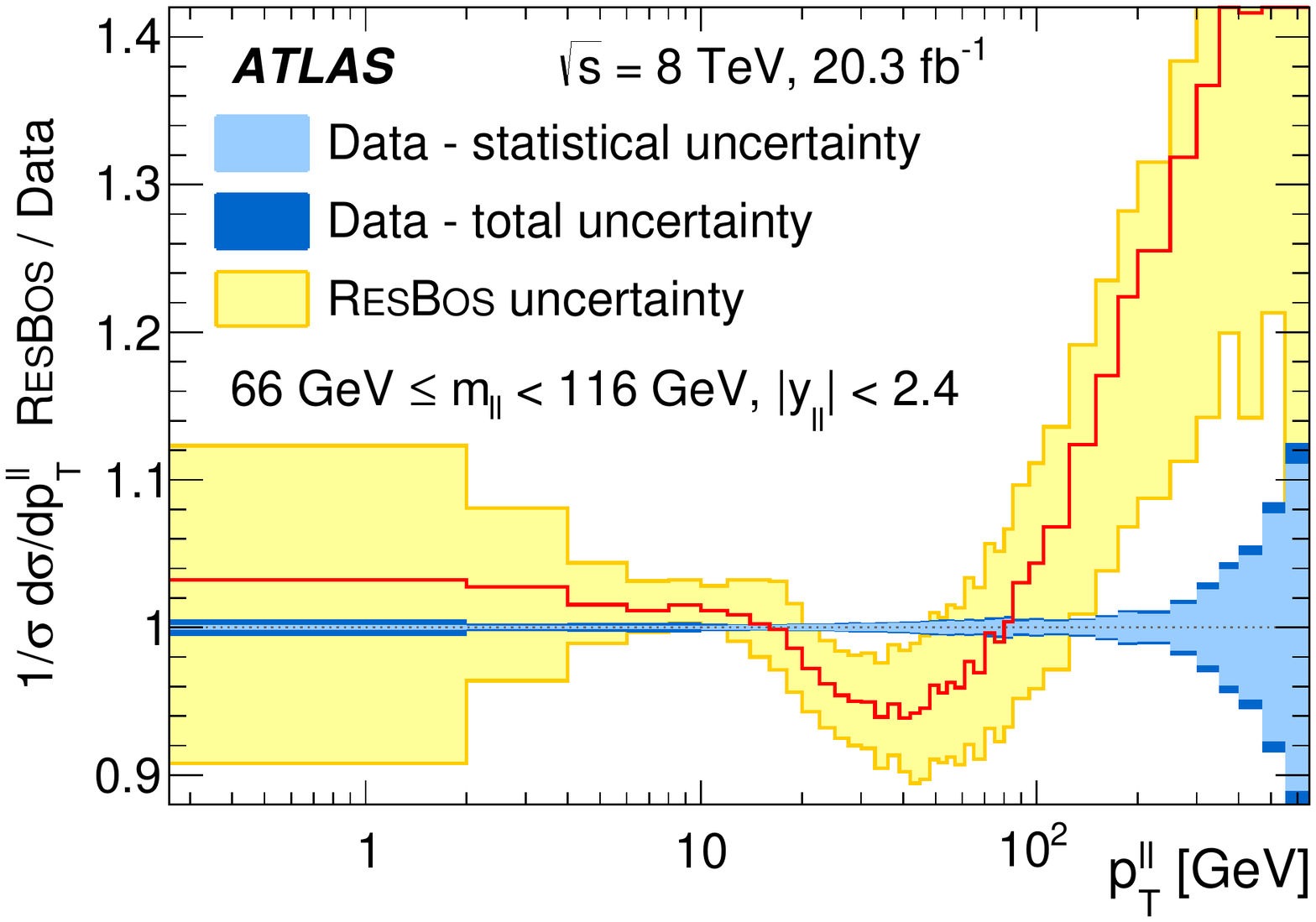}\\
  \caption{The ratio of the predictions of \Resbos\ for the $Z$-boson mass peak and for \mody~$<$~2.4 to the
combined Born-level data for  $(1/\sigma)\,\mathrm{d}\sigma / \mathrm{d}\phistar$ (top) and  $(1/\sigma)\,\mathrm{d}\sigma / \mathrm{d}\Zpt$ (bottom).
  The light-green (light-blue) band represents the statistical uncertainty on the data for \phistar\ (\Zpt) and
  the dark-green (dark-blue) band represents
  the total uncertainty (statistical and systematic) on the data.
  The yellow band represents the uncertainty in the \Resbos{} calculation arising from varying~\cite{ScaleVariations} the QCD scales, 
  the non-perturbative parameter $a_Z$, and PDFs.
  }
  \label{fig:Theory_Phistar_Zpt_Comparison}
\end{figure} 

The \phistar{} measurements are compared in detail to predictions from \Resbos\ in Section~\ref{sec:ResbosComparison}.
In Section~\ref{sec:MCComparison} the normalised \Zpt{} measurements are compared to the predictions from a number of MC generators that use the parton-shower approach.
The fixed-order predictions from \textsc{Dynnlo1.3}~\cite{Catani:2009sm} are compared to the absolute \Zpt{} differential cross sections in Section~\ref{sec:electroweakcorrections}.

\clearpage
\newpage
\subsection{\label{sec:ResbosComparison}Comparison to resummed calculations}

The predictions of $(1/\sigma)\,\mathrm{d}\sigma / \mathrm{d}\phistar$ from \Resbos{} are compared to the Born-level measurements in Figures~\ref{fig:Theory_BR_Phi_Mass} to~\ref{fig:Theory_MC_PhiStarM1_Mass_Ratio}.
As described above, \phistar{} provides particularly precise measurements in the region sensitive to the effects of soft-gluon resummation and
non-perturbative effects and therefore is the observable used to test the predictions from \Resbos{}.
Figure~\ref{fig:Theory_BR_Phi_Mass} shows the ratio of $(1/\sigma)\,\mathrm{d}\sigma / \mathrm{d}\phistar$ as predicted by
\Resbos{} to the combined Born-level data for the six \mody{} regions at the \Zboson{}-boson mass peak.
Figure~\ref{fig:Theory_BR_Phi_OffPeakMass} shows the same comparison for the three \mody{} regions in
the two \Mll\ regions adjacent to the $Z$-boson mass peak.
Also shown in these figures are the statistical and total uncertainties on the data, as well as the uncertainty in the
\Resbos{} calculation arising from varying~\cite{ScaleVariations} the QCD scales, 
  the non-perturbative parameter $a_Z$, and PDFs.
  
For values of $\phistar < 2$  for the \Mll\ region around the \Zboson{}-boson mass peak the predictions from
\Resbos\ are generally consistent with the (much more precise) data within the assigned theoretical uncertainties.
However, at larger values of \phistar\ this is not the case.
For the region of \Mll\ above the $Z$-boson mass peak the predictions from
\Resbos\ are consistent with the data within uncertainties for all values of \phistar.
For the region of \Mll\ from $46 \GeV$ to $66 \GeV$ the predictions from
\Resbos\ lie below the data for $\phistar > 0.4$.
In this context it may be noted that a known deficiency of the \Resbos{} prediction is the lack of
NNLO QCD corrections for the contributions from $\gamma^{*}$ and from $Z/\gamma^{*}$ interference.
Similar deviations from the data in the mass region below the $Z$ peak were observed in the D0 measurement in Ref.~\cite{Abazov:2014mza}.

\begin{figure}[htb]
  \centering
  \includegraphics[width=0.95\textwidth]{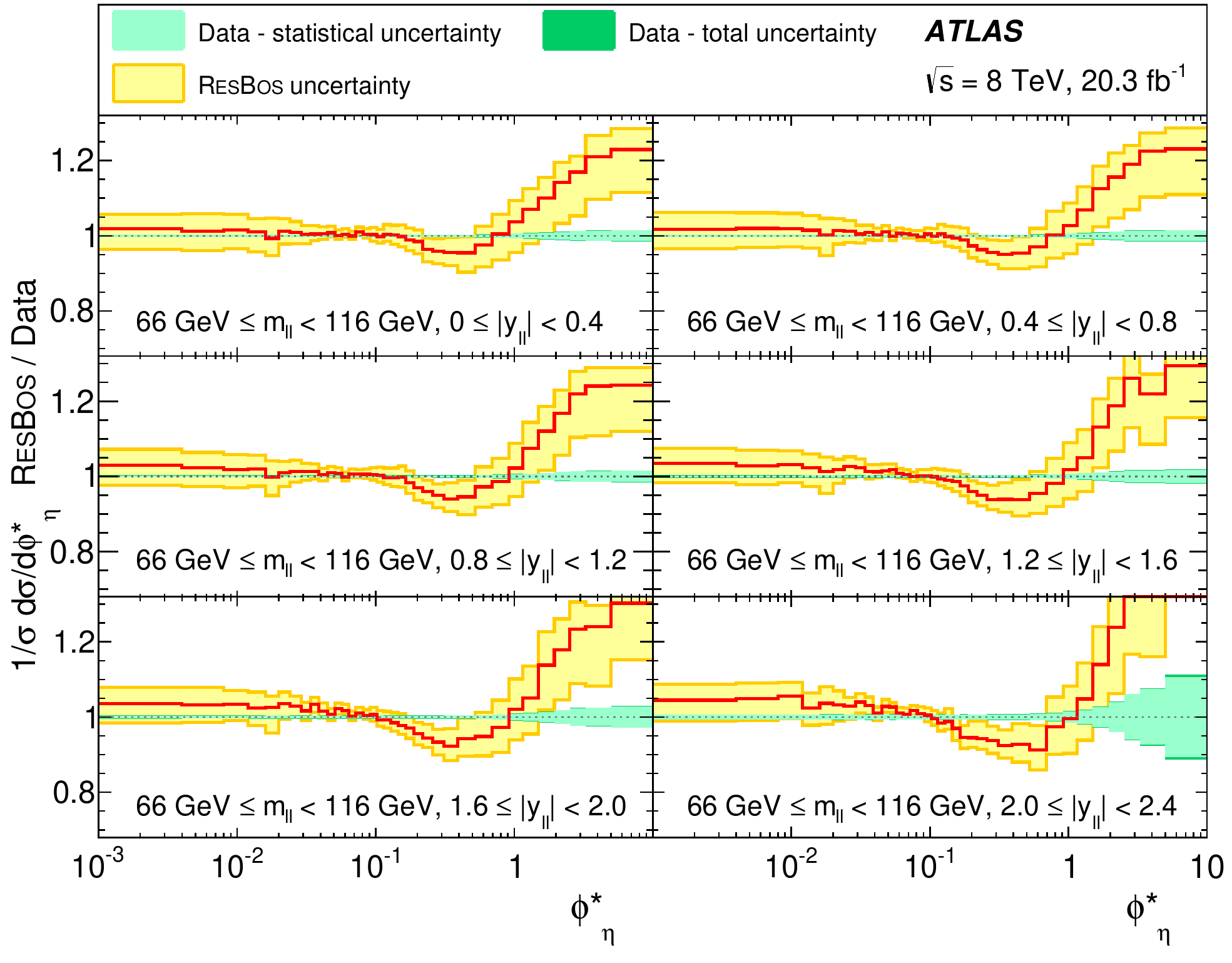}\\
  \caption{The ratio of $(1/\sigma)\,\mathrm{d}\sigma / \mathrm{d}\phistar$ as predicted by \Resbos{} to the combined Born-level data,
  for the six \mody{} regions at the \Zboson{}-boson mass peak.
  The light-green band represents the statistical uncertainty on the data and the dark-green band represents
  the total uncertainty (statistical and systematic) on the data.
  The yellow band represents the uncertainty in the \Resbos{} calculation arising from varying~\cite{ScaleVariations} the QCD scales, 
  the non-perturbative parameter $a_Z$, and PDFs.
  }
  \label{fig:Theory_BR_Phi_Mass}
\end{figure} 

\begin{figure}[htb]
  \centering
  \includegraphics[width=0.95\textwidth]{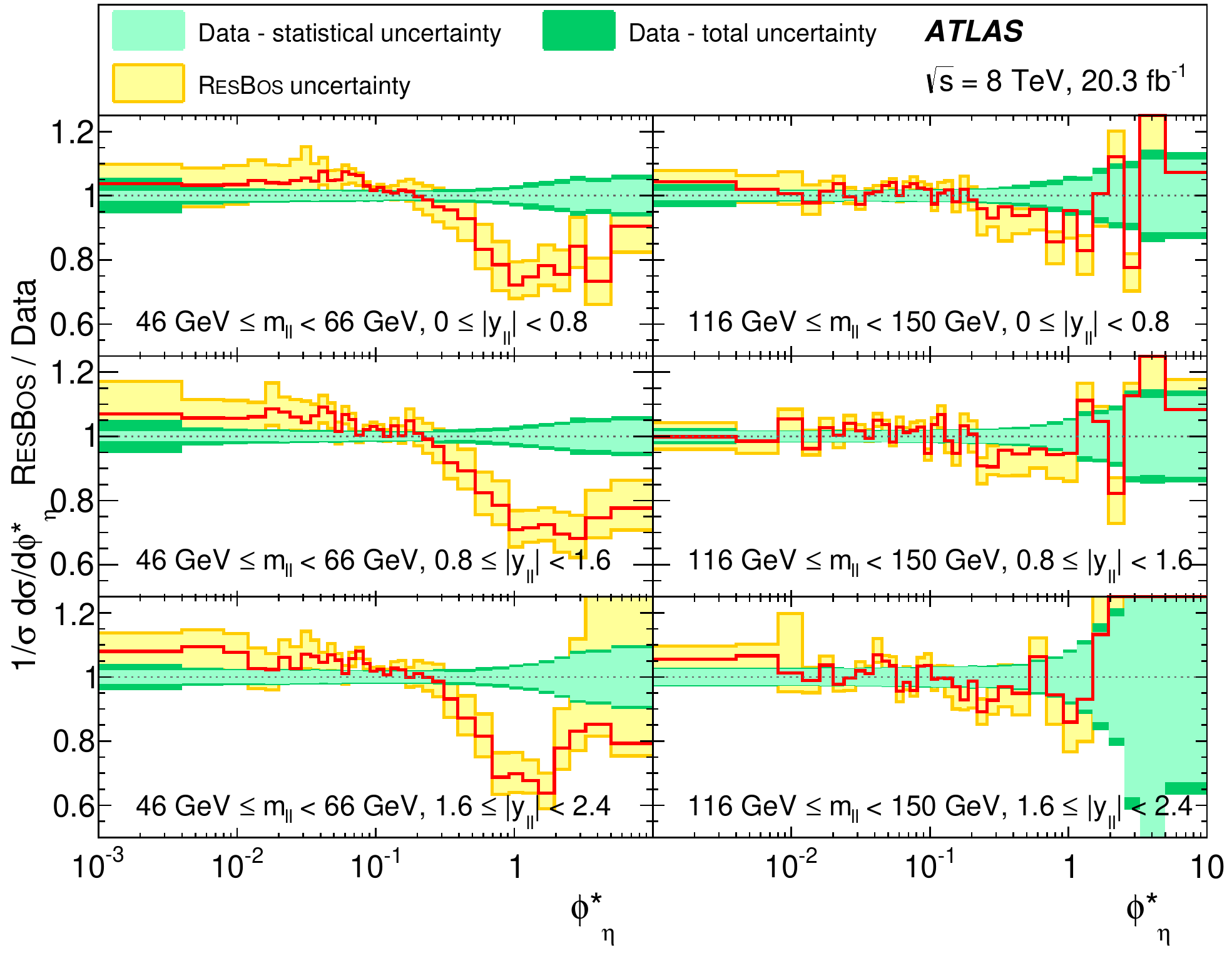}\\
  \caption{The ratio of $(1/\sigma)\,\mathrm{d}\sigma / \mathrm{d}\phistar$ as predicted by \Resbos{} to the combined Born-level data,
  for the three \mody{} regions in the two \Mll\ regions adjacent to the $Z$-boson mass peak.
  The light-green band represents the statistical uncertainty on the data and the dark-green band represents
  the total uncertainty (statistical and systematic) on the data.
  The yellow band represents the uncertainty in the \Resbos{} calculation arising from varying~\cite{ScaleVariations} the QCD scales, 
  the non-perturbative parameter $a_Z$, and PDFs.
  }
  \label{fig:Theory_BR_Phi_OffPeakMass}
\end{figure}

The theoretical uncertainties are highly correlated between different kinematic regions
and therefore, as pointed out in Ref.~\cite{Abazov:2014mza}, the ratio of  $(1/\sigma)\,\mathrm{d}\sigma/ \mathrm{d}\phistar$  in different kinematic regions enables a
more precise comparison of the predictions with data.
For example, the question of whether or not the non-perturbative contribution to \Zpt{} varies with parton
momentum fraction, $x$, or four-momentum transfer, $Q^2$, may be investigated by examining how the shape of  $(1/\sigma)\, \mathrm{d}\sigma / \mathrm{d}\phistar$ evolves with \mody{} and \Mll{} at low \phistar{}.

Figure~\ref{fig:Theory_MC_PhiStarM2_Rapidity_Ratio} shows the ratio of the distribution of $(1/\sigma)\,\mathrm{d}\sigma/ \mathrm{d}\phistar$ in each region of
\mody{} to the distribution in the central region ($\mody<0.4$), for events in the \Mll\ region around
the \Zboson{}-boson mass peak.
The distributions are shown for data (with associated statistical and total uncertainties) as well as for \Resbos{}.
It can be seen that the uncertainties on the \Resbos{} predictions, arising from varying~\cite{ScaleVariations} the QCD scales, 
  the non-perturbative parameter $a_Z$, and PDFs, are of a comparable size to
  the uncertainties on the corrected data.
  The predictions from
\Resbos\ are consistent with the data within the assigned uncertainties.
Figure~\ref{fig:Theory_MC_PhiStarM3_Rapidity_Ratio} shows equivalent comparisons for the \Mll\ regions
from $46 \GeV$ to $66 \GeV$ and from $116 \GeV$ to $150 \GeV$.
It can be seen that the predictions from \Resbos{} are again consistent with the data within the assigned uncertainties.
Therefore it can be concluded that \Resbos{} describes the evolution with \mody\ of the shape of the
$(1/\sigma)\,\mathrm{d}\sigma/ \mathrm{d}\phistar$ measurements well, and rather better than it describes the basic shape
of the data (Figures~\ref{fig:Theory_BR_Phi_Mass} and~\ref{fig:Theory_BR_Phi_OffPeakMass}).

\begin{figure}[htb]
  \centering
  \includegraphics[width=0.95\textwidth]{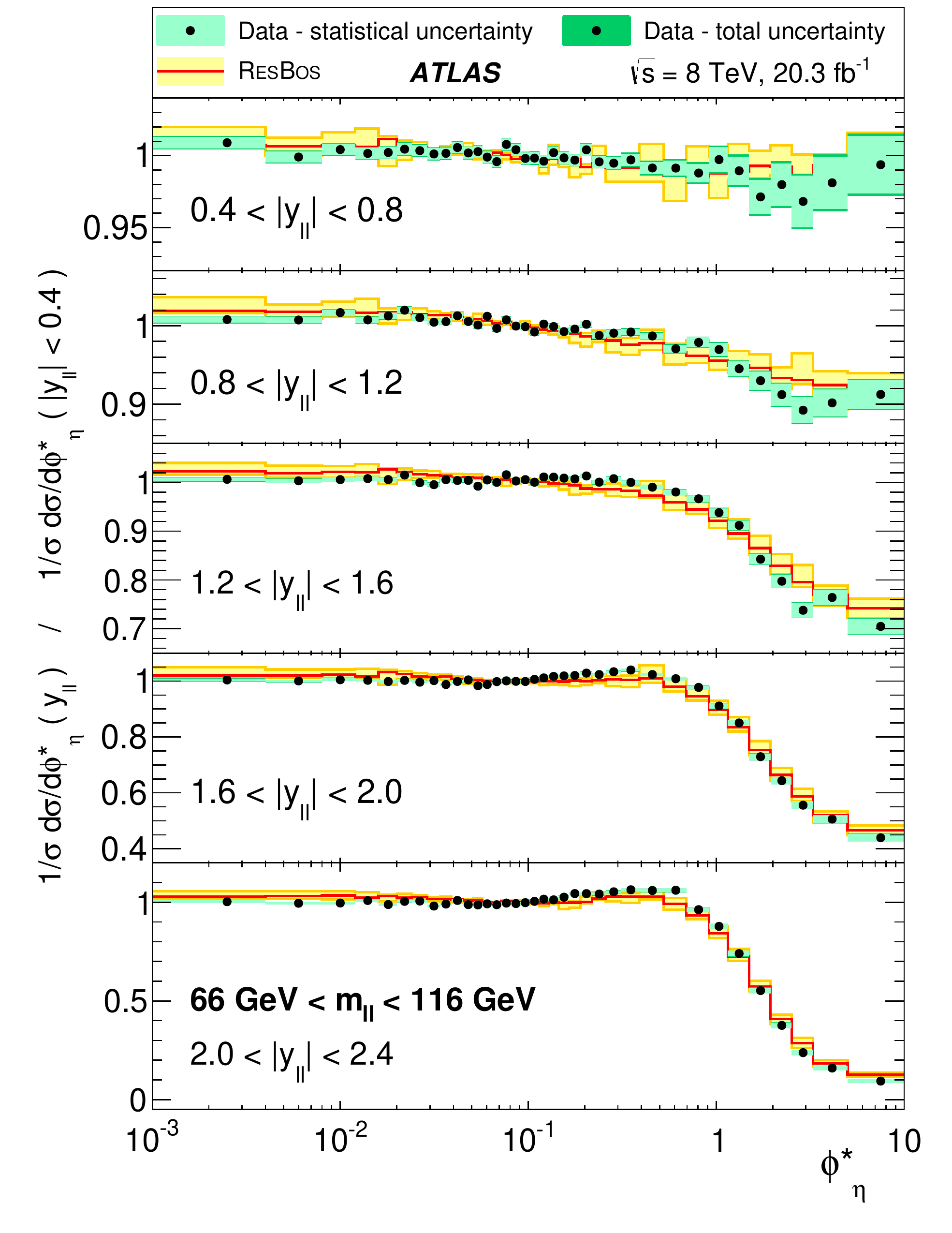}\\
  \caption{The distribution of $(1/\sigma)\,\mathrm{d}\sigma/ \mathrm{d}\phistar$ at Born level in each region of \mody, shown as a ratio
  to the central rapidity region ($\mody<0.4$), for events at the \Zboson{}-boson mass peak.
  The data, shown as points, are compared to the predictions of \Resbos{}.
  The light-green band represents the statistical uncertainty on the data and the dark-green band represents the total uncertainty on the data (treating systematic uncertainties as 
  uncorrelated between regions of \mody).
  The yellow band represents the uncertainty in the \Resbos{} calculation arising from varying~\cite{ScaleVariations} the QCD scales, 
  the non-perturbative parameter $a_Z$, and PDFs.
  }
  \label{fig:Theory_MC_PhiStarM2_Rapidity_Ratio}
\end{figure}

\begin{figure}[htb]
  \centering
   \includegraphics[width=0.80\textwidth]{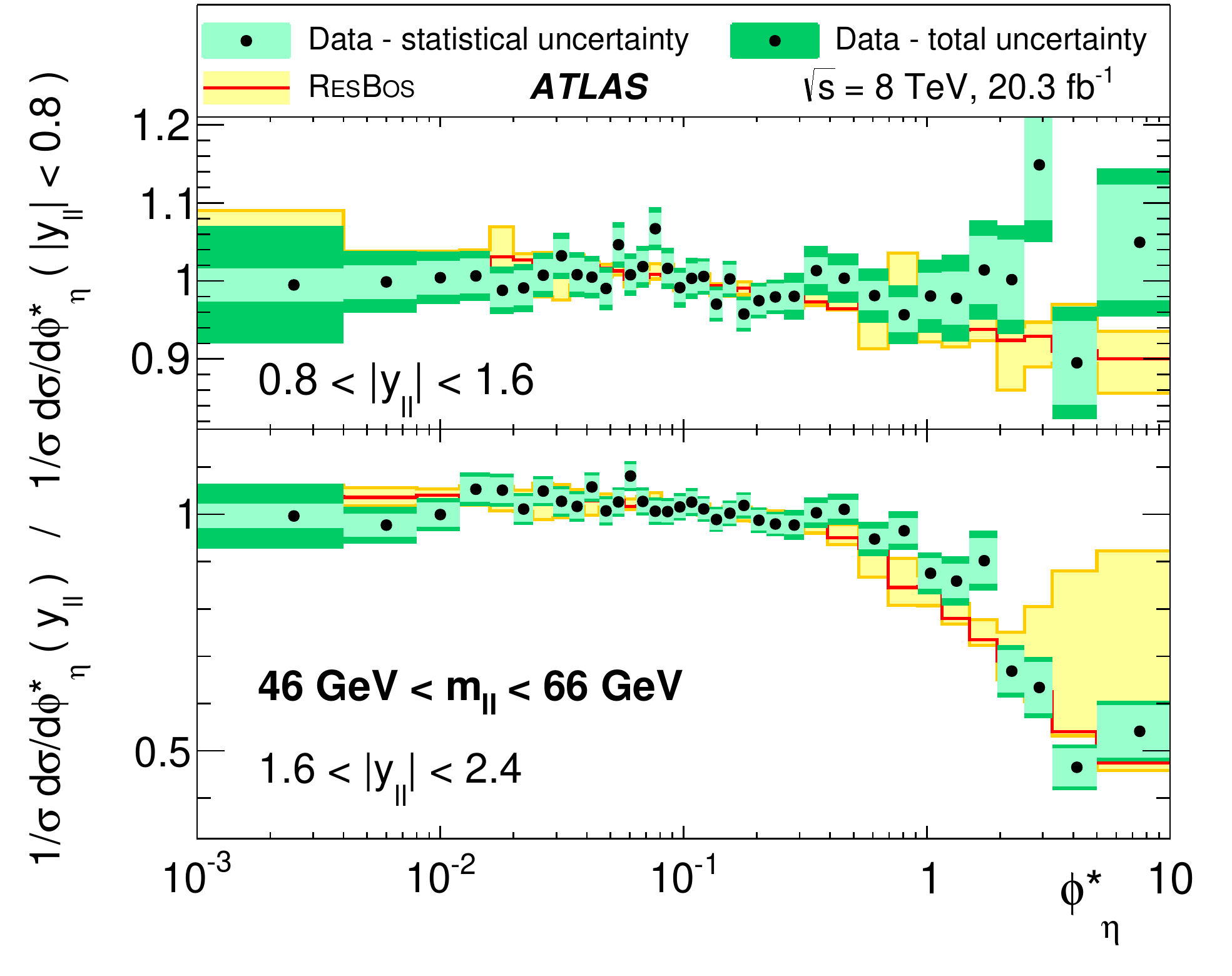}
  \includegraphics[width=0.80\textwidth]{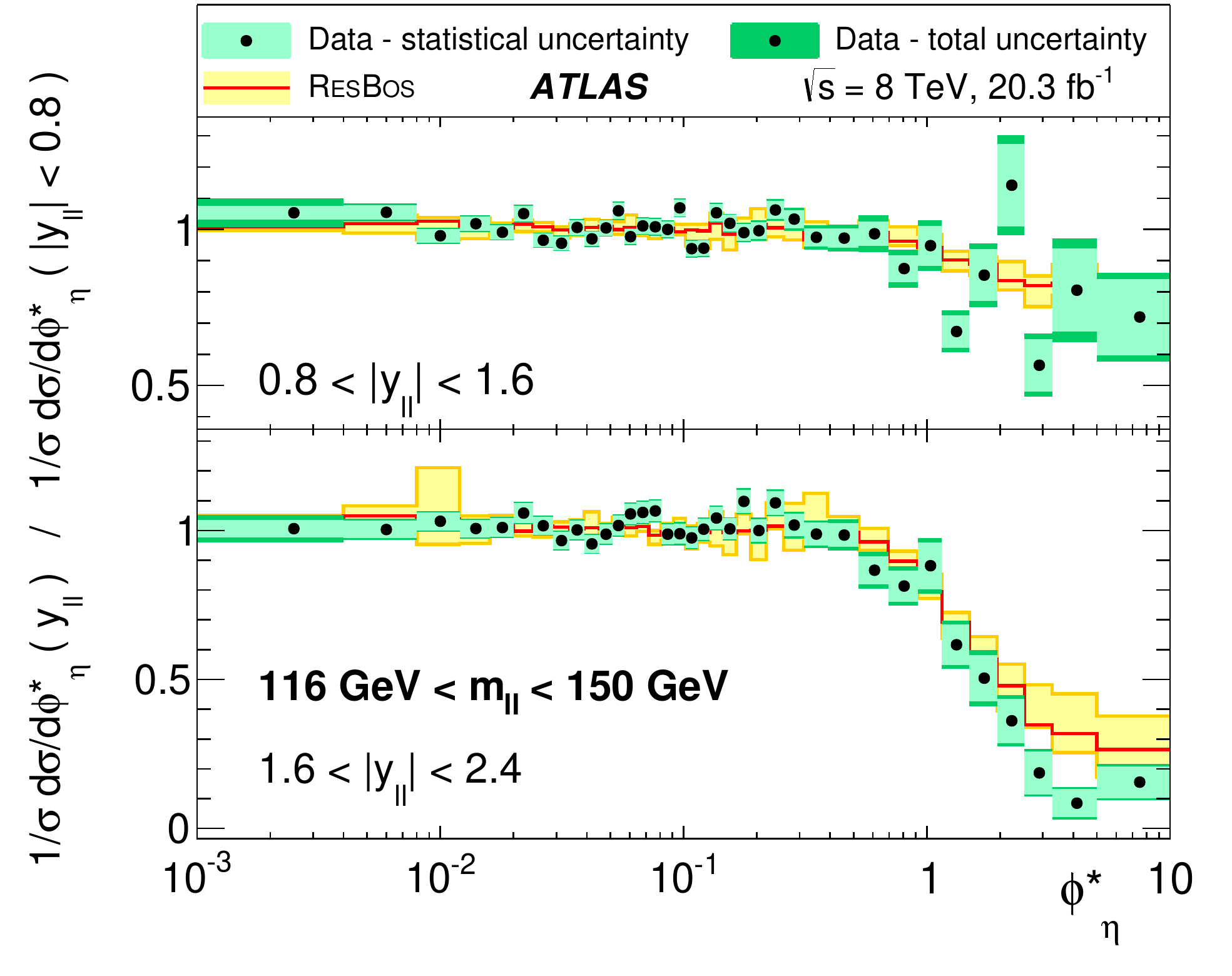}
  \caption{The distribution of $(1/\sigma)\,\mathrm{d}\sigma/ \mathrm{d}\phistar$ at Born level in each region of \mody, shown as a ratio
  to the central rapidity region  ($\mody<0.8$),
   for events with \Mll{} between $46 \GeV$ to $66 \GeV$ (upper plots) and $116 \GeV$ to $150 \GeV$ (lower plots).
  The data, shown as points, are compared to the predictions of \Resbos{}.
  The light-green band represents the statistical uncertainty on the data and the dark-green band represents the total uncertainty on the data (treating systematic uncertainties as
  uncorrelated between regions of \mody).
  The yellow band represents the uncertainty in the \Resbos{} calculation arising from varying~\cite{ScaleVariations} the QCD scales, 
  the non-perturbative parameter $a_Z$, and PDFs.
   }
  \label{fig:Theory_MC_PhiStarM3_Rapidity_Ratio}
\end{figure}

Figure~\ref{fig:Theory_MC_PhiStarM1_Mass_Ratio} shows the ratio of $(1/\sigma)\,\mathrm{d}\sigma/ \mathrm{d}\phistar$ in the \Mll{} region from $116 \GeV$ to $150 \GeV$ to that in the \Mll{} region from $46 \GeV$ to $66 \GeV$,
for the three divisions of \mody{}. The ratio is shown for data (with associated statistical and total uncertainties) as well as for \Resbos{}.
It can again be seen that the uncertainties on the \Resbos{} predictions, arising from varying~\cite{ScaleVariations} the QCD scales, 
  the non-perturbative parameter $a_Z$, and PDFs, and shown as a yellow band, are of a comparable size to
  the uncertainties on the corrected data.
  For values of $\phistar < 0.5$ the predictions from
\Resbos\ are consistent with the data within the assigned theoretical uncertainties showing
that \Resbos{} is able to describe the evolution of the \phistar{} distribution with \Mll.
However, at larger values of \phistar\ this is not the case.

\begin{figure}[htb]
  \centering
  \includegraphics[width=0.95\textwidth]{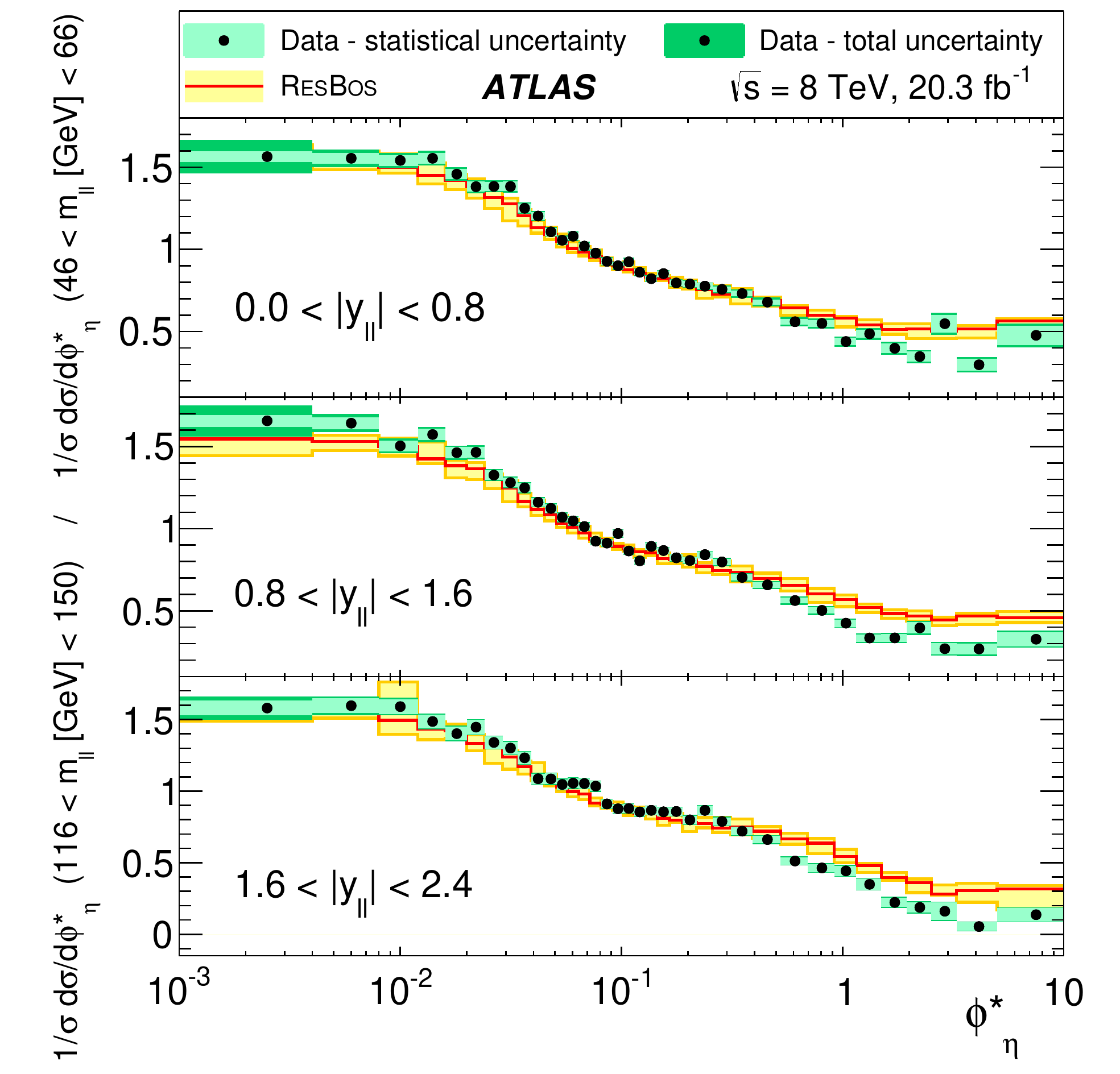}\\
  \caption{The ratio of $(1/\sigma)\,\mathrm{d}\sigma/ \mathrm{d}\phistar$ in the \Mll{} region from $116 \GeV$ to $150 \GeV$ to that in the \Mll{}
  region from $46 \GeV$ to $66 \GeV$, for three regions of \mody.
  The data, shown as points, are compared to the predictions of \Resbos{}.
  The light-green band represents the statistical uncertainty on the data and the dark-green band represents the total uncertainty on the data (treating systematic uncertainties as
 uncorrelated between the mass regions).
  The yellow band represents the uncertainty in the \Resbos{} calculation arising from varying~\cite{ScaleVariations} the QCD scales, 
  the non-perturbative parameter $a_Z$, and PDFs.
  }
  \label{fig:Theory_MC_PhiStarM1_Mass_Ratio}
\end{figure}

\clearpage
\newpage
\subsection{\label{sec:MCComparison}Comparison to parton-shower approaches} 

Figures~\ref{fig:Theory_MC_Zpt_Mass} to~\ref{fig:Theory_MC_Zpt_Rapidity_Ratio} show the comparison of the
$(1/\sigma)\,\mathrm{d}\sigma / \mathrm{d}\Zpt$  distributions to the predictions of
MC generators using the parton-shower approach: \Powheg{}+\Pythia{} (with both the AU2~\cite{ATLAS:2012uec} and AZNLO~\cite{Aad:2014xaa} tunes), \Powheg{}+\Herwig{} (only shown for the \Mll{} region around the \Zboson{} peak) and \Sherpa{}.
Figure~\ref{fig:Theory_MC_Zpt_Mass} shows the ratio of $(1/\sigma)\,\mathrm{d}\sigma / \mathrm{d}\Zpt$ as predicted by the MC generators,
to the combined Born-level data in each of the six \Mll\ regions for $\mody < 2.4$.
Figure~\ref{fig:Theory_MC_Zpt_Rapidity} shows the ratio for each of the six \mody{} regions at the \Zboson{}-boson mass peak.
Between \Zpt{} values of approximately $5 \GeV$ and $100 \GeV$ for $\Mll{} > 46 \GeV$ the MC generators describe the shape of the data to within 10\%. However, outside this range, and in the regions with very low \Mll{}, the agreement worsens.
For values of $\Zpt < 50 \GeV$ for the \Mll\ region around the $Z$-boson mass peak the best description is provided
by \Powheg{}+\Pythia{} (AZNLO), which was
tuned to exactly this kinematic region in the $7 \TeV$ data~\cite{Aad:2014xaa}.
However, at high values of \Zpt\ around the $Z$-boson mass peak and in other \Mll\ regions this MC tune does not describe the data well and also does not outperform the \Powheg{}+\Pythia{} AU2 tune.
The differences between \Sherpa{} and the data are generally of a similar magnitude, but of opposite sign, to those
seen for \Powheg{}+\Pythia{}.

\begin{figure}[htb]
  \centering
  \includegraphics[width=0.95\textwidth]{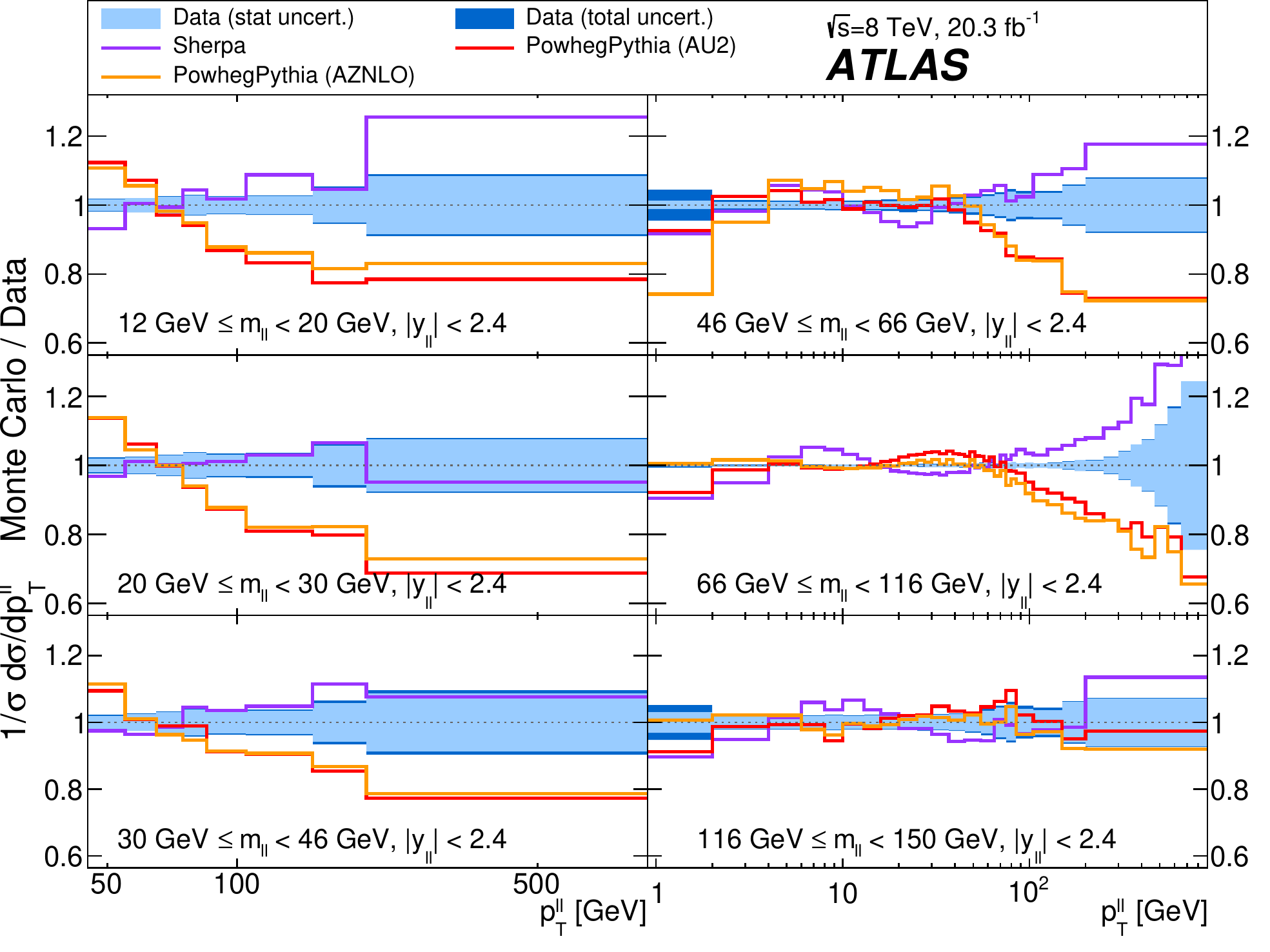}\\
  \caption{The ratio of $(1/\sigma)\,\mathrm{d}\sigma / \mathrm{d}\Zpt$ as predicted by various MC generators to the combined Born-level data,
  in six different regions of \Mll\ for $\mody < 2.4$.
  The light-blue band represents the statistical uncertainty on the data and the dark-blue band represents
  the total uncertainty (statistical and systematic) on the data.
  }
  \label{fig:Theory_MC_Zpt_Mass}
\end{figure}     
  
\begin{figure}[htb]
  \centering
  \includegraphics[width=0.95\textwidth]{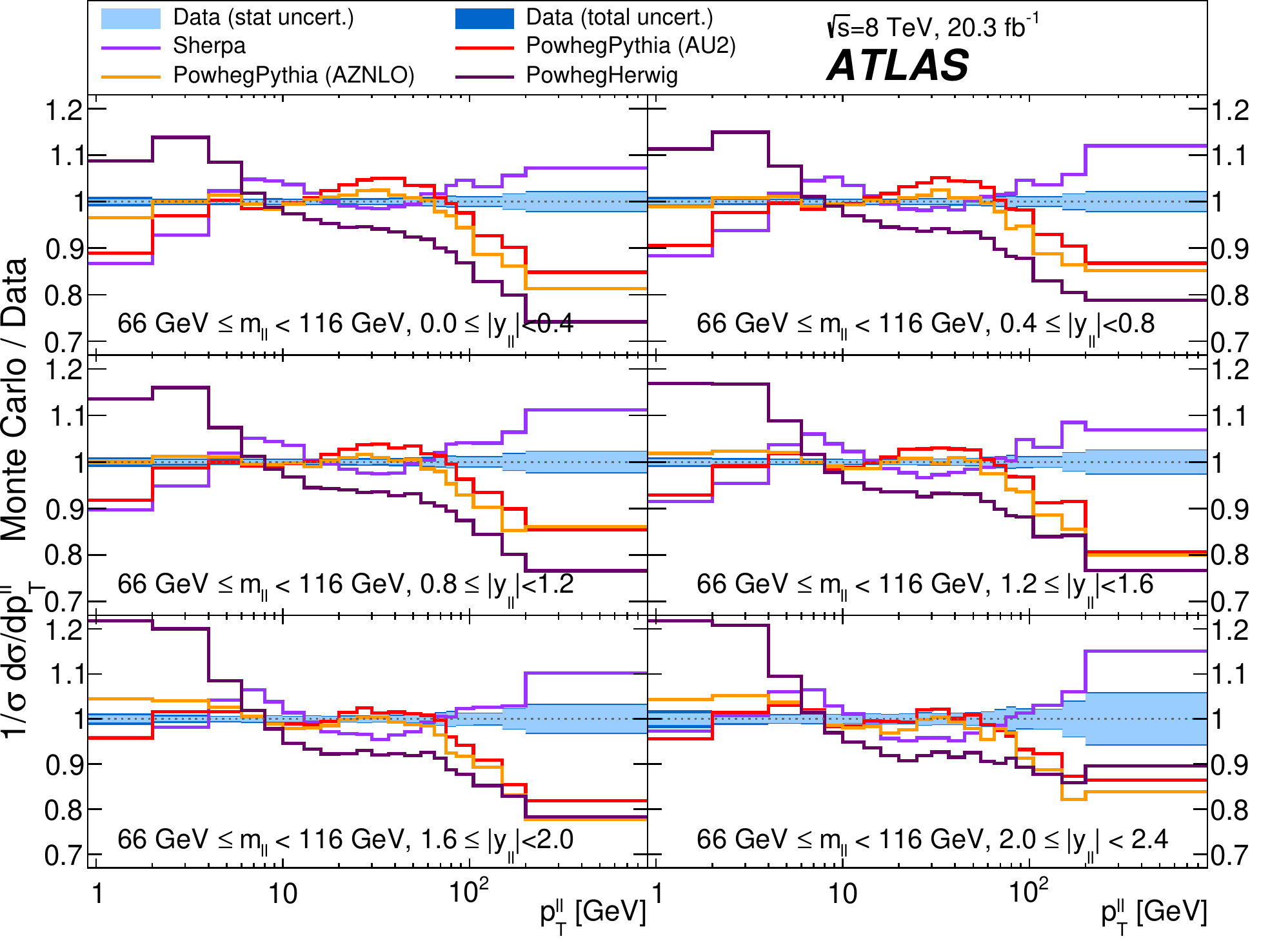}\\
  \caption{The ratio of $(1/\sigma)\,\mathrm{d}\sigma / \mathrm{d}\Zpt$ as predicted by various MC generators to the combined Born-level data,
   in different \mody\ ranges for events at the \Zboson{}-boson mass peak.
  The light-blue band represents the statistical uncertainty on the data and the dark-blue band represents
  the total uncertainty (statistical and systematic) on the data.
  }
  \label{fig:Theory_MC_Zpt_Rapidity}
\end{figure}

Figure~\ref{fig:Theory_MC_Zpt_Rapidity_Ratio} shows the ratio of the distribution of $(1/\sigma)\,\mathrm{d}\sigma/ \mathrm{d}\Zpt$ in each region of
\mody{} to the distribution in the central region ($\mody<0.4$), for events in the \Mll\ region around
the \Zboson{}-boson mass peak.
The distributions are shown for data (with associated statistical and total uncertainties) as well as for predictions from three parton-shower MC generators.
The MC generators describe the data reasonably well over the entire range of \Zpt{} and generally much better than they describe the  $(1/\sigma)\,\mathrm{d}\sigma / \mathrm{d}\Zpt$ distributions (Figures~\ref{fig:Theory_MC_Zpt_Mass} and~\ref{fig:Theory_MC_Zpt_Rapidity}) --- although there are discrepancies of up to 5\% with respect to data for $\Zpt<4 \GeV$.

\begin{figure}[htb]
  \centering
  \includegraphics[width=0.95\textwidth]{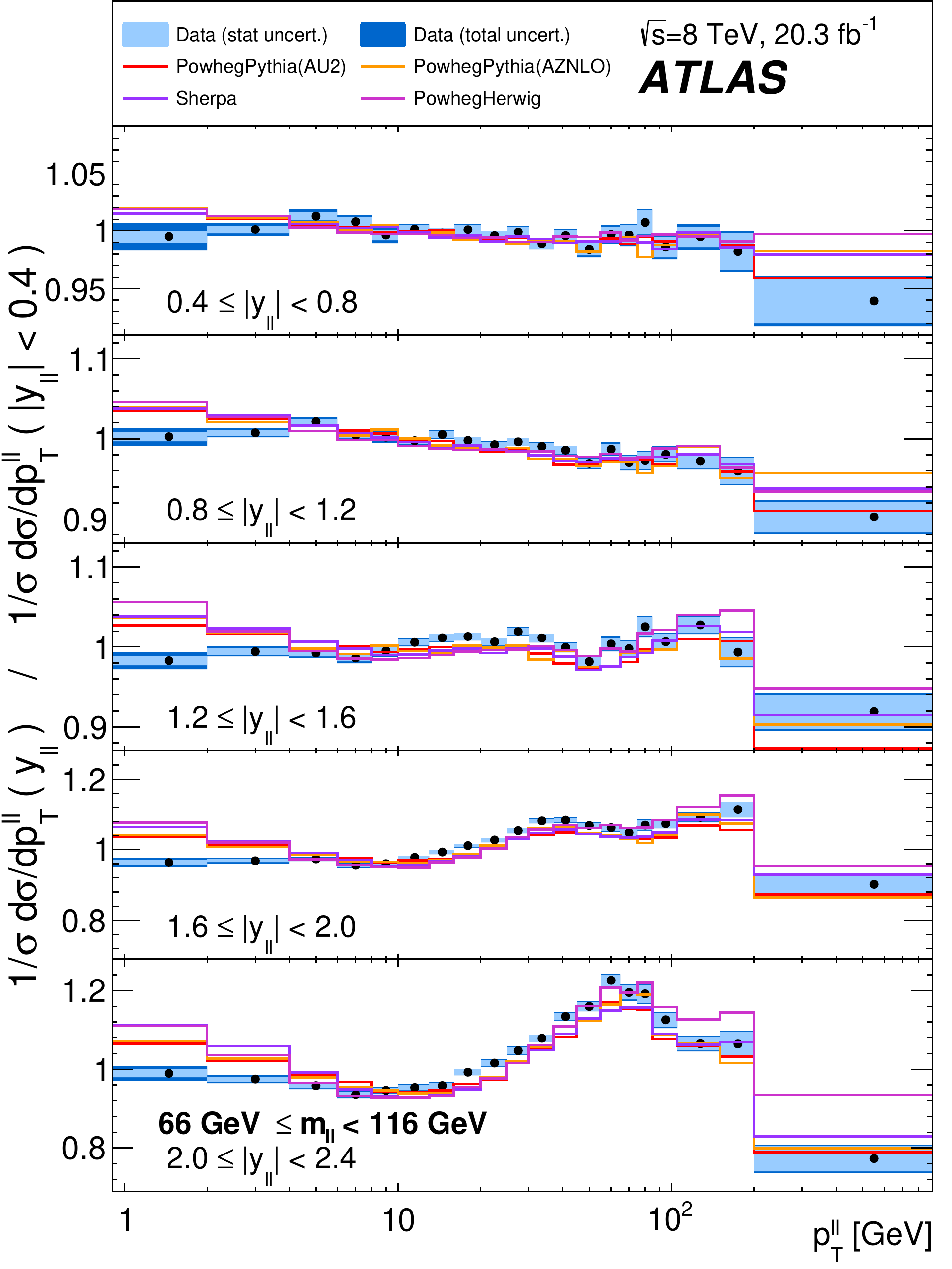}\\
  \caption{The distribution of $(1/\sigma)\,\mathrm{d}\sigma/ \mathrm{d}\Zpt$ at Born level in each region of \mody, shown as a ratio
  to the central rapidity region ($\mody<0.4$), for events at the \Zboson{}-boson mass peak.
  The data, shown as points, are compared to the predictions of various MC generators.
  The light-blue band represents the statistical uncertainty on the data and the dark-blue band represents the total uncertainty on the data (treating systematic uncertainties as
  uncorrelated between regions of \mody).
  }
  \label{fig:Theory_MC_Zpt_Rapidity_Ratio}
\end{figure} 

For comparison with Figure~\ref{fig:Theory_MC_Zpt_Mass}, Figure~\ref{fig:phistar_aznlo} shows the ratio of $(1/\sigma)\,\mathrm{d}\sigma / \mathrm{d}\phistar$ as predicted by the MC generators,
to the combined Born-level data in each of the three \Mll{} regions from $46 \GeV$ to $150 \GeV$ for $\mody < 2.4$.
The differences between MC predictions and data seen in Figure~\ref{fig:phistar_aznlo} are consistent
with those seen in Figure~\ref{fig:Theory_MC_Zpt_Mass}. 

\begin{figure}[htb]
  \centering
  \includegraphics[width=0.95\textwidth]{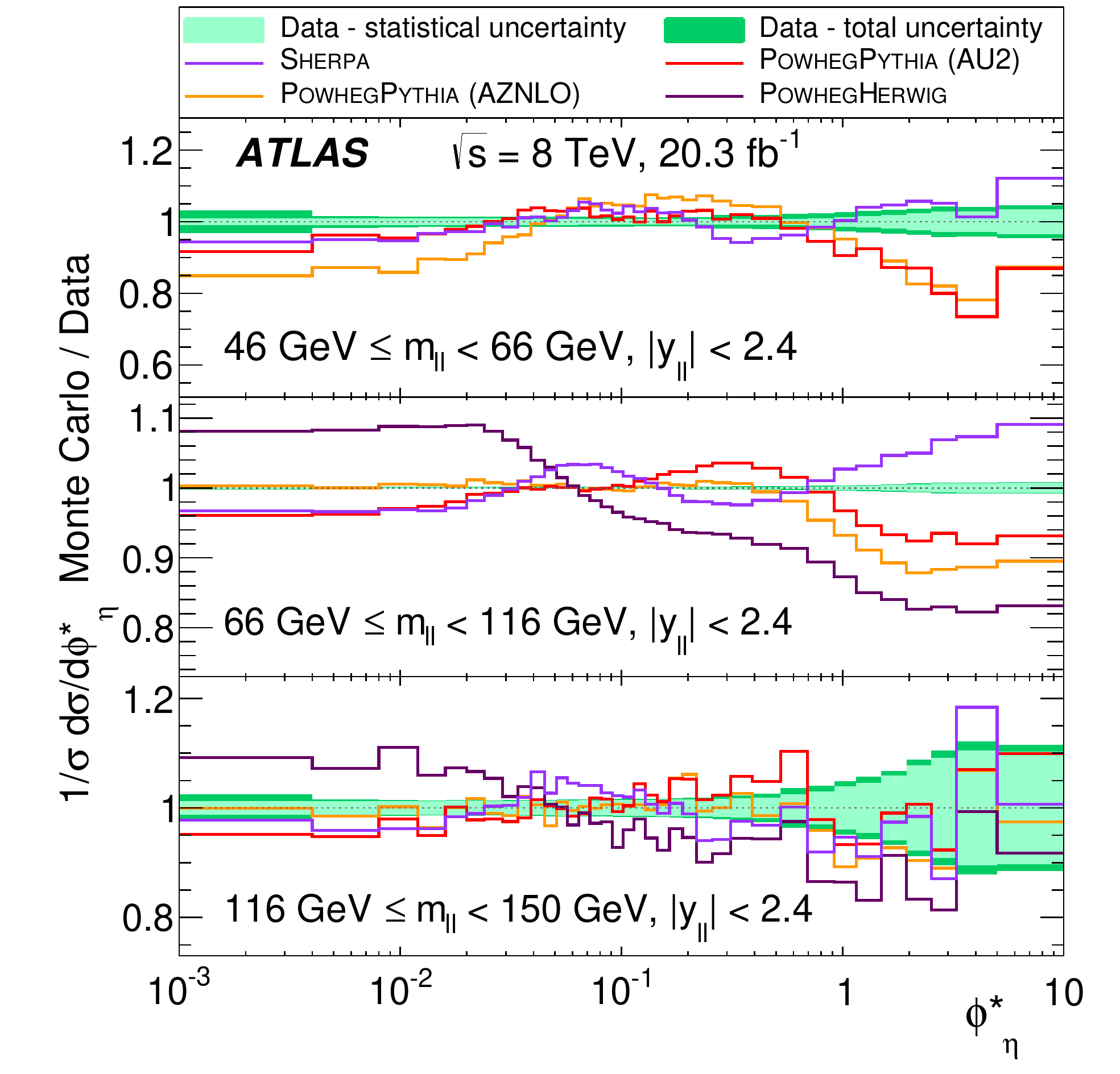}\\
  \caption{The ratio of $(1/\sigma)\,\mathrm{d}\sigma / \mathrm{d}\phistar$ as predicted by various MC generators to the combined Born-level data,
  in three different regions of \Mll\ for $\mody < 2.4$.
  The light-green band represents the statistical uncertainty on the data and the dark-green band represents
  the total uncertainty (statistical and systematic) on the data.
  }
  \label{fig:phistar_aznlo}
\end{figure}

\clearpage
\newpage
\subsection{Fixed-order QCD and electroweak corrections}
\label{sec:electroweakcorrections}

Figure~\ref{fig:Theory_D_Zpt_Mass} shows the ratio of $\mathrm{d}\sigma / \mathrm{d}\Zpt$ as predicted by the fixed-order perturbative QCD predictions of \Dynnlo{} to Born-level data for six regions of \Mll{} from $12 \GeV$ to $150 \GeV$. 
The prediction is shown both with and without NLO EW corrections~\cite{Denner:2011vu}.
The data are shown with their associated statistical and total uncertainties.
The predictions are not expected to describe the shape of the data for lower values of \Zpt{}, where it is known that the effects of soft-gluon emissions become important.
At $\Zpt>30 \GeV$ the shape of the \Zpt{} distribution is described within uncertainties by \Dynnlo{}. However, the prediction is consistently low by about 15\%
compared to the data across all \Mll{} ranges, which is not
covered by the evaluated scale and PDF uncertainties, although a recent calculation suggests the size of order $\alpha_{\mathrm{s}}^3$ corrections to be +(5--10)\% for $\Zpt \gtrsim 60 \GeV$~\cite{Ridder:2015dxa}. The observed behaviour of \Dynnlo{} 
is consistent with the results at $\roots=7 \TeV$ near the $Z$ peak~\cite{Aad:2014xaa}.
The application of NLO EW corrections predicts an approximately 5\%
increase of the cross section below the $Z$-peak region due to effects
of $\gamma^*$ exchange, while a suppression of up to 20\% at highest
\Zpt{} is predicted due to large Sudakov logarithms~\cite{Denner:2011vu}.
The change in the prediction induced by the addition of the EW corrections
is significantly smaller than both the uncertainty on the NNLO QCD prediction and the difference between the prediction and data.
Therefore, no conclusions can be drawn on whether or not their
addition leads to an improvement in agreement between data and theory. 

\begin{figure}[htb]
  \centering
  \includegraphics[width=0.95\textwidth]{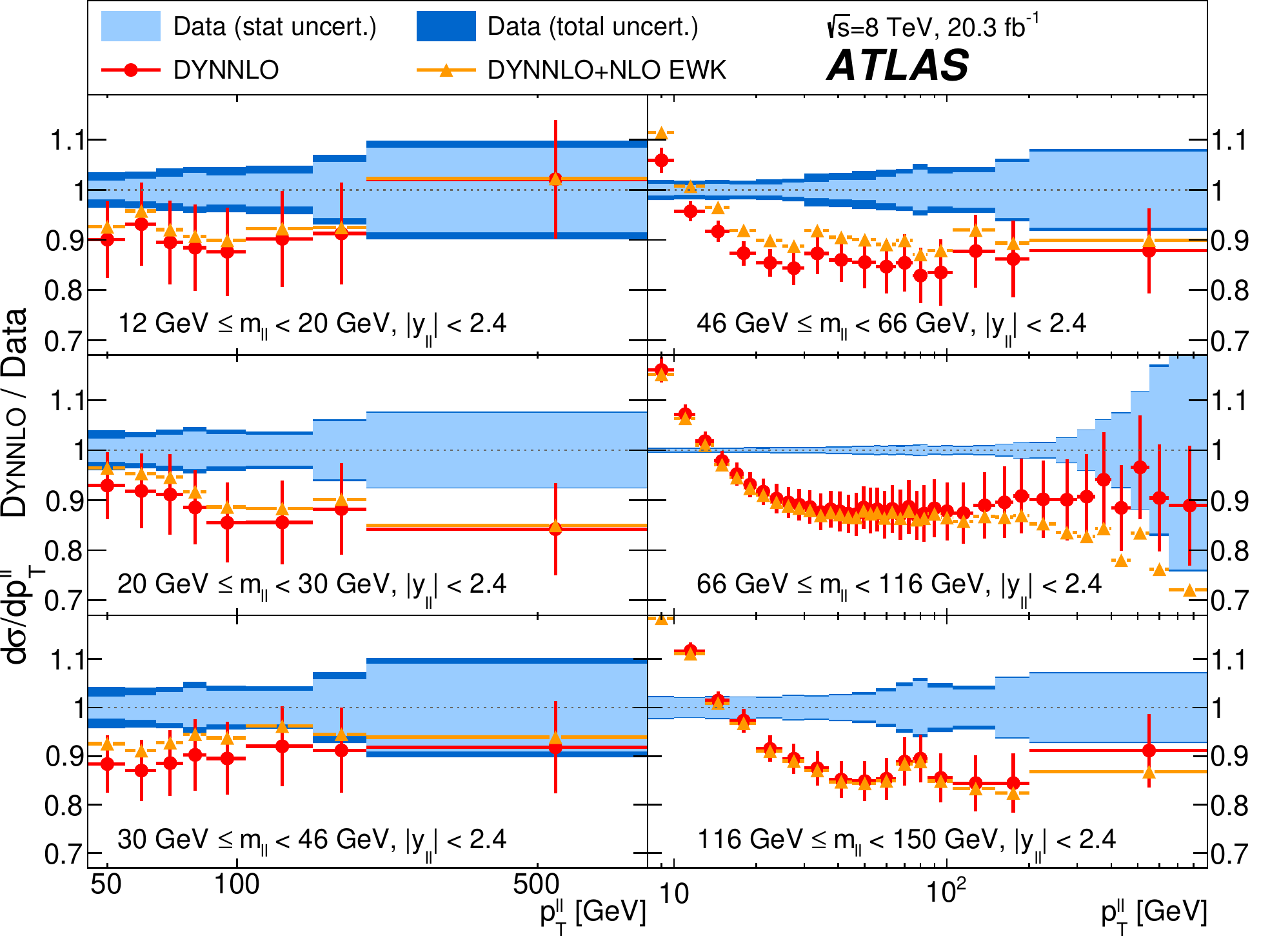}\\
  \caption{The ratio of $\mathrm{d}\sigma / \mathrm{d}\Zpt$ as predicted by the \Dynnlo{} MC generator to the combined Born-level data,
   for six regions of \Mll from $12 \GeV$ to $150 \GeV$.
   Two sets of \Dynnlo{} predictions are shown, one of which includes NLO EW corrections while the other does
  not.
    The error bars on the \Dynnlo\ predictions represent the uncertainty  arising from varying the QCD scales and PDFs.
    Additional uncertainties introduced by the inclusion of the EW
    corrections are at the level of 2--4\% and are always significantly
    smaller than the QCD scale and PDF uncertainties.
   Therefore for clarity these points are shown without uncertainty bars.
   The light-blue band represents the statistical uncertainty on the data and the dark-blue band represents 
  the total uncertainty (statistical and systematic) on the data.
  }
  \label{fig:Theory_D_Zpt_Mass}
\end{figure}    

\FloatBarrier

\section{\label{sec:Conclusion}Conclusion}

Measurements are presented of the \phistar{} and \Zpt{} distributions of Drell--Yan lepton-pair events using 20.3~\ifb{} of
$\sqrt{s}=8 \TeV$ $pp$ collision data collected with the ATLAS detector.
The results presented here expand upon those presented previously by ATLAS at $\sqrt{s}=7 \TeV$,
by providing measurements in regions of \Mll{} above and 
below, as well as on, the \Zboson{}-boson mass peak, and also in finer divisions of \mody\ than were
presented at $\sqrt{s}=7 \TeV$.
Measurements for both the electron- and muon-pair channels are provided corresponding to a variety of
particle-level definitions that differ in the size of the correction for final-state photon radiation.
The results from the two channels at the Born level are combined and compared to a variety of theoretical predictions.
In addition, measurements of the integrated cross section in six bins of \Mll\ are given.

The predictions from \Resbos{}, which include the effects of soft-gluon resummation, are compared to the
normalised \phistar{} distributions  $(1/\sigma)\, \mathrm{d}\sigma / \mathrm{d}\phistar$.
These predictions\ are consistent with the data within the assigned theoretical uncertainties within
certain kinematic regions, especially at low values of \phistar:
$\phistar < 0.4$ for $46 \GeV < \Mll < 66 \GeV$;
$\phistar < 2$  for  $66 \GeV < \Mll < 116 \GeV$; and over the
full range of \phistar\ for $116 \GeV < \Mll < 150 \GeV$.
However, outside these kinematic ranges, i.e., for larger values of \phistar,  the predictions show significant deviations from the data.
The evolution of  $(1/\sigma)\, \mathrm{d}\sigma / \mathrm{d}\phistar$ with \mody\ and \Mll\ (for which the theoretical
uncertainties on the predictions largely cancel) is  generally well described by \Resbos. 

Predictions from MC generators with parton showers are compared to the normalised \Zpt{} distributions in a similar manner.
Between \Zpt{} values of approximately $5 \GeV$ and $100 \GeV$ for $\Mll{} > 46 \GeV$ the MC generators describe the basic shape of the data to within 10\%. However outside this range, and in the very-low regions of \Mll{} the agreement worsens.
The MC generators do though provide a reasonable description of the evolution of the \Zpt{} distributions with \mody{} for the
\Mll\ region around the \Zboson{}-boson mass peak.
Fixed-order predictions from \Dynnlo{} are compared to the absolute \Zpt{} differential cross-section distributions.
The predictions describe the shape of the data within uncertainties for $\Zpt>40 \GeV$ but only describe the absolute values to within 15\%, which is not covered by the evaluated scale and PDF uncertainties.
The data and QCD predictions are not precise enough to be sensitive to the inclusion of EW corrections.

\clearpage
\newpage

\section*{Acknowledgements}

We thank CERN for the very successful operation of the LHC, as well as the
support staff from our institutions without whom ATLAS could not be
operated efficiently.

We acknowledge the support of ANPCyT, Argentina; YerPhI, Armenia; ARC, Australia; BMWFW and FWF, Austria; ANAS, Azerbaijan; SSTC, Belarus; CNPq and FAPESP, Brazil; NSERC, NRC and CFI, Canada; CERN; CONICYT, Chile; CAS, MOST and NSFC, China; COLCIENCIAS, Colombia; MSMT CR, MPO CR and VSC CR, Czech Republic; DNRF and DNSRC, Denmark; IN2P3-CNRS, CEA-DSM/IRFU, France; GNSF, Georgia; BMBF, HGF, and MPG, Germany; GSRT, Greece; RGC, Hong Kong SAR, China; ISF, I-CORE and Benoziyo Center, Israel; INFN, Italy; MEXT and JSPS, Japan; CNRST, Morocco; FOM and NWO, Netherlands; RCN, Norway; MNiSW and NCN, Poland; FCT, Portugal; MNE/IFA, Romania; MES of Russia and NRC KI, Russian Federation; JINR; MESTD, Serbia; MSSR, Slovakia; ARRS and MIZ\v{S}, Slovenia; DST/NRF, South Africa; MINECO, Spain; SRC and Wallenberg Foundation, Sweden; SERI, SNSF and Cantons of Bern and Geneva, Switzerland; MOST, Taiwan; TAEK, Turkey; STFC, United Kingdom; DOE and NSF, United States of America. In addition, individual groups and members have received support from BCKDF, the Canada Council, CANARIE, CRC, Compute Canada, FQRNT, and the Ontario Innovation Trust, Canada; EPLANET, ERC, FP7, Horizon 2020 and Marie Sk{\l}odowska-Curie Actions, European Union; Investissements d'Avenir Labex and Idex, ANR, R{\'e}gion Auvergne and Fondation Partager le Savoir, France; DFG and AvH Foundation, Germany; Herakleitos, Thales and Aristeia programmes co-financed by EU-ESF and the Greek NSRF; BSF, GIF and Minerva, Israel; BRF, Norway; Generalitat de Catalunya, Generalitat Valenciana, Spain; the Royal Society and Leverhulme Trust, United Kingdom.

The crucial computing support from all WLCG partners is acknowledged
gratefully, in particular from CERN and the ATLAS Tier-1 facilities at
TRIUMF (Canada), NDGF (Denmark, Norway, Sweden), CC-IN2P3 (France),
KIT/GridKA (Germany), INFN-CNAF (Italy), NL-T1 (Netherlands), PIC (Spain),
ASGC (Taiwan), RAL (UK) and BNL (USA) and in the Tier-2 facilities
worldwide.

\printbibliography

\clearpage
\newpage
\appendix

\section*{Appendix}
\label{App:crosssection}

In the tables in this appendix the values of $(1/\sigma)\, d\sigma / d\phistar$ and  $(1/\sigma)\,
d\sigma / d\Zpt$  are given for each region of \Mll\ and \mody\ considered.
The electron-pair results are given at the
dressed and Born levels, and the muon-pair results at the bare, dressed and Born levels.
The Born-level combined results are also given. The associated statistical and systematic uncertainties
(both uncorrelated and correlated between bins in \phistar\ or \Zpt) are provided in percentage form.

\begin{table*}
\centering
\caption{The values of $(1/\sigma)\,\mathrm{d}\sigma/\mathrm{d}\phi^*_{\eta}$ in each bin of \PhiStar{} for the electron and muon channels separately (for various particle-level definitions) and for the Born-level combination in the kinematic region $46\ \GeV \leq m_{\ell\ell} < 66\ \GeV,\ 0 \leq |y_{\ell\ell}| < 0.8$. The associated statistical and systematic (both uncorrelated and correlated between bins of \PhiStar{}) are provided in percentage form.} 
\resizebox{\textwidth}{!}{
\begin{tabular}{ ccccccc } \toprule 
Bin & \multicolumn{6}{c}{$(1/\sigma)\,d\sigma/d\phi^*_{\eta}$ $\pm$ Statistical [\%] $\pm$ Uncorrelated systematic [\%] $\pm$ Correlated systematic [\%]} \\ 
 \cmidrule(r){2-7} 
 & \multicolumn{2}{c}{Electron channel} &  \multicolumn{3}{c}{Muon channel} & Combination \\ 
\cmidrule(r){2-3} \cmidrule(r){4-6} \cmidrule(r){7-7} 
 & dressed & Born & bare & dressed & Born & Born \\ 
 0 -- 0.004 & 6.778 $\pm$ 2.4 $\pm$ 0.9 $\pm$ 5.9 & 7.256 $\pm$ 2.4 $\pm$ 1.0 $\pm$ 6.0 & 6.688 $\pm$ 2.0 $\pm$ 0.8 $\pm$ 5.0 & 6.687 $\pm$ 2.0 $\pm$ 0.8 $\pm$ 5.0 & 7.157 $\pm$ 2.0 $\pm$ 0.8 $\pm$ 5.2 & 7.248 $\pm$ 1.5 $\pm$ 0.6 $\pm$ 5.3\\
0.004 -- 0.008 & 6.662 $\pm$ 2.3 $\pm$ 0.9 $\pm$ 2.3 & 7.051 $\pm$ 2.3 $\pm$ 0.9 $\pm$ 2.6 & 7.079 $\pm$ 2.0 $\pm$ 0.8 $\pm$ 1.9 & 7.046 $\pm$ 2.0 $\pm$ 0.8 $\pm$ 1.9 & 7.469 $\pm$ 2.0 $\pm$ 0.8 $\pm$ 2.4 & 7.258 $\pm$ 1.5 $\pm$ 0.6 $\pm$ 2.3\\
0.008 -- 0.012 & 6.781 $\pm$ 2.3 $\pm$ 0.9 $\pm$ 1.5 & 7.179 $\pm$ 2.3 $\pm$ 0.9 $\pm$ 1.9 & 6.747 $\pm$ 2.1 $\pm$ 0.8 $\pm$ 1.4 & 6.704 $\pm$ 2.1 $\pm$ 0.8 $\pm$ 1.4 & 7.169 $\pm$ 2.1 $\pm$ 0.8 $\pm$ 2.0 & 7.141 $\pm$ 1.5 $\pm$ 0.6 $\pm$ 1.7\\
0.012 -- 0.016 & 6.561 $\pm$ 2.3 $\pm$ 0.9 $\pm$ 1.1 & 6.926 $\pm$ 2.3 $\pm$ 0.9 $\pm$ 1.6 & 6.680 $\pm$ 2.0 $\pm$ 0.8 $\pm$ 1.1 & 6.660 $\pm$ 2.1 $\pm$ 0.8 $\pm$ 1.1 & 7.085 $\pm$ 2.0 $\pm$ 0.8 $\pm$ 1.8 & 6.981 $\pm$ 1.5 $\pm$ 0.6 $\pm$ 1.5\\
0.016 -- 0.020 & 6.540 $\pm$ 2.3 $\pm$ 0.9 $\pm$ 1.0 & 6.927 $\pm$ 2.3 $\pm$ 1.0 $\pm$ 1.5 & 6.542 $\pm$ 2.0 $\pm$ 0.8 $\pm$ 0.9 & 6.484 $\pm$ 2.0 $\pm$ 0.8 $\pm$ 0.9 & 6.884 $\pm$ 2.0 $\pm$ 0.8 $\pm$ 1.6 & 6.861 $\pm$ 1.5 $\pm$ 0.6 $\pm$ 1.4\\
0.020 -- 0.024 & 6.327 $\pm$ 2.3 $\pm$ 0.9 $\pm$ 1.1 & 6.714 $\pm$ 2.3 $\pm$ 0.9 $\pm$ 1.6 & 6.437 $\pm$ 2.1 $\pm$ 0.9 $\pm$ 0.9 & 6.415 $\pm$ 2.1 $\pm$ 0.9 $\pm$ 0.9 & 6.755 $\pm$ 2.1 $\pm$ 0.9 $\pm$ 1.7 & 6.693 $\pm$ 1.6 $\pm$ 0.6 $\pm$ 1.4\\
0.024 -- 0.029 & 6.102 $\pm$ 2.1 $\pm$ 0.8 $\pm$ 0.9 & 6.408 $\pm$ 2.1 $\pm$ 0.9 $\pm$ 1.4 & 6.072 $\pm$ 1.9 $\pm$ 0.8 $\pm$ 0.9 & 6.075 $\pm$ 1.9 $\pm$ 0.8 $\pm$ 0.9 & 6.472 $\pm$ 1.9 $\pm$ 0.8 $\pm$ 1.5 & 6.398 $\pm$ 1.4 $\pm$ 0.6 $\pm$ 1.2\\
0.029 -- 0.034 & 5.682 $\pm$ 2.2 $\pm$ 0.8 $\pm$ 0.8 & 5.957 $\pm$ 2.2 $\pm$ 0.9 $\pm$ 1.4 & 5.877 $\pm$ 2.0 $\pm$ 0.8 $\pm$ 0.8 & 5.904 $\pm$ 2.0 $\pm$ 0.8 $\pm$ 0.8 & 6.214 $\pm$ 2.0 $\pm$ 0.8 $\pm$ 1.4 & 6.062 $\pm$ 1.5 $\pm$ 0.6 $\pm$ 1.2\\
0.034 -- 0.039 & 5.868 $\pm$ 2.2 $\pm$ 0.8 $\pm$ 0.9 & 6.185 $\pm$ 2.2 $\pm$ 0.9 $\pm$ 1.4 & 5.468 $\pm$ 2.1 $\pm$ 0.8 $\pm$ 0.8 & 5.482 $\pm$ 2.1 $\pm$ 0.8 $\pm$ 0.8 & 5.798 $\pm$ 2.1 $\pm$ 0.8 $\pm$ 1.4 & 5.919 $\pm$ 1.5 $\pm$ 0.6 $\pm$ 1.2\\
0.039 -- 0.045 & 5.263 $\pm$ 2.1 $\pm$ 0.8 $\pm$ 0.5 & 5.485 $\pm$ 2.1 $\pm$ 0.9 $\pm$ 1.3 & 5.428 $\pm$ 1.9 $\pm$ 0.8 $\pm$ 0.8 & 5.449 $\pm$ 1.9 $\pm$ 0.8 $\pm$ 0.8 & 5.669 $\pm$ 1.9 $\pm$ 0.8 $\pm$ 1.4 & 5.544 $\pm$ 1.4 $\pm$ 0.6 $\pm$ 1.1\\
0.045 -- 0.051 & 5.032 $\pm$ 2.1 $\pm$ 0.8 $\pm$ 0.6 & 5.274 $\pm$ 2.1 $\pm$ 0.9 $\pm$ 1.3 & 5.182 $\pm$ 1.9 $\pm$ 0.8 $\pm$ 0.9 & 5.210 $\pm$ 1.9 $\pm$ 0.8 $\pm$ 0.9 & 5.505 $\pm$ 1.9 $\pm$ 0.8 $\pm$ 1.5 & 5.351 $\pm$ 1.4 $\pm$ 0.6 $\pm$ 1.1\\
0.051 -- 0.057 & 4.796 $\pm$ 2.2 $\pm$ 0.8 $\pm$ 0.6 & 4.964 $\pm$ 2.2 $\pm$ 0.8 $\pm$ 1.3 & 4.862 $\pm$ 2.0 $\pm$ 0.8 $\pm$ 0.8 & 4.897 $\pm$ 2.0 $\pm$ 0.8 $\pm$ 0.8 & 5.111 $\pm$ 2.0 $\pm$ 0.8 $\pm$ 1.4 & 5.003 $\pm$ 1.5 $\pm$ 0.6 $\pm$ 1.1\\
0.057 -- 0.064 & 4.443 $\pm$ 2.1 $\pm$ 0.8 $\pm$ 0.7 & 4.603 $\pm$ 2.1 $\pm$ 0.8 $\pm$ 1.3 & 4.430 $\pm$ 1.9 $\pm$ 0.8 $\pm$ 0.5 & 4.443 $\pm$ 1.9 $\pm$ 0.8 $\pm$ 0.5 & 4.663 $\pm$ 1.9 $\pm$ 0.8 $\pm$ 1.5 & 4.597 $\pm$ 1.4 $\pm$ 0.6 $\pm$ 1.2\\
0.064 -- 0.072 & 4.113 $\pm$ 2.0 $\pm$ 0.8 $\pm$ 0.6 & 4.271 $\pm$ 2.0 $\pm$ 0.8 $\pm$ 1.3 & 4.052 $\pm$ 1.9 $\pm$ 0.8 $\pm$ 0.4 & 4.082 $\pm$ 1.9 $\pm$ 0.8 $\pm$ 0.4 & 4.256 $\pm$ 1.9 $\pm$ 0.8 $\pm$ 1.4 & 4.245 $\pm$ 1.4 $\pm$ 0.6 $\pm$ 1.2\\
0.072 -- 0.081 & 3.766 $\pm$ 2.0 $\pm$ 0.7 $\pm$ 0.7 & 3.876 $\pm$ 2.0 $\pm$ 0.8 $\pm$ 1.3 & 3.759 $\pm$ 1.8 $\pm$ 0.7 $\pm$ 0.5 & 3.787 $\pm$ 1.8 $\pm$ 0.7 $\pm$ 0.5 & 3.912 $\pm$ 1.8 $\pm$ 0.7 $\pm$ 1.5 & 3.866 $\pm$ 1.3 $\pm$ 0.5 $\pm$ 1.2\\
0.081 -- 0.091 & 3.400 $\pm$ 2.0 $\pm$ 1.2 $\pm$ 1.0 & 3.495 $\pm$ 2.0 $\pm$ 1.2 $\pm$ 1.5 & 3.517 $\pm$ 1.8 $\pm$ 0.7 $\pm$ 0.5 & 3.521 $\pm$ 1.8 $\pm$ 0.7 $\pm$ 0.5 & 3.665 $\pm$ 1.8 $\pm$ 0.7 $\pm$ 1.5 & 3.580 $\pm$ 1.3 $\pm$ 0.7 $\pm$ 1.2\\
0.091 -- 0.102 & 3.231 $\pm$ 2.0 $\pm$ 0.7 $\pm$ 0.8 & 3.318 $\pm$ 2.0 $\pm$ 0.8 $\pm$ 1.4 & 3.107 $\pm$ 1.8 $\pm$ 0.7 $\pm$ 0.5 & 3.130 $\pm$ 1.8 $\pm$ 0.7 $\pm$ 0.5 & 3.224 $\pm$ 1.8 $\pm$ 0.7 $\pm$ 1.5 & 3.240 $\pm$ 1.3 $\pm$ 0.5 $\pm$ 1.2\\
0.102 -- 0.114 & 2.833 $\pm$ 2.0 $\pm$ 0.7 $\pm$ 0.8 & 2.848 $\pm$ 2.0 $\pm$ 0.8 $\pm$ 1.4 & 2.814 $\pm$ 1.8 $\pm$ 0.7 $\pm$ 0.5 & 2.822 $\pm$ 1.8 $\pm$ 0.7 $\pm$ 0.5 & 2.882 $\pm$ 1.8 $\pm$ 0.7 $\pm$ 1.5 & 2.844 $\pm$ 1.3 $\pm$ 0.5 $\pm$ 1.2\\
0.114 -- 0.128 & 2.555 $\pm$ 2.0 $\pm$ 0.7 $\pm$ 0.7 & 2.596 $\pm$ 2.0 $\pm$ 0.8 $\pm$ 1.3 & 2.477 $\pm$ 1.8 $\pm$ 0.7 $\pm$ 0.7 & 2.487 $\pm$ 1.8 $\pm$ 0.7 $\pm$ 0.7 & 2.518 $\pm$ 1.8 $\pm$ 0.7 $\pm$ 0.9 & 2.535 $\pm$ 1.3 $\pm$ 0.5 $\pm$ 0.9\\
0.128 -- 0.145 & 2.206 $\pm$ 1.9 $\pm$ 0.7 $\pm$ 0.7 & 2.204 $\pm$ 1.9 $\pm$ 0.7 $\pm$ 1.4 & 2.175 $\pm$ 1.7 $\pm$ 0.7 $\pm$ 0.5 & 2.170 $\pm$ 1.7 $\pm$ 0.7 $\pm$ 0.5 & 2.160 $\pm$ 1.7 $\pm$ 0.7 $\pm$ 0.8 & 2.173 $\pm$ 1.3 $\pm$ 0.5 $\pm$ 0.9\\
0.145 -- 0.165 & 1.830 $\pm$ 1.9 $\pm$ 0.7 $\pm$ 0.7 & 1.799 $\pm$ 1.9 $\pm$ 0.8 $\pm$ 1.4 & 1.846 $\pm$ 1.7 $\pm$ 0.7 $\pm$ 0.6 & 1.850 $\pm$ 1.7 $\pm$ 0.7 $\pm$ 0.6 & 1.836 $\pm$ 1.7 $\pm$ 0.7 $\pm$ 0.8 & 1.811 $\pm$ 1.3 $\pm$ 0.5 $\pm$ 0.9\\
0.165 -- 0.189 & 1.545 $\pm$ 1.9 $\pm$ 0.7 $\pm$ 0.8 & 1.519 $\pm$ 1.9 $\pm$ 0.8 $\pm$ 1.4 & 1.535 $\pm$ 1.7 $\pm$ 0.7 $\pm$ 0.5 & 1.538 $\pm$ 1.7 $\pm$ 0.7 $\pm$ 0.5 & 1.497 $\pm$ 1.7 $\pm$ 0.7 $\pm$ 0.8 & 1.497 $\pm$ 1.3 $\pm$ 0.5 $\pm$ 1.0\\
0.189 -- 0.219 & 1.235 $\pm$ 1.9 $\pm$ 1.1 $\pm$ 1.0 & 1.185 $\pm$ 1.9 $\pm$ 1.2 $\pm$ 1.5 & 1.292 $\pm$ 1.7 $\pm$ 0.7 $\pm$ 0.6 & 1.292 $\pm$ 1.7 $\pm$ 0.7 $\pm$ 0.6 & 1.240 $\pm$ 1.7 $\pm$ 0.7 $\pm$ 0.8 & 1.214 $\pm$ 1.3 $\pm$ 0.6 $\pm$ 0.9\\
0.219 -- 0.258 & 1.008 $\pm$ 1.8 $\pm$ 0.7 $\pm$ 0.9 & 0.949 $\pm$ 1.8 $\pm$ 0.7 $\pm$ 1.5 & 1.003 $\pm$ 1.7 $\pm$ 0.7 $\pm$ 0.6 & 1.001 $\pm$ 1.7 $\pm$ 0.7 $\pm$ 0.6 & 0.944 $\pm$ 1.7 $\pm$ 0.7 $\pm$ 0.9 & 0.943 $\pm$ 1.2 $\pm$ 0.5 $\pm$ 1.0\\
0.258 -- 0.312 & 0.767 $\pm$ 1.8 $\pm$ 0.7 $\pm$ 0.9 & 0.707 $\pm$ 1.8 $\pm$ 0.8 $\pm$ 2.2 & 0.772 $\pm$ 1.6 $\pm$ 0.7 $\pm$ 0.7 & 0.771 $\pm$ 1.6 $\pm$ 0.7 $\pm$ 0.7 & 0.702 $\pm$ 1.6 $\pm$ 0.7 $\pm$ 1.9 & 0.697 $\pm$ 1.2 $\pm$ 0.5 $\pm$ 1.7\\
0.312 -- 0.391 & 0.545 $\pm$ 1.8 $\pm$ 0.8 $\pm$ 0.9 & 0.488 $\pm$ 1.8 $\pm$ 0.9 $\pm$ 2.2 & 0.530 $\pm$ 1.6 $\pm$ 0.7 $\pm$ 0.6 & 0.531 $\pm$ 1.6 $\pm$ 0.7 $\pm$ 0.6 & 0.472 $\pm$ 1.6 $\pm$ 0.7 $\pm$ 1.8 & 0.477 $\pm$ 1.2 $\pm$ 0.5 $\pm$ 1.7\\
0.391 -- 0.524 & 0.337 $\pm$ 1.8 $\pm$ 0.7 $\pm$ 1.0 & 0.299 $\pm$ 1.8 $\pm$ 0.7 $\pm$ 2.2 & 0.336 $\pm$ 1.6 $\pm$ 0.6 $\pm$ 1.1 & 0.335 $\pm$ 1.6 $\pm$ 0.6 $\pm$ 1.1 & 0.293 $\pm$ 1.6 $\pm$ 0.6 $\pm$ 2.0 & 0.295 $\pm$ 1.2 $\pm$ 0.5 $\pm$ 1.7\\
0.524 -- 0.695 & 0.201 $\pm$ 2.0 $\pm$ 0.8 $\pm$ 1.8 & 0.183 $\pm$ 2.0 $\pm$ 0.8 $\pm$ 2.7 & 0.194 $\pm$ 1.8 $\pm$ 0.8 $\pm$ 1.9 & 0.193 $\pm$ 1.8 $\pm$ 0.8 $\pm$ 1.9 & 0.170 $\pm$ 1.8 $\pm$ 0.8 $\pm$ 2.5 & 0.176 $\pm$ 1.4 $\pm$ 0.6 $\pm$ 1.9\\
0.695 -- 0.918 & 0.105 $\pm$ 2.5 $\pm$ 1.0 $\pm$ 1.5 & 0.0978 $\pm$ 2.5 $\pm$ 1.0 $\pm$ 2.5 & 0.113 $\pm$ 2.2 $\pm$ 1.0 $\pm$ 2.1 & 0.112 $\pm$ 2.2 $\pm$ 1.0 $\pm$ 2.1 & 0.102 $\pm$ 2.2 $\pm$ 1.0 $\pm$ 2.7 & 0.101 $\pm$ 1.6 $\pm$ 0.7 $\pm$ 2.0\\
0.918 -- 1.153 & 0.0647 $\pm$ 3.2 $\pm$ 1.3 $\pm$ 1.9 & 0.0623 $\pm$ 3.2 $\pm$ 1.4 $\pm$ 2.7 & 0.0613 $\pm$ 3.0 $\pm$ 1.3 $\pm$ 2.9 & 0.0609 $\pm$ 3.0 $\pm$ 1.3 $\pm$ 2.9 & 0.0569 $\pm$ 3.0 $\pm$ 1.3 $\pm$ 3.4 & 0.0598 $\pm$ 2.2 $\pm$ 0.9 $\pm$ 2.3\\
1.153 -- 1.496 & 0.0342 $\pm$ 3.9 $\pm$ 2.7 $\pm$ 2.9 & 0.0330 $\pm$ 3.9 $\pm$ 2.8 $\pm$ 3.6 & 0.0333 $\pm$ 3.2 $\pm$ 1.7 $\pm$ 4.2 & 0.0328 $\pm$ 3.2 $\pm$ 1.7 $\pm$ 4.2 & 0.0315 $\pm$ 3.2 $\pm$ 1.7 $\pm$ 4.8 & 0.0330 $\pm$ 2.5 $\pm$ 1.5 $\pm$ 3.2\\
1.496 -- 1.947 & 0.0184 $\pm$ 4.7 $\pm$ 2.2 $\pm$ 3.0 & 0.0181 $\pm$ 4.7 $\pm$ 2.2 $\pm$ 3.6 & 0.0169 $\pm$ 4.1 $\pm$ 2.1 $\pm$ 3.3 & 0.0167 $\pm$ 4.1 $\pm$ 2.1 $\pm$ 3.3 & 0.0160 $\pm$ 4.1 $\pm$ 2.1 $\pm$ 4.0 & 0.0170 $\pm$ 3.1 $\pm$ 1.5 $\pm$ 3.2\\
1.947 -- 2.522 & 0.00907 $\pm$ 6.1 $\pm$ 3.0 $\pm$ 3.7 & 0.00885 $\pm$ 6.1 $\pm$ 3.1 $\pm$ 4.3 & 0.00989 $\pm$ 4.7 $\pm$ 2.2 $\pm$ 2.9 & 0.00975 $\pm$ 4.7 $\pm$ 2.2 $\pm$ 2.9 & 0.00950 $\pm$ 4.7 $\pm$ 2.2 $\pm$ 3.6 & 0.00939 $\pm$ 3.7 $\pm$ 1.8 $\pm$ 3.2\\
2.522 -- 3.277 & 0.00454 $\pm$ 7.5 $\pm$ 4.6 $\pm$ 3.1 & 0.00445 $\pm$ 7.5 $\pm$ 4.7 $\pm$ 3.8 & 0.00447 $\pm$ 6.1 $\pm$ 2.7 $\pm$ 4.2 & 0.00441 $\pm$ 6.1 $\pm$ 2.7 $\pm$ 4.2 & 0.00430 $\pm$ 6.1 $\pm$ 2.7 $\pm$ 4.8 & 0.00446 $\pm$ 4.7 $\pm$ 2.4 $\pm$ 3.4\\
3.277 -- 5.000 & 0.00252 $\pm$ 6.3 $\pm$ 2.8 $\pm$ 4.0 & 0.00252 $\pm$ 6.3 $\pm$ 2.8 $\pm$ 4.6 & 0.00220 $\pm$ 5.7 $\pm$ 2.6 $\pm$ 3.8 & 0.00219 $\pm$ 5.7 $\pm$ 2.6 $\pm$ 3.8 & 0.00214 $\pm$ 5.7 $\pm$ 2.6 $\pm$ 4.4 & 0.00232 $\pm$ 4.2 $\pm$ 1.9 $\pm$ 3.3\\
5.000 -- 10.000 & 0.000525 $\pm$ 8.6 $\pm$ 3.9 $\pm$ 3.4 & 0.000510 $\pm$ 8.6 $\pm$ 3.9 $\pm$ 4.0 & 0.000585 $\pm$ 6.4 $\pm$ 2.9 $\pm$ 4.0 & 0.000577 $\pm$ 6.4 $\pm$ 2.9 $\pm$ 4.0 & 0.000545 $\pm$ 6.4 $\pm$ 2.9 $\pm$ 4.6 & 0.000542 $\pm$ 5.1 $\pm$ 2.3 $\pm$ 3.3\\
\bottomrule 
\end{tabular} 
} 
\label{tab:CombPhiStarlowmass_y0008}
\end{table*}

\begin{table*}
\centering
\caption{The values of $(1/\sigma)\,\mathrm{d}\sigma/\mathrm{d}\phi^*_{\eta}$ in each bin of \PhiStar{} for the electron and muon channels separately (for various particle-level definitions) and for the Born-level combination in the kinematic region $46\ \GeV \leq m_{\ell\ell} < 66\ \GeV,\ 0.8 \leq |y_{\ell\ell}| < 1.6$. The associated statistical and systematic (both uncorrelated and correlated between bins of \PhiStar{}) are provided in percentage form.} 
\resizebox{\textwidth}{!}{
\begin{tabular}{ ccccccc } \toprule 
Bin & \multicolumn{6}{c}{$(1/\sigma)\,d\sigma/d\phi^*_{\eta}$ $\pm$ Statistical [\%] $\pm$ Uncorrelated systematic [\%] $\pm$ Correlated systematic [\%]} \\ 
 \cmidrule(r){2-7} 
 & \multicolumn{2}{c}{Electron channel} &  \multicolumn{3}{c}{Muon channel} & Combination \\ 
\cmidrule(r){2-3} \cmidrule(r){4-6} \cmidrule(r){7-7} 
 & dressed & Born & bare & dressed & Born & Born \\ 
 0.000 -- 0.004 & 7.182 $\pm$ 2.7 $\pm$ 1.1 $\pm$ 5.3 & 7.754 $\pm$ 2.7 $\pm$ 1.1 $\pm$ 5.3 & 6.713 $\pm$ 1.9 $\pm$ 0.8 $\pm$ 4.7 & 6.704 $\pm$ 1.9 $\pm$ 0.8 $\pm$ 4.7 & 7.109 $\pm$ 1.9 $\pm$ 0.8 $\pm$ 5.0 & 7.311 $\pm$ 1.5 $\pm$ 0.6 $\pm$ 4.9\\
0.004 -- 0.008 & 7.048 $\pm$ 2.6 $\pm$ 1.1 $\pm$ 2.5 & 7.465 $\pm$ 2.6 $\pm$ 1.1 $\pm$ 2.7 & 6.917 $\pm$ 1.9 $\pm$ 0.8 $\pm$ 1.9 & 6.884 $\pm$ 1.9 $\pm$ 0.8 $\pm$ 1.9 & 7.255 $\pm$ 1.9 $\pm$ 0.8 $\pm$ 2.5 & 7.349 $\pm$ 1.5 $\pm$ 0.6 $\pm$ 2.3\\
0.008 -- 0.012 & 6.842 $\pm$ 2.6 $\pm$ 1.1 $\pm$ 1.7 & 7.254 $\pm$ 2.6 $\pm$ 1.1 $\pm$ 2.0 & 6.898 $\pm$ 1.9 $\pm$ 0.8 $\pm$ 1.3 & 6.851 $\pm$ 1.9 $\pm$ 0.8 $\pm$ 1.3 & 7.247 $\pm$ 1.9 $\pm$ 0.8 $\pm$ 2.0 & 7.270 $\pm$ 1.5 $\pm$ 0.6 $\pm$ 1.7\\
0.012 -- 0.016 & 6.408 $\pm$ 2.7 $\pm$ 1.1 $\pm$ 1.1 & 6.812 $\pm$ 2.7 $\pm$ 1.2 $\pm$ 1.6 & 6.758 $\pm$ 1.9 $\pm$ 0.8 $\pm$ 1.0 & 6.757 $\pm$ 1.9 $\pm$ 0.8 $\pm$ 1.0 & 7.249 $\pm$ 1.9 $\pm$ 0.8 $\pm$ 1.9 & 7.123 $\pm$ 1.6 $\pm$ 0.6 $\pm$ 1.5\\
0.016 -- 0.020 & 6.302 $\pm$ 2.7 $\pm$ 1.1 $\pm$ 1.4 & 6.619 $\pm$ 2.7 $\pm$ 1.1 $\pm$ 1.8 & 6.625 $\pm$ 2.0 $\pm$ 0.8 $\pm$ 1.1 & 6.630 $\pm$ 2.0 $\pm$ 0.8 $\pm$ 1.1 & 7.000 $\pm$ 2.0 $\pm$ 0.8 $\pm$ 1.9 & 6.872 $\pm$ 1.6 $\pm$ 0.7 $\pm$ 1.4\\
0.020 -- 0.024 & 6.328 $\pm$ 2.7 $\pm$ 1.1 $\pm$ 1.1 & 6.654 $\pm$ 2.7 $\pm$ 1.1 $\pm$ 1.6 & 6.326 $\pm$ 2.0 $\pm$ 0.8 $\pm$ 0.9 & 6.334 $\pm$ 2.0 $\pm$ 0.8 $\pm$ 0.9 & 6.722 $\pm$ 2.0 $\pm$ 0.8 $\pm$ 1.8 & 6.725 $\pm$ 1.6 $\pm$ 0.7 $\pm$ 1.4\\
0.024 -- 0.029 & 6.279 $\pm$ 2.4 $\pm$ 0.9 $\pm$ 0.9 & 6.570 $\pm$ 2.4 $\pm$ 1.0 $\pm$ 1.4 & 6.161 $\pm$ 1.8 $\pm$ 0.7 $\pm$ 0.8 & 6.141 $\pm$ 1.8 $\pm$ 0.7 $\pm$ 0.8 & 6.497 $\pm$ 1.8 $\pm$ 0.7 $\pm$ 1.7 & 6.535 $\pm$ 1.4 $\pm$ 0.6 $\pm$ 1.3\\
0.029 -- 0.034 & 6.047 $\pm$ 2.4 $\pm$ 1.0 $\pm$ 0.8 & 6.369 $\pm$ 2.4 $\pm$ 1.0 $\pm$ 1.4 & 5.931 $\pm$ 1.9 $\pm$ 0.8 $\pm$ 0.7 & 5.944 $\pm$ 1.9 $\pm$ 0.8 $\pm$ 0.7 & 6.311 $\pm$ 1.9 $\pm$ 0.8 $\pm$ 1.6 & 6.344 $\pm$ 1.5 $\pm$ 0.6 $\pm$ 1.2\\
0.034 -- 0.039 & 5.803 $\pm$ 2.6 $\pm$ 1.1 $\pm$ 1.1 & 6.074 $\pm$ 2.6 $\pm$ 1.1 $\pm$ 1.5 & 5.684 $\pm$ 1.9 $\pm$ 0.8 $\pm$ 0.7 & 5.664 $\pm$ 1.9 $\pm$ 0.8 $\pm$ 0.7 & 6.004 $\pm$ 1.9 $\pm$ 0.8 $\pm$ 1.6 & 6.049 $\pm$ 1.5 $\pm$ 0.6 $\pm$ 1.2\\
0.039 -- 0.045 & 5.295 $\pm$ 2.4 $\pm$ 0.9 $\pm$ 0.6 & 5.522 $\pm$ 2.4 $\pm$ 1.0 $\pm$ 1.3 & 5.417 $\pm$ 1.8 $\pm$ 0.7 $\pm$ 0.7 & 5.413 $\pm$ 1.8 $\pm$ 0.7 $\pm$ 0.7 & 5.695 $\pm$ 1.8 $\pm$ 0.7 $\pm$ 1.6 & 5.648 $\pm$ 1.4 $\pm$ 0.6 $\pm$ 1.2\\
0.045 -- 0.051 & 5.149 $\pm$ 2.4 $\pm$ 1.0 $\pm$ 0.9 & 5.351 $\pm$ 2.4 $\pm$ 1.0 $\pm$ 1.5 & 5.158 $\pm$ 1.8 $\pm$ 0.7 $\pm$ 1.0 & 5.189 $\pm$ 1.8 $\pm$ 0.7 $\pm$ 1.0 & 5.398 $\pm$ 1.8 $\pm$ 0.7 $\pm$ 1.8 & 5.373 $\pm$ 1.5 $\pm$ 0.6 $\pm$ 1.2\\
0.051 -- 0.057 & 4.906 $\pm$ 2.5 $\pm$ 1.0 $\pm$ 0.7 & 5.112 $\pm$ 2.5 $\pm$ 1.0 $\pm$ 1.3 & 5.115 $\pm$ 1.8 $\pm$ 0.7 $\pm$ 0.8 & 5.139 $\pm$ 1.8 $\pm$ 0.7 $\pm$ 0.8 & 5.394 $\pm$ 1.8 $\pm$ 0.7 $\pm$ 1.7 & 5.308 $\pm$ 1.5 $\pm$ 0.6 $\pm$ 1.2\\
0.057 -- 0.064 & 4.396 $\pm$ 2.4 $\pm$ 0.9 $\pm$ 0.7 & 4.555 $\pm$ 2.4 $\pm$ 1.0 $\pm$ 1.3 & 4.535 $\pm$ 1.8 $\pm$ 0.7 $\pm$ 0.8 & 4.569 $\pm$ 1.8 $\pm$ 0.7 $\pm$ 0.8 & 4.775 $\pm$ 1.8 $\pm$ 0.7 $\pm$ 1.4 & 4.698 $\pm$ 1.4 $\pm$ 0.6 $\pm$ 1.1\\
0.064 -- 0.072 & 4.289 $\pm$ 2.3 $\pm$ 0.9 $\pm$ 1.0 & 4.427 $\pm$ 2.3 $\pm$ 1.0 $\pm$ 1.5 & 4.180 $\pm$ 1.7 $\pm$ 0.7 $\pm$ 0.7 & 4.198 $\pm$ 1.7 $\pm$ 0.7 $\pm$ 0.7 & 4.360 $\pm$ 1.7 $\pm$ 0.7 $\pm$ 1.4 & 4.383 $\pm$ 1.4 $\pm$ 0.6 $\pm$ 1.1\\
0.072 -- 0.081 & 3.884 $\pm$ 2.3 $\pm$ 0.9 $\pm$ 0.8 & 3.993 $\pm$ 2.3 $\pm$ 0.9 $\pm$ 1.4 & 4.079 $\pm$ 1.6 $\pm$ 0.7 $\pm$ 0.7 & 4.110 $\pm$ 1.6 $\pm$ 0.7 $\pm$ 0.7 & 4.273 $\pm$ 1.6 $\pm$ 0.7 $\pm$ 1.3 & 4.182 $\pm$ 1.3 $\pm$ 0.5 $\pm$ 1.1\\
0.081 -- 0.091 & 3.645 $\pm$ 2.2 $\pm$ 0.9 $\pm$ 0.8 & 3.749 $\pm$ 2.2 $\pm$ 0.9 $\pm$ 1.4 & 3.465 $\pm$ 1.7 $\pm$ 0.7 $\pm$ 0.4 & 3.486 $\pm$ 1.7 $\pm$ 0.7 $\pm$ 0.4 & 3.611 $\pm$ 1.7 $\pm$ 0.7 $\pm$ 1.2 & 3.688 $\pm$ 1.3 $\pm$ 0.5 $\pm$ 1.0\\
0.091 -- 0.102 & 3.172 $\pm$ 2.3 $\pm$ 0.9 $\pm$ 0.7 & 3.241 $\pm$ 2.3 $\pm$ 1.0 $\pm$ 1.3 & 3.153 $\pm$ 1.7 $\pm$ 0.7 $\pm$ 0.5 & 3.165 $\pm$ 1.7 $\pm$ 0.7 $\pm$ 0.5 & 3.232 $\pm$ 1.7 $\pm$ 0.7 $\pm$ 1.2 & 3.256 $\pm$ 1.3 $\pm$ 0.6 $\pm$ 1.0\\
0.102 -- 0.114 & 2.869 $\pm$ 2.3 $\pm$ 0.9 $\pm$ 0.7 & 2.927 $\pm$ 2.3 $\pm$ 1.0 $\pm$ 1.4 & 2.795 $\pm$ 1.7 $\pm$ 0.7 $\pm$ 0.5 & 2.802 $\pm$ 1.7 $\pm$ 0.7 $\pm$ 0.5 & 2.850 $\pm$ 1.7 $\pm$ 0.7 $\pm$ 1.2 & 2.893 $\pm$ 1.4 $\pm$ 0.6 $\pm$ 1.0\\
0.114 -- 0.128 & 2.520 $\pm$ 2.3 $\pm$ 1.0 $\pm$ 0.8 & 2.540 $\pm$ 2.3 $\pm$ 1.0 $\pm$ 1.4 & 2.538 $\pm$ 1.6 $\pm$ 0.7 $\pm$ 0.5 & 2.550 $\pm$ 1.6 $\pm$ 0.7 $\pm$ 0.5 & 2.585 $\pm$ 1.6 $\pm$ 0.7 $\pm$ 0.6 & 2.585 $\pm$ 1.3 $\pm$ 0.6 $\pm$ 0.7\\
0.128 -- 0.145 & 2.092 $\pm$ 2.2 $\pm$ 0.9 $\pm$ 0.8 & 2.091 $\pm$ 2.2 $\pm$ 1.0 $\pm$ 1.4 & 2.158 $\pm$ 1.6 $\pm$ 0.6 $\pm$ 0.6 & 2.154 $\pm$ 1.6 $\pm$ 0.6 $\pm$ 0.6 & 2.151 $\pm$ 1.6 $\pm$ 0.6 $\pm$ 0.7 & 2.137 $\pm$ 1.3 $\pm$ 0.5 $\pm$ 0.8\\
0.145 -- 0.165 & 1.806 $\pm$ 2.3 $\pm$ 0.9 $\pm$ 0.8 & 1.768 $\pm$ 2.3 $\pm$ 0.9 $\pm$ 1.4 & 1.891 $\pm$ 1.6 $\pm$ 0.6 $\pm$ 0.6 & 1.889 $\pm$ 1.6 $\pm$ 0.6 $\pm$ 0.6 & 1.868 $\pm$ 1.6 $\pm$ 0.6 $\pm$ 0.7 & 1.841 $\pm$ 1.3 $\pm$ 0.5 $\pm$ 0.8\\
0.165 -- 0.189 & 1.462 $\pm$ 2.3 $\pm$ 0.9 $\pm$ 0.7 & 1.419 $\pm$ 2.3 $\pm$ 1.0 $\pm$ 1.4 & 1.497 $\pm$ 1.6 $\pm$ 0.7 $\pm$ 0.5 & 1.496 $\pm$ 1.6 $\pm$ 0.7 $\pm$ 0.5 & 1.458 $\pm$ 1.6 $\pm$ 0.7 $\pm$ 0.6 & 1.453 $\pm$ 1.3 $\pm$ 0.5 $\pm$ 0.7\\
0.189 -- 0.219 & 1.216 $\pm$ 2.2 $\pm$ 1.1 $\pm$ 0.7 & 1.162 $\pm$ 2.2 $\pm$ 1.1 $\pm$ 1.4 & 1.268 $\pm$ 1.6 $\pm$ 0.6 $\pm$ 0.6 & 1.268 $\pm$ 1.6 $\pm$ 0.6 $\pm$ 0.6 & 1.209 $\pm$ 1.6 $\pm$ 0.6 $\pm$ 0.7 & 1.199 $\pm$ 1.3 $\pm$ 0.6 $\pm$ 0.7\\
0.219 -- 0.258 & 0.989 $\pm$ 2.2 $\pm$ 0.8 $\pm$ 0.8 & 0.937 $\pm$ 2.2 $\pm$ 0.9 $\pm$ 1.4 & 0.982 $\pm$ 1.6 $\pm$ 0.6 $\pm$ 0.6 & 0.991 $\pm$ 1.6 $\pm$ 0.6 $\pm$ 0.6 & 0.928 $\pm$ 1.6 $\pm$ 0.6 $\pm$ 0.7 & 0.937 $\pm$ 1.3 $\pm$ 0.5 $\pm$ 0.8\\
0.258 -- 0.312 & 0.738 $\pm$ 2.1 $\pm$ 0.9 $\pm$ 0.9 & 0.669 $\pm$ 2.1 $\pm$ 1.0 $\pm$ 2.2 & 0.755 $\pm$ 1.5 $\pm$ 0.6 $\pm$ 0.5 & 0.756 $\pm$ 1.5 $\pm$ 0.6 $\pm$ 0.5 & 0.686 $\pm$ 1.5 $\pm$ 0.6 $\pm$ 2.2 & 0.693 $\pm$ 1.2 $\pm$ 0.5 $\pm$ 1.7\\
0.312 -- 0.391 & 0.554 $\pm$ 2.0 $\pm$ 0.9 $\pm$ 1.2 & 0.499 $\pm$ 2.0 $\pm$ 1.0 $\pm$ 2.4 & 0.526 $\pm$ 1.5 $\pm$ 0.6 $\pm$ 1.2 & 0.524 $\pm$ 1.5 $\pm$ 0.6 $\pm$ 1.2 & 0.465 $\pm$ 1.5 $\pm$ 0.6 $\pm$ 2.4 & 0.490 $\pm$ 1.2 $\pm$ 0.5 $\pm$ 1.8\\
0.391 -- 0.524 & 0.325 $\pm$ 2.1 $\pm$ 0.8 $\pm$ 1.3 & 0.288 $\pm$ 2.1 $\pm$ 0.9 $\pm$ 2.4 & 0.331 $\pm$ 1.5 $\pm$ 0.7 $\pm$ 1.6 & 0.329 $\pm$ 1.5 $\pm$ 0.7 $\pm$ 1.6 & 0.291 $\pm$ 1.5 $\pm$ 0.7 $\pm$ 2.6 & 0.300 $\pm$ 1.2 $\pm$ 0.6 $\pm$ 1.9\\
0.524 -- 0.695 & 0.197 $\pm$ 2.4 $\pm$ 1.3 $\pm$ 1.9 & 0.179 $\pm$ 2.4 $\pm$ 1.3 $\pm$ 2.8 & 0.187 $\pm$ 1.7 $\pm$ 0.7 $\pm$ 2.2 & 0.186 $\pm$ 1.7 $\pm$ 0.7 $\pm$ 2.2 & 0.163 $\pm$ 1.7 $\pm$ 0.7 $\pm$ 3.1 & 0.176 $\pm$ 1.4 $\pm$ 0.6 $\pm$ 2.1\\
0.695 -- 0.918 & 0.102 $\pm$ 2.9 $\pm$ 1.8 $\pm$ 1.9 & 0.0961 $\pm$ 2.9 $\pm$ 1.8 $\pm$ 2.8 & 0.101 $\pm$ 2.1 $\pm$ 0.9 $\pm$ 2.5 & 0.0998 $\pm$ 2.1 $\pm$ 0.9 $\pm$ 2.5 & 0.0921 $\pm$ 2.1 $\pm$ 0.9 $\pm$ 3.3 & 0.0978 $\pm$ 1.7 $\pm$ 0.8 $\pm$ 2.2\\
0.918 -- 1.153 & 0.0605 $\pm$ 3.8 $\pm$ 1.8 $\pm$ 2.3 & 0.0584 $\pm$ 3.8 $\pm$ 1.9 $\pm$ 3.1 & 0.0594 $\pm$ 2.6 $\pm$ 1.1 $\pm$ 3.4 & 0.0588 $\pm$ 2.6 $\pm$ 1.1 $\pm$ 3.4 & 0.0547 $\pm$ 2.6 $\pm$ 1.1 $\pm$ 4.0 & 0.0594 $\pm$ 2.1 $\pm$ 1.0 $\pm$ 2.5\\
1.153 -- 1.496 & 0.0352 $\pm$ 4.2 $\pm$ 3.4 $\pm$ 2.6 & 0.0345 $\pm$ 4.2 $\pm$ 3.4 $\pm$ 3.4 & 0.0315 $\pm$ 3.1 $\pm$ 1.6 $\pm$ 2.7 & 0.0310 $\pm$ 3.1 $\pm$ 1.6 $\pm$ 2.7 & 0.0300 $\pm$ 3.1 $\pm$ 1.6 $\pm$ 3.2 & 0.0327 $\pm$ 2.5 $\pm$ 1.5 $\pm$ 2.4\\
1.496 -- 1.947 & 0.0174 $\pm$ 5.2 $\pm$ 3.3 $\pm$ 3.7 & 0.0171 $\pm$ 5.2 $\pm$ 3.4 $\pm$ 4.3 & 0.0177 $\pm$ 3.8 $\pm$ 1.8 $\pm$ 2.2 & 0.0175 $\pm$ 3.8 $\pm$ 1.8 $\pm$ 2.2 & 0.0167 $\pm$ 3.8 $\pm$ 1.8 $\pm$ 2.8 & 0.0175 $\pm$ 3.0 $\pm$ 1.6 $\pm$ 2.4\\
1.947 -- 2.522 & 0.00950 $\pm$ 6.2 $\pm$ 3.6 $\pm$ 3.9 & 0.00936 $\pm$ 6.2 $\pm$ 3.7 $\pm$ 4.5 & 0.00938 $\pm$ 4.4 $\pm$ 2.0 $\pm$ 2.4 & 0.00923 $\pm$ 4.4 $\pm$ 2.0 $\pm$ 2.4 & 0.00905 $\pm$ 4.4 $\pm$ 2.0 $\pm$ 2.9 & 0.00953 $\pm$ 3.5 $\pm$ 1.8 $\pm$ 2.4\\
2.522 -- 3.277 & 0.00567 $\pm$ 6.3 $\pm$ 3.9 $\pm$ 11 & 0.00574 $\pm$ 6.3 $\pm$ 3.9 $\pm$ 11 & 0.00493 $\pm$ 5.4 $\pm$ 3.3 $\pm$ 2.0 & 0.00494 $\pm$ 5.4 $\pm$ 3.3 $\pm$ 2.0 & 0.00480 $\pm$ 5.4 $\pm$ 3.3 $\pm$ 2.6 & 0.00519 $\pm$ 4.1 $\pm$ 2.5 $\pm$ 3.3\\
3.277 -- 5.000 & 0.00208 $\pm$ 7.3 $\pm$ 4.5 $\pm$ 4.1 & 0.00204 $\pm$ 7.3 $\pm$ 4.5 $\pm$ 4.6 & 0.00206 $\pm$ 5.5 $\pm$ 3.4 $\pm$ 3.6 & 0.00204 $\pm$ 5.5 $\pm$ 3.4 $\pm$ 3.6 & 0.00200 $\pm$ 5.5 $\pm$ 3.4 $\pm$ 3.9 & 0.00211 $\pm$ 4.3 $\pm$ 2.7 $\pm$ 2.7\\
5.000 -- 10.000 & 0.000603 $\pm$ 7.9 $\pm$ 4.4 $\pm$ 3.3 & 0.000599 $\pm$ 7.9 $\pm$ 4.4 $\pm$ 3.9 & 0.000536 $\pm$ 6.4 $\pm$ 3.8 $\pm$ 3.4 & 0.000529 $\pm$ 6.4 $\pm$ 3.8 $\pm$ 3.4 & 0.000521 $\pm$ 6.4 $\pm$ 3.8 $\pm$ 3.8 & 0.000577 $\pm$ 4.9 $\pm$ 2.9 $\pm$ 2.7\\
\bottomrule 
\end{tabular} 
} 
\label{tab:CombPhiStarlowmass_y0816}
\end{table*}

\begin{table*}
\centering
\caption{The values of $(1/\sigma)\,\mathrm{d}\sigma/\mathrm{d}\phi^*_{\eta}$ in each bin of \PhiStar{} for the electron and muon channels separately (for various particle-level definitions) and for the Born-level combination in the kinematic region $46\ \GeV \leq m_{\ell\ell} < 66\ \GeV,\ 1.6 \leq |y_{\ell\ell}| < 2.4$. The associated statistical and systematic (both uncorrelated and correlated between bins of \PhiStar{}) are provided in percentage form.} 
\resizebox{\textwidth}{!}{
\begin{tabular}{ ccccccc } \toprule 
Bin & \multicolumn{6}{c}{$(1/\sigma)\,d\sigma/d\phi^*_{\eta}$ $\pm$ Statistical [\%] $\pm$ Uncorrelated systematic [\%] $\pm$ Correlated systematic [\%]} \\ 
 \cmidrule(r){2-7} 
 & \multicolumn{2}{c}{Electron channel} &  \multicolumn{3}{c}{Muon channel} & Combination \\ 
\cmidrule(r){2-3} \cmidrule(r){4-6} \cmidrule(r){7-7} 
 & dressed & Born & bare & dressed & Born & Born \\ 
 0.000 -- 0.004 & 6.724 $\pm$ 3.7 $\pm$ 1.7 $\pm$ 4.0 & 7.096 $\pm$ 3.7 $\pm$ 1.7 $\pm$ 4.2 & 6.844 $\pm$ 2.6 $\pm$ 1.1 $\pm$ 3.2 & 6.783 $\pm$ 2.6 $\pm$ 1.1 $\pm$ 3.2 & 7.249 $\pm$ 2.6 $\pm$ 1.1 $\pm$ 3.3 & 7.260 $\pm$ 2.1 $\pm$ 0.9 $\pm$ 3.5\\
0.004 -- 0.008 & 6.620 $\pm$ 3.7 $\pm$ 1.8 $\pm$ 1.9 & 6.958 $\pm$ 3.7 $\pm$ 1.8 $\pm$ 2.2 & 6.899 $\pm$ 2.4 $\pm$ 1.0 $\pm$ 1.5 & 6.752 $\pm$ 2.4 $\pm$ 1.0 $\pm$ 1.5 & 7.155 $\pm$ 2.4 $\pm$ 1.0 $\pm$ 1.7 & 7.129 $\pm$ 2.0 $\pm$ 0.9 $\pm$ 1.7\\
0.008 -- 0.012 & 6.546 $\pm$ 3.6 $\pm$ 1.8 $\pm$ 1.2 & 6.826 $\pm$ 3.6 $\pm$ 1.8 $\pm$ 1.7 & 7.009 $\pm$ 2.5 $\pm$ 1.0 $\pm$ 1.2 & 6.942 $\pm$ 2.5 $\pm$ 1.0 $\pm$ 1.2 & 7.338 $\pm$ 2.5 $\pm$ 1.0 $\pm$ 1.4 & 7.174 $\pm$ 2.1 $\pm$ 0.9 $\pm$ 1.3\\
0.012 -- 0.016 & 6.493 $\pm$ 3.6 $\pm$ 1.4 $\pm$ 1.6 & 6.877 $\pm$ 3.6 $\pm$ 1.5 $\pm$ 1.9 & 7.341 $\pm$ 2.5 $\pm$ 1.1 $\pm$ 0.9 & 7.206 $\pm$ 2.5 $\pm$ 1.1 $\pm$ 0.9 & 7.625 $\pm$ 2.5 $\pm$ 1.1 $\pm$ 1.2 & 7.389 $\pm$ 2.0 $\pm$ 0.9 $\pm$ 1.2\\
0.016 -- 0.020 & 6.632 $\pm$ 3.6 $\pm$ 1.5 $\pm$ 2.0 & 6.939 $\pm$ 3.6 $\pm$ 1.5 $\pm$ 2.3 & 6.958 $\pm$ 2.5 $\pm$ 1.0 $\pm$ 0.9 & 6.958 $\pm$ 2.5 $\pm$ 1.0 $\pm$ 0.9 & 7.457 $\pm$ 2.5 $\pm$ 1.0 $\pm$ 1.3 & 7.249 $\pm$ 2.1 $\pm$ 0.9 $\pm$ 1.2\\
0.020 -- 0.024 & 6.727 $\pm$ 3.6 $\pm$ 1.5 $\pm$ 0.9 & 7.004 $\pm$ 3.6 $\pm$ 1.5 $\pm$ 1.5 & 6.505 $\pm$ 2.7 $\pm$ 1.1 $\pm$ 0.7 & 6.504 $\pm$ 2.7 $\pm$ 1.1 $\pm$ 0.7 & 6.724 $\pm$ 2.7 $\pm$ 1.1 $\pm$ 1.1 & 6.803 $\pm$ 2.1 $\pm$ 0.9 $\pm$ 1.0\\
0.024 -- 0.029 & 6.531 $\pm$ 3.3 $\pm$ 1.3 $\pm$ 1.1 & 6.896 $\pm$ 3.3 $\pm$ 1.4 $\pm$ 1.6 & 6.452 $\pm$ 2.4 $\pm$ 1.0 $\pm$ 0.7 & 6.378 $\pm$ 2.4 $\pm$ 1.0 $\pm$ 0.7 & 6.721 $\pm$ 2.4 $\pm$ 1.0 $\pm$ 1.1 & 6.747 $\pm$ 1.9 $\pm$ 0.8 $\pm$ 1.0\\
0.029 -- 0.034 & 6.023 $\pm$ 3.4 $\pm$ 1.4 $\pm$ 1.0 & 6.297 $\pm$ 3.4 $\pm$ 1.4 $\pm$ 1.5 & 5.969 $\pm$ 2.5 $\pm$ 1.0 $\pm$ 0.7 & 5.980 $\pm$ 2.5 $\pm$ 1.0 $\pm$ 0.7 & 6.272 $\pm$ 2.5 $\pm$ 1.0 $\pm$ 1.1 & 6.261 $\pm$ 2.0 $\pm$ 0.8 $\pm$ 1.0\\
0.034 -- 0.039 & 6.204 $\pm$ 3.3 $\pm$ 1.4 $\pm$ 0.9 & 6.499 $\pm$ 3.3 $\pm$ 1.4 $\pm$ 1.5 & 5.529 $\pm$ 2.6 $\pm$ 1.1 $\pm$ 0.6 & 5.509 $\pm$ 2.6 $\pm$ 1.1 $\pm$ 0.6 & 5.840 $\pm$ 2.6 $\pm$ 1.1 $\pm$ 1.0 & 6.050 $\pm$ 2.0 $\pm$ 0.8 $\pm$ 1.0\\
0.039 -- 0.045 & 5.461 $\pm$ 3.2 $\pm$ 1.8 $\pm$ 0.9 & 5.675 $\pm$ 3.2 $\pm$ 1.8 $\pm$ 1.4 & 5.619 $\pm$ 2.3 $\pm$ 0.9 $\pm$ 0.7 & 5.625 $\pm$ 2.3 $\pm$ 0.9 $\pm$ 0.7 & 5.999 $\pm$ 2.3 $\pm$ 0.9 $\pm$ 1.1 & 5.895 $\pm$ 1.9 $\pm$ 0.9 $\pm$ 1.0\\
0.045 -- 0.051 & 5.384 $\pm$ 3.3 $\pm$ 1.8 $\pm$ 1.1 & 5.592 $\pm$ 3.3 $\pm$ 1.8 $\pm$ 1.5 & 5.153 $\pm$ 2.4 $\pm$ 1.0 $\pm$ 0.8 & 5.170 $\pm$ 2.4 $\pm$ 1.0 $\pm$ 0.8 & 5.355 $\pm$ 2.4 $\pm$ 1.0 $\pm$ 1.2 & 5.418 $\pm$ 1.9 $\pm$ 0.9 $\pm$ 1.0\\
0.051 -- 0.057 & 5.016 $\pm$ 3.4 $\pm$ 1.4 $\pm$ 1.3 & 5.256 $\pm$ 3.4 $\pm$ 1.5 $\pm$ 1.7 & 4.916 $\pm$ 2.4 $\pm$ 1.0 $\pm$ 0.6 & 4.891 $\pm$ 2.4 $\pm$ 1.0 $\pm$ 0.6 & 5.109 $\pm$ 2.4 $\pm$ 1.0 $\pm$ 1.0 & 5.161 $\pm$ 2.0 $\pm$ 0.8 $\pm$ 1.0\\
0.057 -- 0.064 & 4.760 $\pm$ 3.2 $\pm$ 1.4 $\pm$ 0.8 & 4.923 $\pm$ 3.2 $\pm$ 1.4 $\pm$ 1.4 & 4.848 $\pm$ 2.3 $\pm$ 0.9 $\pm$ 0.4 & 4.878 $\pm$ 2.3 $\pm$ 0.9 $\pm$ 0.4 & 5.037 $\pm$ 2.3 $\pm$ 0.9 $\pm$ 0.8 & 4.995 $\pm$ 1.9 $\pm$ 0.8 $\pm$ 0.8\\
0.064 -- 0.072 & 4.364 $\pm$ 3.2 $\pm$ 2.0 $\pm$ 1.0 & 4.489 $\pm$ 3.2 $\pm$ 2.0 $\pm$ 1.5 & 4.181 $\pm$ 2.2 $\pm$ 0.9 $\pm$ 0.4 & 4.203 $\pm$ 2.2 $\pm$ 0.9 $\pm$ 0.4 & 4.359 $\pm$ 2.2 $\pm$ 0.9 $\pm$ 0.8 & 4.383 $\pm$ 1.8 $\pm$ 0.9 $\pm$ 0.8\\
0.072 -- 0.081 & 3.733 $\pm$ 3.2 $\pm$ 2.0 $\pm$ 0.9 & 3.810 $\pm$ 3.2 $\pm$ 2.1 $\pm$ 1.5 & 3.849 $\pm$ 2.2 $\pm$ 0.9 $\pm$ 0.5 & 3.856 $\pm$ 2.2 $\pm$ 0.9 $\pm$ 0.5 & 3.958 $\pm$ 2.2 $\pm$ 0.9 $\pm$ 0.8 & 3.911 $\pm$ 1.8 $\pm$ 0.9 $\pm$ 0.8\\
0.081 -- 0.091 & 3.600 $\pm$ 3.1 $\pm$ 1.2 $\pm$ 0.6 & 3.679 $\pm$ 3.1 $\pm$ 1.2 $\pm$ 1.3 & 3.482 $\pm$ 2.2 $\pm$ 0.9 $\pm$ 0.4 & 3.505 $\pm$ 2.2 $\pm$ 0.9 $\pm$ 0.4 & 3.594 $\pm$ 2.2 $\pm$ 0.9 $\pm$ 0.8 & 3.620 $\pm$ 1.8 $\pm$ 0.7 $\pm$ 0.8\\
0.091 -- 0.102 & 3.201 $\pm$ 3.1 $\pm$ 1.4 $\pm$ 0.9 & 3.247 $\pm$ 3.1 $\pm$ 1.5 $\pm$ 1.5 & 3.279 $\pm$ 2.3 $\pm$ 0.9 $\pm$ 0.5 & 3.283 $\pm$ 2.3 $\pm$ 0.9 $\pm$ 0.5 & 3.344 $\pm$ 2.3 $\pm$ 0.9 $\pm$ 0.8 & 3.307 $\pm$ 1.8 $\pm$ 0.8 $\pm$ 0.9\\
0.102 -- 0.114 & 2.927 $\pm$ 3.1 $\pm$ 1.4 $\pm$ 1.3 & 2.942 $\pm$ 3.1 $\pm$ 1.5 $\pm$ 1.7 & 2.898 $\pm$ 2.2 $\pm$ 0.9 $\pm$ 0.3 & 2.901 $\pm$ 2.2 $\pm$ 0.9 $\pm$ 0.3 & 2.921 $\pm$ 2.2 $\pm$ 0.9 $\pm$ 0.7 & 2.932 $\pm$ 1.8 $\pm$ 0.8 $\pm$ 0.8\\
0.114 -- 0.128 & 2.599 $\pm$ 3.0 $\pm$ 1.9 $\pm$ 0.9 & 2.618 $\pm$ 3.0 $\pm$ 2.0 $\pm$ 1.5 & 2.526 $\pm$ 2.2 $\pm$ 0.9 $\pm$ 0.5 & 2.528 $\pm$ 2.2 $\pm$ 0.9 $\pm$ 0.5 & 2.560 $\pm$ 2.2 $\pm$ 0.9 $\pm$ 0.8 & 2.577 $\pm$ 1.8 $\pm$ 0.9 $\pm$ 0.8\\
0.128 -- 0.145 & 1.987 $\pm$ 3.1 $\pm$ 2.0 $\pm$ 1.1 & 1.954 $\pm$ 3.1 $\pm$ 2.0 $\pm$ 1.6 & 2.236 $\pm$ 2.1 $\pm$ 0.8 $\pm$ 0.4 & 2.241 $\pm$ 2.1 $\pm$ 0.8 $\pm$ 0.4 & 2.234 $\pm$ 2.1 $\pm$ 0.8 $\pm$ 0.8 & 2.160 $\pm$ 1.7 $\pm$ 0.8 $\pm$ 0.8\\
0.145 -- 0.165 & 1.840 $\pm$ 3.1 $\pm$ 1.3 $\pm$ 1.3 & 1.829 $\pm$ 3.1 $\pm$ 1.3 $\pm$ 1.7 & 1.862 $\pm$ 2.2 $\pm$ 0.9 $\pm$ 0.4 & 1.860 $\pm$ 2.2 $\pm$ 0.9 $\pm$ 0.4 & 1.824 $\pm$ 2.2 $\pm$ 0.9 $\pm$ 0.8 & 1.825 $\pm$ 1.8 $\pm$ 0.7 $\pm$ 0.9\\
0.165 -- 0.189 & 1.544 $\pm$ 3.0 $\pm$ 1.7 $\pm$ 1.1 & 1.500 $\pm$ 3.0 $\pm$ 1.8 $\pm$ 1.6 & 1.594 $\pm$ 2.1 $\pm$ 0.9 $\pm$ 0.5 & 1.610 $\pm$ 2.1 $\pm$ 0.9 $\pm$ 0.5 & 1.546 $\pm$ 2.1 $\pm$ 0.9 $\pm$ 0.8 & 1.533 $\pm$ 1.7 $\pm$ 0.8 $\pm$ 0.9\\
0.189 -- 0.219 & 1.302 $\pm$ 3.0 $\pm$ 2.0 $\pm$ 1.0 & 1.253 $\pm$ 3.0 $\pm$ 2.1 $\pm$ 1.6 & 1.232 $\pm$ 2.2 $\pm$ 0.9 $\pm$ 0.5 & 1.233 $\pm$ 2.2 $\pm$ 0.9 $\pm$ 0.5 & 1.179 $\pm$ 2.2 $\pm$ 0.9 $\pm$ 0.8 & 1.205 $\pm$ 1.8 $\pm$ 0.9 $\pm$ 0.8\\
0.219 -- 0.258 & 1.000 $\pm$ 2.9 $\pm$ 1.6 $\pm$ 0.9 & 0.939 $\pm$ 2.9 $\pm$ 1.6 $\pm$ 1.6 & 0.977 $\pm$ 2.2 $\pm$ 0.9 $\pm$ 0.6 & 0.985 $\pm$ 2.2 $\pm$ 0.9 $\pm$ 0.6 & 0.925 $\pm$ 2.2 $\pm$ 0.9 $\pm$ 0.9 & 0.929 $\pm$ 1.7 $\pm$ 0.8 $\pm$ 0.9\\
0.258 -- 0.312 & 0.728 $\pm$ 2.9 $\pm$ 1.6 $\pm$ 1.5 & 0.669 $\pm$ 2.9 $\pm$ 1.6 $\pm$ 2.4 & 0.749 $\pm$ 2.0 $\pm$ 0.8 $\pm$ 0.6 & 0.747 $\pm$ 2.0 $\pm$ 0.8 $\pm$ 0.6 & 0.690 $\pm$ 2.0 $\pm$ 0.8 $\pm$ 1.9 & 0.685 $\pm$ 1.6 $\pm$ 0.7 $\pm$ 1.7\\
0.312 -- 0.391 & 0.538 $\pm$ 2.8 $\pm$ 1.3 $\pm$ 1.1 & 0.488 $\pm$ 2.8 $\pm$ 1.4 $\pm$ 2.3 & 0.530 $\pm$ 2.0 $\pm$ 0.8 $\pm$ 1.1 & 0.534 $\pm$ 2.0 $\pm$ 0.8 $\pm$ 1.1 & 0.473 $\pm$ 2.0 $\pm$ 0.8 $\pm$ 2.1 & 0.481 $\pm$ 1.6 $\pm$ 0.7 $\pm$ 1.8\\
0.391 -- 0.524 & 0.347 $\pm$ 2.7 $\pm$ 1.0 $\pm$ 1.4 & 0.315 $\pm$ 2.7 $\pm$ 1.1 $\pm$ 2.4 & 0.326 $\pm$ 2.0 $\pm$ 0.8 $\pm$ 1.2 & 0.327 $\pm$ 2.0 $\pm$ 0.8 $\pm$ 1.2 & 0.290 $\pm$ 2.0 $\pm$ 0.8 $\pm$ 2.1 & 0.299 $\pm$ 1.6 $\pm$ 0.7 $\pm$ 1.8\\
0.524 -- 0.695 & 0.191 $\pm$ 3.3 $\pm$ 3.0 $\pm$ 1.7 & 0.176 $\pm$ 3.3 $\pm$ 3.0 $\pm$ 2.6 & 0.181 $\pm$ 2.3 $\pm$ 0.9 $\pm$ 2.0 & 0.182 $\pm$ 2.3 $\pm$ 0.9 $\pm$ 2.0 & 0.163 $\pm$ 2.3 $\pm$ 0.9 $\pm$ 2.7 & 0.168 $\pm$ 1.9 $\pm$ 1.0 $\pm$ 2.1\\
0.695 -- 0.918 & 0.103 $\pm$ 3.9 $\pm$ 2.7 $\pm$ 1.7 & 0.0993 $\pm$ 3.9 $\pm$ 2.7 $\pm$ 2.6 & 0.102 $\pm$ 2.7 $\pm$ 1.2 $\pm$ 2.2 & 0.102 $\pm$ 2.7 $\pm$ 1.2 $\pm$ 2.2 & 0.0953 $\pm$ 2.7 $\pm$ 1.2 $\pm$ 2.8 & 0.0978 $\pm$ 2.2 $\pm$ 1.1 $\pm$ 2.2\\
0.918 -- 1.153 & 0.0543 $\pm$ 5.2 $\pm$ 4.2 $\pm$ 3.4 & 0.0529 $\pm$ 5.2 $\pm$ 4.2 $\pm$ 3.9 & 0.0543 $\pm$ 3.6 $\pm$ 1.5 $\pm$ 1.0 & 0.0539 $\pm$ 3.6 $\pm$ 1.5 $\pm$ 1.0 & 0.0519 $\pm$ 3.6 $\pm$ 1.5 $\pm$ 2.0 & 0.0526 $\pm$ 3.0 $\pm$ 1.5 $\pm$ 1.9\\
1.153 -- 1.496 & 0.0285 $\pm$ 6.0 $\pm$ 3.9 $\pm$ 3.8 & 0.0282 $\pm$ 6.0 $\pm$ 4.0 $\pm$ 4.4 & 0.0300 $\pm$ 4.2 $\pm$ 2.2 $\pm$ 1.4 & 0.0295 $\pm$ 4.2 $\pm$ 2.2 $\pm$ 1.4 & 0.0285 $\pm$ 4.2 $\pm$ 2.2 $\pm$ 1.6 & 0.0285 $\pm$ 3.4 $\pm$ 2.0 $\pm$ 1.7\\
1.496 -- 1.947 & 0.0152 $\pm$ 6.9 $\pm$ 3.6 $\pm$ 3.5 & 0.0152 $\pm$ 6.9 $\pm$ 3.6 $\pm$ 4.1 & 0.0150 $\pm$ 5.7 $\pm$ 2.8 $\pm$ 1.8 & 0.0152 $\pm$ 5.7 $\pm$ 2.8 $\pm$ 1.8 & 0.0153 $\pm$ 5.7 $\pm$ 2.8 $\pm$ 2.0 & 0.0154 $\pm$ 4.4 $\pm$ 2.2 $\pm$ 2.0\\
1.947 -- 2.522 & 0.00628 $\pm$ 9.3 $\pm$ 4.2 $\pm$ 3.9 & 0.00634 $\pm$ 9.3 $\pm$ 4.3 $\pm$ 4.4 & 0.00625 $\pm$ 7.0 $\pm$ 3.3 $\pm$ 1.9 & 0.00634 $\pm$ 7.0 $\pm$ 3.3 $\pm$ 1.9 & 0.00619 $\pm$ 7.0 $\pm$ 3.3 $\pm$ 2.0 & 0.00631 $\pm$ 5.6 $\pm$ 2.6 $\pm$ 2.1\\
2.522 -- 3.277 & 0.00274 $\pm$ 13 $\pm$ 4.4 $\pm$ 4.7 & 0.00272 $\pm$ 13 $\pm$ 4.5 $\pm$ 5.2 & 0.00292 $\pm$ 8.8 $\pm$ 3.8 $\pm$ 1.9 & 0.00294 $\pm$ 8.8 $\pm$ 3.8 $\pm$ 1.9 & 0.00293 $\pm$ 8.8 $\pm$ 3.8 $\pm$ 2.0 & 0.00284 $\pm$ 7.2 $\pm$ 2.9 $\pm$ 2.2\\
3.277 -- 5.000 & 0.00115 $\pm$ 12 $\pm$ 4.3 $\pm$ 5.4 & 0.00118 $\pm$ 12 $\pm$ 4.4 $\pm$ 5.9 & 0.00102 $\pm$ 9.8 $\pm$ 4.3 $\pm$ 1.8 & 0.00106 $\pm$ 9.8 $\pm$ 4.3 $\pm$ 1.8 & 0.00104 $\pm$ 9.8 $\pm$ 4.3 $\pm$ 2.0 & 0.00109 $\pm$ 7.6 $\pm$ 3.1 $\pm$ 2.4\\
5.000 -- 10.000 & 0.000213 $\pm$ 17 $\pm$ 5.6 $\pm$ 5.1 & 0.000209 $\pm$ 17 $\pm$ 5.7 $\pm$ 5.4 & 0.000309 $\pm$ 10 $\pm$ 4.5 $\pm$ 2.2 & 0.000321 $\pm$ 10 $\pm$ 4.5 $\pm$ 2.2 & 0.000354 $\pm$ 10 $\pm$ 4.5 $\pm$ 2.3 & 0.000295 $\pm$ 9.0 $\pm$ 3.5 $\pm$ 2.3\\
\bottomrule 
\end{tabular} 
} 
\label{tab:CombPhiStarlowmass_y1624}
\end{table*}

\begin{table*}
\centering
\caption{The values of $(1/\sigma)\,\mathrm{d}\sigma/\mathrm{d}\phi^*_{\eta}$ in each bin of \PhiStar{} for the electron and muon channels separately (for various particle-level definitions) and for the Born-level combination in the kinematic region $66\ \GeV \leq m_{\ell\ell} < 116\ \GeV,\ 0 \leq |y_{\ell\ell}| < 0.4$. The associated statistical and systematic (both uncorrelated and correlated between bins of \PhiStar{}) are provided in percentage form.} 
\resizebox{\textwidth}{!}{
\begin{tabular}{ ccccccc } \toprule 
Bin & \multicolumn{6}{c}{$(1/\sigma)\,d\sigma/d\phi^*_{\eta}$ $\pm$ Statistical [\%] $\pm$ Uncorrelated systematic [\%] $\pm$ Correlated systematic [\%]} \\ 
 \cmidrule(r){2-7} 
 & \multicolumn{2}{c}{Electron channel} &  \multicolumn{3}{c}{Muon channel} & Combination \\ 
\cmidrule(r){2-3} \cmidrule(r){4-6} \cmidrule(r){7-7} 
 & dressed & Born & bare & dressed & Born & Born \\ 
 0.000 -- 0.004 & 9.252 $\pm$ 0.4 $\pm$ 0.1 $\pm$ 0.2 & 9.331 $\pm$ 0.4 $\pm$ 0.1 $\pm$ 0.2 & 9.359 $\pm$ 0.4 $\pm$ 0.1 $\pm$ 0.1 & 9.359 $\pm$ 0.4 $\pm$ 0.1 $\pm$ 0.1 & 9.437 $\pm$ 0.4 $\pm$ 0.1 $\pm$ 0.2 & 9.386 $\pm$ 0.3 $\pm$ 0.1 $\pm$ 0.2\\
0.004 -- 0.008 & 9.264 $\pm$ 0.4 $\pm$ 0.1 $\pm$ 0.1 & 9.353 $\pm$ 0.4 $\pm$ 0.1 $\pm$ 0.2 & 9.279 $\pm$ 0.4 $\pm$ 0.1 $\pm$ 0.1 & 9.270 $\pm$ 0.4 $\pm$ 0.1 $\pm$ 0.1 & 9.347 $\pm$ 0.4 $\pm$ 0.1 $\pm$ 0.1 & 9.346 $\pm$ 0.3 $\pm$ 0.1 $\pm$ 0.1\\
0.008 -- 0.012 & 9.001 $\pm$ 0.4 $\pm$ 0.1 $\pm$ 0.1 & 9.067 $\pm$ 0.4 $\pm$ 0.1 $\pm$ 0.1 & 9.079 $\pm$ 0.4 $\pm$ 0.1 $\pm$ 0.1 & 9.074 $\pm$ 0.4 $\pm$ 0.1 $\pm$ 0.1 & 9.161 $\pm$ 0.4 $\pm$ 0.1 $\pm$ 0.1 & 9.115 $\pm$ 0.3 $\pm$ 0.1 $\pm$ 0.1\\
0.012 -- 0.016 & 8.810 $\pm$ 0.4 $\pm$ 0.1 $\pm$ 0.1 & 8.881 $\pm$ 0.4 $\pm$ 0.1 $\pm$ 0.1 & 8.878 $\pm$ 0.4 $\pm$ 0.1 $\pm$ 0.1 & 8.875 $\pm$ 0.4 $\pm$ 0.1 $\pm$ 0.1 & 8.933 $\pm$ 0.4 $\pm$ 0.1 $\pm$ 0.1 & 8.908 $\pm$ 0.3 $\pm$ 0.1 $\pm$ 0.1\\
0.016 -- 0.020 & 8.627 $\pm$ 0.4 $\pm$ 0.1 $\pm$ 0.1 & 8.697 $\pm$ 0.4 $\pm$ 0.1 $\pm$ 0.1 & 8.479 $\pm$ 0.4 $\pm$ 0.1 $\pm$ 0.1 & 8.474 $\pm$ 0.4 $\pm$ 0.1 $\pm$ 0.1 & 8.534 $\pm$ 0.4 $\pm$ 0.1 $\pm$ 0.1 & 8.607 $\pm$ 0.3 $\pm$ 0.1 $\pm$ 0.1\\
0.020 -- 0.024 & 8.112 $\pm$ 0.4 $\pm$ 0.1 $\pm$ 0.1 & 8.162 $\pm$ 0.4 $\pm$ 0.1 $\pm$ 0.1 & 8.163 $\pm$ 0.4 $\pm$ 0.1 $\pm$ 0.1 & 8.164 $\pm$ 0.4 $\pm$ 0.1 $\pm$ 0.1 & 8.223 $\pm$ 0.4 $\pm$ 0.1 $\pm$ 0.1 & 8.196 $\pm$ 0.3 $\pm$ 0.1 $\pm$ 0.1\\
0.024 -- 0.029 & 7.778 $\pm$ 0.4 $\pm$ 0.1 $\pm$ 0.1 & 7.817 $\pm$ 0.4 $\pm$ 0.1 $\pm$ 0.1 & 7.823 $\pm$ 0.4 $\pm$ 0.1 $\pm$ 0.0 & 7.822 $\pm$ 0.4 $\pm$ 0.1 $\pm$ 0.0 & 7.864 $\pm$ 0.4 $\pm$ 0.1 $\pm$ 0.1 & 7.840 $\pm$ 0.3 $\pm$ 0.1 $\pm$ 0.1\\
0.029 -- 0.034 & 7.344 $\pm$ 0.4 $\pm$ 0.1 $\pm$ 0.1 & 7.384 $\pm$ 0.4 $\pm$ 0.1 $\pm$ 0.1 & 7.401 $\pm$ 0.4 $\pm$ 0.1 $\pm$ 0.0 & 7.395 $\pm$ 0.4 $\pm$ 0.1 $\pm$ 0.0 & 7.431 $\pm$ 0.4 $\pm$ 0.1 $\pm$ 0.1 & 7.407 $\pm$ 0.3 $\pm$ 0.1 $\pm$ 0.1\\
0.034 -- 0.039 & 6.884 $\pm$ 0.4 $\pm$ 0.1 $\pm$ 0.1 & 6.909 $\pm$ 0.4 $\pm$ 0.1 $\pm$ 0.1 & 6.861 $\pm$ 0.4 $\pm$ 0.1 $\pm$ 0.0 & 6.863 $\pm$ 0.4 $\pm$ 0.1 $\pm$ 0.0 & 6.898 $\pm$ 0.4 $\pm$ 0.1 $\pm$ 0.1 & 6.903 $\pm$ 0.3 $\pm$ 0.1 $\pm$ 0.1\\
0.039 -- 0.045 & 6.367 $\pm$ 0.4 $\pm$ 0.1 $\pm$ 0.1 & 6.375 $\pm$ 0.4 $\pm$ 0.1 $\pm$ 0.1 & 6.392 $\pm$ 0.4 $\pm$ 0.1 $\pm$ 0.0 & 6.392 $\pm$ 0.4 $\pm$ 0.1 $\pm$ 0.0 & 6.416 $\pm$ 0.4 $\pm$ 0.1 $\pm$ 0.1 & 6.396 $\pm$ 0.3 $\pm$ 0.1 $\pm$ 0.1\\
0.045 -- 0.051 & 5.865 $\pm$ 0.4 $\pm$ 0.1 $\pm$ 0.1 & 5.873 $\pm$ 0.4 $\pm$ 0.1 $\pm$ 0.1 & 5.877 $\pm$ 0.4 $\pm$ 0.1 $\pm$ 0.0 & 5.872 $\pm$ 0.4 $\pm$ 0.1 $\pm$ 0.0 & 5.878 $\pm$ 0.4 $\pm$ 0.1 $\pm$ 0.1 & 5.875 $\pm$ 0.3 $\pm$ 0.1 $\pm$ 0.1\\
0.051 -- 0.057 & 5.438 $\pm$ 0.4 $\pm$ 0.1 $\pm$ 0.1 & 5.441 $\pm$ 0.4 $\pm$ 0.1 $\pm$ 0.1 & 5.430 $\pm$ 0.4 $\pm$ 0.1 $\pm$ 0.0 & 5.434 $\pm$ 0.4 $\pm$ 0.1 $\pm$ 0.0 & 5.436 $\pm$ 0.4 $\pm$ 0.1 $\pm$ 0.1 & 5.438 $\pm$ 0.3 $\pm$ 0.1 $\pm$ 0.1\\
0.057 -- 0.064 & 4.954 $\pm$ 0.4 $\pm$ 0.1 $\pm$ 0.1 & 4.952 $\pm$ 0.4 $\pm$ 0.1 $\pm$ 0.1 & 4.970 $\pm$ 0.4 $\pm$ 0.1 $\pm$ 0.1 & 4.966 $\pm$ 0.4 $\pm$ 0.1 $\pm$ 0.1 & 4.962 $\pm$ 0.4 $\pm$ 0.1 $\pm$ 0.2 & 4.957 $\pm$ 0.3 $\pm$ 0.1 $\pm$ 0.1\\
0.064 -- 0.072 & 4.522 $\pm$ 0.4 $\pm$ 0.1 $\pm$ 0.1 & 4.514 $\pm$ 0.4 $\pm$ 0.1 $\pm$ 0.1 & 4.514 $\pm$ 0.4 $\pm$ 0.1 $\pm$ 0.1 & 4.514 $\pm$ 0.4 $\pm$ 0.1 $\pm$ 0.1 & 4.503 $\pm$ 0.4 $\pm$ 0.1 $\pm$ 0.2 & 4.507 $\pm$ 0.3 $\pm$ 0.1 $\pm$ 0.1\\
0.072 -- 0.081 & 4.021 $\pm$ 0.4 $\pm$ 0.1 $\pm$ 0.1 & 4.011 $\pm$ 0.4 $\pm$ 0.1 $\pm$ 0.1 & 3.984 $\pm$ 0.4 $\pm$ 0.1 $\pm$ 0.1 & 3.983 $\pm$ 0.4 $\pm$ 0.1 $\pm$ 0.1 & 3.970 $\pm$ 0.4 $\pm$ 0.1 $\pm$ 0.2 & 3.988 $\pm$ 0.3 $\pm$ 0.1 $\pm$ 0.1\\
0.081 -- 0.091 & 3.572 $\pm$ 0.4 $\pm$ 0.1 $\pm$ 0.1 & 3.558 $\pm$ 0.4 $\pm$ 0.1 $\pm$ 0.1 & 3.576 $\pm$ 0.4 $\pm$ 0.1 $\pm$ 0.1 & 3.576 $\pm$ 0.4 $\pm$ 0.1 $\pm$ 0.1 & 3.567 $\pm$ 0.4 $\pm$ 0.1 $\pm$ 0.2 & 3.561 $\pm$ 0.3 $\pm$ 0.1 $\pm$ 0.1\\
0.091 -- 0.102 & 3.145 $\pm$ 0.4 $\pm$ 0.1 $\pm$ 0.1 & 3.132 $\pm$ 0.4 $\pm$ 0.1 $\pm$ 0.1 & 3.165 $\pm$ 0.4 $\pm$ 0.1 $\pm$ 0.1 & 3.165 $\pm$ 0.4 $\pm$ 0.1 $\pm$ 0.1 & 3.145 $\pm$ 0.4 $\pm$ 0.1 $\pm$ 0.2 & 3.138 $\pm$ 0.3 $\pm$ 0.1 $\pm$ 0.1\\
0.102 -- 0.114 & 2.764 $\pm$ 0.4 $\pm$ 0.1 $\pm$ 0.1 & 2.752 $\pm$ 0.4 $\pm$ 0.1 $\pm$ 0.1 & 2.774 $\pm$ 0.4 $\pm$ 0.1 $\pm$ 0.1 & 2.773 $\pm$ 0.4 $\pm$ 0.1 $\pm$ 0.1 & 2.763 $\pm$ 0.4 $\pm$ 0.1 $\pm$ 0.2 & 2.757 $\pm$ 0.3 $\pm$ 0.1 $\pm$ 0.1\\
0.114 -- 0.128 & 2.394 $\pm$ 0.4 $\pm$ 0.1 $\pm$ 0.1 & 2.382 $\pm$ 0.4 $\pm$ 0.1 $\pm$ 0.1 & 2.379 $\pm$ 0.4 $\pm$ 0.1 $\pm$ 0.1 & 2.378 $\pm$ 0.4 $\pm$ 0.1 $\pm$ 0.1 & 2.363 $\pm$ 0.4 $\pm$ 0.1 $\pm$ 0.1 & 2.371 $\pm$ 0.3 $\pm$ 0.1 $\pm$ 0.1\\
0.128 -- 0.145 & 2.023 $\pm$ 0.4 $\pm$ 0.1 $\pm$ 0.1 & 2.009 $\pm$ 0.4 $\pm$ 0.1 $\pm$ 0.1 & 2.037 $\pm$ 0.4 $\pm$ 0.1 $\pm$ 0.1 & 2.038 $\pm$ 0.4 $\pm$ 0.1 $\pm$ 0.1 & 2.026 $\pm$ 0.4 $\pm$ 0.1 $\pm$ 0.1 & 2.017 $\pm$ 0.3 $\pm$ 0.1 $\pm$ 0.1\\
0.145 -- 0.165 & 1.697 $\pm$ 0.4 $\pm$ 0.1 $\pm$ 0.1 & 1.686 $\pm$ 0.4 $\pm$ 0.1 $\pm$ 0.1 & 1.698 $\pm$ 0.4 $\pm$ 0.1 $\pm$ 0.1 & 1.698 $\pm$ 0.4 $\pm$ 0.1 $\pm$ 0.1 & 1.688 $\pm$ 0.4 $\pm$ 0.1 $\pm$ 0.1 & 1.687 $\pm$ 0.3 $\pm$ 0.1 $\pm$ 0.1\\
0.165 -- 0.189 & 1.396 $\pm$ 0.4 $\pm$ 0.1 $\pm$ 0.1 & 1.388 $\pm$ 0.4 $\pm$ 0.1 $\pm$ 0.1 & 1.391 $\pm$ 0.4 $\pm$ 0.1 $\pm$ 0.1 & 1.391 $\pm$ 0.4 $\pm$ 0.1 $\pm$ 0.1 & 1.382 $\pm$ 0.4 $\pm$ 0.1 $\pm$ 0.2 & 1.384 $\pm$ 0.3 $\pm$ 0.1 $\pm$ 0.1\\
0.189 -- 0.219 & 1.111 $\pm$ 0.4 $\pm$ 0.1 $\pm$ 0.1 & 1.104 $\pm$ 0.4 $\pm$ 0.1 $\pm$ 0.1 & 1.105 $\pm$ 0.4 $\pm$ 0.1 $\pm$ 0.1 & 1.106 $\pm$ 0.4 $\pm$ 0.1 $\pm$ 0.1 & 1.098 $\pm$ 0.4 $\pm$ 0.1 $\pm$ 0.2 & 1.101 $\pm$ 0.3 $\pm$ 0.1 $\pm$ 0.1\\
0.219 -- 0.258 & 0.851 $\pm$ 0.4 $\pm$ 0.1 $\pm$ 0.1 & 0.846 $\pm$ 0.4 $\pm$ 0.1 $\pm$ 0.1 & 0.856 $\pm$ 0.4 $\pm$ 0.1 $\pm$ 0.1 & 0.857 $\pm$ 0.4 $\pm$ 0.1 $\pm$ 0.1 & 0.852 $\pm$ 0.4 $\pm$ 0.1 $\pm$ 0.2 & 0.849 $\pm$ 0.3 $\pm$ 0.1 $\pm$ 0.1\\
0.258 -- 0.312 & 0.618 $\pm$ 0.4 $\pm$ 0.1 $\pm$ 0.2 & 0.615 $\pm$ 0.4 $\pm$ 0.1 $\pm$ 0.3 & 0.616 $\pm$ 0.4 $\pm$ 0.1 $\pm$ 0.1 & 0.616 $\pm$ 0.4 $\pm$ 0.1 $\pm$ 0.1 & 0.613 $\pm$ 0.4 $\pm$ 0.1 $\pm$ 0.2 & 0.614 $\pm$ 0.3 $\pm$ 0.1 $\pm$ 0.2\\
0.312 -- 0.391 & 0.411 $\pm$ 0.4 $\pm$ 0.1 $\pm$ 0.3 & 0.410 $\pm$ 0.4 $\pm$ 0.1 $\pm$ 0.3 & 0.412 $\pm$ 0.4 $\pm$ 0.1 $\pm$ 0.2 & 0.412 $\pm$ 0.4 $\pm$ 0.1 $\pm$ 0.2 & 0.410 $\pm$ 0.4 $\pm$ 0.1 $\pm$ 0.2 & 0.410 $\pm$ 0.3 $\pm$ 0.1 $\pm$ 0.2\\
0.391 -- 0.524 & 0.241 $\pm$ 0.4 $\pm$ 0.1 $\pm$ 0.3 & 0.240 $\pm$ 0.4 $\pm$ 0.1 $\pm$ 0.3 & 0.239 $\pm$ 0.4 $\pm$ 0.1 $\pm$ 0.2 & 0.239 $\pm$ 0.4 $\pm$ 0.1 $\pm$ 0.2 & 0.239 $\pm$ 0.4 $\pm$ 0.1 $\pm$ 0.2 & 0.239 $\pm$ 0.3 $\pm$ 0.1 $\pm$ 0.2\\
0.524 -- 0.695 & 0.126 $\pm$ 0.5 $\pm$ 0.2 $\pm$ 0.3 & 0.126 $\pm$ 0.5 $\pm$ 0.2 $\pm$ 0.3 & 0.124 $\pm$ 0.5 $\pm$ 0.1 $\pm$ 0.2 & 0.124 $\pm$ 0.5 $\pm$ 0.1 $\pm$ 0.2 & 0.124 $\pm$ 0.5 $\pm$ 0.1 $\pm$ 0.3 & 0.125 $\pm$ 0.4 $\pm$ 0.1 $\pm$ 0.2\\
0.695 -- 0.918 & 0.0635 $\pm$ 0.7 $\pm$ 0.2 $\pm$ 0.3 & 0.0633 $\pm$ 0.7 $\pm$ 0.2 $\pm$ 0.3 & 0.0631 $\pm$ 0.6 $\pm$ 0.2 $\pm$ 0.3 & 0.0631 $\pm$ 0.6 $\pm$ 0.2 $\pm$ 0.3 & 0.0629 $\pm$ 0.6 $\pm$ 0.2 $\pm$ 0.3 & 0.0631 $\pm$ 0.5 $\pm$ 0.1 $\pm$ 0.3\\
0.918 -- 1.153 & 0.0335 $\pm$ 0.9 $\pm$ 0.3 $\pm$ 0.3 & 0.0334 $\pm$ 0.9 $\pm$ 0.3 $\pm$ 0.3 & 0.0328 $\pm$ 0.8 $\pm$ 0.2 $\pm$ 0.4 & 0.0329 $\pm$ 0.8 $\pm$ 0.2 $\pm$ 0.4 & 0.0328 $\pm$ 0.8 $\pm$ 0.2 $\pm$ 0.4 & 0.0331 $\pm$ 0.6 $\pm$ 0.2 $\pm$ 0.3\\
1.153 -- 1.496 & 0.0178 $\pm$ 1.0 $\pm$ 0.2 $\pm$ 0.4 & 0.0177 $\pm$ 1.0 $\pm$ 0.2 $\pm$ 0.4 & 0.0173 $\pm$ 1.0 $\pm$ 0.2 $\pm$ 0.4 & 0.0173 $\pm$ 1.0 $\pm$ 0.2 $\pm$ 0.4 & 0.0173 $\pm$ 1.0 $\pm$ 0.2 $\pm$ 0.5 & 0.0175 $\pm$ 0.7 $\pm$ 0.2 $\pm$ 0.4\\
1.496 -- 1.947 & 0.00883 $\pm$ 1.2 $\pm$ 0.3 $\pm$ 0.4 & 0.00880 $\pm$ 1.2 $\pm$ 0.3 $\pm$ 0.4 & 0.00875 $\pm$ 1.2 $\pm$ 0.3 $\pm$ 0.4 & 0.00878 $\pm$ 1.2 $\pm$ 0.3 $\pm$ 0.4 & 0.00877 $\pm$ 1.2 $\pm$ 0.3 $\pm$ 0.5 & 0.00879 $\pm$ 0.9 $\pm$ 0.2 $\pm$ 0.4\\
1.947 -- 2.522 & 0.00451 $\pm$ 1.6 $\pm$ 0.4 $\pm$ 0.6 & 0.00449 $\pm$ 1.6 $\pm$ 0.4 $\pm$ 0.6 & 0.00444 $\pm$ 1.4 $\pm$ 0.4 $\pm$ 0.5 & 0.00445 $\pm$ 1.4 $\pm$ 0.4 $\pm$ 0.5 & 0.00443 $\pm$ 1.4 $\pm$ 0.4 $\pm$ 0.6 & 0.00446 $\pm$ 1.1 $\pm$ 0.3 $\pm$ 0.4\\
2.522 -- 3.277 & 0.00239 $\pm$ 1.9 $\pm$ 0.4 $\pm$ 0.6 & 0.00238 $\pm$ 1.9 $\pm$ 0.4 $\pm$ 0.6 & 0.00227 $\pm$ 1.7 $\pm$ 0.5 $\pm$ 0.6 & 0.00228 $\pm$ 1.7 $\pm$ 0.5 $\pm$ 0.6 & 0.00227 $\pm$ 1.7 $\pm$ 0.5 $\pm$ 0.6 & 0.00232 $\pm$ 1.3 $\pm$ 0.3 $\pm$ 0.4\\
3.277 -- 5.000 & 0.00103 $\pm$ 1.9 $\pm$ 0.4 $\pm$ 0.7 & 0.00103 $\pm$ 1.9 $\pm$ 0.4 $\pm$ 0.7 & 0.00102 $\pm$ 1.7 $\pm$ 0.4 $\pm$ 0.6 & 0.00102 $\pm$ 1.7 $\pm$ 0.4 $\pm$ 0.6 & 0.00102 $\pm$ 1.7 $\pm$ 0.4 $\pm$ 0.6 & 0.00102 $\pm$ 1.3 $\pm$ 0.3 $\pm$ 0.5\\
5.000 -- 10.000 & 0.000306 $\pm$ 2.1 $\pm$ 0.5 $\pm$ 0.7 & 0.000306 $\pm$ 2.1 $\pm$ 0.5 $\pm$ 0.7 & 0.000300 $\pm$ 1.9 $\pm$ 0.5 $\pm$ 0.6 & 0.000301 $\pm$ 1.9 $\pm$ 0.5 $\pm$ 0.6 & 0.000301 $\pm$ 1.9 $\pm$ 0.5 $\pm$ 0.7 & 0.000303 $\pm$ 1.4 $\pm$ 0.3 $\pm$ 0.5\\
\bottomrule 
\end{tabular} 
} 
\label{tab:CombPhiStarpeakmass_y0004}
\end{table*}

\begin{table*}
\centering
\caption{The values of $(1/\sigma)\,\mathrm{d}\sigma/\mathrm{d}\phi^*_{\eta}$ in each bin of \PhiStar{} for the electron and muon channels separately (for various particle-level definitions) and for the Born-level combination in the kinematic region $66\ \GeV \leq m_{\ell\ell} < 116\ \GeV,\ 0.4 \leq |y_{\ell\ell}| < 0.8$. The associated statistical and systematic (both uncorrelated and correlated between bins of \PhiStar{}) are provided in percentage form.} 
\resizebox{\textwidth}{!}{
\begin{tabular}{ ccccccc } \toprule 
Bin & \multicolumn{6}{c}{$(1/\sigma)\,d\sigma/d\phi^*_{\eta}$ $\pm$ Statistical [\%] $\pm$ Uncorrelated systematic [\%] $\pm$ Correlated systematic [\%]} \\ 
 \cmidrule(r){2-7} 
 & \multicolumn{2}{c}{Electron channel} &  \multicolumn{3}{c}{Muon channel} & Combination \\ 
\cmidrule(r){2-3} \cmidrule(r){4-6} \cmidrule(r){7-7} 
 & dressed & Born & bare & dressed & Born & Born \\ 
 0.000 -- 0.004 & 9.405 $\pm$ 0.4 $\pm$ 0.1 $\pm$ 0.2 & 9.503 $\pm$ 0.4 $\pm$ 0.1 $\pm$ 0.2 & 9.386 $\pm$ 0.4 $\pm$ 0.1 $\pm$ 0.1 & 9.383 $\pm$ 0.4 $\pm$ 0.1 $\pm$ 0.1 & 9.452 $\pm$ 0.4 $\pm$ 0.1 $\pm$ 0.2 & 9.465 $\pm$ 0.3 $\pm$ 0.1 $\pm$ 0.2\\
0.004 -- 0.008 & 9.230 $\pm$ 0.4 $\pm$ 0.1 $\pm$ 0.1 & 9.300 $\pm$ 0.4 $\pm$ 0.1 $\pm$ 0.1 & 9.290 $\pm$ 0.4 $\pm$ 0.1 $\pm$ 0.1 & 9.289 $\pm$ 0.4 $\pm$ 0.1 $\pm$ 0.1 & 9.372 $\pm$ 0.4 $\pm$ 0.1 $\pm$ 0.1 & 9.332 $\pm$ 0.3 $\pm$ 0.1 $\pm$ 0.1\\
0.008 -- 0.012 & 9.055 $\pm$ 0.4 $\pm$ 0.1 $\pm$ 0.1 & 9.140 $\pm$ 0.4 $\pm$ 0.1 $\pm$ 0.1 & 9.100 $\pm$ 0.4 $\pm$ 0.1 $\pm$ 0.0 & 9.102 $\pm$ 0.4 $\pm$ 0.1 $\pm$ 0.0 & 9.173 $\pm$ 0.4 $\pm$ 0.1 $\pm$ 0.1 & 9.148 $\pm$ 0.3 $\pm$ 0.1 $\pm$ 0.1\\
0.012 -- 0.016 & 8.813 $\pm$ 0.4 $\pm$ 0.1 $\pm$ 0.1 & 8.877 $\pm$ 0.4 $\pm$ 0.1 $\pm$ 0.1 & 8.888 $\pm$ 0.4 $\pm$ 0.1 $\pm$ 0.0 & 8.890 $\pm$ 0.4 $\pm$ 0.1 $\pm$ 0.0 & 8.960 $\pm$ 0.4 $\pm$ 0.1 $\pm$ 0.1 & 8.917 $\pm$ 0.3 $\pm$ 0.1 $\pm$ 0.1\\
0.016 -- 0.020 & 8.530 $\pm$ 0.4 $\pm$ 0.1 $\pm$ 0.1 & 8.592 $\pm$ 0.4 $\pm$ 0.1 $\pm$ 0.1 & 8.595 $\pm$ 0.4 $\pm$ 0.1 $\pm$ 0.0 & 8.590 $\pm$ 0.4 $\pm$ 0.1 $\pm$ 0.0 & 8.659 $\pm$ 0.4 $\pm$ 0.1 $\pm$ 0.1 & 8.621 $\pm$ 0.3 $\pm$ 0.1 $\pm$ 0.1\\
0.020 -- 0.024 & 8.163 $\pm$ 0.5 $\pm$ 0.1 $\pm$ 0.1 & 8.209 $\pm$ 0.5 $\pm$ 0.1 $\pm$ 0.1 & 8.210 $\pm$ 0.4 $\pm$ 0.1 $\pm$ 0.0 & 8.207 $\pm$ 0.4 $\pm$ 0.1 $\pm$ 0.0 & 8.257 $\pm$ 0.4 $\pm$ 0.1 $\pm$ 0.1 & 8.228 $\pm$ 0.3 $\pm$ 0.1 $\pm$ 0.1\\
0.024 -- 0.029 & 7.849 $\pm$ 0.4 $\pm$ 0.1 $\pm$ 0.1 & 7.898 $\pm$ 0.4 $\pm$ 0.1 $\pm$ 0.1 & 7.813 $\pm$ 0.4 $\pm$ 0.1 $\pm$ 0.1 & 7.809 $\pm$ 0.4 $\pm$ 0.1 $\pm$ 0.1 & 7.853 $\pm$ 0.4 $\pm$ 0.1 $\pm$ 0.2 & 7.863 $\pm$ 0.3 $\pm$ 0.1 $\pm$ 0.1\\
0.029 -- 0.034 & 7.382 $\pm$ 0.4 $\pm$ 0.1 $\pm$ 0.1 & 7.411 $\pm$ 0.4 $\pm$ 0.1 $\pm$ 0.1 & 7.393 $\pm$ 0.4 $\pm$ 0.1 $\pm$ 0.1 & 7.388 $\pm$ 0.4 $\pm$ 0.1 $\pm$ 0.1 & 7.427 $\pm$ 0.4 $\pm$ 0.1 $\pm$ 0.2 & 7.412 $\pm$ 0.3 $\pm$ 0.1 $\pm$ 0.1\\
0.034 -- 0.039 & 6.874 $\pm$ 0.4 $\pm$ 0.1 $\pm$ 0.1 & 6.895 $\pm$ 0.4 $\pm$ 0.1 $\pm$ 0.1 & 6.909 $\pm$ 0.4 $\pm$ 0.1 $\pm$ 0.1 & 6.907 $\pm$ 0.4 $\pm$ 0.1 $\pm$ 0.1 & 6.933 $\pm$ 0.4 $\pm$ 0.1 $\pm$ 0.2 & 6.910 $\pm$ 0.3 $\pm$ 0.1 $\pm$ 0.1\\
0.039 -- 0.045 & 6.410 $\pm$ 0.4 $\pm$ 0.1 $\pm$ 0.1 & 6.425 $\pm$ 0.4 $\pm$ 0.1 $\pm$ 0.1 & 6.425 $\pm$ 0.4 $\pm$ 0.1 $\pm$ 0.1 & 6.417 $\pm$ 0.4 $\pm$ 0.1 $\pm$ 0.1 & 6.443 $\pm$ 0.4 $\pm$ 0.1 $\pm$ 0.2 & 6.428 $\pm$ 0.3 $\pm$ 0.1 $\pm$ 0.1\\
0.045 -- 0.051 & 5.850 $\pm$ 0.4 $\pm$ 0.1 $\pm$ 0.1 & 5.856 $\pm$ 0.4 $\pm$ 0.1 $\pm$ 0.1 & 5.903 $\pm$ 0.4 $\pm$ 0.1 $\pm$ 0.1 & 5.906 $\pm$ 0.4 $\pm$ 0.1 $\pm$ 0.1 & 5.913 $\pm$ 0.4 $\pm$ 0.1 $\pm$ 0.2 & 5.882 $\pm$ 0.3 $\pm$ 0.1 $\pm$ 0.1\\
0.051 -- 0.057 & 5.427 $\pm$ 0.5 $\pm$ 0.1 $\pm$ 0.1 & 5.429 $\pm$ 0.5 $\pm$ 0.1 $\pm$ 0.1 & 5.477 $\pm$ 0.4 $\pm$ 0.1 $\pm$ 0.1 & 5.475 $\pm$ 0.4 $\pm$ 0.1 $\pm$ 0.1 & 5.477 $\pm$ 0.4 $\pm$ 0.1 $\pm$ 0.2 & 5.450 $\pm$ 0.3 $\pm$ 0.1 $\pm$ 0.1\\
0.057 -- 0.064 & 4.933 $\pm$ 0.4 $\pm$ 0.1 $\pm$ 0.1 & 4.930 $\pm$ 0.4 $\pm$ 0.1 $\pm$ 0.1 & 4.979 $\pm$ 0.4 $\pm$ 0.1 $\pm$ 0.1 & 4.977 $\pm$ 0.4 $\pm$ 0.1 $\pm$ 0.1 & 4.972 $\pm$ 0.4 $\pm$ 0.1 $\pm$ 0.2 & 4.949 $\pm$ 0.3 $\pm$ 0.1 $\pm$ 0.1\\
0.064 -- 0.072 & 4.503 $\pm$ 0.4 $\pm$ 0.1 $\pm$ 0.1 & 4.499 $\pm$ 0.4 $\pm$ 0.1 $\pm$ 0.1 & 4.490 $\pm$ 0.4 $\pm$ 0.1 $\pm$ 0.1 & 4.492 $\pm$ 0.4 $\pm$ 0.1 $\pm$ 0.1 & 4.484 $\pm$ 0.4 $\pm$ 0.1 $\pm$ 0.2 & 4.485 $\pm$ 0.3 $\pm$ 0.1 $\pm$ 0.1\\
0.072 -- 0.081 & 4.020 $\pm$ 0.4 $\pm$ 0.1 $\pm$ 0.1 & 4.011 $\pm$ 0.4 $\pm$ 0.1 $\pm$ 0.1 & 4.037 $\pm$ 0.4 $\pm$ 0.1 $\pm$ 0.1 & 4.035 $\pm$ 0.4 $\pm$ 0.1 $\pm$ 0.1 & 4.027 $\pm$ 0.4 $\pm$ 0.1 $\pm$ 0.2 & 4.016 $\pm$ 0.3 $\pm$ 0.1 $\pm$ 0.1\\
0.081 -- 0.091 & 3.585 $\pm$ 0.4 $\pm$ 0.1 $\pm$ 0.1 & 3.573 $\pm$ 0.4 $\pm$ 0.1 $\pm$ 0.1 & 3.586 $\pm$ 0.4 $\pm$ 0.1 $\pm$ 0.1 & 3.587 $\pm$ 0.4 $\pm$ 0.1 $\pm$ 0.1 & 3.579 $\pm$ 0.4 $\pm$ 0.1 $\pm$ 0.2 & 3.574 $\pm$ 0.3 $\pm$ 0.1 $\pm$ 0.1\\
0.091 -- 0.102 & 3.146 $\pm$ 0.4 $\pm$ 0.1 $\pm$ 0.1 & 3.132 $\pm$ 0.4 $\pm$ 0.1 $\pm$ 0.1 & 3.146 $\pm$ 0.4 $\pm$ 0.1 $\pm$ 0.1 & 3.144 $\pm$ 0.4 $\pm$ 0.1 $\pm$ 0.1 & 3.132 $\pm$ 0.4 $\pm$ 0.1 $\pm$ 0.2 & 3.130 $\pm$ 0.3 $\pm$ 0.1 $\pm$ 0.1\\
0.102 -- 0.114 & 2.786 $\pm$ 0.4 $\pm$ 0.1 $\pm$ 0.1 & 2.772 $\pm$ 0.4 $\pm$ 0.1 $\pm$ 0.1 & 2.753 $\pm$ 0.4 $\pm$ 0.1 $\pm$ 0.1 & 2.752 $\pm$ 0.4 $\pm$ 0.1 $\pm$ 0.1 & 2.737 $\pm$ 0.4 $\pm$ 0.1 $\pm$ 0.2 & 2.750 $\pm$ 0.3 $\pm$ 0.1 $\pm$ 0.1\\
0.114 -- 0.128 & 2.380 $\pm$ 0.4 $\pm$ 0.1 $\pm$ 0.1 & 2.365 $\pm$ 0.4 $\pm$ 0.1 $\pm$ 0.1 & 2.370 $\pm$ 0.4 $\pm$ 0.1 $\pm$ 0.1 & 2.369 $\pm$ 0.4 $\pm$ 0.1 $\pm$ 0.1 & 2.358 $\pm$ 0.4 $\pm$ 0.1 $\pm$ 0.1 & 2.361 $\pm$ 0.3 $\pm$ 0.1 $\pm$ 0.1\\
0.128 -- 0.145 & 2.034 $\pm$ 0.4 $\pm$ 0.1 $\pm$ 0.1 & 2.022 $\pm$ 0.4 $\pm$ 0.1 $\pm$ 0.1 & 2.034 $\pm$ 0.4 $\pm$ 0.1 $\pm$ 0.1 & 2.034 $\pm$ 0.4 $\pm$ 0.1 $\pm$ 0.1 & 2.021 $\pm$ 0.4 $\pm$ 0.1 $\pm$ 0.1 & 2.021 $\pm$ 0.3 $\pm$ 0.1 $\pm$ 0.1\\
0.145 -- 0.165 & 1.694 $\pm$ 0.4 $\pm$ 0.1 $\pm$ 0.1 & 1.683 $\pm$ 0.4 $\pm$ 0.1 $\pm$ 0.1 & 1.695 $\pm$ 0.4 $\pm$ 0.1 $\pm$ 0.1 & 1.695 $\pm$ 0.4 $\pm$ 0.1 $\pm$ 0.1 & 1.684 $\pm$ 0.4 $\pm$ 0.1 $\pm$ 0.1 & 1.683 $\pm$ 0.3 $\pm$ 0.1 $\pm$ 0.1\\
0.165 -- 0.189 & 1.396 $\pm$ 0.4 $\pm$ 0.1 $\pm$ 0.1 & 1.388 $\pm$ 0.4 $\pm$ 0.1 $\pm$ 0.1 & 1.381 $\pm$ 0.4 $\pm$ 0.1 $\pm$ 0.1 & 1.381 $\pm$ 0.4 $\pm$ 0.1 $\pm$ 0.1 & 1.373 $\pm$ 0.4 $\pm$ 0.1 $\pm$ 0.1 & 1.379 $\pm$ 0.3 $\pm$ 0.1 $\pm$ 0.1\\
0.189 -- 0.219 & 1.115 $\pm$ 0.4 $\pm$ 0.1 $\pm$ 0.1 & 1.108 $\pm$ 0.4 $\pm$ 0.1 $\pm$ 0.1 & 1.109 $\pm$ 0.4 $\pm$ 0.1 $\pm$ 0.1 & 1.110 $\pm$ 0.4 $\pm$ 0.1 $\pm$ 0.1 & 1.102 $\pm$ 0.4 $\pm$ 0.1 $\pm$ 0.1 & 1.105 $\pm$ 0.3 $\pm$ 0.1 $\pm$ 0.1\\
0.219 -- 0.258 & 0.853 $\pm$ 0.4 $\pm$ 0.1 $\pm$ 0.2 & 0.849 $\pm$ 0.4 $\pm$ 0.1 $\pm$ 0.2 & 0.848 $\pm$ 0.4 $\pm$ 0.1 $\pm$ 0.1 & 0.848 $\pm$ 0.4 $\pm$ 0.1 $\pm$ 0.1 & 0.843 $\pm$ 0.4 $\pm$ 0.1 $\pm$ 0.1 & 0.845 $\pm$ 0.3 $\pm$ 0.1 $\pm$ 0.1\\
0.258 -- 0.312 & 0.616 $\pm$ 0.4 $\pm$ 0.1 $\pm$ 0.2 & 0.613 $\pm$ 0.4 $\pm$ 0.1 $\pm$ 0.2 & 0.612 $\pm$ 0.4 $\pm$ 0.1 $\pm$ 0.1 & 0.612 $\pm$ 0.4 $\pm$ 0.1 $\pm$ 0.1 & 0.609 $\pm$ 0.4 $\pm$ 0.1 $\pm$ 0.2 & 0.610 $\pm$ 0.3 $\pm$ 0.1 $\pm$ 0.2\\
0.312 -- 0.391 & 0.410 $\pm$ 0.4 $\pm$ 0.1 $\pm$ 0.3 & 0.408 $\pm$ 0.4 $\pm$ 0.1 $\pm$ 0.3 & 0.410 $\pm$ 0.4 $\pm$ 0.1 $\pm$ 0.2 & 0.410 $\pm$ 0.4 $\pm$ 0.1 $\pm$ 0.2 & 0.409 $\pm$ 0.4 $\pm$ 0.1 $\pm$ 0.2 & 0.408 $\pm$ 0.3 $\pm$ 0.1 $\pm$ 0.2\\
0.391 -- 0.524 & 0.240 $\pm$ 0.5 $\pm$ 0.1 $\pm$ 0.3 & 0.240 $\pm$ 0.5 $\pm$ 0.1 $\pm$ 0.3 & 0.236 $\pm$ 0.4 $\pm$ 0.1 $\pm$ 0.2 & 0.236 $\pm$ 0.4 $\pm$ 0.1 $\pm$ 0.2 & 0.235 $\pm$ 0.4 $\pm$ 0.1 $\pm$ 0.2 & 0.237 $\pm$ 0.3 $\pm$ 0.1 $\pm$ 0.2\\
0.524 -- 0.695 & 0.124 $\pm$ 0.6 $\pm$ 0.2 $\pm$ 0.3 & 0.124 $\pm$ 0.6 $\pm$ 0.2 $\pm$ 0.3 & 0.124 $\pm$ 0.5 $\pm$ 0.1 $\pm$ 0.2 & 0.124 $\pm$ 0.5 $\pm$ 0.1 $\pm$ 0.2 & 0.124 $\pm$ 0.5 $\pm$ 0.1 $\pm$ 0.2 & 0.124 $\pm$ 0.4 $\pm$ 0.1 $\pm$ 0.2\\
0.695 -- 0.918 & 0.0629 $\pm$ 0.7 $\pm$ 0.2 $\pm$ 0.3 & 0.0627 $\pm$ 0.7 $\pm$ 0.2 $\pm$ 0.3 & 0.0623 $\pm$ 0.6 $\pm$ 0.1 $\pm$ 0.3 & 0.0623 $\pm$ 0.6 $\pm$ 0.1 $\pm$ 0.3 & 0.0621 $\pm$ 0.6 $\pm$ 0.1 $\pm$ 0.3 & 0.0623 $\pm$ 0.5 $\pm$ 0.1 $\pm$ 0.2\\
0.918 -- 1.153 & 0.0330 $\pm$ 0.9 $\pm$ 0.3 $\pm$ 0.3 & 0.0329 $\pm$ 0.9 $\pm$ 0.3 $\pm$ 0.3 & 0.0331 $\pm$ 0.8 $\pm$ 0.2 $\pm$ 0.3 & 0.0332 $\pm$ 0.8 $\pm$ 0.2 $\pm$ 0.3 & 0.0331 $\pm$ 0.8 $\pm$ 0.2 $\pm$ 0.3 & 0.0330 $\pm$ 0.6 $\pm$ 0.2 $\pm$ 0.2\\
1.153 -- 1.496 & 0.0175 $\pm$ 1.0 $\pm$ 0.3 $\pm$ 0.5 & 0.0174 $\pm$ 1.0 $\pm$ 0.3 $\pm$ 0.5 & 0.0174 $\pm$ 1.0 $\pm$ 0.2 $\pm$ 0.3 & 0.0174 $\pm$ 1.0 $\pm$ 0.2 $\pm$ 0.3 & 0.0173 $\pm$ 1.0 $\pm$ 0.2 $\pm$ 0.5 & 0.0173 $\pm$ 0.7 $\pm$ 0.2 $\pm$ 0.4\\
1.496 -- 1.947 & 0.00850 $\pm$ 1.3 $\pm$ 0.3 $\pm$ 0.5 & 0.00847 $\pm$ 1.3 $\pm$ 0.3 $\pm$ 0.5 & 0.00861 $\pm$ 1.2 $\pm$ 0.2 $\pm$ 0.4 & 0.00862 $\pm$ 1.2 $\pm$ 0.2 $\pm$ 0.4 & 0.00862 $\pm$ 1.2 $\pm$ 0.2 $\pm$ 0.5 & 0.00853 $\pm$ 0.9 $\pm$ 0.2 $\pm$ 0.4\\
1.947 -- 2.522 & 0.00440 $\pm$ 1.6 $\pm$ 0.4 $\pm$ 0.6 & 0.00439 $\pm$ 1.6 $\pm$ 0.4 $\pm$ 0.6 & 0.00438 $\pm$ 1.5 $\pm$ 0.4 $\pm$ 0.3 & 0.00439 $\pm$ 1.5 $\pm$ 0.4 $\pm$ 0.3 & 0.00437 $\pm$ 1.5 $\pm$ 0.4 $\pm$ 0.5 & 0.00437 $\pm$ 1.1 $\pm$ 0.3 $\pm$ 0.4\\
2.522 -- 3.277 & 0.00229 $\pm$ 2.0 $\pm$ 0.4 $\pm$ 0.6 & 0.00229 $\pm$ 2.0 $\pm$ 0.5 $\pm$ 0.6 & 0.00224 $\pm$ 1.8 $\pm$ 0.5 $\pm$ 0.4 & 0.00224 $\pm$ 1.8 $\pm$ 0.5 $\pm$ 0.4 & 0.00223 $\pm$ 1.8 $\pm$ 0.5 $\pm$ 0.6 & 0.00225 $\pm$ 1.3 $\pm$ 0.3 $\pm$ 0.4\\
3.277 -- 5.000 & 0.000994 $\pm$ 2.0 $\pm$ 0.6 $\pm$ 0.8 & 0.000993 $\pm$ 2.0 $\pm$ 0.6 $\pm$ 0.8 & 0.00101 $\pm$ 1.8 $\pm$ 0.4 $\pm$ 0.3 & 0.00101 $\pm$ 1.8 $\pm$ 0.4 $\pm$ 0.3 & 0.00102 $\pm$ 1.8 $\pm$ 0.4 $\pm$ 0.5 & 0.00100 $\pm$ 1.3 $\pm$ 0.4 $\pm$ 0.4\\
5.000 -- 10.000 & 0.000309 $\pm$ 2.1 $\pm$ 0.5 $\pm$ 0.8 & 0.000308 $\pm$ 2.1 $\pm$ 0.5 $\pm$ 0.8 & 0.000297 $\pm$ 1.9 $\pm$ 0.5 $\pm$ 0.4 & 0.000297 $\pm$ 1.9 $\pm$ 0.5 $\pm$ 0.4 & 0.000297 $\pm$ 1.9 $\pm$ 0.5 $\pm$ 0.6 & 0.000301 $\pm$ 1.4 $\pm$ 0.4 $\pm$ 0.5\\
\bottomrule 
\end{tabular} 
} 
\label{tab:CombPhiStarpeakmass_y0408}
\end{table*}

\begin{table*}
\centering
\caption{The values of $(1/\sigma)\,\mathrm{d}\sigma/\mathrm{d}\phi^*_{\eta}$ in each bin of \PhiStar{} for the electron and muon channels separately (for various particle-level definitions) and for the Born-level combination in the kinematic region $66\ \GeV \leq m_{\ell\ell} < 116\ \GeV,\ 0.8 \leq |y_{\ell\ell}| < 1.2$. The associated statistical and systematic (both uncorrelated and correlated between bins of \PhiStar{}) are provided in percentage form.} 
\resizebox{\textwidth}{!}{
\begin{tabular}{ ccccccc } \toprule 
Bin & \multicolumn{6}{c}{$(1/\sigma)\,d\sigma/d\phi^*_{\eta}$ $\pm$ Statistical [\%] $\pm$ Uncorrelated systematic [\%] $\pm$ Correlated systematic [\%]} \\ 
 \cmidrule(r){2-7} 
 & \multicolumn{2}{c}{Electron channel} &  \multicolumn{3}{c}{Muon channel} & Combination \\ 
\cmidrule(r){2-3} \cmidrule(r){4-6} \cmidrule(r){7-7} 
 & dressed & Born & bare & dressed & Born & Born \\ 
 0.000 -- 0.004 & 9.387 $\pm$ 0.4 $\pm$ 0.1 $\pm$ 0.2 & 9.476 $\pm$ 0.4 $\pm$ 0.1 $\pm$ 0.2 & 9.394 $\pm$ 0.4 $\pm$ 0.1 $\pm$ 0.1 & 9.388 $\pm$ 0.4 $\pm$ 0.1 $\pm$ 0.1 & 9.455 $\pm$ 0.4 $\pm$ 0.1 $\pm$ 0.2 & 9.458 $\pm$ 0.3 $\pm$ 0.1 $\pm$ 0.2\\
0.004 -- 0.008 & 9.268 $\pm$ 0.4 $\pm$ 0.1 $\pm$ 0.1 & 9.352 $\pm$ 0.4 $\pm$ 0.1 $\pm$ 0.1 & 9.374 $\pm$ 0.4 $\pm$ 0.1 $\pm$ 0.0 & 9.371 $\pm$ 0.4 $\pm$ 0.1 $\pm$ 0.0 & 9.462 $\pm$ 0.4 $\pm$ 0.1 $\pm$ 0.2 & 9.412 $\pm$ 0.3 $\pm$ 0.1 $\pm$ 0.1\\
0.008 -- 0.012 & 9.252 $\pm$ 0.4 $\pm$ 0.1 $\pm$ 0.1 & 9.330 $\pm$ 0.4 $\pm$ 0.1 $\pm$ 0.1 & 9.160 $\pm$ 0.4 $\pm$ 0.1 $\pm$ 0.0 & 9.155 $\pm$ 0.4 $\pm$ 0.1 $\pm$ 0.0 & 9.231 $\pm$ 0.4 $\pm$ 0.1 $\pm$ 0.2 & 9.265 $\pm$ 0.3 $\pm$ 0.1 $\pm$ 0.1\\
0.012 -- 0.016 & 8.913 $\pm$ 0.5 $\pm$ 0.2 $\pm$ 0.1 & 8.989 $\pm$ 0.5 $\pm$ 0.2 $\pm$ 0.1 & 8.907 $\pm$ 0.4 $\pm$ 0.1 $\pm$ 0.0 & 8.900 $\pm$ 0.4 $\pm$ 0.1 $\pm$ 0.0 & 8.968 $\pm$ 0.4 $\pm$ 0.1 $\pm$ 0.2 & 8.973 $\pm$ 0.3 $\pm$ 0.1 $\pm$ 0.1\\
0.016 -- 0.020 & 8.699 $\pm$ 0.5 $\pm$ 0.2 $\pm$ 0.1 & 8.772 $\pm$ 0.5 $\pm$ 0.2 $\pm$ 0.1 & 8.619 $\pm$ 0.4 $\pm$ 0.1 $\pm$ 0.0 & 8.613 $\pm$ 0.4 $\pm$ 0.1 $\pm$ 0.0 & 8.681 $\pm$ 0.4 $\pm$ 0.1 $\pm$ 0.2 & 8.712 $\pm$ 0.3 $\pm$ 0.1 $\pm$ 0.1\\
0.020 -- 0.024 & 8.264 $\pm$ 0.5 $\pm$ 0.1 $\pm$ 0.1 & 8.312 $\pm$ 0.5 $\pm$ 0.1 $\pm$ 0.1 & 8.352 $\pm$ 0.4 $\pm$ 0.1 $\pm$ 0.0 & 8.348 $\pm$ 0.4 $\pm$ 0.1 $\pm$ 0.0 & 8.394 $\pm$ 0.4 $\pm$ 0.1 $\pm$ 0.2 & 8.356 $\pm$ 0.3 $\pm$ 0.1 $\pm$ 0.1\\
0.024 -- 0.029 & 7.876 $\pm$ 0.4 $\pm$ 0.1 $\pm$ 0.1 & 7.920 $\pm$ 0.4 $\pm$ 0.1 $\pm$ 0.1 & 7.869 $\pm$ 0.4 $\pm$ 0.1 $\pm$ 0.1 & 7.868 $\pm$ 0.4 $\pm$ 0.1 $\pm$ 0.1 & 7.926 $\pm$ 0.4 $\pm$ 0.1 $\pm$ 0.1 & 7.918 $\pm$ 0.3 $\pm$ 0.1 $\pm$ 0.1\\
0.029 -- 0.034 & 7.364 $\pm$ 0.4 $\pm$ 0.2 $\pm$ 0.1 & 7.396 $\pm$ 0.4 $\pm$ 0.2 $\pm$ 0.1 & 7.443 $\pm$ 0.4 $\pm$ 0.1 $\pm$ 0.1 & 7.437 $\pm$ 0.4 $\pm$ 0.1 $\pm$ 0.1 & 7.473 $\pm$ 0.4 $\pm$ 0.1 $\pm$ 0.1 & 7.439 $\pm$ 0.3 $\pm$ 0.1 $\pm$ 0.1\\
0.034 -- 0.039 & 6.923 $\pm$ 0.5 $\pm$ 0.2 $\pm$ 0.1 & 6.950 $\pm$ 0.5 $\pm$ 0.2 $\pm$ 0.1 & 6.915 $\pm$ 0.4 $\pm$ 0.1 $\pm$ 0.0 & 6.906 $\pm$ 0.4 $\pm$ 0.1 $\pm$ 0.0 & 6.934 $\pm$ 0.4 $\pm$ 0.1 $\pm$ 0.1 & 6.937 $\pm$ 0.3 $\pm$ 0.1 $\pm$ 0.1\\
0.039 -- 0.045 & 6.430 $\pm$ 0.4 $\pm$ 0.1 $\pm$ 0.1 & 6.450 $\pm$ 0.4 $\pm$ 0.1 $\pm$ 0.1 & 6.484 $\pm$ 0.4 $\pm$ 0.1 $\pm$ 0.1 & 6.483 $\pm$ 0.4 $\pm$ 0.1 $\pm$ 0.1 & 6.499 $\pm$ 0.4 $\pm$ 0.1 $\pm$ 0.1 & 6.475 $\pm$ 0.3 $\pm$ 0.1 $\pm$ 0.1\\
0.045 -- 0.051 & 5.921 $\pm$ 0.5 $\pm$ 0.1 $\pm$ 0.1 & 5.923 $\pm$ 0.5 $\pm$ 0.2 $\pm$ 0.1 & 5.884 $\pm$ 0.4 $\pm$ 0.1 $\pm$ 0.1 & 5.884 $\pm$ 0.4 $\pm$ 0.1 $\pm$ 0.1 & 5.898 $\pm$ 0.4 $\pm$ 0.1 $\pm$ 0.1 & 5.905 $\pm$ 0.3 $\pm$ 0.1 $\pm$ 0.1\\
0.051 -- 0.057 & 5.410 $\pm$ 0.5 $\pm$ 0.1 $\pm$ 0.1 & 5.410 $\pm$ 0.5 $\pm$ 0.1 $\pm$ 0.1 & 5.469 $\pm$ 0.4 $\pm$ 0.1 $\pm$ 0.1 & 5.470 $\pm$ 0.4 $\pm$ 0.1 $\pm$ 0.1 & 5.466 $\pm$ 0.4 $\pm$ 0.1 $\pm$ 0.1 & 5.441 $\pm$ 0.3 $\pm$ 0.1 $\pm$ 0.1\\
0.057 -- 0.064 & 5.012 $\pm$ 0.5 $\pm$ 0.1 $\pm$ 0.1 & 5.008 $\pm$ 0.5 $\pm$ 0.1 $\pm$ 0.1 & 5.019 $\pm$ 0.4 $\pm$ 0.1 $\pm$ 0.1 & 5.016 $\pm$ 0.4 $\pm$ 0.1 $\pm$ 0.1 & 5.023 $\pm$ 0.4 $\pm$ 0.1 $\pm$ 0.1 & 5.015 $\pm$ 0.3 $\pm$ 0.1 $\pm$ 0.1\\
0.064 -- 0.072 & 4.492 $\pm$ 0.5 $\pm$ 0.1 $\pm$ 0.1 & 4.483 $\pm$ 0.5 $\pm$ 0.1 $\pm$ 0.1 & 4.506 $\pm$ 0.4 $\pm$ 0.1 $\pm$ 0.1 & 4.506 $\pm$ 0.4 $\pm$ 0.1 $\pm$ 0.1 & 4.498 $\pm$ 0.4 $\pm$ 0.1 $\pm$ 0.1 & 4.491 $\pm$ 0.3 $\pm$ 0.1 $\pm$ 0.1\\
0.072 -- 0.081 & 4.019 $\pm$ 0.5 $\pm$ 0.1 $\pm$ 0.1 & 4.009 $\pm$ 0.5 $\pm$ 0.1 $\pm$ 0.1 & 4.038 $\pm$ 0.4 $\pm$ 0.1 $\pm$ 0.1 & 4.037 $\pm$ 0.4 $\pm$ 0.1 $\pm$ 0.1 & 4.024 $\pm$ 0.4 $\pm$ 0.1 $\pm$ 0.1 & 4.016 $\pm$ 0.3 $\pm$ 0.1 $\pm$ 0.1\\
0.081 -- 0.091 & 3.589 $\pm$ 0.5 $\pm$ 0.1 $\pm$ 0.1 & 3.576 $\pm$ 0.5 $\pm$ 0.1 $\pm$ 0.1 & 3.566 $\pm$ 0.4 $\pm$ 0.1 $\pm$ 0.1 & 3.564 $\pm$ 0.4 $\pm$ 0.1 $\pm$ 0.1 & 3.550 $\pm$ 0.4 $\pm$ 0.1 $\pm$ 0.1 & 3.559 $\pm$ 0.3 $\pm$ 0.1 $\pm$ 0.1\\
0.091 -- 0.102 & 3.141 $\pm$ 0.5 $\pm$ 0.1 $\pm$ 0.1 & 3.128 $\pm$ 0.5 $\pm$ 0.1 $\pm$ 0.1 & 3.147 $\pm$ 0.4 $\pm$ 0.1 $\pm$ 0.1 & 3.149 $\pm$ 0.4 $\pm$ 0.1 $\pm$ 0.1 & 3.138 $\pm$ 0.4 $\pm$ 0.1 $\pm$ 0.1 & 3.133 $\pm$ 0.3 $\pm$ 0.1 $\pm$ 0.1\\
0.102 -- 0.114 & 2.755 $\pm$ 0.5 $\pm$ 0.1 $\pm$ 0.1 & 2.740 $\pm$ 0.5 $\pm$ 0.1 $\pm$ 0.1 & 2.746 $\pm$ 0.4 $\pm$ 0.1 $\pm$ 0.1 & 2.745 $\pm$ 0.4 $\pm$ 0.1 $\pm$ 0.1 & 2.732 $\pm$ 0.4 $\pm$ 0.1 $\pm$ 0.1 & 2.734 $\pm$ 0.3 $\pm$ 0.1 $\pm$ 0.1\\
0.114 -- 0.128 & 2.394 $\pm$ 0.5 $\pm$ 0.1 $\pm$ 0.1 & 2.380 $\pm$ 0.5 $\pm$ 0.1 $\pm$ 0.1 & 2.388 $\pm$ 0.4 $\pm$ 0.1 $\pm$ 0.1 & 2.388 $\pm$ 0.4 $\pm$ 0.1 $\pm$ 0.1 & 2.374 $\pm$ 0.4 $\pm$ 0.1 $\pm$ 0.1 & 2.375 $\pm$ 0.3 $\pm$ 0.1 $\pm$ 0.1\\
0.128 -- 0.145 & 2.029 $\pm$ 0.5 $\pm$ 0.1 $\pm$ 0.1 & 2.017 $\pm$ 0.5 $\pm$ 0.1 $\pm$ 0.1 & 2.025 $\pm$ 0.4 $\pm$ 0.1 $\pm$ 0.1 & 2.025 $\pm$ 0.4 $\pm$ 0.1 $\pm$ 0.1 & 2.015 $\pm$ 0.4 $\pm$ 0.1 $\pm$ 0.1 & 2.015 $\pm$ 0.3 $\pm$ 0.1 $\pm$ 0.1\\
0.145 -- 0.165 & 1.695 $\pm$ 0.5 $\pm$ 0.1 $\pm$ 0.1 & 1.684 $\pm$ 0.5 $\pm$ 0.1 $\pm$ 0.1 & 1.678 $\pm$ 0.4 $\pm$ 0.1 $\pm$ 0.1 & 1.679 $\pm$ 0.4 $\pm$ 0.1 $\pm$ 0.1 & 1.667 $\pm$ 0.4 $\pm$ 0.1 $\pm$ 0.1 & 1.673 $\pm$ 0.3 $\pm$ 0.1 $\pm$ 0.1\\
0.165 -- 0.189 & 1.378 $\pm$ 0.5 $\pm$ 0.1 $\pm$ 0.1 & 1.368 $\pm$ 0.5 $\pm$ 0.1 $\pm$ 0.1 & 1.394 $\pm$ 0.4 $\pm$ 0.1 $\pm$ 0.1 & 1.394 $\pm$ 0.4 $\pm$ 0.1 $\pm$ 0.1 & 1.385 $\pm$ 0.4 $\pm$ 0.1 $\pm$ 0.1 & 1.378 $\pm$ 0.3 $\pm$ 0.1 $\pm$ 0.1\\
0.189 -- 0.219 & 1.106 $\pm$ 0.5 $\pm$ 0.1 $\pm$ 0.1 & 1.100 $\pm$ 0.5 $\pm$ 0.1 $\pm$ 0.1 & 1.109 $\pm$ 0.4 $\pm$ 0.1 $\pm$ 0.1 & 1.110 $\pm$ 0.4 $\pm$ 0.1 $\pm$ 0.1 & 1.105 $\pm$ 0.4 $\pm$ 0.1 $\pm$ 0.2 & 1.102 $\pm$ 0.3 $\pm$ 0.1 $\pm$ 0.1\\
0.219 -- 0.258 & 0.845 $\pm$ 0.5 $\pm$ 0.1 $\pm$ 0.1 & 0.840 $\pm$ 0.5 $\pm$ 0.1 $\pm$ 0.1 & 0.843 $\pm$ 0.4 $\pm$ 0.1 $\pm$ 0.1 & 0.843 $\pm$ 0.4 $\pm$ 0.1 $\pm$ 0.1 & 0.838 $\pm$ 0.4 $\pm$ 0.1 $\pm$ 0.1 & 0.839 $\pm$ 0.3 $\pm$ 0.1 $\pm$ 0.1\\
0.258 -- 0.312 & 0.614 $\pm$ 0.5 $\pm$ 0.1 $\pm$ 0.2 & 0.611 $\pm$ 0.5 $\pm$ 0.1 $\pm$ 0.2 & 0.609 $\pm$ 0.4 $\pm$ 0.1 $\pm$ 0.1 & 0.609 $\pm$ 0.4 $\pm$ 0.1 $\pm$ 0.1 & 0.606 $\pm$ 0.4 $\pm$ 0.1 $\pm$ 0.2 & 0.608 $\pm$ 0.3 $\pm$ 0.1 $\pm$ 0.2\\
0.312 -- 0.391 & 0.409 $\pm$ 0.5 $\pm$ 0.1 $\pm$ 0.3 & 0.407 $\pm$ 0.5 $\pm$ 0.1 $\pm$ 0.3 & 0.408 $\pm$ 0.4 $\pm$ 0.1 $\pm$ 0.2 & 0.408 $\pm$ 0.4 $\pm$ 0.1 $\pm$ 0.2 & 0.406 $\pm$ 0.4 $\pm$ 0.1 $\pm$ 0.2 & 0.406 $\pm$ 0.3 $\pm$ 0.1 $\pm$ 0.2\\
0.391 -- 0.524 & 0.238 $\pm$ 0.5 $\pm$ 0.1 $\pm$ 0.3 & 0.237 $\pm$ 0.5 $\pm$ 0.1 $\pm$ 0.3 & 0.236 $\pm$ 0.4 $\pm$ 0.1 $\pm$ 0.2 & 0.236 $\pm$ 0.4 $\pm$ 0.1 $\pm$ 0.2 & 0.235 $\pm$ 0.4 $\pm$ 0.1 $\pm$ 0.2 & 0.236 $\pm$ 0.3 $\pm$ 0.1 $\pm$ 0.2\\
0.524 -- 0.695 & 0.122 $\pm$ 0.6 $\pm$ 0.3 $\pm$ 0.3 & 0.122 $\pm$ 0.6 $\pm$ 0.3 $\pm$ 0.3 & 0.121 $\pm$ 0.5 $\pm$ 0.1 $\pm$ 0.2 & 0.121 $\pm$ 0.5 $\pm$ 0.1 $\pm$ 0.2 & 0.121 $\pm$ 0.5 $\pm$ 0.1 $\pm$ 0.3 & 0.121 $\pm$ 0.4 $\pm$ 0.1 $\pm$ 0.2\\
0.695 -- 0.918 & 0.0624 $\pm$ 0.7 $\pm$ 0.3 $\pm$ 0.3 & 0.0622 $\pm$ 0.7 $\pm$ 0.3 $\pm$ 0.3 & 0.0616 $\pm$ 0.6 $\pm$ 0.1 $\pm$ 0.2 & 0.0617 $\pm$ 0.6 $\pm$ 0.1 $\pm$ 0.2 & 0.0615 $\pm$ 0.6 $\pm$ 0.1 $\pm$ 0.3 & 0.0618 $\pm$ 0.5 $\pm$ 0.2 $\pm$ 0.2\\
0.918 -- 1.153 & 0.0321 $\pm$ 1.0 $\pm$ 0.3 $\pm$ 0.3 & 0.0320 $\pm$ 1.0 $\pm$ 0.3 $\pm$ 0.3 & 0.0323 $\pm$ 0.8 $\pm$ 0.2 $\pm$ 0.2 & 0.0323 $\pm$ 0.8 $\pm$ 0.2 $\pm$ 0.2 & 0.0322 $\pm$ 0.8 $\pm$ 0.2 $\pm$ 0.3 & 0.0321 $\pm$ 0.6 $\pm$ 0.2 $\pm$ 0.2\\
1.153 -- 1.496 & 0.0165 $\pm$ 1.1 $\pm$ 0.3 $\pm$ 0.4 & 0.0165 $\pm$ 1.1 $\pm$ 0.3 $\pm$ 0.4 & 0.0166 $\pm$ 1.0 $\pm$ 0.2 $\pm$ 0.2 & 0.0167 $\pm$ 1.0 $\pm$ 0.2 $\pm$ 0.2 & 0.0166 $\pm$ 1.0 $\pm$ 0.2 $\pm$ 0.3 & 0.0165 $\pm$ 0.7 $\pm$ 0.2 $\pm$ 0.3\\
1.496 -- 1.947 & 0.00801 $\pm$ 1.4 $\pm$ 0.4 $\pm$ 0.4 & 0.00797 $\pm$ 1.4 $\pm$ 0.4 $\pm$ 0.4 & 0.00835 $\pm$ 1.2 $\pm$ 0.3 $\pm$ 0.3 & 0.00838 $\pm$ 1.2 $\pm$ 0.3 $\pm$ 0.3 & 0.00833 $\pm$ 1.2 $\pm$ 0.3 $\pm$ 0.3 & 0.00817 $\pm$ 0.9 $\pm$ 0.2 $\pm$ 0.3\\
1.947 -- 2.522 & 0.00412 $\pm$ 1.7 $\pm$ 0.4 $\pm$ 0.4 & 0.00410 $\pm$ 1.7 $\pm$ 0.4 $\pm$ 0.4 & 0.00406 $\pm$ 1.6 $\pm$ 0.3 $\pm$ 0.4 & 0.00407 $\pm$ 1.6 $\pm$ 0.3 $\pm$ 0.4 & 0.00405 $\pm$ 1.6 $\pm$ 0.3 $\pm$ 0.4 & 0.00407 $\pm$ 1.2 $\pm$ 0.3 $\pm$ 0.3\\
2.522 -- 3.277 & 0.00210 $\pm$ 2.1 $\pm$ 0.5 $\pm$ 0.5 & 0.00209 $\pm$ 2.1 $\pm$ 0.5 $\pm$ 0.5 & 0.00207 $\pm$ 1.9 $\pm$ 0.4 $\pm$ 0.3 & 0.00208 $\pm$ 1.9 $\pm$ 0.4 $\pm$ 0.3 & 0.00207 $\pm$ 1.9 $\pm$ 0.4 $\pm$ 0.4 & 0.00207 $\pm$ 1.4 $\pm$ 0.3 $\pm$ 0.3\\
3.277 -- 5.000 & 0.000942 $\pm$ 2.1 $\pm$ 0.5 $\pm$ 0.8 & 0.000940 $\pm$ 2.1 $\pm$ 0.6 $\pm$ 0.8 & 0.000909 $\pm$ 1.8 $\pm$ 0.4 $\pm$ 0.4 & 0.000909 $\pm$ 1.9 $\pm$ 0.4 $\pm$ 0.4 & 0.000909 $\pm$ 1.9 $\pm$ 0.4 $\pm$ 0.4 & 0.000922 $\pm$ 1.4 $\pm$ 0.3 $\pm$ 0.4\\
5.000 -- 10.000 & 0.000282 $\pm$ 2.3 $\pm$ 0.5 $\pm$ 0.6 & 0.000282 $\pm$ 2.3 $\pm$ 0.5 $\pm$ 0.6 & 0.000274 $\pm$ 2.0 $\pm$ 0.5 $\pm$ 0.4 & 0.000275 $\pm$ 2.0 $\pm$ 0.5 $\pm$ 0.4 & 0.000273 $\pm$ 2.0 $\pm$ 0.5 $\pm$ 0.5 & 0.000276 $\pm$ 1.5 $\pm$ 0.3 $\pm$ 0.4\\
\bottomrule 
\end{tabular} 
} 
\label{tab:CombPhiStarpeakmass_y0812}
\end{table*}

\begin{table*}
\centering
\caption{The values of $(1/\sigma)\,\mathrm{d}\sigma/\mathrm{d}\phi^*_{\eta}$ in each bin of \PhiStar{} for the electron and muon channels separately (for various particle-level definitions) and for the Born-level combination in the kinematic region $66\ \GeV \leq m_{\ell\ell} < 116\ \GeV,\ 1.2 \leq |y_{\ell\ell}| < 1.6$. The associated statistical and systematic (both uncorrelated and correlated between bins of \PhiStar{}) are provided in percentage form.} 
\resizebox{\textwidth}{!}{
\begin{tabular}{ ccccccc } \toprule 
Bin & \multicolumn{6}{c}{$(1/\sigma)\,d\sigma/d\phi^*_{\eta}$ $\pm$ Statistical [\%] $\pm$ Uncorrelated systematic [\%] $\pm$ Correlated systematic [\%]} \\ 
 \cmidrule(r){2-7} 
 & \multicolumn{2}{c}{Electron channel} &  \multicolumn{3}{c}{Muon channel} & Combination \\ 
\cmidrule(r){2-3} \cmidrule(r){4-6} \cmidrule(r){7-7} 
 & dressed & Born & bare & dressed & Born & Born \\ 
 0.000 -- 0.004 & 9.382 $\pm$ 0.5 $\pm$ 0.1 $\pm$ 0.2 & 9.470 $\pm$ 0.5 $\pm$ 0.2 $\pm$ 0.2 & 9.358 $\pm$ 0.4 $\pm$ 0.1 $\pm$ 0.1 & 9.352 $\pm$ 0.4 $\pm$ 0.1 $\pm$ 0.1 & 9.433 $\pm$ 0.4 $\pm$ 0.1 $\pm$ 0.2 & 9.443 $\pm$ 0.3 $\pm$ 0.1 $\pm$ 0.2\\
0.004 -- 0.008 & 9.317 $\pm$ 0.5 $\pm$ 0.2 $\pm$ 0.2 & 9.402 $\pm$ 0.5 $\pm$ 0.2 $\pm$ 0.2 & 9.289 $\pm$ 0.4 $\pm$ 0.1 $\pm$ 0.1 & 9.281 $\pm$ 0.4 $\pm$ 0.1 $\pm$ 0.1 & 9.363 $\pm$ 0.4 $\pm$ 0.1 $\pm$ 0.2 & 9.374 $\pm$ 0.3 $\pm$ 0.1 $\pm$ 0.1\\
0.008 -- 0.012 & 9.095 $\pm$ 0.5 $\pm$ 0.2 $\pm$ 0.2 & 9.169 $\pm$ 0.5 $\pm$ 0.2 $\pm$ 0.2 & 9.101 $\pm$ 0.4 $\pm$ 0.1 $\pm$ 0.0 & 9.093 $\pm$ 0.4 $\pm$ 0.1 $\pm$ 0.0 & 9.167 $\pm$ 0.4 $\pm$ 0.1 $\pm$ 0.2 & 9.165 $\pm$ 0.3 $\pm$ 0.1 $\pm$ 0.1\\
0.012 -- 0.016 & 8.874 $\pm$ 0.5 $\pm$ 0.1 $\pm$ 0.2 & 8.945 $\pm$ 0.5 $\pm$ 0.2 $\pm$ 0.2 & 8.918 $\pm$ 0.4 $\pm$ 0.1 $\pm$ 0.0 & 8.905 $\pm$ 0.4 $\pm$ 0.1 $\pm$ 0.0 & 8.999 $\pm$ 0.4 $\pm$ 0.1 $\pm$ 0.2 & 8.975 $\pm$ 0.3 $\pm$ 0.1 $\pm$ 0.1\\
0.016 -- 0.020 & 8.552 $\pm$ 0.5 $\pm$ 0.2 $\pm$ 0.2 & 8.617 $\pm$ 0.5 $\pm$ 0.2 $\pm$ 0.2 & 8.632 $\pm$ 0.4 $\pm$ 0.1 $\pm$ 0.0 & 8.629 $\pm$ 0.4 $\pm$ 0.1 $\pm$ 0.0 & 8.682 $\pm$ 0.4 $\pm$ 0.1 $\pm$ 0.2 & 8.655 $\pm$ 0.3 $\pm$ 0.1 $\pm$ 0.1\\
0.020 -- 0.024 & 8.242 $\pm$ 0.5 $\pm$ 0.2 $\pm$ 0.2 & 8.299 $\pm$ 0.5 $\pm$ 0.2 $\pm$ 0.2 & 8.265 $\pm$ 0.4 $\pm$ 0.1 $\pm$ 0.0 & 8.264 $\pm$ 0.4 $\pm$ 0.1 $\pm$ 0.0 & 8.332 $\pm$ 0.4 $\pm$ 0.1 $\pm$ 0.2 & 8.317 $\pm$ 0.3 $\pm$ 0.1 $\pm$ 0.1\\
0.024 -- 0.029 & 7.803 $\pm$ 0.5 $\pm$ 0.2 $\pm$ 0.2 & 7.841 $\pm$ 0.5 $\pm$ 0.2 $\pm$ 0.2 & 7.805 $\pm$ 0.4 $\pm$ 0.1 $\pm$ 0.0 & 7.792 $\pm$ 0.4 $\pm$ 0.1 $\pm$ 0.0 & 7.835 $\pm$ 0.4 $\pm$ 0.1 $\pm$ 0.2 & 7.833 $\pm$ 0.3 $\pm$ 0.1 $\pm$ 0.2\\
0.029 -- 0.034 & 7.312 $\pm$ 0.5 $\pm$ 0.1 $\pm$ 0.2 & 7.340 $\pm$ 0.5 $\pm$ 0.2 $\pm$ 0.2 & 7.371 $\pm$ 0.4 $\pm$ 0.1 $\pm$ 0.0 & 7.363 $\pm$ 0.4 $\pm$ 0.1 $\pm$ 0.0 & 7.395 $\pm$ 0.4 $\pm$ 0.1 $\pm$ 0.2 & 7.371 $\pm$ 0.3 $\pm$ 0.1 $\pm$ 0.2\\
0.034 -- 0.039 & 6.920 $\pm$ 0.5 $\pm$ 0.2 $\pm$ 0.2 & 6.938 $\pm$ 0.5 $\pm$ 0.2 $\pm$ 0.2 & 6.907 $\pm$ 0.4 $\pm$ 0.1 $\pm$ 0.0 & 6.914 $\pm$ 0.4 $\pm$ 0.1 $\pm$ 0.0 & 6.944 $\pm$ 0.4 $\pm$ 0.1 $\pm$ 0.2 & 6.939 $\pm$ 0.3 $\pm$ 0.1 $\pm$ 0.2\\
0.039 -- 0.045 & 6.397 $\pm$ 0.5 $\pm$ 0.1 $\pm$ 0.2 & 6.413 $\pm$ 0.5 $\pm$ 0.1 $\pm$ 0.2 & 6.401 $\pm$ 0.4 $\pm$ 0.1 $\pm$ 0.0 & 6.404 $\pm$ 0.4 $\pm$ 0.1 $\pm$ 0.0 & 6.427 $\pm$ 0.4 $\pm$ 0.1 $\pm$ 0.2 & 6.420 $\pm$ 0.3 $\pm$ 0.1 $\pm$ 0.2\\
0.045 -- 0.051 & 5.877 $\pm$ 0.5 $\pm$ 0.2 $\pm$ 0.1 & 5.886 $\pm$ 0.5 $\pm$ 0.2 $\pm$ 0.1 & 5.914 $\pm$ 0.4 $\pm$ 0.1 $\pm$ 0.0 & 5.908 $\pm$ 0.4 $\pm$ 0.1 $\pm$ 0.0 & 5.910 $\pm$ 0.4 $\pm$ 0.1 $\pm$ 0.2 & 5.898 $\pm$ 0.3 $\pm$ 0.1 $\pm$ 0.2\\
0.051 -- 0.057 & 5.366 $\pm$ 0.5 $\pm$ 0.2 $\pm$ 0.1 & 5.364 $\pm$ 0.5 $\pm$ 0.2 $\pm$ 0.1 & 5.417 $\pm$ 0.4 $\pm$ 0.1 $\pm$ 0.0 & 5.415 $\pm$ 0.4 $\pm$ 0.1 $\pm$ 0.0 & 5.416 $\pm$ 0.4 $\pm$ 0.1 $\pm$ 0.2 & 5.394 $\pm$ 0.3 $\pm$ 0.1 $\pm$ 0.2\\
0.057 -- 0.064 & 4.985 $\pm$ 0.5 $\pm$ 0.1 $\pm$ 0.1 & 4.985 $\pm$ 0.5 $\pm$ 0.2 $\pm$ 0.1 & 4.981 $\pm$ 0.4 $\pm$ 0.1 $\pm$ 0.1 & 4.979 $\pm$ 0.4 $\pm$ 0.1 $\pm$ 0.1 & 4.982 $\pm$ 0.4 $\pm$ 0.1 $\pm$ 0.2 & 4.980 $\pm$ 0.3 $\pm$ 0.1 $\pm$ 0.2\\
0.064 -- 0.072 & 4.526 $\pm$ 0.5 $\pm$ 0.1 $\pm$ 0.1 & 4.519 $\pm$ 0.5 $\pm$ 0.1 $\pm$ 0.1 & 4.524 $\pm$ 0.4 $\pm$ 0.1 $\pm$ 0.1 & 4.521 $\pm$ 0.4 $\pm$ 0.1 $\pm$ 0.1 & 4.503 $\pm$ 0.4 $\pm$ 0.1 $\pm$ 0.2 & 4.506 $\pm$ 0.3 $\pm$ 0.1 $\pm$ 0.2\\
0.072 -- 0.081 & 4.049 $\pm$ 0.5 $\pm$ 0.2 $\pm$ 0.1 & 4.037 $\pm$ 0.5 $\pm$ 0.2 $\pm$ 0.1 & 4.071 $\pm$ 0.4 $\pm$ 0.1 $\pm$ 0.1 & 4.071 $\pm$ 0.4 $\pm$ 0.1 $\pm$ 0.1 & 4.059 $\pm$ 0.4 $\pm$ 0.1 $\pm$ 0.2 & 4.049 $\pm$ 0.3 $\pm$ 0.1 $\pm$ 0.1\\
0.081 -- 0.091 & 3.593 $\pm$ 0.5 $\pm$ 0.2 $\pm$ 0.1 & 3.576 $\pm$ 0.5 $\pm$ 0.2 $\pm$ 0.1 & 3.590 $\pm$ 0.4 $\pm$ 0.1 $\pm$ 0.1 & 3.587 $\pm$ 0.4 $\pm$ 0.1 $\pm$ 0.1 & 3.571 $\pm$ 0.4 $\pm$ 0.1 $\pm$ 0.2 & 3.570 $\pm$ 0.3 $\pm$ 0.1 $\pm$ 0.1\\
0.091 -- 0.102 & 3.196 $\pm$ 0.5 $\pm$ 0.2 $\pm$ 0.1 & 3.182 $\pm$ 0.5 $\pm$ 0.2 $\pm$ 0.1 & 3.159 $\pm$ 0.4 $\pm$ 0.1 $\pm$ 0.1 & 3.158 $\pm$ 0.4 $\pm$ 0.1 $\pm$ 0.1 & 3.142 $\pm$ 0.4 $\pm$ 0.1 $\pm$ 0.2 & 3.154 $\pm$ 0.3 $\pm$ 0.1 $\pm$ 0.1\\
0.102 -- 0.114 & 2.756 $\pm$ 0.5 $\pm$ 0.2 $\pm$ 0.1 & 2.742 $\pm$ 0.5 $\pm$ 0.2 $\pm$ 0.1 & 2.781 $\pm$ 0.4 $\pm$ 0.1 $\pm$ 0.1 & 2.784 $\pm$ 0.4 $\pm$ 0.1 $\pm$ 0.1 & 2.769 $\pm$ 0.4 $\pm$ 0.1 $\pm$ 0.2 & 2.757 $\pm$ 0.3 $\pm$ 0.1 $\pm$ 0.1\\
0.114 -- 0.128 & 2.403 $\pm$ 0.5 $\pm$ 0.2 $\pm$ 0.1 & 2.388 $\pm$ 0.5 $\pm$ 0.2 $\pm$ 0.1 & 2.418 $\pm$ 0.4 $\pm$ 0.1 $\pm$ 0.1 & 2.418 $\pm$ 0.4 $\pm$ 0.1 $\pm$ 0.1 & 2.403 $\pm$ 0.4 $\pm$ 0.1 $\pm$ 0.1 & 2.396 $\pm$ 0.3 $\pm$ 0.1 $\pm$ 0.1\\
0.128 -- 0.145 & 2.065 $\pm$ 0.5 $\pm$ 0.2 $\pm$ 0.1 & 2.053 $\pm$ 0.5 $\pm$ 0.2 $\pm$ 0.1 & 2.045 $\pm$ 0.4 $\pm$ 0.1 $\pm$ 0.1 & 2.045 $\pm$ 0.4 $\pm$ 0.1 $\pm$ 0.1 & 2.032 $\pm$ 0.4 $\pm$ 0.1 $\pm$ 0.1 & 2.039 $\pm$ 0.3 $\pm$ 0.1 $\pm$ 0.1\\
0.145 -- 0.165 & 1.715 $\pm$ 0.5 $\pm$ 0.2 $\pm$ 0.1 & 1.705 $\pm$ 0.5 $\pm$ 0.2 $\pm$ 0.1 & 1.709 $\pm$ 0.4 $\pm$ 0.1 $\pm$ 0.1 & 1.711 $\pm$ 0.4 $\pm$ 0.1 $\pm$ 0.1 & 1.700 $\pm$ 0.4 $\pm$ 0.1 $\pm$ 0.1 & 1.701 $\pm$ 0.3 $\pm$ 0.1 $\pm$ 0.1\\
0.165 -- 0.189 & 1.404 $\pm$ 0.5 $\pm$ 0.2 $\pm$ 0.1 & 1.395 $\pm$ 0.5 $\pm$ 0.2 $\pm$ 0.1 & 1.403 $\pm$ 0.4 $\pm$ 0.1 $\pm$ 0.1 & 1.404 $\pm$ 0.4 $\pm$ 0.1 $\pm$ 0.1 & 1.394 $\pm$ 0.4 $\pm$ 0.1 $\pm$ 0.1 & 1.394 $\pm$ 0.3 $\pm$ 0.1 $\pm$ 0.1\\
0.189 -- 0.219 & 1.117 $\pm$ 0.5 $\pm$ 0.3 $\pm$ 0.2 & 1.110 $\pm$ 0.5 $\pm$ 0.3 $\pm$ 0.2 & 1.123 $\pm$ 0.4 $\pm$ 0.1 $\pm$ 0.1 & 1.124 $\pm$ 0.4 $\pm$ 0.1 $\pm$ 0.1 & 1.118 $\pm$ 0.4 $\pm$ 0.1 $\pm$ 0.1 & 1.115 $\pm$ 0.3 $\pm$ 0.1 $\pm$ 0.1\\
0.219 -- 0.258 & 0.851 $\pm$ 0.5 $\pm$ 0.3 $\pm$ 0.2 & 0.846 $\pm$ 0.5 $\pm$ 0.3 $\pm$ 0.2 & 0.856 $\pm$ 0.4 $\pm$ 0.1 $\pm$ 0.1 & 0.856 $\pm$ 0.4 $\pm$ 0.1 $\pm$ 0.1 & 0.852 $\pm$ 0.4 $\pm$ 0.1 $\pm$ 0.1 & 0.850 $\pm$ 0.3 $\pm$ 0.1 $\pm$ 0.1\\
0.258 -- 0.312 & 0.621 $\pm$ 0.5 $\pm$ 0.2 $\pm$ 0.3 & 0.618 $\pm$ 0.5 $\pm$ 0.2 $\pm$ 0.3 & 0.622 $\pm$ 0.4 $\pm$ 0.1 $\pm$ 0.1 & 0.623 $\pm$ 0.4 $\pm$ 0.1 $\pm$ 0.1 & 0.620 $\pm$ 0.4 $\pm$ 0.1 $\pm$ 0.2 & 0.618 $\pm$ 0.3 $\pm$ 0.1 $\pm$ 0.2\\
0.312 -- 0.391 & 0.415 $\pm$ 0.5 $\pm$ 0.2 $\pm$ 0.3 & 0.413 $\pm$ 0.5 $\pm$ 0.2 $\pm$ 0.4 & 0.409 $\pm$ 0.4 $\pm$ 0.1 $\pm$ 0.1 & 0.409 $\pm$ 0.4 $\pm$ 0.1 $\pm$ 0.1 & 0.408 $\pm$ 0.4 $\pm$ 0.1 $\pm$ 0.2 & 0.409 $\pm$ 0.3 $\pm$ 0.1 $\pm$ 0.2\\
0.391 -- 0.524 & 0.238 $\pm$ 0.5 $\pm$ 0.1 $\pm$ 0.3 & 0.238 $\pm$ 0.5 $\pm$ 0.1 $\pm$ 0.3 & 0.237 $\pm$ 0.4 $\pm$ 0.1 $\pm$ 0.1 & 0.237 $\pm$ 0.4 $\pm$ 0.1 $\pm$ 0.1 & 0.237 $\pm$ 0.4 $\pm$ 0.1 $\pm$ 0.2 & 0.237 $\pm$ 0.3 $\pm$ 0.1 $\pm$ 0.2\\
0.524 -- 0.695 & 0.124 $\pm$ 0.7 $\pm$ 0.4 $\pm$ 0.3 & 0.124 $\pm$ 0.7 $\pm$ 0.4 $\pm$ 0.3 & 0.122 $\pm$ 0.5 $\pm$ 0.2 $\pm$ 0.2 & 0.122 $\pm$ 0.5 $\pm$ 0.2 $\pm$ 0.2 & 0.122 $\pm$ 0.5 $\pm$ 0.2 $\pm$ 0.2 & 0.122 $\pm$ 0.4 $\pm$ 0.2 $\pm$ 0.2\\
0.695 -- 0.918 & 0.0615 $\pm$ 0.8 $\pm$ 0.4 $\pm$ 0.4 & 0.0614 $\pm$ 0.8 $\pm$ 0.4 $\pm$ 0.4 & 0.0609 $\pm$ 0.7 $\pm$ 0.2 $\pm$ 0.1 & 0.0610 $\pm$ 0.7 $\pm$ 0.2 $\pm$ 0.1 & 0.0609 $\pm$ 0.7 $\pm$ 0.2 $\pm$ 0.2 & 0.0610 $\pm$ 0.5 $\pm$ 0.2 $\pm$ 0.2\\
0.918 -- 1.153 & 0.0313 $\pm$ 1.1 $\pm$ 0.5 $\pm$ 0.4 & 0.0312 $\pm$ 1.1 $\pm$ 0.5 $\pm$ 0.4 & 0.0310 $\pm$ 0.9 $\pm$ 0.2 $\pm$ 0.2 & 0.0311 $\pm$ 0.9 $\pm$ 0.2 $\pm$ 0.2 & 0.0310 $\pm$ 0.9 $\pm$ 0.2 $\pm$ 0.2 & 0.0310 $\pm$ 0.7 $\pm$ 0.2 $\pm$ 0.2\\
1.153 -- 1.496 & 0.0163 $\pm$ 1.3 $\pm$ 0.3 $\pm$ 0.5 & 0.0163 $\pm$ 1.3 $\pm$ 0.3 $\pm$ 0.5 & 0.0158 $\pm$ 1.1 $\pm$ 0.3 $\pm$ 0.2 & 0.0158 $\pm$ 1.1 $\pm$ 0.3 $\pm$ 0.2 & 0.0158 $\pm$ 1.1 $\pm$ 0.3 $\pm$ 0.4 & 0.0160 $\pm$ 0.8 $\pm$ 0.2 $\pm$ 0.3\\
1.496 -- 1.947 & 0.00752 $\pm$ 1.6 $\pm$ 0.4 $\pm$ 0.5 & 0.00749 $\pm$ 1.6 $\pm$ 0.4 $\pm$ 0.5 & 0.00740 $\pm$ 1.3 $\pm$ 0.3 $\pm$ 0.3 & 0.00741 $\pm$ 1.3 $\pm$ 0.3 $\pm$ 0.3 & 0.00737 $\pm$ 1.3 $\pm$ 0.3 $\pm$ 0.5 & 0.00740 $\pm$ 1.0 $\pm$ 0.2 $\pm$ 0.3\\
1.947 -- 2.522 & 0.00356 $\pm$ 2.0 $\pm$ 0.5 $\pm$ 0.6 & 0.00354 $\pm$ 2.0 $\pm$ 0.5 $\pm$ 0.6 & 0.00356 $\pm$ 1.7 $\pm$ 0.4 $\pm$ 0.4 & 0.00357 $\pm$ 1.7 $\pm$ 0.4 $\pm$ 0.4 & 0.00357 $\pm$ 1.7 $\pm$ 0.4 $\pm$ 0.5 & 0.00355 $\pm$ 1.3 $\pm$ 0.3 $\pm$ 0.4\\
2.522 -- 3.277 & 0.00168 $\pm$ 2.5 $\pm$ 0.6 $\pm$ 0.6 & 0.00168 $\pm$ 2.5 $\pm$ 0.6 $\pm$ 0.6 & 0.00175 $\pm$ 2.1 $\pm$ 0.4 $\pm$ 0.6 & 0.00176 $\pm$ 2.1 $\pm$ 0.4 $\pm$ 0.6 & 0.00175 $\pm$ 2.1 $\pm$ 0.4 $\pm$ 0.7 & 0.00171 $\pm$ 1.6 $\pm$ 0.4 $\pm$ 0.4\\
3.277 -- 5.000 & 0.000769 $\pm$ 2.4 $\pm$ 0.6 $\pm$ 0.6 & 0.000768 $\pm$ 2.4 $\pm$ 0.6 $\pm$ 0.6 & 0.000792 $\pm$ 2.1 $\pm$ 0.7 $\pm$ 0.5 & 0.000796 $\pm$ 2.1 $\pm$ 0.7 $\pm$ 0.5 & 0.000795 $\pm$ 2.1 $\pm$ 0.7 $\pm$ 0.6 & 0.000781 $\pm$ 1.6 $\pm$ 0.5 $\pm$ 0.4\\
5.000 -- 10.000 & 0.000215 $\pm$ 2.7 $\pm$ 0.8 $\pm$ 0.7 & 0.000215 $\pm$ 2.7 $\pm$ 0.8 $\pm$ 0.7 & 0.000213 $\pm$ 2.4 $\pm$ 0.5 $\pm$ 0.4 & 0.000213 $\pm$ 2.4 $\pm$ 0.5 $\pm$ 0.4 & 0.000213 $\pm$ 2.4 $\pm$ 0.5 $\pm$ 0.6 & 0.000213 $\pm$ 1.8 $\pm$ 0.4 $\pm$ 0.4\\
\bottomrule 
\end{tabular} 
} 
\label{tab:CombPhiStarpeakmass_y1216}
\end{table*}

\begin{table*}
\centering
\caption{The values of $(1/\sigma)\,\mathrm{d}\sigma/\mathrm{d}\phi^*_{\eta}$ in each bin of \PhiStar{} for the electron and muon channels separately (for various particle-level definitions) and for the Born-level combination in the kinematic region $66\ \GeV \leq m_{\ell\ell} < 116\ \GeV,\ 1.6 \leq |y_{\ell\ell}| < 2.0$. The associated statistical and systematic (both uncorrelated and correlated between bins of \PhiStar{}) are provided in percentage form.} 
\resizebox{\textwidth}{!}{
\begin{tabular}{ ccccccc } \toprule 
Bin & \multicolumn{6}{c}{$(1/\sigma)\,d\sigma/d\phi^*_{\eta}$ $\pm$ Statistical [\%] $\pm$ Uncorrelated systematic [\%] $\pm$ Correlated systematic [\%]} \\ 
 \cmidrule(r){2-7} 
 & \multicolumn{2}{c}{Electron channel} &  \multicolumn{3}{c}{Muon channel} & Combination \\ 
\cmidrule(r){2-3} \cmidrule(r){4-6} \cmidrule(r){7-7} 
 & dressed & Born & bare & dressed & Born & Born \\ 
 0.000 -- 0.004 & 9.378 $\pm$ 0.6 $\pm$ 0.3 $\pm$ 0.3 & 9.472 $\pm$ 0.6 $\pm$ 0.3 $\pm$ 0.3 & 9.338 $\pm$ 0.5 $\pm$ 0.1 $\pm$ 0.1 & 9.325 $\pm$ 0.5 $\pm$ 0.1 $\pm$ 0.1 & 9.405 $\pm$ 0.5 $\pm$ 0.1 $\pm$ 0.2 & 9.418 $\pm$ 0.4 $\pm$ 0.1 $\pm$ 0.1\\
0.004 -- 0.008 & 9.264 $\pm$ 0.6 $\pm$ 0.2 $\pm$ 0.2 & 9.368 $\pm$ 0.6 $\pm$ 0.2 $\pm$ 0.2 & 9.273 $\pm$ 0.5 $\pm$ 0.1 $\pm$ 0.0 & 9.274 $\pm$ 0.5 $\pm$ 0.1 $\pm$ 0.0 & 9.347 $\pm$ 0.5 $\pm$ 0.1 $\pm$ 0.1 & 9.345 $\pm$ 0.4 $\pm$ 0.1 $\pm$ 0.1\\
0.008 -- 0.012 & 9.014 $\pm$ 0.6 $\pm$ 0.2 $\pm$ 0.2 & 9.077 $\pm$ 0.6 $\pm$ 0.2 $\pm$ 0.3 & 9.131 $\pm$ 0.5 $\pm$ 0.1 $\pm$ 0.0 & 9.130 $\pm$ 0.5 $\pm$ 0.1 $\pm$ 0.0 & 9.199 $\pm$ 0.5 $\pm$ 0.1 $\pm$ 0.1 & 9.153 $\pm$ 0.4 $\pm$ 0.1 $\pm$ 0.1\\
0.012 -- 0.016 & 8.875 $\pm$ 0.7 $\pm$ 0.2 $\pm$ 0.2 & 8.951 $\pm$ 0.7 $\pm$ 0.2 $\pm$ 0.2 & 8.864 $\pm$ 0.5 $\pm$ 0.1 $\pm$ 0.0 & 8.855 $\pm$ 0.5 $\pm$ 0.1 $\pm$ 0.0 & 8.930 $\pm$ 0.5 $\pm$ 0.1 $\pm$ 0.1 & 8.928 $\pm$ 0.4 $\pm$ 0.1 $\pm$ 0.1\\
0.016 -- 0.020 & 8.592 $\pm$ 0.7 $\pm$ 0.2 $\pm$ 0.2 & 8.657 $\pm$ 0.7 $\pm$ 0.2 $\pm$ 0.2 & 8.503 $\pm$ 0.5 $\pm$ 0.1 $\pm$ 0.0 & 8.493 $\pm$ 0.5 $\pm$ 0.1 $\pm$ 0.0 & 8.561 $\pm$ 0.5 $\pm$ 0.1 $\pm$ 0.1 & 8.588 $\pm$ 0.4 $\pm$ 0.1 $\pm$ 0.1\\
0.020 -- 0.024 & 8.207 $\pm$ 0.7 $\pm$ 0.2 $\pm$ 0.2 & 8.259 $\pm$ 0.7 $\pm$ 0.2 $\pm$ 0.2 & 8.149 $\pm$ 0.5 $\pm$ 0.1 $\pm$ 0.1 & 8.143 $\pm$ 0.5 $\pm$ 0.1 $\pm$ 0.1 & 8.195 $\pm$ 0.5 $\pm$ 0.1 $\pm$ 0.2 & 8.211 $\pm$ 0.4 $\pm$ 0.1 $\pm$ 0.1\\
0.024 -- 0.029 & 7.721 $\pm$ 0.6 $\pm$ 0.2 $\pm$ 0.2 & 7.755 $\pm$ 0.6 $\pm$ 0.2 $\pm$ 0.2 & 7.776 $\pm$ 0.5 $\pm$ 0.1 $\pm$ 0.0 & 7.769 $\pm$ 0.5 $\pm$ 0.1 $\pm$ 0.0 & 7.842 $\pm$ 0.5 $\pm$ 0.1 $\pm$ 0.3 & 7.804 $\pm$ 0.4 $\pm$ 0.1 $\pm$ 0.2\\
0.029 -- 0.034 & 7.439 $\pm$ 0.6 $\pm$ 0.2 $\pm$ 0.2 & 7.484 $\pm$ 0.6 $\pm$ 0.2 $\pm$ 0.3 & 7.343 $\pm$ 0.5 $\pm$ 0.1 $\pm$ 0.0 & 7.345 $\pm$ 0.5 $\pm$ 0.1 $\pm$ 0.0 & 7.398 $\pm$ 0.5 $\pm$ 0.1 $\pm$ 0.3 & 7.416 $\pm$ 0.4 $\pm$ 0.1 $\pm$ 0.2\\
0.034 -- 0.039 & 6.755 $\pm$ 0.7 $\pm$ 0.2 $\pm$ 0.2 & 6.765 $\pm$ 0.7 $\pm$ 0.3 $\pm$ 0.3 & 6.832 $\pm$ 0.5 $\pm$ 0.1 $\pm$ 0.1 & 6.824 $\pm$ 0.5 $\pm$ 0.1 $\pm$ 0.1 & 6.848 $\pm$ 0.5 $\pm$ 0.1 $\pm$ 0.3 & 6.813 $\pm$ 0.4 $\pm$ 0.1 $\pm$ 0.2\\
0.039 -- 0.045 & 6.394 $\pm$ 0.6 $\pm$ 0.2 $\pm$ 0.2 & 6.414 $\pm$ 0.6 $\pm$ 0.2 $\pm$ 0.2 & 6.377 $\pm$ 0.5 $\pm$ 0.1 $\pm$ 0.0 & 6.375 $\pm$ 0.5 $\pm$ 0.1 $\pm$ 0.0 & 6.383 $\pm$ 0.5 $\pm$ 0.1 $\pm$ 0.3 & 6.383 $\pm$ 0.4 $\pm$ 0.1 $\pm$ 0.2\\
0.045 -- 0.051 & 5.878 $\pm$ 0.7 $\pm$ 0.2 $\pm$ 0.2 & 5.881 $\pm$ 0.7 $\pm$ 0.2 $\pm$ 0.2 & 5.906 $\pm$ 0.5 $\pm$ 0.1 $\pm$ 0.1 & 5.903 $\pm$ 0.5 $\pm$ 0.1 $\pm$ 0.1 & 5.916 $\pm$ 0.5 $\pm$ 0.1 $\pm$ 0.3 & 5.895 $\pm$ 0.4 $\pm$ 0.1 $\pm$ 0.2\\
0.051 -- 0.057 & 5.309 $\pm$ 0.7 $\pm$ 0.2 $\pm$ 0.2 & 5.302 $\pm$ 0.7 $\pm$ 0.2 $\pm$ 0.2 & 5.371 $\pm$ 0.5 $\pm$ 0.1 $\pm$ 0.0 & 5.370 $\pm$ 0.5 $\pm$ 0.1 $\pm$ 0.0 & 5.374 $\pm$ 0.5 $\pm$ 0.1 $\pm$ 0.3 & 5.344 $\pm$ 0.4 $\pm$ 0.1 $\pm$ 0.2\\
0.057 -- 0.064 & 4.899 $\pm$ 0.7 $\pm$ 0.2 $\pm$ 0.2 & 4.896 $\pm$ 0.7 $\pm$ 0.2 $\pm$ 0.2 & 4.909 $\pm$ 0.5 $\pm$ 0.1 $\pm$ 0.1 & 4.907 $\pm$ 0.5 $\pm$ 0.1 $\pm$ 0.1 & 4.900 $\pm$ 0.5 $\pm$ 0.1 $\pm$ 0.2 & 4.893 $\pm$ 0.4 $\pm$ 0.1 $\pm$ 0.1\\
0.064 -- 0.072 & 4.508 $\pm$ 0.6 $\pm$ 0.2 $\pm$ 0.2 & 4.503 $\pm$ 0.6 $\pm$ 0.2 $\pm$ 0.2 & 4.511 $\pm$ 0.5 $\pm$ 0.1 $\pm$ 0.1 & 4.507 $\pm$ 0.5 $\pm$ 0.1 $\pm$ 0.1 & 4.498 $\pm$ 0.5 $\pm$ 0.1 $\pm$ 0.2 & 4.495 $\pm$ 0.4 $\pm$ 0.1 $\pm$ 0.1\\
0.072 -- 0.081 & 4.011 $\pm$ 0.6 $\pm$ 0.4 $\pm$ 0.2 & 3.996 $\pm$ 0.6 $\pm$ 0.4 $\pm$ 0.2 & 4.008 $\pm$ 0.5 $\pm$ 0.1 $\pm$ 0.1 & 4.005 $\pm$ 0.5 $\pm$ 0.1 $\pm$ 0.1 & 3.990 $\pm$ 0.5 $\pm$ 0.1 $\pm$ 0.2 & 3.988 $\pm$ 0.4 $\pm$ 0.1 $\pm$ 0.1\\
0.081 -- 0.091 & 3.574 $\pm$ 0.7 $\pm$ 0.4 $\pm$ 0.2 & 3.561 $\pm$ 0.7 $\pm$ 0.4 $\pm$ 0.2 & 3.574 $\pm$ 0.5 $\pm$ 0.1 $\pm$ 0.1 & 3.575 $\pm$ 0.5 $\pm$ 0.1 $\pm$ 0.1 & 3.558 $\pm$ 0.5 $\pm$ 0.1 $\pm$ 0.2 & 3.556 $\pm$ 0.4 $\pm$ 0.1 $\pm$ 0.1\\
0.091 -- 0.102 & 3.159 $\pm$ 0.7 $\pm$ 0.2 $\pm$ 0.2 & 3.142 $\pm$ 0.7 $\pm$ 0.2 $\pm$ 0.2 & 3.157 $\pm$ 0.5 $\pm$ 0.1 $\pm$ 0.1 & 3.155 $\pm$ 0.5 $\pm$ 0.1 $\pm$ 0.1 & 3.130 $\pm$ 0.5 $\pm$ 0.1 $\pm$ 0.2 & 3.131 $\pm$ 0.4 $\pm$ 0.1 $\pm$ 0.1\\
0.102 -- 0.114 & 2.806 $\pm$ 0.7 $\pm$ 0.2 $\pm$ 0.1 & 2.793 $\pm$ 0.7 $\pm$ 0.2 $\pm$ 0.1 & 2.774 $\pm$ 0.5 $\pm$ 0.1 $\pm$ 0.1 & 2.776 $\pm$ 0.5 $\pm$ 0.1 $\pm$ 0.1 & 2.764 $\pm$ 0.5 $\pm$ 0.1 $\pm$ 0.2 & 2.772 $\pm$ 0.4 $\pm$ 0.1 $\pm$ 0.1\\
0.114 -- 0.128 & 2.407 $\pm$ 0.7 $\pm$ 0.2 $\pm$ 0.1 & 2.392 $\pm$ 0.7 $\pm$ 0.2 $\pm$ 0.1 & 2.416 $\pm$ 0.5 $\pm$ 0.1 $\pm$ 0.1 & 2.416 $\pm$ 0.5 $\pm$ 0.1 $\pm$ 0.1 & 2.401 $\pm$ 0.5 $\pm$ 0.1 $\pm$ 0.2 & 2.396 $\pm$ 0.4 $\pm$ 0.1 $\pm$ 0.1\\
0.128 -- 0.145 & 2.072 $\pm$ 0.7 $\pm$ 0.2 $\pm$ 0.1 & 2.058 $\pm$ 0.7 $\pm$ 0.2 $\pm$ 0.1 & 2.061 $\pm$ 0.5 $\pm$ 0.1 $\pm$ 0.1 & 2.061 $\pm$ 0.5 $\pm$ 0.1 $\pm$ 0.1 & 2.048 $\pm$ 0.5 $\pm$ 0.1 $\pm$ 0.2 & 2.050 $\pm$ 0.4 $\pm$ 0.1 $\pm$ 0.1\\
0.145 -- 0.165 & 1.730 $\pm$ 0.7 $\pm$ 0.2 $\pm$ 0.2 & 1.716 $\pm$ 0.7 $\pm$ 0.2 $\pm$ 0.2 & 1.729 $\pm$ 0.5 $\pm$ 0.1 $\pm$ 0.1 & 1.730 $\pm$ 0.5 $\pm$ 0.1 $\pm$ 0.1 & 1.718 $\pm$ 0.5 $\pm$ 0.1 $\pm$ 0.2 & 1.717 $\pm$ 0.4 $\pm$ 0.1 $\pm$ 0.1\\
0.165 -- 0.189 & 1.417 $\pm$ 0.7 $\pm$ 0.2 $\pm$ 0.2 & 1.408 $\pm$ 0.7 $\pm$ 0.2 $\pm$ 0.2 & 1.423 $\pm$ 0.5 $\pm$ 0.1 $\pm$ 0.1 & 1.422 $\pm$ 0.5 $\pm$ 0.1 $\pm$ 0.1 & 1.414 $\pm$ 0.5 $\pm$ 0.1 $\pm$ 0.2 & 1.411 $\pm$ 0.4 $\pm$ 0.1 $\pm$ 0.1\\
0.189 -- 0.219 & 1.137 $\pm$ 0.7 $\pm$ 0.4 $\pm$ 0.3 & 1.130 $\pm$ 0.7 $\pm$ 0.4 $\pm$ 0.3 & 1.138 $\pm$ 0.5 $\pm$ 0.1 $\pm$ 0.1 & 1.139 $\pm$ 0.5 $\pm$ 0.1 $\pm$ 0.1 & 1.133 $\pm$ 0.5 $\pm$ 0.1 $\pm$ 0.2 & 1.131 $\pm$ 0.4 $\pm$ 0.1 $\pm$ 0.2\\
0.219 -- 0.258 & 0.871 $\pm$ 0.7 $\pm$ 0.4 $\pm$ 0.3 & 0.866 $\pm$ 0.7 $\pm$ 0.4 $\pm$ 0.3 & 0.875 $\pm$ 0.5 $\pm$ 0.1 $\pm$ 0.1 & 0.876 $\pm$ 0.5 $\pm$ 0.1 $\pm$ 0.1 & 0.870 $\pm$ 0.5 $\pm$ 0.1 $\pm$ 0.2 & 0.869 $\pm$ 0.4 $\pm$ 0.1 $\pm$ 0.2\\
0.258 -- 0.312 & 0.643 $\pm$ 0.7 $\pm$ 0.2 $\pm$ 0.4 & 0.641 $\pm$ 0.7 $\pm$ 0.3 $\pm$ 0.4 & 0.634 $\pm$ 0.5 $\pm$ 0.1 $\pm$ 0.1 & 0.635 $\pm$ 0.5 $\pm$ 0.1 $\pm$ 0.1 & 0.631 $\pm$ 0.5 $\pm$ 0.1 $\pm$ 0.2 & 0.634 $\pm$ 0.4 $\pm$ 0.1 $\pm$ 0.2\\
0.312 -- 0.391 & 0.428 $\pm$ 0.7 $\pm$ 0.3 $\pm$ 0.4 & 0.426 $\pm$ 0.7 $\pm$ 0.3 $\pm$ 0.4 & 0.427 $\pm$ 0.5 $\pm$ 0.1 $\pm$ 0.1 & 0.427 $\pm$ 0.5 $\pm$ 0.1 $\pm$ 0.1 & 0.425 $\pm$ 0.5 $\pm$ 0.1 $\pm$ 0.2 & 0.425 $\pm$ 0.4 $\pm$ 0.1 $\pm$ 0.2\\
0.391 -- 0.524 & 0.244 $\pm$ 0.7 $\pm$ 0.3 $\pm$ 0.5 & 0.244 $\pm$ 0.7 $\pm$ 0.3 $\pm$ 0.5 & 0.245 $\pm$ 0.5 $\pm$ 0.1 $\pm$ 0.1 & 0.246 $\pm$ 0.5 $\pm$ 0.1 $\pm$ 0.1 & 0.245 $\pm$ 0.5 $\pm$ 0.1 $\pm$ 0.2 & 0.245 $\pm$ 0.4 $\pm$ 0.1 $\pm$ 0.2\\
0.524 -- 0.695 & 0.125 $\pm$ 0.8 $\pm$ 0.4 $\pm$ 0.6 & 0.124 $\pm$ 0.8 $\pm$ 0.4 $\pm$ 0.6 & 0.126 $\pm$ 0.6 $\pm$ 0.1 $\pm$ 0.2 & 0.127 $\pm$ 0.6 $\pm$ 0.1 $\pm$ 0.2 & 0.126 $\pm$ 0.6 $\pm$ 0.1 $\pm$ 0.2 & 0.126 $\pm$ 0.5 $\pm$ 0.2 $\pm$ 0.2\\
0.695 -- 0.918 & 0.0621 $\pm$ 1.0 $\pm$ 0.4 $\pm$ 0.7 & 0.0620 $\pm$ 1.0 $\pm$ 0.4 $\pm$ 0.7 & 0.0613 $\pm$ 0.8 $\pm$ 0.2 $\pm$ 0.2 & 0.0615 $\pm$ 0.8 $\pm$ 0.2 $\pm$ 0.2 & 0.0615 $\pm$ 0.8 $\pm$ 0.2 $\pm$ 0.3 & 0.0616 $\pm$ 0.6 $\pm$ 0.2 $\pm$ 0.3\\
0.918 -- 1.153 & 0.0305 $\pm$ 1.4 $\pm$ 0.5 $\pm$ 0.7 & 0.0304 $\pm$ 1.4 $\pm$ 0.5 $\pm$ 0.7 & 0.0301 $\pm$ 1.1 $\pm$ 0.2 $\pm$ 0.2 & 0.0301 $\pm$ 1.1 $\pm$ 0.2 $\pm$ 0.2 & 0.0300 $\pm$ 1.1 $\pm$ 0.2 $\pm$ 0.3 & 0.0301 $\pm$ 0.9 $\pm$ 0.2 $\pm$ 0.3\\
1.153 -- 1.496 & 0.0148 $\pm$ 1.6 $\pm$ 0.6 $\pm$ 0.8 & 0.0148 $\pm$ 1.6 $\pm$ 0.6 $\pm$ 0.8 & 0.0149 $\pm$ 1.3 $\pm$ 0.2 $\pm$ 0.3 & 0.0150 $\pm$ 1.3 $\pm$ 0.2 $\pm$ 0.3 & 0.0149 $\pm$ 1.3 $\pm$ 0.2 $\pm$ 0.6 & 0.0149 $\pm$ 1.0 $\pm$ 0.2 $\pm$ 0.4\\
1.496 -- 1.947 & 0.00643 $\pm$ 2.1 $\pm$ 0.6 $\pm$ 1.0 & 0.00641 $\pm$ 2.1 $\pm$ 0.6 $\pm$ 1.0 & 0.00638 $\pm$ 1.8 $\pm$ 0.3 $\pm$ 0.4 & 0.00641 $\pm$ 1.8 $\pm$ 0.3 $\pm$ 0.4 & 0.00642 $\pm$ 1.8 $\pm$ 0.3 $\pm$ 0.7 & 0.00641 $\pm$ 1.4 $\pm$ 0.3 $\pm$ 0.5\\
1.947 -- 2.522 & 0.00274 $\pm$ 2.9 $\pm$ 0.7 $\pm$ 1.0 & 0.00272 $\pm$ 2.9 $\pm$ 0.7 $\pm$ 1.1 & 0.00299 $\pm$ 2.3 $\pm$ 0.4 $\pm$ 0.4 & 0.00299 $\pm$ 2.3 $\pm$ 0.4 $\pm$ 0.4 & 0.00297 $\pm$ 2.3 $\pm$ 0.4 $\pm$ 0.7 & 0.00287 $\pm$ 1.8 $\pm$ 0.4 $\pm$ 0.5\\
2.522 -- 3.277 & 0.00131 $\pm$ 3.6 $\pm$ 0.7 $\pm$ 1.2 & 0.00130 $\pm$ 3.6 $\pm$ 0.8 $\pm$ 1.2 & 0.00128 $\pm$ 3.0 $\pm$ 0.6 $\pm$ 0.4 & 0.00128 $\pm$ 3.0 $\pm$ 0.6 $\pm$ 0.4 & 0.00129 $\pm$ 3.0 $\pm$ 0.6 $\pm$ 0.7 & 0.00129 $\pm$ 2.3 $\pm$ 0.5 $\pm$ 0.5\\
3.277 -- 5.000 & 0.000510 $\pm$ 3.8 $\pm$ 1.1 $\pm$ 1.4 & 0.000509 $\pm$ 3.8 $\pm$ 1.1 $\pm$ 1.4 & 0.000519 $\pm$ 3.1 $\pm$ 0.6 $\pm$ 0.7 & 0.000524 $\pm$ 3.1 $\pm$ 0.6 $\pm$ 0.7 & 0.000525 $\pm$ 3.1 $\pm$ 0.6 $\pm$ 0.9 & 0.000517 $\pm$ 2.4 $\pm$ 0.5 $\pm$ 0.7\\
5.000 -- 10.000 & 0.000141 $\pm$ 4.2 $\pm$ 0.9 $\pm$ 1.3 & 0.000141 $\pm$ 4.2 $\pm$ 0.9 $\pm$ 1.3 & 0.000127 $\pm$ 3.8 $\pm$ 0.6 $\pm$ 0.5 & 0.000128 $\pm$ 3.8 $\pm$ 0.6 $\pm$ 0.5 & 0.000128 $\pm$ 3.8 $\pm$ 0.6 $\pm$ 0.7 & 0.000133 $\pm$ 2.8 $\pm$ 0.5 $\pm$ 0.6\\
\bottomrule 
\end{tabular} 
} 
\label{tab:CombPhiStarpeakmass_y1620}
\end{table*}

\begin{table*}
\centering
\caption{The values of $(1/\sigma)\,\mathrm{d}\sigma/\mathrm{d}\phi^*_{\eta}$ in each bin of \PhiStar{} for the electron and muon channels separately (for various particle-level definitions) and for the Born-level combination in the kinematic region $66\ \GeV \leq m_{\ell\ell} < 116\ \GeV,\ 2.0 \leq |y_{\ell\ell}| < 2.4$. The associated statistical and systematic (both uncorrelated and correlated between bins of \PhiStar{}) are provided in percentage form.} 
\resizebox{\textwidth}{!}{
\begin{tabular}{ ccccccc } \toprule 
Bin & \multicolumn{6}{c}{$(1/\sigma)\,d\sigma/d\phi^*_{\eta}$ $\pm$ Statistical [\%] $\pm$ Uncorrelated systematic [\%] $\pm$ Correlated systematic [\%]} \\ 
 \cmidrule(r){2-7} 
 & \multicolumn{2}{c}{Electron channel} &  \multicolumn{3}{c}{Muon channel} & Combination \\ 
\cmidrule(r){2-3} \cmidrule(r){4-6} \cmidrule(r){7-7} 
 & dressed & Born & bare & dressed & Born & Born \\ 
 0.000 -- 0.004 & 9.324 $\pm$ 1.0 $\pm$ 0.4 $\pm$ 0.3 & 9.417 $\pm$ 1.0 $\pm$ 0.4 $\pm$ 0.3 & 9.348 $\pm$ 0.8 $\pm$ 0.1 $\pm$ 0.1 & 9.347 $\pm$ 0.8 $\pm$ 0.1 $\pm$ 0.1 & 9.417 $\pm$ 0.8 $\pm$ 0.1 $\pm$ 0.4 & 9.417 $\pm$ 0.6 $\pm$ 0.2 $\pm$ 0.2\\
0.004 -- 0.008 & 9.101 $\pm$ 1.0 $\pm$ 0.4 $\pm$ 0.3 & 9.182 $\pm$ 1.0 $\pm$ 0.4 $\pm$ 0.3 & 9.294 $\pm$ 0.9 $\pm$ 0.2 $\pm$ 0.1 & 9.288 $\pm$ 0.9 $\pm$ 0.2 $\pm$ 0.1 & 9.372 $\pm$ 0.9 $\pm$ 0.2 $\pm$ 0.4 & 9.303 $\pm$ 0.7 $\pm$ 0.2 $\pm$ 0.2\\
0.008 -- 0.012 & 9.083 $\pm$ 1.0 $\pm$ 0.4 $\pm$ 0.3 & 9.179 $\pm$ 1.0 $\pm$ 0.4 $\pm$ 0.3 & 8.969 $\pm$ 0.9 $\pm$ 0.2 $\pm$ 0.1 & 8.959 $\pm$ 0.9 $\pm$ 0.2 $\pm$ 0.1 & 9.023 $\pm$ 0.9 $\pm$ 0.2 $\pm$ 0.4 & 9.084 $\pm$ 0.7 $\pm$ 0.2 $\pm$ 0.2\\
0.012 -- 0.016 & 8.825 $\pm$ 1.0 $\pm$ 0.4 $\pm$ 0.3 & 8.895 $\pm$ 1.0 $\pm$ 0.4 $\pm$ 0.3 & 9.015 $\pm$ 0.9 $\pm$ 0.2 $\pm$ 0.1 & 8.993 $\pm$ 0.9 $\pm$ 0.2 $\pm$ 0.1 & 9.050 $\pm$ 0.9 $\pm$ 0.2 $\pm$ 0.4 & 8.991 $\pm$ 0.7 $\pm$ 0.2 $\pm$ 0.2\\
0.016 -- 0.020 & 8.401 $\pm$ 1.0 $\pm$ 0.4 $\pm$ 0.3 & 8.462 $\pm$ 1.0 $\pm$ 0.4 $\pm$ 0.3 & 8.469 $\pm$ 0.9 $\pm$ 0.2 $\pm$ 0.1 & 8.456 $\pm$ 0.9 $\pm$ 0.2 $\pm$ 0.1 & 8.540 $\pm$ 0.9 $\pm$ 0.2 $\pm$ 0.4 & 8.511 $\pm$ 0.7 $\pm$ 0.2 $\pm$ 0.2\\
0.020 -- 0.024 & 8.047 $\pm$ 1.1 $\pm$ 0.5 $\pm$ 0.3 & 8.094 $\pm$ 1.1 $\pm$ 0.5 $\pm$ 0.3 & 8.242 $\pm$ 0.9 $\pm$ 0.2 $\pm$ 0.1 & 8.241 $\pm$ 0.9 $\pm$ 0.2 $\pm$ 0.1 & 8.305 $\pm$ 0.9 $\pm$ 0.2 $\pm$ 0.4 & 8.236 $\pm$ 0.7 $\pm$ 0.2 $\pm$ 0.2\\
0.024 -- 0.029 & 7.986 $\pm$ 1.0 $\pm$ 0.5 $\pm$ 0.3 & 8.065 $\pm$ 1.0 $\pm$ 0.5 $\pm$ 0.4 & 7.748 $\pm$ 0.8 $\pm$ 0.2 $\pm$ 0.0 & 7.743 $\pm$ 0.8 $\pm$ 0.2 $\pm$ 0.0 & 7.799 $\pm$ 0.8 $\pm$ 0.2 $\pm$ 0.4 & 7.891 $\pm$ 0.6 $\pm$ 0.2 $\pm$ 0.2\\
0.029 -- 0.034 & 7.168 $\pm$ 1.0 $\pm$ 0.5 $\pm$ 0.3 & 7.173 $\pm$ 1.0 $\pm$ 0.5 $\pm$ 0.3 & 7.276 $\pm$ 0.9 $\pm$ 0.1 $\pm$ 0.0 & 7.270 $\pm$ 0.9 $\pm$ 0.1 $\pm$ 0.0 & 7.310 $\pm$ 0.9 $\pm$ 0.1 $\pm$ 0.4 & 7.267 $\pm$ 0.7 $\pm$ 0.2 $\pm$ 0.2\\
0.034 -- 0.039 & 6.833 $\pm$ 1.0 $\pm$ 0.5 $\pm$ 0.3 & 6.857 $\pm$ 1.0 $\pm$ 0.5 $\pm$ 0.3 & 6.781 $\pm$ 0.9 $\pm$ 0.2 $\pm$ 0.1 & 6.783 $\pm$ 0.9 $\pm$ 0.2 $\pm$ 0.1 & 6.828 $\pm$ 0.9 $\pm$ 0.2 $\pm$ 0.4 & 6.843 $\pm$ 0.7 $\pm$ 0.2 $\pm$ 0.2\\
0.039 -- 0.045 & 6.468 $\pm$ 1.0 $\pm$ 0.4 $\pm$ 0.3 & 6.490 $\pm$ 1.0 $\pm$ 0.4 $\pm$ 0.3 & 6.415 $\pm$ 0.8 $\pm$ 0.1 $\pm$ 0.0 & 6.408 $\pm$ 0.8 $\pm$ 0.1 $\pm$ 0.0 & 6.441 $\pm$ 0.8 $\pm$ 0.1 $\pm$ 0.4 & 6.457 $\pm$ 0.6 $\pm$ 0.2 $\pm$ 0.2\\
0.045 -- 0.051 & 5.717 $\pm$ 1.0 $\pm$ 0.4 $\pm$ 0.2 & 5.715 $\pm$ 1.0 $\pm$ 0.5 $\pm$ 0.3 & 5.865 $\pm$ 0.9 $\pm$ 0.2 $\pm$ 0.0 & 5.870 $\pm$ 0.9 $\pm$ 0.2 $\pm$ 0.0 & 5.854 $\pm$ 0.9 $\pm$ 0.2 $\pm$ 0.4 & 5.807 $\pm$ 0.7 $\pm$ 0.2 $\pm$ 0.2\\
0.051 -- 0.057 & 5.413 $\pm$ 1.1 $\pm$ 0.4 $\pm$ 0.2 & 5.410 $\pm$ 1.1 $\pm$ 0.4 $\pm$ 0.2 & 5.333 $\pm$ 0.9 $\pm$ 0.2 $\pm$ 0.1 & 5.330 $\pm$ 0.9 $\pm$ 0.2 $\pm$ 0.1 & 5.336 $\pm$ 0.9 $\pm$ 0.2 $\pm$ 0.4 & 5.366 $\pm$ 0.7 $\pm$ 0.2 $\pm$ 0.2\\
0.057 -- 0.064 & 4.935 $\pm$ 1.0 $\pm$ 0.4 $\pm$ 0.2 & 4.934 $\pm$ 1.0 $\pm$ 0.4 $\pm$ 0.2 & 4.923 $\pm$ 0.9 $\pm$ 0.2 $\pm$ 0.0 & 4.915 $\pm$ 0.9 $\pm$ 0.2 $\pm$ 0.0 & 4.906 $\pm$ 0.9 $\pm$ 0.2 $\pm$ 0.2 & 4.918 $\pm$ 0.7 $\pm$ 0.2 $\pm$ 0.1\\
0.064 -- 0.072 & 4.502 $\pm$ 1.0 $\pm$ 0.4 $\pm$ 0.2 & 4.491 $\pm$ 1.0 $\pm$ 0.4 $\pm$ 0.2 & 4.434 $\pm$ 0.9 $\pm$ 0.2 $\pm$ 0.1 & 4.435 $\pm$ 0.9 $\pm$ 0.2 $\pm$ 0.1 & 4.427 $\pm$ 0.9 $\pm$ 0.2 $\pm$ 0.2 & 4.452 $\pm$ 0.7 $\pm$ 0.2 $\pm$ 0.1\\
0.072 -- 0.081 & 3.994 $\pm$ 1.0 $\pm$ 0.5 $\pm$ 0.2 & 3.977 $\pm$ 1.0 $\pm$ 0.5 $\pm$ 0.2 & 3.978 $\pm$ 0.9 $\pm$ 0.1 $\pm$ 0.1 & 3.982 $\pm$ 0.9 $\pm$ 0.1 $\pm$ 0.1 & 3.968 $\pm$ 0.9 $\pm$ 0.1 $\pm$ 0.2 & 3.972 $\pm$ 0.7 $\pm$ 0.2 $\pm$ 0.2\\
0.081 -- 0.091 & 3.517 $\pm$ 1.0 $\pm$ 0.5 $\pm$ 0.1 & 3.496 $\pm$ 1.0 $\pm$ 0.5 $\pm$ 0.2 & 3.572 $\pm$ 0.9 $\pm$ 0.1 $\pm$ 0.1 & 3.569 $\pm$ 0.9 $\pm$ 0.1 $\pm$ 0.1 & 3.563 $\pm$ 0.9 $\pm$ 0.1 $\pm$ 0.2 & 3.539 $\pm$ 0.7 $\pm$ 0.2 $\pm$ 0.1\\
0.091 -- 0.102 & 3.161 $\pm$ 1.0 $\pm$ 0.6 $\pm$ 0.1 & 3.148 $\pm$ 1.0 $\pm$ 0.6 $\pm$ 0.2 & 3.146 $\pm$ 0.8 $\pm$ 0.2 $\pm$ 0.0 & 3.144 $\pm$ 0.8 $\pm$ 0.2 $\pm$ 0.0 & 3.122 $\pm$ 0.8 $\pm$ 0.2 $\pm$ 0.2 & 3.132 $\pm$ 0.7 $\pm$ 0.2 $\pm$ 0.1\\
0.102 -- 0.114 & 2.793 $\pm$ 1.1 $\pm$ 0.6 $\pm$ 0.2 & 2.784 $\pm$ 1.1 $\pm$ 0.6 $\pm$ 0.2 & 2.775 $\pm$ 0.9 $\pm$ 0.1 $\pm$ 0.1 & 2.779 $\pm$ 0.9 $\pm$ 0.1 $\pm$ 0.1 & 2.762 $\pm$ 0.9 $\pm$ 0.1 $\pm$ 0.2 & 2.770 $\pm$ 0.7 $\pm$ 0.2 $\pm$ 0.1\\
0.114 -- 0.128 & 2.404 $\pm$ 1.1 $\pm$ 0.6 $\pm$ 0.1 & 2.385 $\pm$ 1.1 $\pm$ 0.6 $\pm$ 0.1 & 2.441 $\pm$ 0.9 $\pm$ 0.1 $\pm$ 0.1 & 2.438 $\pm$ 0.9 $\pm$ 0.1 $\pm$ 0.1 & 2.420 $\pm$ 0.9 $\pm$ 0.1 $\pm$ 0.3 & 2.408 $\pm$ 0.7 $\pm$ 0.2 $\pm$ 0.2\\
0.128 -- 0.145 & 2.036 $\pm$ 1.0 $\pm$ 0.6 $\pm$ 0.1 & 2.021 $\pm$ 1.0 $\pm$ 0.6 $\pm$ 0.1 & 2.074 $\pm$ 0.8 $\pm$ 0.2 $\pm$ 0.1 & 2.074 $\pm$ 0.8 $\pm$ 0.2 $\pm$ 0.1 & 2.053 $\pm$ 0.8 $\pm$ 0.2 $\pm$ 0.3 & 2.043 $\pm$ 0.7 $\pm$ 0.2 $\pm$ 0.2\\
0.145 -- 0.165 & 1.743 $\pm$ 1.0 $\pm$ 0.6 $\pm$ 0.2 & 1.732 $\pm$ 1.0 $\pm$ 0.6 $\pm$ 0.2 & 1.736 $\pm$ 0.9 $\pm$ 0.1 $\pm$ 0.1 & 1.737 $\pm$ 0.9 $\pm$ 0.1 $\pm$ 0.1 & 1.730 $\pm$ 0.9 $\pm$ 0.1 $\pm$ 0.3 & 1.730 $\pm$ 0.7 $\pm$ 0.2 $\pm$ 0.2\\
0.165 -- 0.189 & 1.447 $\pm$ 1.0 $\pm$ 0.6 $\pm$ 0.2 & 1.439 $\pm$ 1.0 $\pm$ 0.6 $\pm$ 0.2 & 1.453 $\pm$ 0.9 $\pm$ 0.1 $\pm$ 0.1 & 1.455 $\pm$ 0.9 $\pm$ 0.1 $\pm$ 0.1 & 1.449 $\pm$ 0.9 $\pm$ 0.1 $\pm$ 0.3 & 1.445 $\pm$ 0.7 $\pm$ 0.2 $\pm$ 0.2\\
0.189 -- 0.219 & 1.175 $\pm$ 1.0 $\pm$ 0.4 $\pm$ 0.2 & 1.167 $\pm$ 1.0 $\pm$ 0.4 $\pm$ 0.2 & 1.147 $\pm$ 0.9 $\pm$ 0.1 $\pm$ 0.1 & 1.147 $\pm$ 0.9 $\pm$ 0.1 $\pm$ 0.1 & 1.139 $\pm$ 0.9 $\pm$ 0.1 $\pm$ 0.3 & 1.149 $\pm$ 0.7 $\pm$ 0.2 $\pm$ 0.2\\
0.219 -- 0.258 & 0.894 $\pm$ 1.0 $\pm$ 0.5 $\pm$ 0.3 & 0.889 $\pm$ 1.0 $\pm$ 0.5 $\pm$ 0.3 & 0.888 $\pm$ 0.8 $\pm$ 0.2 $\pm$ 0.1 & 0.889 $\pm$ 0.8 $\pm$ 0.2 $\pm$ 0.1 & 0.883 $\pm$ 0.8 $\pm$ 0.2 $\pm$ 0.3 & 0.885 $\pm$ 0.7 $\pm$ 0.2 $\pm$ 0.2\\
0.258 -- 0.312 & 0.656 $\pm$ 1.0 $\pm$ 0.6 $\pm$ 0.4 & 0.653 $\pm$ 1.0 $\pm$ 0.6 $\pm$ 0.4 & 0.646 $\pm$ 0.9 $\pm$ 0.1 $\pm$ 0.2 & 0.646 $\pm$ 0.9 $\pm$ 0.1 $\pm$ 0.2 & 0.643 $\pm$ 0.9 $\pm$ 0.1 $\pm$ 0.4 & 0.646 $\pm$ 0.7 $\pm$ 0.2 $\pm$ 0.3\\
0.312 -- 0.391 & 0.438 $\pm$ 1.1 $\pm$ 0.6 $\pm$ 0.5 & 0.436 $\pm$ 1.1 $\pm$ 0.6 $\pm$ 0.5 & 0.436 $\pm$ 0.9 $\pm$ 0.1 $\pm$ 0.1 & 0.437 $\pm$ 0.9 $\pm$ 0.1 $\pm$ 0.1 & 0.435 $\pm$ 0.9 $\pm$ 0.1 $\pm$ 0.4 & 0.436 $\pm$ 0.7 $\pm$ 0.2 $\pm$ 0.3\\
0.391 -- 0.524 & 0.255 $\pm$ 1.1 $\pm$ 0.6 $\pm$ 0.6 & 0.255 $\pm$ 1.1 $\pm$ 0.6 $\pm$ 0.6 & 0.253 $\pm$ 0.9 $\pm$ 0.1 $\pm$ 0.2 & 0.253 $\pm$ 0.9 $\pm$ 0.1 $\pm$ 0.2 & 0.253 $\pm$ 0.9 $\pm$ 0.1 $\pm$ 0.4 & 0.254 $\pm$ 0.7 $\pm$ 0.2 $\pm$ 0.3\\
0.524 -- 0.695 & 0.131 $\pm$ 1.3 $\pm$ 0.4 $\pm$ 0.7 & 0.130 $\pm$ 1.3 $\pm$ 0.4 $\pm$ 0.7 & 0.134 $\pm$ 1.1 $\pm$ 0.2 $\pm$ 0.2 & 0.134 $\pm$ 1.1 $\pm$ 0.2 $\pm$ 0.2 & 0.134 $\pm$ 1.1 $\pm$ 0.2 $\pm$ 0.4 & 0.132 $\pm$ 0.8 $\pm$ 0.2 $\pm$ 0.3\\
0.695 -- 0.918 & 0.0612 $\pm$ 1.7 $\pm$ 0.5 $\pm$ 0.9 & 0.0611 $\pm$ 1.7 $\pm$ 0.5 $\pm$ 0.9 & 0.0608 $\pm$ 1.4 $\pm$ 0.2 $\pm$ 0.3 & 0.0609 $\pm$ 1.4 $\pm$ 0.2 $\pm$ 0.3 & 0.0607 $\pm$ 1.4 $\pm$ 0.2 $\pm$ 0.4 & 0.0608 $\pm$ 1.1 $\pm$ 0.3 $\pm$ 0.4\\
0.918 -- 1.153 & 0.0292 $\pm$ 2.3 $\pm$ 0.5 $\pm$ 1.1 & 0.0291 $\pm$ 2.3 $\pm$ 0.5 $\pm$ 1.1 & 0.0288 $\pm$ 2.0 $\pm$ 0.3 $\pm$ 0.3 & 0.0290 $\pm$ 2.0 $\pm$ 0.3 $\pm$ 0.3 & 0.0291 $\pm$ 2.0 $\pm$ 0.3 $\pm$ 0.4 & 0.0291 $\pm$ 1.5 $\pm$ 0.3 $\pm$ 0.4\\
1.153 -- 1.496 & 0.0132 $\pm$ 2.8 $\pm$ 0.6 $\pm$ 1.3 & 0.0132 $\pm$ 2.8 $\pm$ 0.6 $\pm$ 1.3 & 0.0129 $\pm$ 2.4 $\pm$ 0.5 $\pm$ 0.3 & 0.0129 $\pm$ 2.4 $\pm$ 0.5 $\pm$ 0.3 & 0.0128 $\pm$ 2.4 $\pm$ 0.5 $\pm$ 1.0 & 0.0130 $\pm$ 1.8 $\pm$ 0.4 $\pm$ 0.6\\
1.496 -- 1.947 & 0.00492 $\pm$ 4.0 $\pm$ 0.8 $\pm$ 1.6 & 0.00493 $\pm$ 4.0 $\pm$ 0.8 $\pm$ 1.6 & 0.00483 $\pm$ 3.5 $\pm$ 0.5 $\pm$ 0.4 & 0.00486 $\pm$ 3.5 $\pm$ 0.5 $\pm$ 0.4 & 0.00483 $\pm$ 3.5 $\pm$ 0.5 $\pm$ 1.0 & 0.00486 $\pm$ 2.6 $\pm$ 0.4 $\pm$ 0.7\\
1.947 -- 2.522 & 0.00191 $\pm$ 5.7 $\pm$ 1.1 $\pm$ 2.0 & 0.00189 $\pm$ 5.7 $\pm$ 1.1 $\pm$ 1.9 & 0.00153 $\pm$ 5.4 $\pm$ 0.9 $\pm$ 0.7 & 0.00154 $\pm$ 5.4 $\pm$ 0.9 $\pm$ 0.7 & 0.00153 $\pm$ 5.4 $\pm$ 0.9 $\pm$ 1.2 & 0.00168 $\pm$ 3.9 $\pm$ 0.7 $\pm$ 0.9\\
2.522 -- 3.277 & 0.000583 $\pm$ 8.9 $\pm$ 1.7 $\pm$ 2.3 & 0.000582 $\pm$ 8.9 $\pm$ 1.8 $\pm$ 2.3 & 0.000535 $\pm$ 8.3 $\pm$ 1.1 $\pm$ 0.7 & 0.000536 $\pm$ 8.3 $\pm$ 1.1 $\pm$ 0.7 & 0.000535 $\pm$ 8.3 $\pm$ 1.1 $\pm$ 1.2 & 0.000553 $\pm$ 6.1 $\pm$ 1.0 $\pm$ 1.0\\
3.277 -- 5.000 & 0.000139 $\pm$ 12 $\pm$ 2.4 $\pm$ 3.1 & 0.000138 $\pm$ 12 $\pm$ 2.5 $\pm$ 3.1 & 0.000182 $\pm$ 9.3 $\pm$ 1.3 $\pm$ 1.4 & 0.000183 $\pm$ 9.3 $\pm$ 1.3 $\pm$ 1.4 & 0.000183 $\pm$ 9.3 $\pm$ 1.3 $\pm$ 1.7 & 0.000163 $\pm$ 7.4 $\pm$ 1.3 $\pm$ 1.4\\
5.000 -- 10.000 & 0.0000247 $\pm$ 18 $\pm$ 5.4 $\pm$ 8.4 & 0.0000248 $\pm$ 18 $\pm$ 5.4 $\pm$ 8.5 & 0.0000308 $\pm$ 13 $\pm$ 2.4 $\pm$ 1.8 & 0.0000311 $\pm$ 13 $\pm$ 2.4 $\pm$ 1.8 & 0.0000314 $\pm$ 13 $\pm$ 2.4 $\pm$ 2.0 & 0.0000285 $\pm$ 11 $\pm$ 2.6 $\pm$ 3.2\\
\bottomrule 
\end{tabular} 
} 
\label{tab:CombPhiStarpeakmass_y2024}
\end{table*}

\begin{table*}
\centering
\caption{The values of $(1/\sigma)\,\mathrm{d}\sigma/\mathrm{d}\phi^*_{\eta}$ in each bin of \PhiStar{} for the electron and muon channels separately (for various particle-level definitions) and for the Born-level combination in the kinematic region $116\ \GeV \leq m_{\ell\ell} < 150\ \GeV,\ 0 \leq |y_{\ell\ell}| < 0.8$. The associated statistical and systematic (both uncorrelated and correlated between bins of \PhiStar{}) are provided in percentage form.} 
\resizebox{\textwidth}{!}{
\begin{tabular}{ ccccccc } \toprule 
Bin & \multicolumn{6}{c}{$(1/\sigma)\,d\sigma/d\phi^*_{\eta}$ $\pm$ Statistical [\%] $\pm$ Uncorrelated systematic [\%] $\pm$ Correlated systematic [\%]} \\ 
 \cmidrule(r){2-7} 
 & \multicolumn{2}{c}{Electron channel} &  \multicolumn{3}{c}{Muon channel} & Combination \\ 
\cmidrule(r){2-3} \cmidrule(r){4-6} \cmidrule(r){7-7} 
 & dressed & Born & bare & dressed & Born & Born \\ 
 0.000 -- 0.004 & 11.489 $\pm$ 2.7 $\pm$ 0.9 $\pm$ 2.9 & 11.576 $\pm$ 2.7 $\pm$ 0.9 $\pm$ 2.9 & 11.151 $\pm$ 1.8 $\pm$ 0.7 $\pm$ 3.4 & 11.135 $\pm$ 1.8 $\pm$ 0.7 $\pm$ 3.4 & 11.235 $\pm$ 1.8 $\pm$ 0.7 $\pm$ 3.5 & 11.412 $\pm$ 1.5 $\pm$ 0.6 $\pm$ 3.2\\
0.004 -- 0.008 & 11.375 $\pm$ 2.7 $\pm$ 0.7 $\pm$ 1.3 & 11.449 $\pm$ 2.7 $\pm$ 0.8 $\pm$ 1.3 & 11.398 $\pm$ 1.7 $\pm$ 0.8 $\pm$ 1.5 & 11.361 $\pm$ 1.7 $\pm$ 0.8 $\pm$ 1.5 & 11.340 $\pm$ 1.7 $\pm$ 0.8 $\pm$ 1.6 & 11.355 $\pm$ 1.5 $\pm$ 0.6 $\pm$ 1.4\\
0.008 -- 0.012 & 11.051 $\pm$ 2.7 $\pm$ 0.8 $\pm$ 0.9 & 11.091 $\pm$ 2.7 $\pm$ 0.8 $\pm$ 0.9 & 11.056 $\pm$ 1.8 $\pm$ 0.8 $\pm$ 1.2 & 10.980 $\pm$ 1.8 $\pm$ 0.8 $\pm$ 1.2 & 11.131 $\pm$ 1.8 $\pm$ 0.8 $\pm$ 1.4 & 11.077 $\pm$ 1.5 $\pm$ 0.6 $\pm$ 1.0\\
0.012 -- 0.016 & 10.804 $\pm$ 2.7 $\pm$ 0.7 $\pm$ 1.2 & 10.904 $\pm$ 2.7 $\pm$ 0.8 $\pm$ 1.2 & 11.077 $\pm$ 1.8 $\pm$ 0.8 $\pm$ 0.9 & 11.006 $\pm$ 1.8 $\pm$ 0.8 $\pm$ 0.9 & 10.970 $\pm$ 1.8 $\pm$ 0.8 $\pm$ 1.2 & 10.920 $\pm$ 1.5 $\pm$ 0.6 $\pm$ 0.9\\
0.016 -- 0.020 & 10.038 $\pm$ 2.8 $\pm$ 0.7 $\pm$ 0.8 & 10.105 $\pm$ 2.8 $\pm$ 0.8 $\pm$ 0.7 & 10.116 $\pm$ 1.9 $\pm$ 0.9 $\pm$ 0.9 & 10.039 $\pm$ 1.9 $\pm$ 0.9 $\pm$ 0.9 & 10.107 $\pm$ 1.9 $\pm$ 0.9 $\pm$ 1.2 & 10.068 $\pm$ 1.6 $\pm$ 0.6 $\pm$ 0.8\\
0.020 -- 0.024 & 9.378 $\pm$ 2.9 $\pm$ 0.8 $\pm$ 0.7 & 9.372 $\pm$ 2.9 $\pm$ 0.8 $\pm$ 0.8 & 9.319 $\pm$ 1.9 $\pm$ 0.8 $\pm$ 1.0 & 9.293 $\pm$ 1.9 $\pm$ 0.8 $\pm$ 1.0 & 9.330 $\pm$ 1.9 $\pm$ 0.8 $\pm$ 1.2 & 9.307 $\pm$ 1.6 $\pm$ 0.6 $\pm$ 0.8\\
0.024 -- 0.029 & 9.075 $\pm$ 2.7 $\pm$ 0.7 $\pm$ 0.6 & 9.117 $\pm$ 2.7 $\pm$ 0.8 $\pm$ 0.6 & 8.794 $\pm$ 1.8 $\pm$ 0.8 $\pm$ 0.5 & 8.773 $\pm$ 1.8 $\pm$ 0.8 $\pm$ 0.5 & 8.847 $\pm$ 1.8 $\pm$ 0.8 $\pm$ 0.7 & 8.907 $\pm$ 1.5 $\pm$ 0.6 $\pm$ 0.6\\
0.029 -- 0.034 & 8.348 $\pm$ 2.7 $\pm$ 0.7 $\pm$ 0.6 & 8.376 $\pm$ 2.7 $\pm$ 0.8 $\pm$ 0.6 & 8.466 $\pm$ 1.8 $\pm$ 0.8 $\pm$ 0.9 & 8.532 $\pm$ 1.8 $\pm$ 0.8 $\pm$ 0.9 & 8.557 $\pm$ 1.8 $\pm$ 0.8 $\pm$ 1.0 & 8.437 $\pm$ 1.5 $\pm$ 0.6 $\pm$ 0.7\\
0.034 -- 0.039 & 6.798 $\pm$ 3.1 $\pm$ 0.9 $\pm$ 0.7 & 6.776 $\pm$ 3.1 $\pm$ 0.9 $\pm$ 0.7 & 7.793 $\pm$ 1.9 $\pm$ 0.8 $\pm$ 0.9 & 7.781 $\pm$ 1.9 $\pm$ 0.8 $\pm$ 0.9 & 7.815 $\pm$ 1.9 $\pm$ 0.8 $\pm$ 1.0 & 7.449 $\pm$ 1.6 $\pm$ 0.6 $\pm$ 0.7\\
0.039 -- 0.045 & 6.684 $\pm$ 2.8 $\pm$ 0.9 $\pm$ 0.7 & 6.689 $\pm$ 2.8 $\pm$ 0.9 $\pm$ 0.7 & 6.810 $\pm$ 1.9 $\pm$ 0.8 $\pm$ 1.1 & 6.826 $\pm$ 1.9 $\pm$ 0.8 $\pm$ 1.1 & 6.806 $\pm$ 1.9 $\pm$ 0.8 $\pm$ 1.1 & 6.711 $\pm$ 1.6 $\pm$ 0.6 $\pm$ 0.8\\
0.045 -- 0.051 & 6.001 $\pm$ 2.9 $\pm$ 0.8 $\pm$ 0.7 & 5.999 $\pm$ 2.9 $\pm$ 0.9 $\pm$ 0.7 & 5.993 $\pm$ 2.0 $\pm$ 0.8 $\pm$ 1.0 & 5.965 $\pm$ 2.0 $\pm$ 0.8 $\pm$ 1.0 & 6.017 $\pm$ 2.0 $\pm$ 0.8 $\pm$ 1.1 & 5.961 $\pm$ 1.6 $\pm$ 0.6 $\pm$ 0.7\\
0.051 -- 0.057 & 5.433 $\pm$ 3.1 $\pm$ 0.9 $\pm$ 0.9 & 5.384 $\pm$ 3.1 $\pm$ 0.9 $\pm$ 1.0 & 5.455 $\pm$ 2.1 $\pm$ 0.9 $\pm$ 1.1 & 5.439 $\pm$ 2.1 $\pm$ 0.9 $\pm$ 1.1 & 5.355 $\pm$ 2.1 $\pm$ 0.9 $\pm$ 1.2 & 5.316 $\pm$ 1.7 $\pm$ 0.7 $\pm$ 0.8\\
0.057 -- 0.064 & 5.175 $\pm$ 2.9 $\pm$ 0.9 $\pm$ 0.7 & 5.210 $\pm$ 2.9 $\pm$ 0.9 $\pm$ 0.7 & 4.958 $\pm$ 2.0 $\pm$ 0.8 $\pm$ 0.8 & 4.949 $\pm$ 2.0 $\pm$ 0.8 $\pm$ 0.8 & 4.933 $\pm$ 2.0 $\pm$ 0.8 $\pm$ 0.8 & 4.999 $\pm$ 1.7 $\pm$ 0.6 $\pm$ 0.7\\
0.064 -- 0.072 & 4.099 $\pm$ 3.1 $\pm$ 0.9 $\pm$ 0.7 & 4.049 $\pm$ 3.1 $\pm$ 0.9 $\pm$ 0.7 & 4.509 $\pm$ 2.0 $\pm$ 0.9 $\pm$ 0.8 & 4.528 $\pm$ 2.0 $\pm$ 0.9 $\pm$ 0.8 & 4.531 $\pm$ 2.0 $\pm$ 0.9 $\pm$ 0.9 & 4.356 $\pm$ 1.7 $\pm$ 0.7 $\pm$ 0.7\\
0.072 -- 0.081 & 4.003 $\pm$ 3.0 $\pm$ 0.8 $\pm$ 0.7 & 4.009 $\pm$ 3.0 $\pm$ 0.8 $\pm$ 0.7 & 3.750 $\pm$ 2.0 $\pm$ 0.8 $\pm$ 0.9 & 3.744 $\pm$ 2.0 $\pm$ 0.8 $\pm$ 0.9 & 3.731 $\pm$ 2.0 $\pm$ 0.8 $\pm$ 0.9 & 3.802 $\pm$ 1.7 $\pm$ 0.6 $\pm$ 0.7\\
0.081 -- 0.091 & 3.423 $\pm$ 3.1 $\pm$ 0.9 $\pm$ 0.9 & 3.437 $\pm$ 3.1 $\pm$ 1.0 $\pm$ 0.9 & 3.303 $\pm$ 2.1 $\pm$ 0.9 $\pm$ 0.6 & 3.302 $\pm$ 2.1 $\pm$ 0.9 $\pm$ 0.6 & 3.303 $\pm$ 2.1 $\pm$ 0.9 $\pm$ 0.7 & 3.339 $\pm$ 1.7 $\pm$ 0.7 $\pm$ 0.7\\
0.091 -- 0.102 & 2.992 $\pm$ 3.1 $\pm$ 0.9 $\pm$ 0.9 & 2.983 $\pm$ 3.1 $\pm$ 1.0 $\pm$ 0.9 & 2.946 $\pm$ 2.1 $\pm$ 1.0 $\pm$ 0.7 & 2.958 $\pm$ 2.1 $\pm$ 1.0 $\pm$ 0.7 & 2.916 $\pm$ 2.1 $\pm$ 1.0 $\pm$ 0.8 & 2.934 $\pm$ 1.7 $\pm$ 0.7 $\pm$ 0.7\\
0.102 -- 0.114 & 2.495 $\pm$ 3.3 $\pm$ 1.9 $\pm$ 2.8 & 2.489 $\pm$ 3.3 $\pm$ 1.9 $\pm$ 2.7 & 2.694 $\pm$ 2.1 $\pm$ 0.9 $\pm$ 0.7 & 2.697 $\pm$ 2.1 $\pm$ 0.9 $\pm$ 0.7 & 2.710 $\pm$ 2.1 $\pm$ 0.9 $\pm$ 0.8 & 2.644 $\pm$ 1.8 $\pm$ 0.8 $\pm$ 0.9\\
0.114 -- 0.128 & 2.134 $\pm$ 3.3 $\pm$ 0.9 $\pm$ 0.7 & 2.125 $\pm$ 3.3 $\pm$ 0.9 $\pm$ 0.7 & 2.240 $\pm$ 2.1 $\pm$ 0.9 $\pm$ 0.9 & 2.246 $\pm$ 2.1 $\pm$ 0.9 $\pm$ 0.9 & 2.227 $\pm$ 2.1 $\pm$ 0.9 $\pm$ 0.9 & 2.195 $\pm$ 1.8 $\pm$ 0.7 $\pm$ 0.7\\
0.128 -- 0.145 & 1.857 $\pm$ 3.3 $\pm$ 1.1 $\pm$ 1.1 & 1.843 $\pm$ 3.3 $\pm$ 1.1 $\pm$ 1.2 & 1.771 $\pm$ 2.2 $\pm$ 0.9 $\pm$ 0.8 & 1.785 $\pm$ 2.2 $\pm$ 0.9 $\pm$ 0.8 & 1.786 $\pm$ 2.2 $\pm$ 0.9 $\pm$ 0.9 & 1.796 $\pm$ 1.8 $\pm$ 0.7 $\pm$ 0.8\\
0.145 -- 0.165 & 1.507 $\pm$ 3.3 $\pm$ 0.9 $\pm$ 0.8 & 1.505 $\pm$ 3.3 $\pm$ 1.0 $\pm$ 0.9 & 1.579 $\pm$ 2.2 $\pm$ 1.0 $\pm$ 0.8 & 1.582 $\pm$ 2.2 $\pm$ 1.0 $\pm$ 0.8 & 1.581 $\pm$ 2.2 $\pm$ 1.0 $\pm$ 0.8 & 1.553 $\pm$ 1.8 $\pm$ 0.7 $\pm$ 0.7\\
0.165 -- 0.189 & 1.212 $\pm$ 3.5 $\pm$ 1.5 $\pm$ 1.5 & 1.200 $\pm$ 3.5 $\pm$ 1.5 $\pm$ 1.5 & 1.203 $\pm$ 2.3 $\pm$ 0.9 $\pm$ 0.8 & 1.211 $\pm$ 2.3 $\pm$ 0.9 $\pm$ 0.8 & 1.193 $\pm$ 2.3 $\pm$ 0.9 $\pm$ 0.8 & 1.199 $\pm$ 1.9 $\pm$ 0.8 $\pm$ 0.8\\
0.189 -- 0.219 & 0.979 $\pm$ 3.4 $\pm$ 1.2 $\pm$ 1.2 & 0.978 $\pm$ 3.4 $\pm$ 1.2 $\pm$ 1.1 & 0.933 $\pm$ 2.3 $\pm$ 0.9 $\pm$ 1.9 & 0.933 $\pm$ 2.3 $\pm$ 0.9 $\pm$ 1.9 & 0.939 $\pm$ 2.3 $\pm$ 0.9 $\pm$ 1.9 & 0.964 $\pm$ 1.9 $\pm$ 0.7 $\pm$ 1.1\\
0.219 -- 0.258 & 0.705 $\pm$ 3.6 $\pm$ 1.4 $\pm$ 1.2 & 0.701 $\pm$ 3.6 $\pm$ 1.4 $\pm$ 1.2 & 0.741 $\pm$ 2.2 $\pm$ 1.0 $\pm$ 2.1 & 0.740 $\pm$ 2.2 $\pm$ 1.0 $\pm$ 2.1 & 0.736 $\pm$ 2.2 $\pm$ 1.0 $\pm$ 2.1 & 0.736 $\pm$ 1.9 $\pm$ 0.8 $\pm$ 1.2\\
0.258 -- 0.312 & 0.526 $\pm$ 3.7 $\pm$ 2.2 $\pm$ 2.0 & 0.524 $\pm$ 3.7 $\pm$ 2.2 $\pm$ 2.0 & 0.524 $\pm$ 2.3 $\pm$ 1.1 $\pm$ 1.6 & 0.524 $\pm$ 2.3 $\pm$ 1.1 $\pm$ 1.6 & 0.522 $\pm$ 2.3 $\pm$ 1.1 $\pm$ 1.8 & 0.531 $\pm$ 2.0 $\pm$ 1.0 $\pm$ 1.3\\
0.312 -- 0.391 & 0.354 $\pm$ 3.8 $\pm$ 1.7 $\pm$ 2.3 & 0.354 $\pm$ 3.8 $\pm$ 1.7 $\pm$ 2.3 & 0.340 $\pm$ 2.4 $\pm$ 1.2 $\pm$ 2.2 & 0.341 $\pm$ 2.4 $\pm$ 1.2 $\pm$ 2.2 & 0.340 $\pm$ 2.4 $\pm$ 1.2 $\pm$ 2.4 & 0.351 $\pm$ 2.0 $\pm$ 1.0 $\pm$ 1.7\\
0.391 -- 0.524 & 0.199 $\pm$ 4.1 $\pm$ 2.0 $\pm$ 3.3 & 0.199 $\pm$ 4.1 $\pm$ 2.0 $\pm$ 3.3 & 0.196 $\pm$ 2.6 $\pm$ 1.4 $\pm$ 3.0 & 0.197 $\pm$ 2.6 $\pm$ 1.4 $\pm$ 3.0 & 0.196 $\pm$ 2.6 $\pm$ 1.4 $\pm$ 3.1 & 0.201 $\pm$ 2.1 $\pm$ 1.2 $\pm$ 2.3\\
0.524 -- 0.695 & 0.103 $\pm$ 5.3 $\pm$ 2.8 $\pm$ 3.8 & 0.103 $\pm$ 5.3 $\pm$ 2.8 $\pm$ 3.8 & 0.0918 $\pm$ 3.3 $\pm$ 2.7 $\pm$ 4.1 & 0.0920 $\pm$ 3.3 $\pm$ 2.7 $\pm$ 4.1 & 0.0910 $\pm$ 3.3 $\pm$ 2.7 $\pm$ 4.1 & 0.0994 $\pm$ 2.8 $\pm$ 2.0 $\pm$ 3.2\\
0.695 -- 0.918 & 0.0540 $\pm$ 6.3 $\pm$ 2.9 $\pm$ 3.5 & 0.0538 $\pm$ 6.3 $\pm$ 2.9 $\pm$ 3.5 & 0.0527 $\pm$ 4.0 $\pm$ 2.9 $\pm$ 4.3 & 0.0529 $\pm$ 4.0 $\pm$ 2.9 $\pm$ 4.3 & 0.0528 $\pm$ 4.0 $\pm$ 2.9 $\pm$ 4.3 & 0.0558 $\pm$ 3.3 $\pm$ 2.1 $\pm$ 3.2\\
0.918 -- 1.153 & 0.0258 $\pm$ 8.8 $\pm$ 6.1 $\pm$ 6.2 & 0.0258 $\pm$ 8.8 $\pm$ 6.1 $\pm$ 6.2 & 0.0259 $\pm$ 5.6 $\pm$ 3.3 $\pm$ 2.8 & 0.0262 $\pm$ 5.6 $\pm$ 3.3 $\pm$ 2.8 & 0.0263 $\pm$ 5.6 $\pm$ 3.3 $\pm$ 2.9 & 0.0264 $\pm$ 4.7 $\pm$ 2.9 $\pm$ 3.0\\
1.153 -- 1.496 & 0.0180 $\pm$ 8.2 $\pm$ 3.2 $\pm$ 2.6 & 0.0181 $\pm$ 8.2 $\pm$ 3.2 $\pm$ 2.6 & 0.0135 $\pm$ 5.8 $\pm$ 3.0 $\pm$ 4.9 & 0.0137 $\pm$ 5.8 $\pm$ 3.0 $\pm$ 4.9 & 0.0141 $\pm$ 5.8 $\pm$ 3.0 $\pm$ 5.2 & 0.0161 $\pm$ 4.6 $\pm$ 2.3 $\pm$ 3.2\\
1.496 -- 1.947 & 0.00804 $\pm$ 11 $\pm$ 4.5 $\pm$ 5.0 & 0.00804 $\pm$ 11 $\pm$ 4.5 $\pm$ 5.1 & 0.00542 $\pm$ 9.2 $\pm$ 4.8 $\pm$ 7.9 & 0.00546 $\pm$ 9.2 $\pm$ 4.8 $\pm$ 7.9 & 0.00537 $\pm$ 9.2 $\pm$ 4.8 $\pm$ 8.1 & 0.00681 $\pm$ 6.7 $\pm$ 3.5 $\pm$ 4.2\\
1.947 -- 2.522 & 0.00285 $\pm$ 19 $\pm$ 14 $\pm$ 8.4 & 0.00281 $\pm$ 19 $\pm$ 14 $\pm$ 8.4 & 0.00312 $\pm$ 9.5 $\pm$ 4.7 $\pm$ 4.0 & 0.00320 $\pm$ 9.5 $\pm$ 4.7 $\pm$ 4.0 & 0.00319 $\pm$ 9.5 $\pm$ 4.7 $\pm$ 4.3 & 0.00328 $\pm$ 8.3 $\pm$ 4.6 $\pm$ 3.7\\
2.522 -- 3.277 & 0.00136 $\pm$ 24 $\pm$ 12 $\pm$ 10 & 0.00134 $\pm$ 24 $\pm$ 12 $\pm$ 10.0 & 0.00260 $\pm$ 9.4 $\pm$ 5.5 $\pm$ 5.6 & 0.00262 $\pm$ 9.4 $\pm$ 5.5 $\pm$ 5.6 & 0.00259 $\pm$ 9.4 $\pm$ 5.5 $\pm$ 5.9 & 0.00246 $\pm$ 8.5 $\pm$ 5.0 $\pm$ 4.5\\
3.277 -- 5.000 & 0.000850 $\pm$ 19 $\pm$ 9.6 $\pm$ 6.8 & 0.000843 $\pm$ 19 $\pm$ 9.6 $\pm$ 6.8 & 0.000517 $\pm$ 15 $\pm$ 8.7 $\pm$ 10 & 0.000532 $\pm$ 15 $\pm$ 8.7 $\pm$ 10 & 0.000530 $\pm$ 15 $\pm$ 8.7 $\pm$ 10 & 0.000696 $\pm$ 11 $\pm$ 6.6 $\pm$ 5.9\\
5.000 -- 10.000 & 0.000285 $\pm$ 21 $\pm$ 9.8 $\pm$ 5.3 & 0.000280 $\pm$ 21 $\pm$ 9.8 $\pm$ 5.3 & 0.000215 $\pm$ 14 $\pm$ 7.2 $\pm$ 7.5 & 0.000218 $\pm$ 14 $\pm$ 7.2 $\pm$ 7.5 & 0.000226 $\pm$ 14 $\pm$ 7.2 $\pm$ 7.7 & 0.000260 $\pm$ 11 $\pm$ 5.8 $\pm$ 4.7\\
\bottomrule 
\end{tabular} 
} 
\label{tab:CombPhiStarhighmass_y0008}
\end{table*}

\begin{table*}
\centering
\caption{The values of $(1/\sigma)\,\mathrm{d}\sigma/\mathrm{d}\phi^*_{\eta}$ in each bin of \PhiStar{} for the electron and muon channels separately (for various particle-level definitions) and for the Born-level combination in the kinematic region $116\ \GeV \leq m_{\ell\ell} < 150\ \GeV,\ 0.8 \leq |y_{\ell\ell}| < 1.6$. The associated statistical and systematic (both uncorrelated and correlated between bins of \PhiStar{}) are provided in percentage form.} 
\resizebox{\textwidth}{!}{
\begin{tabular}{ ccccccc } \toprule 
Bin & \multicolumn{6}{c}{$(1/\sigma)\,d\sigma/d\phi^*_{\eta}$ $\pm$ Statistical [\%] $\pm$ Uncorrelated systematic [\%] $\pm$ Correlated systematic [\%]} \\ 
 \cmidrule(r){2-7} 
 & \multicolumn{2}{c}{Electron channel} &  \multicolumn{3}{c}{Muon channel} & Combination \\ 
\cmidrule(r){2-3} \cmidrule(r){4-6} \cmidrule(r){7-7} 
 & dressed & Born & bare & dressed & Born & Born \\ 
 0.000 -- 0.004 & 10.972 $\pm$ 3.1 $\pm$ 1.0 $\pm$ 2.3 & 10.997 $\pm$ 3.1 $\pm$ 1.0 $\pm$ 2.3 & 12.244 $\pm$ 1.8 $\pm$ 0.8 $\pm$ 1.8 & 12.206 $\pm$ 1.8 $\pm$ 0.8 $\pm$ 1.8 & 12.288 $\pm$ 1.8 $\pm$ 0.8 $\pm$ 1.8 & 11.982 $\pm$ 1.6 $\pm$ 0.6 $\pm$ 1.9\\
0.004 -- 0.008 & 11.989 $\pm$ 3.0 $\pm$ 0.9 $\pm$ 1.0 & 12.122 $\pm$ 3.0 $\pm$ 0.9 $\pm$ 1.0 & 11.775 $\pm$ 1.8 $\pm$ 0.8 $\pm$ 0.9 & 11.843 $\pm$ 1.8 $\pm$ 0.8 $\pm$ 0.9 & 11.904 $\pm$ 1.8 $\pm$ 0.8 $\pm$ 0.9 & 11.940 $\pm$ 1.6 $\pm$ 0.6 $\pm$ 0.8\\
0.008 -- 0.012 & 10.628 $\pm$ 3.1 $\pm$ 0.9 $\pm$ 1.3 & 10.637 $\pm$ 3.1 $\pm$ 0.9 $\pm$ 1.3 & 10.892 $\pm$ 1.9 $\pm$ 0.8 $\pm$ 0.7 & 10.805 $\pm$ 1.9 $\pm$ 0.8 $\pm$ 0.7 & 10.871 $\pm$ 1.9 $\pm$ 0.8 $\pm$ 0.7 & 10.818 $\pm$ 1.6 $\pm$ 0.7 $\pm$ 0.7\\
0.012 -- 0.016 & 10.718 $\pm$ 3.1 $\pm$ 1.0 $\pm$ 0.8 & 10.832 $\pm$ 3.1 $\pm$ 1.0 $\pm$ 0.8 & 11.265 $\pm$ 1.9 $\pm$ 0.8 $\pm$ 0.8 & 11.237 $\pm$ 1.9 $\pm$ 0.8 $\pm$ 0.8 & 11.211 $\pm$ 1.9 $\pm$ 0.8 $\pm$ 0.8 & 11.085 $\pm$ 1.6 $\pm$ 0.7 $\pm$ 0.7\\
0.016 -- 0.020 & 9.871 $\pm$ 3.2 $\pm$ 0.9 $\pm$ 1.1 & 9.849 $\pm$ 3.2 $\pm$ 0.9 $\pm$ 1.1 & 9.966 $\pm$ 2.0 $\pm$ 0.9 $\pm$ 0.6 & 9.860 $\pm$ 2.0 $\pm$ 0.9 $\pm$ 0.6 & 10.009 $\pm$ 2.0 $\pm$ 0.9 $\pm$ 0.7 & 9.945 $\pm$ 1.7 $\pm$ 0.7 $\pm$ 0.6\\
0.020 -- 0.024 & 9.754 $\pm$ 3.3 $\pm$ 1.1 $\pm$ 0.9 & 9.811 $\pm$ 3.3 $\pm$ 1.1 $\pm$ 0.9 & 9.756 $\pm$ 2.0 $\pm$ 0.9 $\pm$ 0.7 & 9.737 $\pm$ 2.0 $\pm$ 0.9 $\pm$ 0.7 & 9.765 $\pm$ 2.0 $\pm$ 0.9 $\pm$ 0.7 & 9.747 $\pm$ 1.7 $\pm$ 0.7 $\pm$ 0.6\\
0.024 -- 0.029 & 8.577 $\pm$ 3.1 $\pm$ 1.0 $\pm$ 0.7 & 8.597 $\pm$ 3.1 $\pm$ 1.0 $\pm$ 0.7 & 8.647 $\pm$ 1.9 $\pm$ 0.8 $\pm$ 0.5 & 8.611 $\pm$ 1.9 $\pm$ 0.8 $\pm$ 0.5 & 8.609 $\pm$ 1.9 $\pm$ 0.8 $\pm$ 0.6 & 8.572 $\pm$ 1.6 $\pm$ 0.7 $\pm$ 0.5\\
0.029 -- 0.034 & 7.516 $\pm$ 3.4 $\pm$ 1.4 $\pm$ 1.7 & 7.495 $\pm$ 3.4 $\pm$ 1.4 $\pm$ 1.7 & 8.261 $\pm$ 2.0 $\pm$ 0.9 $\pm$ 0.5 & 8.268 $\pm$ 2.0 $\pm$ 0.9 $\pm$ 0.5 & 8.305 $\pm$ 2.0 $\pm$ 0.9 $\pm$ 0.6 & 8.038 $\pm$ 1.7 $\pm$ 0.7 $\pm$ 0.6\\
0.034 -- 0.039 & 7.643 $\pm$ 3.3 $\pm$ 1.0 $\pm$ 0.8 & 7.677 $\pm$ 3.3 $\pm$ 1.1 $\pm$ 0.8 & 7.411 $\pm$ 2.1 $\pm$ 0.9 $\pm$ 0.5 & 7.404 $\pm$ 2.1 $\pm$ 0.9 $\pm$ 0.5 & 7.432 $\pm$ 2.1 $\pm$ 0.9 $\pm$ 0.6 & 7.473 $\pm$ 1.8 $\pm$ 0.7 $\pm$ 0.5\\
0.039 -- 0.045 & 6.365 $\pm$ 3.3 $\pm$ 1.0 $\pm$ 0.9 & 6.356 $\pm$ 3.3 $\pm$ 1.0 $\pm$ 0.9 & 6.576 $\pm$ 2.0 $\pm$ 0.9 $\pm$ 0.4 & 6.583 $\pm$ 2.0 $\pm$ 0.9 $\pm$ 0.4 & 6.552 $\pm$ 2.0 $\pm$ 0.9 $\pm$ 0.5 & 6.489 $\pm$ 1.7 $\pm$ 0.7 $\pm$ 0.5\\
0.045 -- 0.051 & 6.080 $\pm$ 3.4 $\pm$ 1.0 $\pm$ 0.6 & 6.080 $\pm$ 3.4 $\pm$ 1.0 $\pm$ 0.6 & 5.937 $\pm$ 2.1 $\pm$ 0.9 $\pm$ 0.4 & 5.931 $\pm$ 2.1 $\pm$ 0.9 $\pm$ 0.4 & 5.947 $\pm$ 2.1 $\pm$ 0.9 $\pm$ 0.5 & 5.971 $\pm$ 1.8 $\pm$ 0.7 $\pm$ 0.5\\
0.051 -- 0.057 & 5.577 $\pm$ 3.5 $\pm$ 1.0 $\pm$ 0.6 & 5.578 $\pm$ 3.5 $\pm$ 1.0 $\pm$ 0.6 & 5.636 $\pm$ 2.1 $\pm$ 0.9 $\pm$ 0.4 & 5.657 $\pm$ 2.1 $\pm$ 0.9 $\pm$ 0.4 & 5.652 $\pm$ 2.1 $\pm$ 0.9 $\pm$ 0.6 & 5.616 $\pm$ 1.8 $\pm$ 0.7 $\pm$ 0.5\\
0.057 -- 0.064 & 4.794 $\pm$ 3.6 $\pm$ 1.0 $\pm$ 0.7 & 4.748 $\pm$ 3.6 $\pm$ 1.0 $\pm$ 0.7 & 4.936 $\pm$ 2.1 $\pm$ 0.9 $\pm$ 0.5 & 4.918 $\pm$ 2.1 $\pm$ 0.9 $\pm$ 0.5 & 4.927 $\pm$ 2.1 $\pm$ 0.9 $\pm$ 0.5 & 4.868 $\pm$ 1.8 $\pm$ 0.7 $\pm$ 0.5\\
0.064 -- 0.072 & 4.645 $\pm$ 3.4 $\pm$ 1.7 $\pm$ 0.6 & 4.637 $\pm$ 3.4 $\pm$ 1.8 $\pm$ 0.6 & 4.375 $\pm$ 2.1 $\pm$ 0.9 $\pm$ 0.6 & 4.370 $\pm$ 2.1 $\pm$ 0.9 $\pm$ 0.6 & 4.333 $\pm$ 2.1 $\pm$ 0.9 $\pm$ 0.7 & 4.391 $\pm$ 1.8 $\pm$ 0.8 $\pm$ 0.6\\
0.072 -- 0.081 & 4.082 $\pm$ 3.4 $\pm$ 1.1 $\pm$ 0.7 & 4.078 $\pm$ 3.4 $\pm$ 1.2 $\pm$ 0.7 & 3.774 $\pm$ 2.1 $\pm$ 0.9 $\pm$ 0.6 & 3.773 $\pm$ 2.1 $\pm$ 0.9 $\pm$ 0.6 & 3.727 $\pm$ 2.1 $\pm$ 0.9 $\pm$ 0.6 & 3.822 $\pm$ 1.8 $\pm$ 0.7 $\pm$ 0.6\\
0.081 -- 0.091 & 3.474 $\pm$ 3.5 $\pm$ 1.0 $\pm$ 0.8 & 3.457 $\pm$ 3.5 $\pm$ 1.0 $\pm$ 0.8 & 3.262 $\pm$ 2.2 $\pm$ 0.9 $\pm$ 0.7 & 3.238 $\pm$ 2.2 $\pm$ 0.9 $\pm$ 0.7 & 3.269 $\pm$ 2.2 $\pm$ 0.9 $\pm$ 0.7 & 3.329 $\pm$ 1.9 $\pm$ 0.7 $\pm$ 0.6\\
0.091 -- 0.102 & 3.205 $\pm$ 3.5 $\pm$ 1.1 $\pm$ 0.9 & 3.212 $\pm$ 3.5 $\pm$ 1.2 $\pm$ 0.9 & 3.119 $\pm$ 2.1 $\pm$ 0.9 $\pm$ 0.5 & 3.125 $\pm$ 2.1 $\pm$ 0.9 $\pm$ 0.5 & 3.116 $\pm$ 2.1 $\pm$ 0.9 $\pm$ 0.6 & 3.126 $\pm$ 1.8 $\pm$ 0.7 $\pm$ 0.5\\
0.102 -- 0.114 & 2.435 $\pm$ 3.9 $\pm$ 1.2 $\pm$ 0.9 & 2.420 $\pm$ 3.9 $\pm$ 1.3 $\pm$ 0.9 & 2.528 $\pm$ 2.3 $\pm$ 0.9 $\pm$ 0.6 & 2.533 $\pm$ 2.3 $\pm$ 0.9 $\pm$ 0.6 & 2.509 $\pm$ 2.3 $\pm$ 0.9 $\pm$ 0.6 & 2.473 $\pm$ 2.0 $\pm$ 0.8 $\pm$ 0.5\\
0.114 -- 0.128 & 2.051 $\pm$ 3.9 $\pm$ 1.2 $\pm$ 0.8 & 2.041 $\pm$ 3.9 $\pm$ 1.3 $\pm$ 0.8 & 2.071 $\pm$ 2.3 $\pm$ 1.1 $\pm$ 0.6 & 2.075 $\pm$ 2.3 $\pm$ 1.1 $\pm$ 0.6 & 2.066 $\pm$ 2.3 $\pm$ 1.1 $\pm$ 0.6 & 2.057 $\pm$ 2.0 $\pm$ 0.9 $\pm$ 0.5\\
0.128 -- 0.145 & 1.828 $\pm$ 3.8 $\pm$ 1.5 $\pm$ 1.7 & 1.823 $\pm$ 3.8 $\pm$ 1.5 $\pm$ 1.8 & 1.907 $\pm$ 2.2 $\pm$ 0.9 $\pm$ 0.6 & 1.902 $\pm$ 2.2 $\pm$ 0.9 $\pm$ 0.6 & 1.905 $\pm$ 2.2 $\pm$ 0.9 $\pm$ 0.6 & 1.886 $\pm$ 1.9 $\pm$ 0.8 $\pm$ 0.6\\
0.145 -- 0.165 & 1.608 $\pm$ 3.8 $\pm$ 1.2 $\pm$ 1.4 & 1.597 $\pm$ 3.8 $\pm$ 1.3 $\pm$ 1.4 & 1.579 $\pm$ 2.2 $\pm$ 1.1 $\pm$ 0.5 & 1.584 $\pm$ 2.2 $\pm$ 1.1 $\pm$ 0.5 & 1.580 $\pm$ 2.2 $\pm$ 1.1 $\pm$ 0.5 & 1.579 $\pm$ 1.9 $\pm$ 0.9 $\pm$ 0.5\\
0.165 -- 0.189 & 1.211 $\pm$ 3.9 $\pm$ 1.3 $\pm$ 1.4 & 1.207 $\pm$ 3.9 $\pm$ 1.3 $\pm$ 1.5 & 1.177 $\pm$ 2.4 $\pm$ 1.0 $\pm$ 0.8 & 1.177 $\pm$ 2.4 $\pm$ 1.0 $\pm$ 0.8 & 1.173 $\pm$ 2.4 $\pm$ 1.0 $\pm$ 0.8 & 1.183 $\pm$ 2.0 $\pm$ 0.8 $\pm$ 0.7\\
0.189 -- 0.219 & 0.940 $\pm$ 4.0 $\pm$ 1.8 $\pm$ 1.2 & 0.938 $\pm$ 4.0 $\pm$ 1.8 $\pm$ 1.3 & 0.970 $\pm$ 2.3 $\pm$ 1.3 $\pm$ 0.7 & 0.972 $\pm$ 2.3 $\pm$ 1.3 $\pm$ 0.7 & 0.968 $\pm$ 2.3 $\pm$ 1.3 $\pm$ 0.7 & 0.957 $\pm$ 2.0 $\pm$ 1.1 $\pm$ 0.6\\
0.219 -- 0.258 & 0.774 $\pm$ 4.0 $\pm$ 1.7 $\pm$ 1.1 & 0.773 $\pm$ 4.0 $\pm$ 1.7 $\pm$ 1.0 & 0.784 $\pm$ 2.3 $\pm$ 1.3 $\pm$ 0.5 & 0.789 $\pm$ 2.3 $\pm$ 1.3 $\pm$ 0.5 & 0.786 $\pm$ 2.3 $\pm$ 1.3 $\pm$ 0.5 & 0.780 $\pm$ 2.0 $\pm$ 1.1 $\pm$ 0.5\\
0.258 -- 0.312 & 0.547 $\pm$ 4.1 $\pm$ 1.6 $\pm$ 2.4 & 0.546 $\pm$ 4.1 $\pm$ 1.6 $\pm$ 2.4 & 0.543 $\pm$ 2.3 $\pm$ 0.9 $\pm$ 1.6 & 0.545 $\pm$ 2.3 $\pm$ 0.9 $\pm$ 1.6 & 0.542 $\pm$ 2.3 $\pm$ 0.9 $\pm$ 1.6 & 0.547 $\pm$ 2.0 $\pm$ 0.8 $\pm$ 1.2\\
0.312 -- 0.391 & 0.355 $\pm$ 4.3 $\pm$ 1.8 $\pm$ 2.3 & 0.354 $\pm$ 4.3 $\pm$ 1.8 $\pm$ 2.3 & 0.329 $\pm$ 2.6 $\pm$ 1.1 $\pm$ 2.0 & 0.330 $\pm$ 2.6 $\pm$ 1.1 $\pm$ 2.0 & 0.330 $\pm$ 2.6 $\pm$ 1.1 $\pm$ 2.0 & 0.341 $\pm$ 2.2 $\pm$ 0.9 $\pm$ 1.5\\
0.391 -- 0.524 & 0.212 $\pm$ 4.4 $\pm$ 2.0 $\pm$ 2.1 & 0.212 $\pm$ 4.4 $\pm$ 2.0 $\pm$ 2.1 & 0.186 $\pm$ 2.6 $\pm$ 1.1 $\pm$ 2.0 & 0.188 $\pm$ 2.6 $\pm$ 1.1 $\pm$ 2.0 & 0.188 $\pm$ 2.6 $\pm$ 1.1 $\pm$ 2.0 & 0.195 $\pm$ 2.3 $\pm$ 1.0 $\pm$ 1.7\\
0.524 -- 0.695 & 0.101 $\pm$ 5.9 $\pm$ 2.8 $\pm$ 3.0 & 0.101 $\pm$ 5.9 $\pm$ 2.8 $\pm$ 3.0 & 0.0959 $\pm$ 3.4 $\pm$ 1.9 $\pm$ 1.8 & 0.0964 $\pm$ 3.4 $\pm$ 1.9 $\pm$ 1.8 & 0.0969 $\pm$ 3.4 $\pm$ 1.9 $\pm$ 1.8 & 0.0977 $\pm$ 2.9 $\pm$ 1.6 $\pm$ 1.8\\
0.695 -- 0.918 & 0.0504 $\pm$ 7.3 $\pm$ 3.3 $\pm$ 3.1 & 0.0507 $\pm$ 7.3 $\pm$ 3.3 $\pm$ 3.1 & 0.0475 $\pm$ 4.3 $\pm$ 2.2 $\pm$ 2.5 & 0.0477 $\pm$ 4.3 $\pm$ 2.2 $\pm$ 2.5 & 0.0473 $\pm$ 4.3 $\pm$ 2.2 $\pm$ 2.5 & 0.0486 $\pm$ 3.7 $\pm$ 1.9 $\pm$ 2.3\\
0.918 -- 1.153 & 0.0254 $\pm$ 9.5 $\pm$ 3.9 $\pm$ 5.0 & 0.0252 $\pm$ 9.5 $\pm$ 4.0 $\pm$ 5.0 & 0.0247 $\pm$ 5.6 $\pm$ 2.7 $\pm$ 2.2 & 0.0251 $\pm$ 5.6 $\pm$ 2.7 $\pm$ 2.2 & 0.0251 $\pm$ 5.6 $\pm$ 2.7 $\pm$ 2.2 & 0.0250 $\pm$ 4.9 $\pm$ 2.2 $\pm$ 2.2\\
1.153 -- 1.496 & 0.0114 $\pm$ 12 $\pm$ 12 $\pm$ 25 & 0.0115 $\pm$ 12 $\pm$ 12 $\pm$ 25 & 0.0109 $\pm$ 7.1 $\pm$ 3.5 $\pm$ 2.5 & 0.0111 $\pm$ 7.1 $\pm$ 3.5 $\pm$ 2.5 & 0.0107 $\pm$ 7.1 $\pm$ 3.5 $\pm$ 2.6 & 0.0108 $\pm$ 6.3 $\pm$ 3.6 $\pm$ 3.4\\
1.496 -- 1.947 & 0.00523 $\pm$ 15 $\pm$ 6.7 $\pm$ 5.6 & 0.00520 $\pm$ 15 $\pm$ 6.7 $\pm$ 5.6 & 0.00588 $\pm$ 7.9 $\pm$ 4.1 $\pm$ 2.4 & 0.00597 $\pm$ 7.9 $\pm$ 4.1 $\pm$ 2.4 & 0.00592 $\pm$ 7.9 $\pm$ 4.1 $\pm$ 2.5 & 0.00579 $\pm$ 7.0 $\pm$ 3.5 $\pm$ 2.5\\
1.947 -- 2.522 & 0.00236 $\pm$ 20 $\pm$ 8.6 $\pm$ 8.2 & 0.00237 $\pm$ 20 $\pm$ 8.6 $\pm$ 8.2 & 0.00407 $\pm$ 8.9 $\pm$ 4.3 $\pm$ 4.6 & 0.00407 $\pm$ 8.9 $\pm$ 4.3 $\pm$ 4.6 & 0.00403 $\pm$ 8.9 $\pm$ 4.3 $\pm$ 4.7 & 0.00374 $\pm$ 8.0 $\pm$ 3.9 $\pm$ 3.6\\
2.522 -- 3.277 & 0.00147 $\pm$ 22 $\pm$ 9.3 $\pm$ 10.0 & 0.00153 $\pm$ 22 $\pm$ 9.4 $\pm$ 10 & 0.00124 $\pm$ 15 $\pm$ 7.3 $\pm$ 4.2 & 0.00128 $\pm$ 15 $\pm$ 7.3 $\pm$ 4.2 & 0.00129 $\pm$ 15 $\pm$ 7.3 $\pm$ 4.2 & 0.00138 $\pm$ 12 $\pm$ 5.8 $\pm$ 4.1\\
3.277 -- 5.000 & 0.000654 $\pm$ 22 $\pm$ 13 $\pm$ 9.8 & 0.000644 $\pm$ 22 $\pm$ 13 $\pm$ 9.9 & 0.000505 $\pm$ 15 $\pm$ 7.9 $\pm$ 6.5 & 0.000511 $\pm$ 15 $\pm$ 7.9 $\pm$ 6.5 & 0.000506 $\pm$ 15 $\pm$ 7.9 $\pm$ 6.5 & 0.000559 $\pm$ 12 $\pm$ 6.8 $\pm$ 5.0\\
5.000 -- 10.000 & 0.000228 $\pm$ 21 $\pm$ 8.2 $\pm$ 9.0 & 0.000230 $\pm$ 21 $\pm$ 8.3 $\pm$ 8.7 & 0.000168 $\pm$ 16 $\pm$ 8.2 $\pm$ 2.4 & 0.000171 $\pm$ 16 $\pm$ 8.2 $\pm$ 2.4 & 0.000169 $\pm$ 16 $\pm$ 8.2 $\pm$ 2.5 & 0.000187 $\pm$ 13 $\pm$ 6.2 $\pm$ 3.0\\
\bottomrule 
\end{tabular} 
} 
\label{tab:CombPhiStarhighmass_y0816}
\end{table*}

\begin{table*}
\centering
\caption{The values of $(1/\sigma)\,\mathrm{d}\sigma/\mathrm{d}\phi^*_{\eta}$ in each bin of \PhiStar{} for the electron and muon channels separately (for various particle-level definitions) and for the Born-level combination in the kinematic region $116\ \GeV \leq m_{\ell\ell} < 150\ \GeV,\ 1.6 \leq |y_{\ell\ell}| < 2.4$. The associated statistical and systematic (both uncorrelated and correlated between bins of \PhiStar{}) are provided in percentage form.} 
\resizebox{\textwidth}{!}{
\begin{tabular}{ ccccccc } \toprule 
Bin & \multicolumn{6}{c}{$(1/\sigma)\,d\sigma/d\phi^*_{\eta}$ $\pm$ Statistical [\%] $\pm$ Uncorrelated systematic [\%] $\pm$ Correlated systematic [\%]} \\ 
 \cmidrule(r){2-7} 
 & \multicolumn{2}{c}{Electron channel} &  \multicolumn{3}{c}{Muon channel} & Combination \\ 
\cmidrule(r){2-3} \cmidrule(r){4-6} \cmidrule(r){7-7} 
 & dressed & Born & bare & dressed & Born & Born \\ 
 0.000 -- 0.004 & 11.892 $\pm$ 5.1 $\pm$ 1.6 $\pm$ 1.5 & 12.002 $\pm$ 5.1 $\pm$ 1.7 $\pm$ 1.5 & 11.291 $\pm$ 2.9 $\pm$ 1.2 $\pm$ 1.3 & 11.291 $\pm$ 2.9 $\pm$ 1.2 $\pm$ 1.3 & 11.263 $\pm$ 2.9 $\pm$ 1.2 $\pm$ 1.4 & 11.481 $\pm$ 2.5 $\pm$ 1.0 $\pm$ 1.2\\
0.004 -- 0.008 & 11.648 $\pm$ 5.2 $\pm$ 2.7 $\pm$ 1.8 & 11.765 $\pm$ 5.2 $\pm$ 2.8 $\pm$ 1.8 & 11.245 $\pm$ 2.9 $\pm$ 1.2 $\pm$ 0.7 & 11.207 $\pm$ 2.9 $\pm$ 1.2 $\pm$ 0.7 & 11.271 $\pm$ 2.9 $\pm$ 1.2 $\pm$ 0.9 & 11.391 $\pm$ 2.6 $\pm$ 1.1 $\pm$ 0.8\\
0.008 -- 0.012 & 11.455 $\pm$ 5.0 $\pm$ 1.5 $\pm$ 1.0 & 11.528 $\pm$ 5.0 $\pm$ 1.6 $\pm$ 1.0 & 11.166 $\pm$ 2.9 $\pm$ 1.2 $\pm$ 0.6 & 11.164 $\pm$ 2.9 $\pm$ 1.2 $\pm$ 0.6 & 11.356 $\pm$ 2.9 $\pm$ 1.2 $\pm$ 0.8 & 11.419 $\pm$ 2.5 $\pm$ 1.0 $\pm$ 0.7\\
0.012 -- 0.016 & 11.171 $\pm$ 5.1 $\pm$ 1.6 $\pm$ 1.0 & 11.159 $\pm$ 5.1 $\pm$ 1.7 $\pm$ 1.0 & 10.965 $\pm$ 2.9 $\pm$ 1.2 $\pm$ 0.8 & 10.902 $\pm$ 2.9 $\pm$ 1.2 $\pm$ 0.8 & 10.904 $\pm$ 2.9 $\pm$ 1.2 $\pm$ 1.0 & 10.991 $\pm$ 2.6 $\pm$ 1.0 $\pm$ 0.8\\
0.016 -- 0.020 & 9.714 $\pm$ 5.6 $\pm$ 1.9 $\pm$ 1.2 & 9.678 $\pm$ 5.6 $\pm$ 2.0 $\pm$ 1.2 & 10.150 $\pm$ 3.1 $\pm$ 1.3 $\pm$ 0.8 & 10.200 $\pm$ 3.1 $\pm$ 1.3 $\pm$ 0.8 & 10.303 $\pm$ 3.1 $\pm$ 1.3 $\pm$ 1.0 & 10.167 $\pm$ 2.7 $\pm$ 1.1 $\pm$ 0.8\\
0.020 -- 0.024 & 9.337 $\pm$ 5.7 $\pm$ 2.5 $\pm$ 1.9 & 9.362 $\pm$ 5.7 $\pm$ 2.5 $\pm$ 1.8 & 10.034 $\pm$ 3.1 $\pm$ 1.3 $\pm$ 0.5 & 10.084 $\pm$ 3.1 $\pm$ 1.3 $\pm$ 0.5 & 9.997 $\pm$ 3.1 $\pm$ 1.3 $\pm$ 0.7 & 9.850 $\pm$ 2.7 $\pm$ 1.1 $\pm$ 0.7\\
0.024 -- 0.029 & 9.075 $\pm$ 5.1 $\pm$ 2.4 $\pm$ 1.7 & 9.100 $\pm$ 5.1 $\pm$ 2.4 $\pm$ 1.7 & 9.051 $\pm$ 2.9 $\pm$ 1.2 $\pm$ 0.3 & 9.019 $\pm$ 2.9 $\pm$ 1.2 $\pm$ 0.3 & 9.027 $\pm$ 2.9 $\pm$ 1.2 $\pm$ 0.4 & 9.049 $\pm$ 2.5 $\pm$ 1.1 $\pm$ 0.5\\
0.029 -- 0.034 & 7.757 $\pm$ 5.6 $\pm$ 1.9 $\pm$ 1.1 & 7.764 $\pm$ 5.6 $\pm$ 1.9 $\pm$ 1.1 & 8.218 $\pm$ 3.1 $\pm$ 1.2 $\pm$ 0.6 & 8.239 $\pm$ 3.1 $\pm$ 1.2 $\pm$ 0.6 & 8.263 $\pm$ 3.1 $\pm$ 1.2 $\pm$ 0.7 & 8.151 $\pm$ 2.7 $\pm$ 1.1 $\pm$ 0.6\\
0.034 -- 0.039 & 7.263 $\pm$ 5.7 $\pm$ 1.9 $\pm$ 1.6 & 7.270 $\pm$ 5.7 $\pm$ 2.0 $\pm$ 1.6 & 7.469 $\pm$ 3.2 $\pm$ 1.3 $\pm$ 0.3 & 7.356 $\pm$ 3.2 $\pm$ 1.3 $\pm$ 0.3 & 7.515 $\pm$ 3.2 $\pm$ 1.3 $\pm$ 0.5 & 7.462 $\pm$ 2.8 $\pm$ 1.1 $\pm$ 0.5\\
0.039 -- 0.045 & 6.336 $\pm$ 5.6 $\pm$ 1.8 $\pm$ 1.3 & 6.324 $\pm$ 5.6 $\pm$ 1.8 $\pm$ 1.3 & 6.501 $\pm$ 3.2 $\pm$ 1.3 $\pm$ 0.5 & 6.463 $\pm$ 3.2 $\pm$ 1.3 $\pm$ 0.5 & 6.425 $\pm$ 3.2 $\pm$ 1.3 $\pm$ 0.6 & 6.409 $\pm$ 2.7 $\pm$ 1.1 $\pm$ 0.5\\
0.045 -- 0.051 & 6.574 $\pm$ 5.5 $\pm$ 1.9 $\pm$ 1.4 & 6.582 $\pm$ 5.5 $\pm$ 1.9 $\pm$ 1.4 & 5.615 $\pm$ 3.3 $\pm$ 1.3 $\pm$ 0.9 & 5.675 $\pm$ 3.3 $\pm$ 1.3 $\pm$ 0.9 & 5.648 $\pm$ 3.3 $\pm$ 1.3 $\pm$ 1.0 & 5.886 $\pm$ 2.9 $\pm$ 1.1 $\pm$ 0.8\\
0.051 -- 0.057 & 5.317 $\pm$ 6.0 $\pm$ 2.0 $\pm$ 1.3 & 5.312 $\pm$ 6.0 $\pm$ 2.0 $\pm$ 1.3 & 5.470 $\pm$ 3.3 $\pm$ 1.3 $\pm$ 0.6 & 5.445 $\pm$ 3.3 $\pm$ 1.3 $\pm$ 0.6 & 5.424 $\pm$ 3.3 $\pm$ 1.3 $\pm$ 0.6 & 5.405 $\pm$ 2.9 $\pm$ 1.1 $\pm$ 0.6\\
0.057 -- 0.064 & 4.755 $\pm$ 6.0 $\pm$ 2.0 $\pm$ 1.5 & 4.707 $\pm$ 6.0 $\pm$ 2.1 $\pm$ 1.6 & 5.408 $\pm$ 3.2 $\pm$ 1.3 $\pm$ 0.9 & 5.424 $\pm$ 3.2 $\pm$ 1.3 $\pm$ 0.9 & 5.460 $\pm$ 3.2 $\pm$ 1.3 $\pm$ 1.0 & 5.277 $\pm$ 2.8 $\pm$ 1.1 $\pm$ 0.8\\
0.064 -- 0.072 & 4.983 $\pm$ 5.5 $\pm$ 1.8 $\pm$ 1.6 & 5.002 $\pm$ 5.5 $\pm$ 1.8 $\pm$ 1.6 & 4.468 $\pm$ 3.4 $\pm$ 1.4 $\pm$ 0.6 & 4.460 $\pm$ 3.4 $\pm$ 1.4 $\pm$ 0.6 & 4.476 $\pm$ 3.4 $\pm$ 1.4 $\pm$ 0.7 & 4.619 $\pm$ 2.9 $\pm$ 1.1 $\pm$ 0.7\\
0.072 -- 0.081 & 3.911 $\pm$ 5.8 $\pm$ 2.4 $\pm$ 1.6 & 3.870 $\pm$ 5.8 $\pm$ 2.4 $\pm$ 1.5 & 4.120 $\pm$ 3.2 $\pm$ 1.3 $\pm$ 0.8 & 4.103 $\pm$ 3.2 $\pm$ 1.3 $\pm$ 0.8 & 4.090 $\pm$ 3.2 $\pm$ 1.3 $\pm$ 0.9 & 4.051 $\pm$ 2.8 $\pm$ 1.1 $\pm$ 0.8\\
0.081 -- 0.091 & 3.070 $\pm$ 6.3 $\pm$ 2.4 $\pm$ 1.5 & 3.030 $\pm$ 6.3 $\pm$ 2.5 $\pm$ 1.5 & 3.441 $\pm$ 3.4 $\pm$ 1.4 $\pm$ 0.6 & 3.426 $\pm$ 3.4 $\pm$ 1.4 $\pm$ 0.6 & 3.388 $\pm$ 3.4 $\pm$ 1.4 $\pm$ 0.7 & 3.298 $\pm$ 3.0 $\pm$ 1.2 $\pm$ 0.6\\
0.091 -- 0.102 & 3.005 $\pm$ 6.1 $\pm$ 1.8 $\pm$ 4.8 & 3.011 $\pm$ 6.1 $\pm$ 1.9 $\pm$ 4.8 & 2.876 $\pm$ 3.5 $\pm$ 1.6 $\pm$ 0.7 & 2.874 $\pm$ 3.5 $\pm$ 1.6 $\pm$ 0.7 & 2.852 $\pm$ 3.5 $\pm$ 1.6 $\pm$ 0.8 & 2.902 $\pm$ 3.0 $\pm$ 1.3 $\pm$ 1.0\\
0.102 -- 0.114 & 2.530 $\pm$ 6.3 $\pm$ 1.9 $\pm$ 2.2 & 2.520 $\pm$ 6.3 $\pm$ 2.0 $\pm$ 2.2 & 2.595 $\pm$ 3.6 $\pm$ 1.4 $\pm$ 0.3 & 2.590 $\pm$ 3.6 $\pm$ 1.4 $\pm$ 0.3 & 2.595 $\pm$ 3.6 $\pm$ 1.4 $\pm$ 0.5 & 2.578 $\pm$ 3.1 $\pm$ 1.2 $\pm$ 0.5\\
0.114 -- 0.128 & 2.350 $\pm$ 6.1 $\pm$ 1.8 $\pm$ 2.2 & 2.358 $\pm$ 6.1 $\pm$ 1.9 $\pm$ 2.2 & 2.160 $\pm$ 3.6 $\pm$ 1.4 $\pm$ 0.6 & 2.157 $\pm$ 3.6 $\pm$ 1.4 $\pm$ 0.6 & 2.152 $\pm$ 3.6 $\pm$ 1.4 $\pm$ 0.7 & 2.205 $\pm$ 3.1 $\pm$ 1.2 $\pm$ 0.7\\
0.128 -- 0.145 & 1.975 $\pm$ 5.9 $\pm$ 1.6 $\pm$ 1.2 & 1.985 $\pm$ 5.9 $\pm$ 1.7 $\pm$ 1.2 & 1.806 $\pm$ 3.6 $\pm$ 1.6 $\pm$ 0.4 & 1.814 $\pm$ 3.6 $\pm$ 1.6 $\pm$ 0.4 & 1.825 $\pm$ 3.6 $\pm$ 1.6 $\pm$ 0.5 & 1.871 $\pm$ 3.1 $\pm$ 1.2 $\pm$ 0.5\\
0.145 -- 0.165 & 1.409 $\pm$ 6.5 $\pm$ 1.9 $\pm$ 0.8 & 1.406 $\pm$ 6.5 $\pm$ 1.9 $\pm$ 0.8 & 1.626 $\pm$ 3.5 $\pm$ 1.4 $\pm$ 0.8 & 1.634 $\pm$ 3.5 $\pm$ 1.4 $\pm$ 0.8 & 1.617 $\pm$ 3.5 $\pm$ 1.4 $\pm$ 0.8 & 1.562 $\pm$ 3.1 $\pm$ 1.2 $\pm$ 0.6\\
0.165 -- 0.189 & 1.252 $\pm$ 6.4 $\pm$ 2.4 $\pm$ 3.5 & 1.252 $\pm$ 6.4 $\pm$ 2.5 $\pm$ 3.5 & 1.333 $\pm$ 3.4 $\pm$ 1.4 $\pm$ 0.9 & 1.343 $\pm$ 3.4 $\pm$ 1.4 $\pm$ 0.9 & 1.327 $\pm$ 3.4 $\pm$ 1.4 $\pm$ 0.9 & 1.316 $\pm$ 3.0 $\pm$ 1.2 $\pm$ 0.9\\
0.189 -- 0.219 & 0.970 $\pm$ 6.6 $\pm$ 4.5 $\pm$ 4.3 & 0.970 $\pm$ 6.6 $\pm$ 4.6 $\pm$ 4.3 & 0.970 $\pm$ 3.7 $\pm$ 1.5 $\pm$ 1.0 & 0.963 $\pm$ 3.7 $\pm$ 1.5 $\pm$ 1.0 & 0.962 $\pm$ 3.7 $\pm$ 1.5 $\pm$ 1.0 & 0.963 $\pm$ 3.3 $\pm$ 1.5 $\pm$ 1.0\\
0.219 -- 0.258 & 0.856 $\pm$ 6.1 $\pm$ 2.3 $\pm$ 2.3 & 0.859 $\pm$ 6.1 $\pm$ 2.3 $\pm$ 2.3 & 0.789 $\pm$ 3.6 $\pm$ 1.4 $\pm$ 0.7 & 0.788 $\pm$ 3.6 $\pm$ 1.4 $\pm$ 0.7 & 0.784 $\pm$ 3.6 $\pm$ 1.4 $\pm$ 0.8 & 0.805 $\pm$ 3.1 $\pm$ 1.2 $\pm$ 0.8\\
0.258 -- 0.312 & 0.538 $\pm$ 6.6 $\pm$ 2.1 $\pm$ 1.1 & 0.535 $\pm$ 6.6 $\pm$ 2.2 $\pm$ 1.1 & 0.541 $\pm$ 3.7 $\pm$ 1.7 $\pm$ 1.0 & 0.543 $\pm$ 3.7 $\pm$ 1.7 $\pm$ 1.0 & 0.540 $\pm$ 3.7 $\pm$ 1.7 $\pm$ 1.2 & 0.541 $\pm$ 3.2 $\pm$ 1.4 $\pm$ 1.0\\
0.312 -- 0.391 & 0.356 $\pm$ 6.8 $\pm$ 2.7 $\pm$ 2.5 & 0.357 $\pm$ 6.8 $\pm$ 2.7 $\pm$ 2.4 & 0.339 $\pm$ 3.9 $\pm$ 1.6 $\pm$ 1.0 & 0.342 $\pm$ 3.9 $\pm$ 1.6 $\pm$ 1.0 & 0.341 $\pm$ 3.9 $\pm$ 1.6 $\pm$ 1.2 & 0.347 $\pm$ 3.3 $\pm$ 1.4 $\pm$ 1.1\\
0.391 -- 0.524 & 0.194 $\pm$ 7.4 $\pm$ 3.2 $\pm$ 2.4 & 0.194 $\pm$ 7.4 $\pm$ 3.2 $\pm$ 2.4 & 0.201 $\pm$ 4.0 $\pm$ 1.7 $\pm$ 1.5 & 0.199 $\pm$ 4.0 $\pm$ 1.7 $\pm$ 1.5 & 0.199 $\pm$ 4.0 $\pm$ 1.7 $\pm$ 1.7 & 0.198 $\pm$ 3.5 $\pm$ 1.5 $\pm$ 1.5\\
0.524 -- 0.695 & 0.0787 $\pm$ 10 $\pm$ 5.9 $\pm$ 8.2 & 0.0779 $\pm$ 10 $\pm$ 6.0 $\pm$ 8.2 & 0.0865 $\pm$ 5.4 $\pm$ 2.2 $\pm$ 1.3 & 0.0878 $\pm$ 5.4 $\pm$ 2.2 $\pm$ 1.3 & 0.0873 $\pm$ 5.4 $\pm$ 2.2 $\pm$ 1.5 & 0.0861 $\pm$ 4.7 $\pm$ 2.1 $\pm$ 1.7\\
0.695 -- 0.918 & 0.0465 $\pm$ 11 $\pm$ 4.1 $\pm$ 4.6 & 0.0464 $\pm$ 11 $\pm$ 4.1 $\pm$ 4.6 & 0.0440 $\pm$ 6.4 $\pm$ 2.6 $\pm$ 2.6 & 0.0448 $\pm$ 6.4 $\pm$ 2.6 $\pm$ 2.6 & 0.0450 $\pm$ 6.4 $\pm$ 2.6 $\pm$ 2.6 & 0.0454 $\pm$ 5.6 $\pm$ 2.2 $\pm$ 2.2\\
0.918 -- 1.153 & 0.0228 $\pm$ 16 $\pm$ 5.2 $\pm$ 4.6 & 0.0227 $\pm$ 16 $\pm$ 5.3 $\pm$ 4.6 & 0.0230 $\pm$ 8.8 $\pm$ 3.7 $\pm$ 2.8 & 0.0238 $\pm$ 8.8 $\pm$ 3.7 $\pm$ 2.8 & 0.0235 $\pm$ 8.8 $\pm$ 3.7 $\pm$ 2.9 & 0.0233 $\pm$ 7.6 $\pm$ 3.0 $\pm$ 2.3\\
1.153 -- 1.496 & 0.00970 $\pm$ 19 $\pm$ 6.6 $\pm$ 6.5 & 0.00955 $\pm$ 19 $\pm$ 6.7 $\pm$ 6.5 & 0.00973 $\pm$ 12 $\pm$ 5.0 $\pm$ 4.7 & 0.00960 $\pm$ 12 $\pm$ 5.0 $\pm$ 4.7 & 0.00981 $\pm$ 12 $\pm$ 5.0 $\pm$ 4.8 & 0.00994 $\pm$ 10 $\pm$ 4.0 $\pm$ 3.6\\
1.496 -- 1.947 & 0.00496 $\pm$ 22 $\pm$ 6.7 $\pm$ 7.6 & 0.00491 $\pm$ 22 $\pm$ 6.8 $\pm$ 8.0 & 0.00262 $\pm$ 19 $\pm$ 7.6 $\pm$ 11 & 0.00272 $\pm$ 19 $\pm$ 7.6 $\pm$ 11 & 0.00277 $\pm$ 19 $\pm$ 7.6 $\pm$ 11 & 0.00343 $\pm$ 14 $\pm$ 5.6 $\pm$ 7.0\\
1.947 -- 2.522 & 0.00155 $\pm$ 34 $\pm$ 9.4 $\pm$ 3.9 & 0.00155 $\pm$ 34 $\pm$ 9.5 $\pm$ 3.9 & 0.000971 $\pm$ 23 $\pm$ 9.6 $\pm$ 10 & 0.000988 $\pm$ 23 $\pm$ 9.6 $\pm$ 10 & 0.00101 $\pm$ 23 $\pm$ 9.6 $\pm$ 10 & 0.00119 $\pm$ 19 $\pm$ 7.5 $\pm$ 6.9\\
2.522 -- 3.277 & 0.000678 $\pm$ 44 $\pm$ 9.9 $\pm$ 9.2 & 0.000631 $\pm$ 44 $\pm$ 11 $\pm$ 9.5 & 0.0000976 $\pm$ 116 $\pm$ 49 $\pm$ 8.2 & 0.000101 $\pm$ 116 $\pm$ 49 $\pm$ 8.2 & 0.000106 $\pm$ 116 $\pm$ 49 $\pm$ 8.2 & 0.000458 $\pm$ 38 $\pm$ 17 $\pm$ 5.4\\
3.277 -- 5.000 & 0.0000936 $\pm$ 101 $\pm$ 83 $\pm$ 84 & 0.0000944 $\pm$ 101 $\pm$ 83 $\pm$ 84 & 0.0000533 $\pm$ 65 $\pm$ 30 $\pm$ 7.0 & 0.0000535 $\pm$ 65 $\pm$ 30 $\pm$ 7.0 & 0.0000513 $\pm$ 65 $\pm$ 30 $\pm$ 7.1 & 0.0000589 $\pm$ 55 $\pm$ 29 $\pm$ 12\\
5.000 -- 10.000 & 0.0000540 $\pm$ 68 $\pm$ 19 $\pm$ 38 & 0.0000506 $\pm$ 68 $\pm$ 19 $\pm$ 38 & 0.0000431 $\pm$ 37 $\pm$ 17 $\pm$ 8.4 & 0.0000416 $\pm$ 37 $\pm$ 17 $\pm$ 8.4 & 0.0000387 $\pm$ 37 $\pm$ 17 $\pm$ 8.5 & 0.0000404 $\pm$ 33 $\pm$ 14 $\pm$ 8.4\\
\bottomrule 
\end{tabular} 
} 
\label{tab:CombPhiStarhighmass_y1624}
\end{table*}

\begin{table*}
\centering
\caption{The values of $(1/\sigma)\,\mathrm{d}\sigma/\mathrm{d}\phi^*_{\eta}$ in each bin of \PhiStar{} for the electron and muon channels separately (for various particle-level definitions) and for the Born-level combination in the kinematic region $46\ \GeV \leq m_{\ell\ell} < 66\ \GeV,\ |y_{\ell\ell}| < 2.4$. The associated statistical and systematic (both uncorrelated and correlated between bins of \PhiStar{}) are provided in percentage form.} 
\resizebox{\textwidth}{!}{
\begin{tabular}{ ccccccc } \toprule 
Bin & \multicolumn{6}{c}{$(1/\sigma)\,d\sigma/d\phi^*_{\eta}$ $\pm$ Statistical [\%] $\pm$ Uncorrelated systematic [\%] $\pm$ Correlated systematic [\%]} \\ 
 \cmidrule(r){2-7} 
 & \multicolumn{2}{c}{Electron channel} &  \multicolumn{3}{c}{Muon channel} & Combination \\ 
\cmidrule(r){2-3} \cmidrule(r){4-6} \cmidrule(r){7-7} 
 & dressed & Born & bare & dressed & Born & Born \\ 
 0.000 -- 0.004 & 6.941 $\pm$ 1.6 $\pm$ 0.7 $\pm$ 5.3 & 7.435 $\pm$ 1.6 $\pm$ 0.8 $\pm$ 5.4 & 6.741 $\pm$ 1.3 $\pm$ 0.3 $\pm$ 4.2 & 6.724 $\pm$ 1.3 $\pm$ 0.3 $\pm$ 4.2 & 7.169 $\pm$ 1.3 $\pm$ 0.3 $\pm$ 4.4 & 7.655 $\pm$ 1.0 $\pm$ 0.3 $\pm$ 3.0\\
0.004 -- 0.008 & 6.819 $\pm$ 1.6 $\pm$ 0.7 $\pm$ 2.3 & 7.209 $\pm$ 1.6 $\pm$ 0.7 $\pm$ 2.5 & 6.967 $\pm$ 1.2 $\pm$ 0.4 $\pm$ 1.2 & 6.909 $\pm$ 1.2 $\pm$ 0.4 $\pm$ 1.2 & 7.307 $\pm$ 1.2 $\pm$ 0.4 $\pm$ 1.8 & 7.439 $\pm$ 1.0 $\pm$ 0.3 $\pm$ 1.3\\
0.008 -- 0.012 & 6.793 $\pm$ 1.6 $\pm$ 0.7 $\pm$ 1.5 & 7.176 $\pm$ 1.6 $\pm$ 0.7 $\pm$ 1.8 & 6.877 $\pm$ 1.2 $\pm$ 0.3 $\pm$ 0.6 & 6.827 $\pm$ 1.2 $\pm$ 0.3 $\pm$ 0.6 & 7.251 $\pm$ 1.2 $\pm$ 0.3 $\pm$ 1.5 & 7.346 $\pm$ 1.0 $\pm$ 0.3 $\pm$ 1.1\\
0.012 -- 0.016 & 6.507 $\pm$ 1.6 $\pm$ 0.7 $\pm$ 1.0 & 6.895 $\pm$ 1.6 $\pm$ 0.7 $\pm$ 1.5 & 6.851 $\pm$ 1.2 $\pm$ 0.3 $\pm$ 0.4 & 6.814 $\pm$ 1.2 $\pm$ 0.3 $\pm$ 0.4 & 7.265 $\pm$ 1.2 $\pm$ 0.3 $\pm$ 1.4 & 7.235 $\pm$ 0.9 $\pm$ 0.3 $\pm$ 1.1\\
0.016 -- 0.020 & 6.498 $\pm$ 1.6 $\pm$ 0.7 $\pm$ 1.2 & 6.844 $\pm$ 1.6 $\pm$ 0.7 $\pm$ 1.6 & 6.648 $\pm$ 1.2 $\pm$ 0.4 $\pm$ 0.3 & 6.627 $\pm$ 1.2 $\pm$ 0.4 $\pm$ 0.3 & 7.035 $\pm$ 1.2 $\pm$ 0.4 $\pm$ 1.4 & 7.039 $\pm$ 1.0 $\pm$ 0.3 $\pm$ 1.1\\
0.020 -- 0.024 & 6.428 $\pm$ 1.6 $\pm$ 0.7 $\pm$ 0.8 & 6.768 $\pm$ 1.6 $\pm$ 0.7 $\pm$ 1.4 & 6.445 $\pm$ 1.3 $\pm$ 0.4 $\pm$ 0.2 & 6.440 $\pm$ 1.3 $\pm$ 0.4 $\pm$ 0.2 & 6.774 $\pm$ 1.3 $\pm$ 0.4 $\pm$ 1.4 & 6.852 $\pm$ 1.0 $\pm$ 0.4 $\pm$ 1.1\\
0.024 -- 0.029 & 6.261 $\pm$ 1.4 $\pm$ 0.6 $\pm$ 0.8 & 6.573 $\pm$ 1.4 $\pm$ 0.6 $\pm$ 1.4 & 6.167 $\pm$ 1.2 $\pm$ 0.3 $\pm$ 0.3 & 6.145 $\pm$ 1.2 $\pm$ 0.3 $\pm$ 0.3 & 6.514 $\pm$ 1.2 $\pm$ 0.3 $\pm$ 1.3 & 6.597 $\pm$ 0.9 $\pm$ 0.3 $\pm$ 1.0\\
0.029 -- 0.034 & 5.900 $\pm$ 1.5 $\pm$ 0.6 $\pm$ 0.7 & 6.193 $\pm$ 1.5 $\pm$ 0.6 $\pm$ 1.3 & 5.940 $\pm$ 1.2 $\pm$ 0.3 $\pm$ 0.3 & 5.959 $\pm$ 1.2 $\pm$ 0.3 $\pm$ 0.3 & 6.288 $\pm$ 1.2 $\pm$ 0.3 $\pm$ 1.3 & 6.316 $\pm$ 0.9 $\pm$ 0.3 $\pm$ 1.0\\
0.034 -- 0.039 & 5.934 $\pm$ 1.5 $\pm$ 0.6 $\pm$ 0.8 & 6.232 $\pm$ 1.5 $\pm$ 0.7 $\pm$ 1.4 & 5.545 $\pm$ 1.2 $\pm$ 0.3 $\pm$ 0.4 & 5.539 $\pm$ 1.2 $\pm$ 0.3 $\pm$ 0.4 & 5.867 $\pm$ 1.2 $\pm$ 0.3 $\pm$ 1.3 & 6.041 $\pm$ 1.0 $\pm$ 0.3 $\pm$ 1.1\\
0.039 -- 0.045 & 5.324 $\pm$ 1.4 $\pm$ 0.7 $\pm$ 0.5 & 5.547 $\pm$ 1.4 $\pm$ 0.7 $\pm$ 1.3 & 5.466 $\pm$ 1.1 $\pm$ 0.3 $\pm$ 0.4 & 5.475 $\pm$ 1.1 $\pm$ 0.3 $\pm$ 0.4 & 5.753 $\pm$ 1.1 $\pm$ 0.3 $\pm$ 1.3 & 5.729 $\pm$ 0.9 $\pm$ 0.3 $\pm$ 1.0\\
0.045 -- 0.051 & 5.159 $\pm$ 1.4 $\pm$ 0.7 $\pm$ 0.6 & 5.379 $\pm$ 1.4 $\pm$ 0.8 $\pm$ 1.3 & 5.181 $\pm$ 1.1 $\pm$ 0.3 $\pm$ 0.6 & 5.208 $\pm$ 1.1 $\pm$ 0.3 $\pm$ 0.6 & 5.446 $\pm$ 1.1 $\pm$ 0.3 $\pm$ 1.4 & 5.451 $\pm$ 0.9 $\pm$ 0.3 $\pm$ 1.1\\
0.051 -- 0.057 & 4.874 $\pm$ 1.5 $\pm$ 0.6 $\pm$ 0.6 & 5.071 $\pm$ 1.5 $\pm$ 0.6 $\pm$ 1.3 & 4.960 $\pm$ 1.1 $\pm$ 0.3 $\pm$ 0.6 & 4.977 $\pm$ 1.1 $\pm$ 0.3 $\pm$ 0.6 & 5.208 $\pm$ 1.1 $\pm$ 0.3 $\pm$ 1.4 & 5.194 $\pm$ 0.9 $\pm$ 0.3 $\pm$ 1.1\\
0.057 -- 0.064 & 4.499 $\pm$ 1.4 $\pm$ 0.6 $\pm$ 0.6 & 4.661 $\pm$ 1.4 $\pm$ 0.6 $\pm$ 1.3 & 4.575 $\pm$ 1.1 $\pm$ 0.3 $\pm$ 0.7 & 4.600 $\pm$ 1.1 $\pm$ 0.3 $\pm$ 0.7 & 4.802 $\pm$ 1.1 $\pm$ 0.3 $\pm$ 1.4 & 4.781 $\pm$ 0.9 $\pm$ 0.3 $\pm$ 1.1\\
0.064 -- 0.072 & 4.231 $\pm$ 1.4 $\pm$ 0.6 $\pm$ 0.6 & 4.374 $\pm$ 1.4 $\pm$ 0.7 $\pm$ 1.3 & 4.124 $\pm$ 1.1 $\pm$ 0.3 $\pm$ 0.7 & 4.148 $\pm$ 1.1 $\pm$ 0.3 $\pm$ 0.7 & 4.313 $\pm$ 1.1 $\pm$ 0.3 $\pm$ 1.4 & 4.369 $\pm$ 0.9 $\pm$ 0.3 $\pm$ 1.1\\
0.072 -- 0.081 & 3.805 $\pm$ 1.4 $\pm$ 0.6 $\pm$ 0.6 & 3.907 $\pm$ 1.4 $\pm$ 0.6 $\pm$ 1.3 & 3.920 $\pm$ 1.1 $\pm$ 0.3 $\pm$ 0.8 & 3.945 $\pm$ 1.1 $\pm$ 0.3 $\pm$ 0.8 & 4.080 $\pm$ 1.1 $\pm$ 0.3 $\pm$ 1.4 & 4.042 $\pm$ 0.8 $\pm$ 0.3 $\pm$ 1.1\\
0.081 -- 0.091 & 3.526 $\pm$ 1.3 $\pm$ 0.7 $\pm$ 0.7 & 3.620 $\pm$ 1.3 $\pm$ 0.7 $\pm$ 1.3 & 3.494 $\pm$ 1.0 $\pm$ 0.3 $\pm$ 0.9 & 3.509 $\pm$ 1.0 $\pm$ 0.3 $\pm$ 0.9 & 3.633 $\pm$ 1.0 $\pm$ 0.3 $\pm$ 1.5 & 3.657 $\pm$ 0.8 $\pm$ 0.3 $\pm$ 1.1\\
0.091 -- 0.102 & 3.202 $\pm$ 1.4 $\pm$ 0.6 $\pm$ 0.6 & 3.273 $\pm$ 1.4 $\pm$ 0.6 $\pm$ 1.3 & 3.169 $\pm$ 1.0 $\pm$ 0.3 $\pm$ 0.9 & 3.184 $\pm$ 1.0 $\pm$ 0.3 $\pm$ 0.9 & 3.260 $\pm$ 1.0 $\pm$ 0.3 $\pm$ 1.5 & 3.289 $\pm$ 0.8 $\pm$ 0.3 $\pm$ 1.1\\
0.102 -- 0.114 & 2.856 $\pm$ 1.4 $\pm$ 0.6 $\pm$ 0.8 & 2.886 $\pm$ 1.4 $\pm$ 0.6 $\pm$ 1.4 & 2.817 $\pm$ 1.1 $\pm$ 0.3 $\pm$ 1.0 & 2.824 $\pm$ 1.1 $\pm$ 0.3 $\pm$ 1.0 & 2.871 $\pm$ 1.1 $\pm$ 0.3 $\pm$ 1.5 & 2.896 $\pm$ 0.8 $\pm$ 0.3 $\pm$ 1.1\\
0.114 -- 0.128 & 2.549 $\pm$ 1.3 $\pm$ 0.6 $\pm$ 0.6 & 2.577 $\pm$ 1.3 $\pm$ 0.6 $\pm$ 1.3 & 2.515 $\pm$ 1.0 $\pm$ 0.2 $\pm$ 1.1 & 2.524 $\pm$ 1.0 $\pm$ 0.2 $\pm$ 1.1 & 2.557 $\pm$ 1.0 $\pm$ 0.2 $\pm$ 1.2 & 2.576 $\pm$ 0.8 $\pm$ 0.2 $\pm$ 0.9\\
0.128 -- 0.145 & 2.122 $\pm$ 1.3 $\pm$ 0.5 $\pm$ 0.6 & 2.114 $\pm$ 1.3 $\pm$ 0.6 $\pm$ 1.3 & 2.183 $\pm$ 1.0 $\pm$ 0.3 $\pm$ 1.0 & 2.180 $\pm$ 1.0 $\pm$ 0.3 $\pm$ 1.0 & 2.174 $\pm$ 1.0 $\pm$ 0.3 $\pm$ 1.2 & 2.163 $\pm$ 0.8 $\pm$ 0.3 $\pm$ 0.9\\
0.145 -- 0.165 & 1.817 $\pm$ 1.3 $\pm$ 0.6 $\pm$ 0.8 & 1.787 $\pm$ 1.3 $\pm$ 0.6 $\pm$ 1.4 & 1.868 $\pm$ 1.0 $\pm$ 0.3 $\pm$ 1.0 & 1.868 $\pm$ 1.0 $\pm$ 0.3 $\pm$ 1.0 & 1.847 $\pm$ 1.0 $\pm$ 0.3 $\pm$ 1.2 & 1.834 $\pm$ 0.8 $\pm$ 0.3 $\pm$ 0.9\\
0.165 -- 0.189 & 1.512 $\pm$ 1.3 $\pm$ 0.6 $\pm$ 0.8 & 1.474 $\pm$ 1.3 $\pm$ 0.6 $\pm$ 1.4 & 1.535 $\pm$ 1.0 $\pm$ 0.3 $\pm$ 1.2 & 1.539 $\pm$ 1.0 $\pm$ 0.3 $\pm$ 1.2 & 1.495 $\pm$ 1.0 $\pm$ 0.3 $\pm$ 1.3 & 1.497 $\pm$ 0.8 $\pm$ 0.3 $\pm$ 0.9\\
0.189 -- 0.219 & 1.240 $\pm$ 1.3 $\pm$ 0.8 $\pm$ 0.8 & 1.188 $\pm$ 1.3 $\pm$ 0.8 $\pm$ 1.4 & 1.269 $\pm$ 1.0 $\pm$ 0.3 $\pm$ 1.1 & 1.269 $\pm$ 1.0 $\pm$ 0.3 $\pm$ 1.1 & 1.214 $\pm$ 1.0 $\pm$ 0.3 $\pm$ 1.3 & 1.213 $\pm$ 0.8 $\pm$ 0.3 $\pm$ 0.9\\
0.219 -- 0.258 & 0.999 $\pm$ 1.3 $\pm$ 0.6 $\pm$ 0.8 & 0.942 $\pm$ 1.3 $\pm$ 0.6 $\pm$ 1.4 & 0.987 $\pm$ 1.0 $\pm$ 0.3 $\pm$ 1.2 & 0.991 $\pm$ 1.0 $\pm$ 0.3 $\pm$ 1.2 & 0.931 $\pm$ 1.0 $\pm$ 0.3 $\pm$ 1.3 & 0.940 $\pm$ 0.8 $\pm$ 0.3 $\pm$ 0.9\\
0.258 -- 0.312 & 0.748 $\pm$ 1.2 $\pm$ 0.6 $\pm$ 0.8 & 0.685 $\pm$ 1.2 $\pm$ 0.7 $\pm$ 2.2 & 0.763 $\pm$ 0.9 $\pm$ 0.2 $\pm$ 1.3 & 0.762 $\pm$ 0.9 $\pm$ 0.2 $\pm$ 1.3 & 0.696 $\pm$ 0.9 $\pm$ 0.2 $\pm$ 2.4 & 0.704 $\pm$ 0.7 $\pm$ 0.3 $\pm$ 1.8\\
0.312 -- 0.391 & 0.545 $\pm$ 1.2 $\pm$ 0.7 $\pm$ 1.0 & 0.490 $\pm$ 1.2 $\pm$ 0.7 $\pm$ 2.2 & 0.528 $\pm$ 1.0 $\pm$ 0.3 $\pm$ 1.7 & 0.529 $\pm$ 1.0 $\pm$ 0.3 $\pm$ 1.7 & 0.469 $\pm$ 1.0 $\pm$ 0.3 $\pm$ 2.6 & 0.487 $\pm$ 0.8 $\pm$ 0.3 $\pm$ 1.8\\
0.391 -- 0.524 & 0.333 $\pm$ 1.2 $\pm$ 0.6 $\pm$ 1.1 & 0.297 $\pm$ 1.2 $\pm$ 0.6 $\pm$ 2.3 & 0.331 $\pm$ 1.0 $\pm$ 0.3 $\pm$ 2.1 & 0.330 $\pm$ 1.0 $\pm$ 0.3 $\pm$ 2.1 & 0.290 $\pm$ 1.0 $\pm$ 0.3 $\pm$ 2.9 & 0.301 $\pm$ 0.7 $\pm$ 0.3 $\pm$ 1.8\\
0.524 -- 0.695 & 0.197 $\pm$ 1.4 $\pm$ 0.9 $\pm$ 1.6 & 0.179 $\pm$ 1.4 $\pm$ 1.0 $\pm$ 2.6 & 0.187 $\pm$ 1.1 $\pm$ 0.3 $\pm$ 2.9 & 0.187 $\pm$ 1.1 $\pm$ 0.3 $\pm$ 2.9 & 0.165 $\pm$ 1.1 $\pm$ 0.3 $\pm$ 3.5 & 0.176 $\pm$ 0.9 $\pm$ 0.3 $\pm$ 2.0\\
0.695 -- 0.918 & 0.103 $\pm$ 1.7 $\pm$ 1.1 $\pm$ 1.5 & 0.0972 $\pm$ 1.7 $\pm$ 1.1 $\pm$ 2.5 & 0.105 $\pm$ 1.3 $\pm$ 0.4 $\pm$ 3.1 & 0.104 $\pm$ 1.3 $\pm$ 0.4 $\pm$ 3.1 & 0.0963 $\pm$ 1.3 $\pm$ 0.4 $\pm$ 3.7 & 0.100 $\pm$ 1.0 $\pm$ 0.4 $\pm$ 2.0\\
0.918 -- 1.153 & 0.0611 $\pm$ 2.2 $\pm$ 1.3 $\pm$ 1.8 & 0.0590 $\pm$ 2.2 $\pm$ 1.3 $\pm$ 2.7 & 0.0591 $\pm$ 1.8 $\pm$ 0.5 $\pm$ 3.5 & 0.0586 $\pm$ 1.8 $\pm$ 0.5 $\pm$ 3.5 & 0.0550 $\pm$ 1.8 $\pm$ 0.5 $\pm$ 4.0 & 0.0586 $\pm$ 1.4 $\pm$ 0.6 $\pm$ 2.2\\
1.153 -- 1.496 & 0.0333 $\pm$ 2.6 $\pm$ 2.1 $\pm$ 2.3 & 0.0324 $\pm$ 2.6 $\pm$ 2.1 $\pm$ 3.1 & 0.0320 $\pm$ 2.1 $\pm$ 1.1 $\pm$ 3.6 & 0.0315 $\pm$ 2.1 $\pm$ 1.1 $\pm$ 3.6 & 0.0303 $\pm$ 2.1 $\pm$ 1.1 $\pm$ 4.1 & 0.0322 $\pm$ 1.6 $\pm$ 1.0 $\pm$ 2.4\\
1.496 -- 1.947 & 0.0174 $\pm$ 3.1 $\pm$ 2.0 $\pm$ 2.5 & 0.0171 $\pm$ 3.1 $\pm$ 2.0 $\pm$ 3.3 & 0.0168 $\pm$ 2.5 $\pm$ 1.1 $\pm$ 3.0 & 0.0167 $\pm$ 2.5 $\pm$ 1.1 $\pm$ 3.0 & 0.0161 $\pm$ 2.5 $\pm$ 1.1 $\pm$ 3.5 & 0.0169 $\pm$ 1.9 $\pm$ 1.0 $\pm$ 2.4\\
1.947 -- 2.522 & 0.00863 $\pm$ 4.0 $\pm$ 2.3 $\pm$ 3.0 & 0.00850 $\pm$ 4.0 $\pm$ 2.3 $\pm$ 3.7 & 0.00885 $\pm$ 3.0 $\pm$ 1.2 $\pm$ 3.0 & 0.00875 $\pm$ 3.0 $\pm$ 1.2 $\pm$ 3.0 & 0.00855 $\pm$ 3.0 $\pm$ 1.2 $\pm$ 3.5 & 0.00880 $\pm$ 2.4 $\pm$ 1.1 $\pm$ 2.5\\
2.522 -- 3.277 & 0.00457 $\pm$ 4.6 $\pm$ 3.0 $\pm$ 5.8 & 0.00456 $\pm$ 4.6 $\pm$ 3.0 $\pm$ 6.2 & 0.00432 $\pm$ 3.8 $\pm$ 1.6 $\pm$ 3.1 & 0.00430 $\pm$ 3.8 $\pm$ 1.6 $\pm$ 3.1 & 0.00420 $\pm$ 3.8 $\pm$ 1.6 $\pm$ 3.6 & 0.00445 $\pm$ 2.9 $\pm$ 1.5 $\pm$ 2.7\\
3.277 -- 5.000 & 0.00207 $\pm$ 4.4 $\pm$ 2.6 $\pm$ 3.0 & 0.00206 $\pm$ 4.4 $\pm$ 2.6 $\pm$ 3.7 & 0.00187 $\pm$ 3.8 $\pm$ 1.6 $\pm$ 4.0 & 0.00187 $\pm$ 3.8 $\pm$ 1.6 $\pm$ 4.0 & 0.00183 $\pm$ 3.8 $\pm$ 1.6 $\pm$ 4.4 & 0.00198 $\pm$ 2.9 $\pm$ 1.4 $\pm$ 2.6\\
5.000 -- 10.000 & 0.000486 $\pm$ 5.6 $\pm$ 3.0 $\pm$ 3.3 & 0.000478 $\pm$ 5.6 $\pm$ 3.0 $\pm$ 3.9 & 0.000501 $\pm$ 4.5 $\pm$ 1.7 $\pm$ 3.9 & 0.000497 $\pm$ 4.5 $\pm$ 1.7 $\pm$ 3.9 & 0.000487 $\pm$ 4.5 $\pm$ 1.7 $\pm$ 4.3 & 0.000502 $\pm$ 3.4 $\pm$ 1.5 $\pm$ 2.7\\
\bottomrule 
\end{tabular} 
} 
\label{tab:CombPhiStarlowmass_allrapidity}
\end{table*}

\begin{table*}
\centering
\caption{The values of $(1/\sigma)\,\mathrm{d}\sigma/\mathrm{d}\phi^*_{\eta}$ in each bin of \PhiStar{} for the electron and muon channels separately (for various particle-level definitions) and for the Born-level combination in the kinematic region $66\ \GeV \leq m_{\ell\ell} < 116\ \GeV,\ |y_{\ell\ell}| < 2.4$. The associated statistical and systematic (both uncorrelated and correlated between bins of \PhiStar{}) are provided in percentage form.} 
\resizebox{\textwidth}{!}{
\begin{tabular}{ ccccccc } \toprule 
Bin & \multicolumn{6}{c}{$(1/\sigma)\,d\sigma/d\phi^*_{\eta}$ $\pm$ Statistical [\%] $\pm$ Uncorrelated systematic [\%] $\pm$ Correlated systematic [\%]} \\ 
 \cmidrule(r){2-7} 
 & \multicolumn{2}{c}{Electron channel} &  \multicolumn{3}{c}{Muon channel} & Combination \\ 
\cmidrule(r){2-3} \cmidrule(r){4-6} \cmidrule(r){7-7} 
 & dressed & Born & bare & dressed & Born & Born \\ 
 0.000 -- 0.004 & 9.362 $\pm$ 0.2 $\pm$ 0.1 $\pm$ 0.2 & 9.451 $\pm$ 0.2 $\pm$ 0.1 $\pm$ 0.2 & 9.364 $\pm$ 0.2 $\pm$ 0.0 $\pm$ 0.1 & 9.359 $\pm$ 0.2 $\pm$ 0.0 $\pm$ 0.1 & 9.433 $\pm$ 0.2 $\pm$ 0.0 $\pm$ 0.1 & 9.441 $\pm$ 0.1 $\pm$ 0.0 $\pm$ 0.1\\
0.004 -- 0.008 & 9.267 $\pm$ 0.2 $\pm$ 0.1 $\pm$ 0.1 & 9.352 $\pm$ 0.2 $\pm$ 0.1 $\pm$ 0.1 & 9.299 $\pm$ 0.2 $\pm$ 0.0 $\pm$ 0.1 & 9.294 $\pm$ 0.2 $\pm$ 0.0 $\pm$ 0.1 & 9.376 $\pm$ 0.2 $\pm$ 0.0 $\pm$ 0.1 & 9.365 $\pm$ 0.1 $\pm$ 0.0 $\pm$ 0.1\\
0.008 -- 0.012 & 9.094 $\pm$ 0.2 $\pm$ 0.1 $\pm$ 0.1 & 9.169 $\pm$ 0.2 $\pm$ 0.1 $\pm$ 0.1 & 9.101 $\pm$ 0.2 $\pm$ 0.0 $\pm$ 0.1 & 9.098 $\pm$ 0.2 $\pm$ 0.0 $\pm$ 0.1 & 9.173 $\pm$ 0.2 $\pm$ 0.0 $\pm$ 0.1 & 9.171 $\pm$ 0.1 $\pm$ 0.0 $\pm$ 0.1\\
0.012 -- 0.016 & 8.855 $\pm$ 0.2 $\pm$ 0.1 $\pm$ 0.1 & 8.926 $\pm$ 0.2 $\pm$ 0.1 $\pm$ 0.1 & 8.894 $\pm$ 0.2 $\pm$ 0.0 $\pm$ 0.0 & 8.888 $\pm$ 0.2 $\pm$ 0.0 $\pm$ 0.0 & 8.959 $\pm$ 0.2 $\pm$ 0.0 $\pm$ 0.1 & 8.945 $\pm$ 0.1 $\pm$ 0.0 $\pm$ 0.1\\
0.016 -- 0.020 & 8.601 $\pm$ 0.2 $\pm$ 0.1 $\pm$ 0.1 & 8.668 $\pm$ 0.2 $\pm$ 0.1 $\pm$ 0.1 & 8.562 $\pm$ 0.2 $\pm$ 0.0 $\pm$ 0.0 & 8.556 $\pm$ 0.2 $\pm$ 0.0 $\pm$ 0.0 & 8.620 $\pm$ 0.2 $\pm$ 0.0 $\pm$ 0.1 & 8.640 $\pm$ 0.1 $\pm$ 0.0 $\pm$ 0.1\\
0.020 -- 0.024 & 8.188 $\pm$ 0.2 $\pm$ 0.1 $\pm$ 0.1 & 8.238 $\pm$ 0.2 $\pm$ 0.1 $\pm$ 0.1 & 8.231 $\pm$ 0.2 $\pm$ 0.0 $\pm$ 0.0 & 8.229 $\pm$ 0.2 $\pm$ 0.0 $\pm$ 0.0 & 8.284 $\pm$ 0.2 $\pm$ 0.0 $\pm$ 0.1 & 8.264 $\pm$ 0.1 $\pm$ 0.0 $\pm$ 0.1\\
0.024 -- 0.029 & 7.825 $\pm$ 0.2 $\pm$ 0.1 $\pm$ 0.1 & 7.868 $\pm$ 0.2 $\pm$ 0.1 $\pm$ 0.1 & 7.816 $\pm$ 0.2 $\pm$ 0.0 $\pm$ 0.0 & 7.811 $\pm$ 0.2 $\pm$ 0.0 $\pm$ 0.0 & 7.861 $\pm$ 0.2 $\pm$ 0.0 $\pm$ 0.1 & 7.863 $\pm$ 0.1 $\pm$ 0.0 $\pm$ 0.1\\
0.029 -- 0.034 & 7.356 $\pm$ 0.2 $\pm$ 0.1 $\pm$ 0.1 & 7.389 $\pm$ 0.2 $\pm$ 0.1 $\pm$ 0.1 & 7.389 $\pm$ 0.2 $\pm$ 0.0 $\pm$ 0.0 & 7.384 $\pm$ 0.2 $\pm$ 0.0 $\pm$ 0.0 & 7.422 $\pm$ 0.2 $\pm$ 0.0 $\pm$ 0.1 & 7.408 $\pm$ 0.1 $\pm$ 0.0 $\pm$ 0.1\\
0.034 -- 0.039 & 6.883 $\pm$ 0.2 $\pm$ 0.1 $\pm$ 0.1 & 6.905 $\pm$ 0.2 $\pm$ 0.1 $\pm$ 0.1 & 6.883 $\pm$ 0.2 $\pm$ 0.0 $\pm$ 0.0 & 6.881 $\pm$ 0.2 $\pm$ 0.0 $\pm$ 0.0 & 6.911 $\pm$ 0.2 $\pm$ 0.0 $\pm$ 0.1 & 6.908 $\pm$ 0.1 $\pm$ 0.0 $\pm$ 0.1\\
0.039 -- 0.045 & 6.403 $\pm$ 0.2 $\pm$ 0.1 $\pm$ 0.1 & 6.419 $\pm$ 0.2 $\pm$ 0.1 $\pm$ 0.1 & 6.419 $\pm$ 0.2 $\pm$ 0.0 $\pm$ 0.1 & 6.417 $\pm$ 0.2 $\pm$ 0.0 $\pm$ 0.1 & 6.438 $\pm$ 0.2 $\pm$ 0.0 $\pm$ 0.1 & 6.429 $\pm$ 0.1 $\pm$ 0.0 $\pm$ 0.1\\
0.045 -- 0.051 & 5.871 $\pm$ 0.2 $\pm$ 0.1 $\pm$ 0.1 & 5.876 $\pm$ 0.2 $\pm$ 0.1 $\pm$ 0.1 & 5.893 $\pm$ 0.2 $\pm$ 0.0 $\pm$ 0.1 & 5.891 $\pm$ 0.2 $\pm$ 0.0 $\pm$ 0.1 & 5.899 $\pm$ 0.2 $\pm$ 0.0 $\pm$ 0.1 & 5.888 $\pm$ 0.1 $\pm$ 0.0 $\pm$ 0.1\\
0.051 -- 0.057 & 5.404 $\pm$ 0.2 $\pm$ 0.1 $\pm$ 0.1 & 5.404 $\pm$ 0.2 $\pm$ 0.1 $\pm$ 0.1 & 5.434 $\pm$ 0.2 $\pm$ 0.0 $\pm$ 0.0 & 5.434 $\pm$ 0.2 $\pm$ 0.0 $\pm$ 0.0 & 5.435 $\pm$ 0.2 $\pm$ 0.0 $\pm$ 0.1 & 5.422 $\pm$ 0.1 $\pm$ 0.0 $\pm$ 0.1\\
0.057 -- 0.064 & 4.960 $\pm$ 0.2 $\pm$ 0.1 $\pm$ 0.1 & 4.957 $\pm$ 0.2 $\pm$ 0.1 $\pm$ 0.1 & 4.975 $\pm$ 0.2 $\pm$ 0.0 $\pm$ 0.1 & 4.972 $\pm$ 0.2 $\pm$ 0.0 $\pm$ 0.1 & 4.971 $\pm$ 0.2 $\pm$ 0.0 $\pm$ 0.1 & 4.964 $\pm$ 0.1 $\pm$ 0.0 $\pm$ 0.1\\
0.064 -- 0.072 & 4.509 $\pm$ 0.2 $\pm$ 0.1 $\pm$ 0.1 & 4.502 $\pm$ 0.2 $\pm$ 0.1 $\pm$ 0.1 & 4.506 $\pm$ 0.2 $\pm$ 0.0 $\pm$ 0.1 & 4.505 $\pm$ 0.2 $\pm$ 0.0 $\pm$ 0.1 & 4.494 $\pm$ 0.2 $\pm$ 0.0 $\pm$ 0.1 & 4.496 $\pm$ 0.1 $\pm$ 0.0 $\pm$ 0.1\\
0.072 -- 0.081 & 4.022 $\pm$ 0.2 $\pm$ 0.1 $\pm$ 0.1 & 4.011 $\pm$ 0.2 $\pm$ 0.1 $\pm$ 0.1 & 4.026 $\pm$ 0.2 $\pm$ 0.0 $\pm$ 0.1 & 4.025 $\pm$ 0.2 $\pm$ 0.0 $\pm$ 0.1 & 4.013 $\pm$ 0.2 $\pm$ 0.0 $\pm$ 0.1 & 4.011 $\pm$ 0.1 $\pm$ 0.0 $\pm$ 0.1\\
0.081 -- 0.091 & 3.580 $\pm$ 0.2 $\pm$ 0.1 $\pm$ 0.1 & 3.566 $\pm$ 0.2 $\pm$ 0.1 $\pm$ 0.1 & 3.577 $\pm$ 0.2 $\pm$ 0.0 $\pm$ 0.1 & 3.577 $\pm$ 0.2 $\pm$ 0.0 $\pm$ 0.1 & 3.565 $\pm$ 0.2 $\pm$ 0.0 $\pm$ 0.1 & 3.564 $\pm$ 0.1 $\pm$ 0.0 $\pm$ 0.1\\
0.091 -- 0.102 & 3.155 $\pm$ 0.2 $\pm$ 0.1 $\pm$ 0.1 & 3.141 $\pm$ 0.2 $\pm$ 0.1 $\pm$ 0.1 & 3.154 $\pm$ 0.2 $\pm$ 0.0 $\pm$ 0.1 & 3.153 $\pm$ 0.2 $\pm$ 0.0 $\pm$ 0.1 & 3.137 $\pm$ 0.2 $\pm$ 0.0 $\pm$ 0.1 & 3.138 $\pm$ 0.1 $\pm$ 0.0 $\pm$ 0.1\\
0.102 -- 0.114 & 2.771 $\pm$ 0.2 $\pm$ 0.1 $\pm$ 0.1 & 2.758 $\pm$ 0.2 $\pm$ 0.1 $\pm$ 0.1 & 2.765 $\pm$ 0.2 $\pm$ 0.0 $\pm$ 0.1 & 2.765 $\pm$ 0.2 $\pm$ 0.0 $\pm$ 0.1 & 2.752 $\pm$ 0.2 $\pm$ 0.0 $\pm$ 0.1 & 2.753 $\pm$ 0.1 $\pm$ 0.0 $\pm$ 0.1\\
0.114 -- 0.128 & 2.394 $\pm$ 0.2 $\pm$ 0.1 $\pm$ 0.1 & 2.380 $\pm$ 0.2 $\pm$ 0.1 $\pm$ 0.1 & 2.394 $\pm$ 0.2 $\pm$ 0.0 $\pm$ 0.1 & 2.394 $\pm$ 0.2 $\pm$ 0.0 $\pm$ 0.1 & 2.380 $\pm$ 0.2 $\pm$ 0.0 $\pm$ 0.1 & 2.379 $\pm$ 0.1 $\pm$ 0.0 $\pm$ 0.1\\
0.128 -- 0.145 & 2.039 $\pm$ 0.2 $\pm$ 0.1 $\pm$ 0.1 & 2.026 $\pm$ 0.2 $\pm$ 0.1 $\pm$ 0.1 & 2.040 $\pm$ 0.2 $\pm$ 0.0 $\pm$ 0.1 & 2.040 $\pm$ 0.2 $\pm$ 0.0 $\pm$ 0.1 & 2.028 $\pm$ 0.2 $\pm$ 0.0 $\pm$ 0.1 & 2.026 $\pm$ 0.1 $\pm$ 0.0 $\pm$ 0.1\\
0.145 -- 0.165 & 1.704 $\pm$ 0.2 $\pm$ 0.1 $\pm$ 0.1 & 1.693 $\pm$ 0.2 $\pm$ 0.1 $\pm$ 0.1 & 1.701 $\pm$ 0.2 $\pm$ 0.0 $\pm$ 0.1 & 1.702 $\pm$ 0.2 $\pm$ 0.0 $\pm$ 0.1 & 1.691 $\pm$ 0.2 $\pm$ 0.0 $\pm$ 0.1 & 1.691 $\pm$ 0.1 $\pm$ 0.0 $\pm$ 0.1\\
0.165 -- 0.189 & 1.398 $\pm$ 0.2 $\pm$ 0.1 $\pm$ 0.1 & 1.389 $\pm$ 0.2 $\pm$ 0.1 $\pm$ 0.1 & 1.398 $\pm$ 0.2 $\pm$ 0.0 $\pm$ 0.1 & 1.399 $\pm$ 0.2 $\pm$ 0.0 $\pm$ 0.1 & 1.390 $\pm$ 0.2 $\pm$ 0.0 $\pm$ 0.1 & 1.389 $\pm$ 0.1 $\pm$ 0.0 $\pm$ 0.1\\
0.189 -- 0.219 & 1.117 $\pm$ 0.2 $\pm$ 0.1 $\pm$ 0.1 & 1.110 $\pm$ 0.2 $\pm$ 0.1 $\pm$ 0.1 & 1.116 $\pm$ 0.2 $\pm$ 0.0 $\pm$ 0.1 & 1.117 $\pm$ 0.2 $\pm$ 0.0 $\pm$ 0.1 & 1.110 $\pm$ 0.2 $\pm$ 0.0 $\pm$ 0.1 & 1.110 $\pm$ 0.1 $\pm$ 0.0 $\pm$ 0.1\\
0.219 -- 0.258 & 0.854 $\pm$ 0.2 $\pm$ 0.1 $\pm$ 0.1 & 0.849 $\pm$ 0.2 $\pm$ 0.1 $\pm$ 0.1 & 0.855 $\pm$ 0.2 $\pm$ 0.0 $\pm$ 0.1 & 0.856 $\pm$ 0.2 $\pm$ 0.0 $\pm$ 0.1 & 0.851 $\pm$ 0.2 $\pm$ 0.0 $\pm$ 0.1 & 0.850 $\pm$ 0.1 $\pm$ 0.0 $\pm$ 0.1\\
0.258 -- 0.312 & 0.621 $\pm$ 0.2 $\pm$ 0.1 $\pm$ 0.3 & 0.618 $\pm$ 0.2 $\pm$ 0.1 $\pm$ 0.3 & 0.618 $\pm$ 0.2 $\pm$ 0.0 $\pm$ 0.1 & 0.619 $\pm$ 0.2 $\pm$ 0.0 $\pm$ 0.1 & 0.615 $\pm$ 0.2 $\pm$ 0.0 $\pm$ 0.2 & 0.616 $\pm$ 0.1 $\pm$ 0.0 $\pm$ 0.2\\
0.312 -- 0.391 & 0.414 $\pm$ 0.2 $\pm$ 0.1 $\pm$ 0.3 & 0.412 $\pm$ 0.2 $\pm$ 0.1 $\pm$ 0.3 & 0.413 $\pm$ 0.2 $\pm$ 0.0 $\pm$ 0.1 & 0.413 $\pm$ 0.2 $\pm$ 0.0 $\pm$ 0.1 & 0.411 $\pm$ 0.2 $\pm$ 0.0 $\pm$ 0.2 & 0.411 $\pm$ 0.1 $\pm$ 0.0 $\pm$ 0.2\\
0.391 -- 0.524 & 0.241 $\pm$ 0.2 $\pm$ 0.1 $\pm$ 0.3 & 0.240 $\pm$ 0.2 $\pm$ 0.1 $\pm$ 0.3 & 0.239 $\pm$ 0.2 $\pm$ 0.0 $\pm$ 0.2 & 0.239 $\pm$ 0.2 $\pm$ 0.0 $\pm$ 0.2 & 0.238 $\pm$ 0.2 $\pm$ 0.0 $\pm$ 0.2 & 0.239 $\pm$ 0.1 $\pm$ 0.0 $\pm$ 0.2\\
0.524 -- 0.695 & 0.124 $\pm$ 0.3 $\pm$ 0.1 $\pm$ 0.3 & 0.124 $\pm$ 0.3 $\pm$ 0.1 $\pm$ 0.3 & 0.124 $\pm$ 0.2 $\pm$ 0.1 $\pm$ 0.2 & 0.124 $\pm$ 0.2 $\pm$ 0.1 $\pm$ 0.2 & 0.124 $\pm$ 0.2 $\pm$ 0.1 $\pm$ 0.2 & 0.124 $\pm$ 0.2 $\pm$ 0.1 $\pm$ 0.2\\
0.695 -- 0.918 & 0.0625 $\pm$ 0.3 $\pm$ 0.1 $\pm$ 0.3 & 0.0623 $\pm$ 0.3 $\pm$ 0.1 $\pm$ 0.3 & 0.0619 $\pm$ 0.3 $\pm$ 0.1 $\pm$ 0.2 & 0.0620 $\pm$ 0.3 $\pm$ 0.1 $\pm$ 0.2 & 0.0619 $\pm$ 0.3 $\pm$ 0.1 $\pm$ 0.2 & 0.0620 $\pm$ 0.2 $\pm$ 0.1 $\pm$ 0.2\\
0.918 -- 1.153 & 0.0322 $\pm$ 0.4 $\pm$ 0.2 $\pm$ 0.4 & 0.0321 $\pm$ 0.4 $\pm$ 0.2 $\pm$ 0.4 & 0.0320 $\pm$ 0.4 $\pm$ 0.1 $\pm$ 0.2 & 0.0320 $\pm$ 0.4 $\pm$ 0.1 $\pm$ 0.2 & 0.0319 $\pm$ 0.4 $\pm$ 0.1 $\pm$ 0.3 & 0.0320 $\pm$ 0.3 $\pm$ 0.1 $\pm$ 0.2\\
1.153 -- 1.496 & 0.0166 $\pm$ 0.5 $\pm$ 0.1 $\pm$ 0.4 & 0.0166 $\pm$ 0.5 $\pm$ 0.1 $\pm$ 0.4 & 0.0165 $\pm$ 0.5 $\pm$ 0.1 $\pm$ 0.2 & 0.0165 $\pm$ 0.5 $\pm$ 0.1 $\pm$ 0.2 & 0.0164 $\pm$ 0.5 $\pm$ 0.1 $\pm$ 0.3 & 0.0165 $\pm$ 0.3 $\pm$ 0.1 $\pm$ 0.3\\
1.496 -- 1.947 & 0.00791 $\pm$ 0.6 $\pm$ 0.2 $\pm$ 0.5 & 0.00789 $\pm$ 0.6 $\pm$ 0.2 $\pm$ 0.5 & 0.00796 $\pm$ 0.6 $\pm$ 0.1 $\pm$ 0.3 & 0.00798 $\pm$ 0.6 $\pm$ 0.1 $\pm$ 0.3 & 0.00796 $\pm$ 0.6 $\pm$ 0.1 $\pm$ 0.3 & 0.00791 $\pm$ 0.4 $\pm$ 0.1 $\pm$ 0.3\\
1.947 -- 2.522 & 0.00392 $\pm$ 0.8 $\pm$ 0.2 $\pm$ 0.5 & 0.00390 $\pm$ 0.8 $\pm$ 0.2 $\pm$ 0.5 & 0.00390 $\pm$ 0.7 $\pm$ 0.2 $\pm$ 0.3 & 0.00391 $\pm$ 0.7 $\pm$ 0.2 $\pm$ 0.3 & 0.00390 $\pm$ 0.7 $\pm$ 0.2 $\pm$ 0.4 & 0.00389 $\pm$ 0.5 $\pm$ 0.1 $\pm$ 0.3\\
2.522 -- 3.277 & 0.00198 $\pm$ 1.0 $\pm$ 0.2 $\pm$ 0.6 & 0.00198 $\pm$ 1.0 $\pm$ 0.2 $\pm$ 0.6 & 0.00194 $\pm$ 0.9 $\pm$ 0.2 $\pm$ 0.3 & 0.00195 $\pm$ 0.9 $\pm$ 0.2 $\pm$ 0.3 & 0.00194 $\pm$ 0.9 $\pm$ 0.2 $\pm$ 0.4 & 0.00195 $\pm$ 0.7 $\pm$ 0.2 $\pm$ 0.3\\
3.277 -- 5.000 & 0.000860 $\pm$ 1.0 $\pm$ 0.3 $\pm$ 0.7 & 0.000859 $\pm$ 1.0 $\pm$ 0.3 $\pm$ 0.7 & 0.000861 $\pm$ 0.9 $\pm$ 0.2 $\pm$ 0.3 & 0.000863 $\pm$ 0.9 $\pm$ 0.2 $\pm$ 0.3 & 0.000864 $\pm$ 0.9 $\pm$ 0.2 $\pm$ 0.3 & 0.000859 $\pm$ 0.7 $\pm$ 0.2 $\pm$ 0.3\\
5.000 -- 10.000 & 0.000255 $\pm$ 1.1 $\pm$ 0.3 $\pm$ 0.7 & 0.000255 $\pm$ 1.1 $\pm$ 0.3 $\pm$ 0.7 & 0.000247 $\pm$ 1.0 $\pm$ 0.2 $\pm$ 0.3 & 0.000247 $\pm$ 1.0 $\pm$ 0.2 $\pm$ 0.3 & 0.000247 $\pm$ 1.0 $\pm$ 0.2 $\pm$ 0.4 & 0.000250 $\pm$ 0.7 $\pm$ 0.2 $\pm$ 0.4\\
\bottomrule 
\end{tabular} 
} 
\label{tab:CombPhiStarpeakmass_allrapidity}
\end{table*}

\begin{table*}
\centering
\caption{The values of $(1/\sigma)\,\mathrm{d}\sigma/\mathrm{d}\phi^*_{\eta}$ in each bin of \PhiStar{} for the electron and muon channels separately (for various particle-level definitions) and for the Born-level combination in the kinematic region $116\ \GeV \leq m_{\ell\ell} < 150\ \GeV,\ |y_{\ell\ell}| < 2.4$. The associated statistical and systematic (both uncorrelated and correlated between bins of \PhiStar{}) are provided in percentage form.} 
\resizebox{\textwidth}{!}{
\begin{tabular}{ ccccccc } \toprule 
Bin & \multicolumn{6}{c}{$(1/\sigma)\,d\sigma/d\phi^*_{\eta}$ $\pm$ Statistical [\%] $\pm$ Uncorrelated systematic [\%] $\pm$ Correlated systematic [\%]} \\ 
 \cmidrule(r){2-7} 
 & \multicolumn{2}{c}{Electron channel} &  \multicolumn{3}{c}{Muon channel} & Combination \\ 
\cmidrule(r){2-3} \cmidrule(r){4-6} \cmidrule(r){7-7} 
 & dressed & Born & bare & dressed & Born & Born \\ 
 0.000 -- 0.004 & 11.378 $\pm$ 1.9 $\pm$ 0.6 $\pm$ 2.4 & 11.447 $\pm$ 1.9 $\pm$ 0.6 $\pm$ 2.4 & 11.623 $\pm$ 1.6 $\pm$ 0.4 $\pm$ 2.1 & 11.601 $\pm$ 1.6 $\pm$ 0.4 $\pm$ 2.1 & 11.676 $\pm$ 1.6 $\pm$ 0.4 $\pm$ 2.2 & 11.634 $\pm$ 1.2 $\pm$ 0.3 $\pm$ 2.1\\
0.004 -- 0.008 & 11.623 $\pm$ 1.8 $\pm$ 0.6 $\pm$ 1.0 & 11.722 $\pm$ 1.8 $\pm$ 0.7 $\pm$ 1.0 & 11.447 $\pm$ 1.6 $\pm$ 0.4 $\pm$ 0.6 & 11.449 $\pm$ 1.6 $\pm$ 0.4 $\pm$ 0.6 & 11.472 $\pm$ 1.6 $\pm$ 0.4 $\pm$ 0.9 & 11.593 $\pm$ 1.2 $\pm$ 0.4 $\pm$ 0.8\\
0.008 -- 0.012 & 10.950 $\pm$ 1.9 $\pm$ 0.6 $\pm$ 0.8 & 10.983 $\pm$ 1.9 $\pm$ 0.6 $\pm$ 0.8 & 10.949 $\pm$ 1.6 $\pm$ 0.4 $\pm$ 0.3 & 10.880 $\pm$ 1.6 $\pm$ 0.4 $\pm$ 0.3 & 11.003 $\pm$ 1.6 $\pm$ 0.4 $\pm$ 0.7 & 10.994 $\pm$ 1.2 $\pm$ 0.3 $\pm$ 0.6\\
0.012 -- 0.016 & 10.819 $\pm$ 1.9 $\pm$ 0.6 $\pm$ 0.7 & 10.908 $\pm$ 1.9 $\pm$ 0.6 $\pm$ 0.7 & 11.071 $\pm$ 1.7 $\pm$ 0.4 $\pm$ 0.2 & 11.018 $\pm$ 1.7 $\pm$ 0.4 $\pm$ 0.2 & 10.992 $\pm$ 1.7 $\pm$ 0.4 $\pm$ 0.6 & 10.948 $\pm$ 1.2 $\pm$ 0.3 $\pm$ 0.5\\
0.016 -- 0.020 & 9.950 $\pm$ 2.0 $\pm$ 0.6 $\pm$ 0.6 & 9.968 $\pm$ 2.0 $\pm$ 0.6 $\pm$ 0.6 & 10.060 $\pm$ 1.7 $\pm$ 0.5 $\pm$ 0.3 & 9.989 $\pm$ 1.7 $\pm$ 0.5 $\pm$ 0.3 & 10.094 $\pm$ 1.7 $\pm$ 0.5 $\pm$ 0.7 & 10.022 $\pm$ 1.3 $\pm$ 0.4 $\pm$ 0.5\\
\bottomrule 
\end{tabular} 
} 
\label{tab:CombPhiStarhighmass_allrapidity}
\end{table*}

\clearpage 

\begin{table*}
\centering
\caption{The values of $(1/\sigma)\,d\sigma/d\Zpt$ in each bin of \Zpt\ for the electron and muon channels separately (for various generator-level definitions) and for the Born-level combination in the kinematic region $ 66 \textrm{ GeV} \leq m_{\ell\ell} < 116 \textrm{ GeV},\ 0.0 \leq |y_{\ell\ell}| < 0.4$. The associated statistical and systematic (both uncorrelated and correlated between bins) are provided in percentage form.}
\resizebox{\textwidth}{!}{
\centering 
\begin{tabular}{ c  cccccc } \toprule  
 Bin & \multicolumn{6}{c}{$(1/\sigma)\,d\sigma/d\Zpt$ $\pm$ Statistical [\%] $\pm$ Uncorrelated systematic [\%] $\pm$ Correlated systematic [\%] } \\ 
 \cmidrule(r){2-7} 
  & 	\multicolumn{2}{c}{Electron channel} & \multicolumn{3}{c}{Muon channel} & Combination \\ 
 \cmidrule(r){2-3} \cmidrule(r){4-6} \cmidrule(r){7-7}
  & Dressed & Born & Bare & Dressed & Born & Born \\ 
 0.0 --  2.0&2.622e-02 $\pm$  0.5 $\pm$  0.2 $\pm$   0.8&2.688e-02 $\pm$  0.5 $\pm$  0.2 $\pm$   0.8&2.586e-02 $\pm$  0.4 $\pm$  0.2 $\pm$   0.9&2.638e-02  $\pm$  0.4 $\pm$  0.2 $\pm$   0.9&2.704e-02 $\pm$  0.4 $\pm$  0.2 $\pm$   1.0&2.704e-02 $\pm$  0.3 $\pm$  0.2 $\pm$   0.7 \\ 
  2.0 --  4.0 &5.438e-02 $\pm$  0.3 $\pm$  0.1 $\pm$   0.4&5.544e-02 $\pm$  0.3 $\pm$  0.1 $\pm$   0.4&5.385e-02 $\pm$  0.3 $\pm$  0.1 $\pm$   0.4&5.472e-02  $\pm$  0.3 $\pm$  0.1 $\pm$   0.4&5.582e-02 $\pm$  0.3 $\pm$  0.1 $\pm$   0.6&5.569e-02 $\pm$  0.2 $\pm$  0.1 $\pm$   0.3 \\ 
  4.0 --  6.0 &5.513e-02 $\pm$  0.3 $\pm$  0.2 $\pm$   0.4&5.562e-02 $\pm$  0.3 $\pm$  0.2 $\pm$   0.4&5.502e-02 $\pm$  0.3 $\pm$  0.2 $\pm$   0.5&5.534e-02  $\pm$  0.3 $\pm$  0.2 $\pm$   0.5&5.584e-02 $\pm$  0.3 $\pm$  0.2 $\pm$   0.6&5.571e-02 $\pm$  0.2 $\pm$  0.1 $\pm$   0.4 \\ 
  6.0 --  8.0 &4.896e-02 $\pm$  0.4 $\pm$  0.2 $\pm$   0.3&4.897e-02 $\pm$  0.4 $\pm$  0.2 $\pm$   0.3&4.887e-02 $\pm$  0.3 $\pm$  0.2 $\pm$   0.3&4.879e-02  $\pm$  0.3 $\pm$  0.2 $\pm$   0.3&4.875e-02 $\pm$  0.3 $\pm$  0.2 $\pm$   0.6&4.883e-02 $\pm$  0.2 $\pm$  0.1 $\pm$   0.3 \\ 
  8.0 -- 10.0 &4.127e-02 $\pm$  0.4 $\pm$  0.2 $\pm$   0.4&4.096e-02 $\pm$  0.4 $\pm$  0.2 $\pm$   0.6&4.139e-02 $\pm$  0.4 $\pm$  0.2 $\pm$   0.4&4.111e-02  $\pm$  0.4 $\pm$  0.2 $\pm$   0.4&4.082e-02 $\pm$  0.4 $\pm$  0.2 $\pm$   0.7&4.086e-02 $\pm$  0.3 $\pm$  0.1 $\pm$   0.4 \\
\bottomrule 
\end{tabular}
}
\label{tab:CombZPtY1U2}
\end{table*}

\begin{table*}
\centering
\caption{The values of $(1/\sigma)\,d\sigma/d\Zpt$ in each bin of \Zpt\ for the electron and muon channels separately (for various generator-level definitions) and for the Born-level combination in the kinematic region $ 66 \textrm{ GeV} \leq m_{\ell\ell} < 116 \textrm{ GeV},\ 0.4 \leq |y_{\ell\ell}| < 0.8$. The associated statistical and systematic (both uncorrelated and correlated between bins) are provided in percentage form.}
\resizebox{\textwidth}{!}{
\centering 
\begin{tabular}{ c  cccccc } \toprule  
 Bin & \multicolumn{6}{c}{$(1/\sigma)\,d\sigma/d\Zpt$ $\pm$ Statistical [\%] $\pm$ Uncorrelated systematic [\%] $\pm$ Correlated systematic [\%] } \\ 
 \cmidrule(r){2-7} 
  & 	\multicolumn{2}{c}{Electron channel} & \multicolumn{3}{c}{Muon channel} & Combination \\ 
 \cmidrule(r){2-3} \cmidrule(r){4-6} \cmidrule(r){7-7}
  & Dressed & Born & Bare & Dressed & Born & Born \\ 
 0.0 --  2.0&2.601e-02 $\pm$  0.6 $\pm$  0.3 $\pm$   1.2&2.663e-02 $\pm$  0.6 $\pm$  0.3 $\pm$   1.2&2.620e-02 $\pm$  0.4 $\pm$  0.2 $\pm$   0.9&2.672e-02  $\pm$  0.4 $\pm$  0.2 $\pm$   0.9&2.739e-02 $\pm$  0.4 $\pm$  0.2 $\pm$   0.9&2.690e-02 $\pm$  0.3 $\pm$  0.2 $\pm$   0.8 \\ 
  2.0 --  4.0 &5.469e-02 $\pm$  0.4 $\pm$  0.2 $\pm$   0.6&5.577e-02 $\pm$  0.4 $\pm$  0.2 $\pm$   0.6&5.405e-02 $\pm$  0.3 $\pm$  0.1 $\pm$   0.3&5.485e-02  $\pm$  0.3 $\pm$  0.1 $\pm$   0.3&5.590e-02 $\pm$  0.3 $\pm$  0.1 $\pm$   0.4&5.575e-02 $\pm$  0.2 $\pm$  0.1 $\pm$   0.3 \\ 
  4.0 --  6.0 &5.582e-02 $\pm$  0.4 $\pm$  0.1 $\pm$   0.4&5.634e-02 $\pm$  0.4 $\pm$  0.1 $\pm$   0.4&5.530e-02 $\pm$  0.3 $\pm$  0.1 $\pm$   0.4&5.569e-02  $\pm$  0.3 $\pm$  0.1 $\pm$   0.4&5.617e-02 $\pm$  0.3 $\pm$  0.1 $\pm$   0.5&5.642e-02 $\pm$  0.2 $\pm$  0.1 $\pm$   0.3 \\ 
  6.0 --  8.0 &4.875e-02 $\pm$  0.4 $\pm$  0.2 $\pm$   0.5&4.874e-02 $\pm$  0.4 $\pm$  0.2 $\pm$   0.6&4.941e-02 $\pm$  0.3 $\pm$  0.2 $\pm$   0.2&4.936e-02  $\pm$  0.3 $\pm$  0.2 $\pm$   0.2&4.933e-02 $\pm$  0.3 $\pm$  0.2 $\pm$   0.5&4.922e-02 $\pm$  0.3 $\pm$  0.1 $\pm$   0.4 \\ 
  8.0 -- 10.0 &4.085e-02 $\pm$  0.4 $\pm$  0.2 $\pm$   0.5&4.054e-02 $\pm$  0.4 $\pm$  0.2 $\pm$   0.5&4.143e-02 $\pm$  0.4 $\pm$  0.2 $\pm$   0.3&4.112e-02  $\pm$  0.4 $\pm$  0.2 $\pm$   0.3&4.085e-02 $\pm$  0.4 $\pm$  0.2 $\pm$   0.5&4.071e-02 $\pm$  0.3 $\pm$  0.1 $\pm$   0.4 \\
\bottomrule 
\end{tabular}
}
\label{tab:CombZPtY2U2}
\end{table*}

\begin{table*}
\centering
\caption{The values of $(1/\sigma)\,d\sigma/d\Zpt$ in each bin of \Zpt\ for the electron and muon channels separately (for various generator-level definitions) and for the Born-level combination in the kinematic region $ 66 \textrm{ GeV} \leq m_{\ell\ell} < 116 \textrm{ GeV},\ 0.8 \leq |y_{\ell\ell}| < 1.2$. The associated statistical and systematic (both uncorrelated and correlated between bins) are provided in percentage form.}
\resizebox{\textwidth}{!}{
\centering 
\begin{tabular}{ c  cccccc } \toprule  
 Bin & \multicolumn{6}{c}{$(1/\sigma)\,d\sigma/d\Zpt$ $\pm$ Statistical [\%] $\pm$ Uncorrelated systematic [\%] $\pm$ Correlated systematic [\%] } \\ 
 \cmidrule(r){2-7} 
  & 	\multicolumn{2}{c}{Electron channel} & \multicolumn{3}{c}{Muon channel} & Combination \\ 
 \cmidrule(r){2-3} \cmidrule(r){4-6} \cmidrule(r){7-7}
  & Dressed & Born & Bare & Dressed & Born & Born \\ 
 0.0 --  2.0&2.640e-02 $\pm$  0.6 $\pm$  0.3 $\pm$   1.0&2.708e-02 $\pm$  0.6 $\pm$  0.3 $\pm$   1.0&2.606e-02 $\pm$  0.5 $\pm$  0.2 $\pm$   0.9&2.662e-02  $\pm$  0.5 $\pm$  0.2 $\pm$   0.9&2.731e-02 $\pm$  0.5 $\pm$  0.2 $\pm$   0.9&2.711e-02 $\pm$  0.4 $\pm$  0.2 $\pm$   0.6 \\ 
  2.0 --  4.0 &5.532e-02 $\pm$  0.4 $\pm$  0.2 $\pm$   0.6&5.639e-02 $\pm$  0.4 $\pm$  0.2 $\pm$   0.6&5.406e-02 $\pm$  0.3 $\pm$  0.2 $\pm$   0.3&5.491e-02  $\pm$  0.3 $\pm$  0.2 $\pm$   0.3&5.597e-02 $\pm$  0.3 $\pm$  0.2 $\pm$   0.4&5.611e-02 $\pm$  0.2 $\pm$  0.1 $\pm$   0.3 \\ 
  4.0 --  6.0 &5.598e-02 $\pm$  0.4 $\pm$  0.2 $\pm$   0.3&5.653e-02 $\pm$  0.4 $\pm$  0.2 $\pm$   0.3&5.608e-02 $\pm$  0.3 $\pm$  0.1 $\pm$   0.4&5.645e-02  $\pm$  0.3 $\pm$  0.1 $\pm$   0.4&5.700e-02 $\pm$  0.3 $\pm$  0.1 $\pm$   0.5&5.691e-02 $\pm$  0.2 $\pm$  0.1 $\pm$   0.3 \\ 
  6.0 --  8.0 &4.889e-02 $\pm$  0.4 $\pm$  0.2 $\pm$   0.4&4.887e-02 $\pm$  0.4 $\pm$  0.2 $\pm$   0.4&4.919e-02 $\pm$  0.3 $\pm$  0.1 $\pm$   0.4&4.913e-02  $\pm$  0.3 $\pm$  0.1 $\pm$   0.4&4.913e-02 $\pm$  0.3 $\pm$  0.1 $\pm$   0.5&4.906e-02 $\pm$  0.3 $\pm$  0.1 $\pm$   0.3 \\ 
  8.0 -- 10.0 &4.126e-02 $\pm$  0.5 $\pm$  0.3 $\pm$   0.5&4.095e-02 $\pm$  0.5 $\pm$  0.3 $\pm$   0.6&4.159e-02 $\pm$  0.4 $\pm$  0.2 $\pm$   0.4&4.132e-02  $\pm$  0.4 $\pm$  0.2 $\pm$   0.4&4.098e-02 $\pm$  0.4 $\pm$  0.2 $\pm$   0.6&4.096e-02 $\pm$  0.3 $\pm$  0.1 $\pm$   0.4 \\
\bottomrule 
\end{tabular}
}
\label{tab:CombZPtY3U2}
\end{table*}

\begin{table*}
\centering
\caption{The values of $(1/\sigma)\,d\sigma/d\Zpt$ in each bin of \Zpt\ for the electron and muon channels separately (for various generator-level definitions) and for the Born-level combination in the kinematic region $ 66 \textrm{ GeV} \leq m_{\ell\ell} < 116 \textrm{ GeV},\ 1.2 \leq |y_{\ell\ell}| < 1.6$. The associated statistical and systematic (both uncorrelated and correlated between bins) are provided in percentage form.}
\resizebox{\textwidth}{!}{
\centering 
\begin{tabular}{ c  cccccc } \toprule  
 Bin & \multicolumn{6}{c}{$(1/\sigma)\,d\sigma/d\Zpt$ $\pm$ Statistical [\%] $\pm$ Uncorrelated systematic [\%] $\pm$ Correlated systematic [\%] } \\ 
 \cmidrule(r){2-7} 
  & 	\multicolumn{2}{c}{Electron channel} & \multicolumn{3}{c}{Muon channel} & Combination \\ 
 \cmidrule(r){2-3} \cmidrule(r){4-6} \cmidrule(r){7-7}
  & Dressed & Born & Bare & Dressed & Born & Born \\ 
 0.0 --  2.0&2.584e-02 $\pm$  0.6 $\pm$  0.3 $\pm$   1.1&2.655e-02 $\pm$  0.6 $\pm$  0.3 $\pm$   1.1&2.512e-02 $\pm$  0.5 $\pm$  0.2 $\pm$   0.9&2.573e-02  $\pm$  0.5 $\pm$  0.2 $\pm$   0.9&2.645e-02 $\pm$  0.5 $\pm$  0.2 $\pm$   0.9&2.658e-02 $\pm$  0.4 $\pm$  0.2 $\pm$   0.7 \\ 
  2.0 --  4.0 &5.405e-02 $\pm$  0.4 $\pm$  0.2 $\pm$   0.8&5.522e-02 $\pm$  0.4 $\pm$  0.2 $\pm$   0.8&5.308e-02 $\pm$  0.3 $\pm$  0.2 $\pm$   0.5&5.405e-02  $\pm$  0.3 $\pm$  0.2 $\pm$   0.5&5.525e-02 $\pm$  0.3 $\pm$  0.2 $\pm$   0.5&5.536e-02 $\pm$  0.3 $\pm$  0.1 $\pm$   0.4 \\ 
  4.0 --  6.0 &5.436e-02 $\pm$  0.4 $\pm$  0.2 $\pm$   0.5&5.494e-02 $\pm$  0.4 $\pm$  0.2 $\pm$   0.5&5.453e-02 $\pm$  0.3 $\pm$  0.2 $\pm$   0.4&5.498e-02  $\pm$  0.3 $\pm$  0.2 $\pm$   0.4&5.554e-02 $\pm$  0.3 $\pm$  0.2 $\pm$   0.4&5.529e-02 $\pm$  0.3 $\pm$  0.2 $\pm$   0.4 \\ 
  6.0 --  8.0 &4.804e-02 $\pm$  0.5 $\pm$  0.3 $\pm$   0.6&4.803e-02 $\pm$  0.5 $\pm$  0.3 $\pm$   0.6&4.836e-02 $\pm$  0.4 $\pm$  0.2 $\pm$   0.3&4.831e-02  $\pm$  0.4 $\pm$  0.2 $\pm$   0.3&4.829e-02 $\pm$  0.4 $\pm$  0.2 $\pm$   0.6&4.816e-02 $\pm$  0.3 $\pm$  0.1 $\pm$   0.5 \\ 
  8.0 -- 10.0 &4.096e-02 $\pm$  0.5 $\pm$  0.2 $\pm$   0.5&4.057e-02 $\pm$  0.5 $\pm$  0.2 $\pm$   0.6&4.138e-02 $\pm$  0.4 $\pm$  0.2 $\pm$   0.4&4.104e-02  $\pm$  0.4 $\pm$  0.2 $\pm$   0.4&4.068e-02 $\pm$  0.4 $\pm$  0.2 $\pm$   0.6&4.066e-02 $\pm$  0.3 $\pm$  0.2 $\pm$   0.5 \\
\bottomrule 
\end{tabular}
}
\label{tab:CombZPtY4U2}
\end{table*}

\begin{table*}
\centering
\caption{The values of $(1/\sigma)\,d\sigma/d\Zpt$ in each bin of \Zpt\ for the electron and muon channels separately (for various generator-level definitions) and for the Born-level combination in the kinematic region $ 66 \textrm{ GeV} \leq m_{\ell\ell} < 116 \textrm{ GeV},\ 1.6 \leq |y_{\ell\ell}| < 2.0$. The associated statistical and systematic (both uncorrelated and correlated between bins) are provided in percentage form.}
\resizebox{\textwidth}{!}{
\centering 
\begin{tabular}{ c  cccccc } \toprule  
 Bin & \multicolumn{6}{c}{$(1/\sigma)\,d\sigma/d\Zpt$ $\pm$ Statistical [\%] $\pm$ Uncorrelated systematic [\%] $\pm$ Correlated systematic [\%] } \\ 
 \cmidrule(r){2-7} 
  & 	\multicolumn{2}{c}{Electron channel} & \multicolumn{3}{c}{Muon channel} & Combination \\ 
 \cmidrule(r){2-3} \cmidrule(r){4-6} \cmidrule(r){7-7}
  & Dressed & Born & Bare & Dressed & Born & Born \\ 
 0.0 --  2.0&2.536e-02 $\pm$  0.8 $\pm$  0.5 $\pm$   1.5&2.616e-02 $\pm$  0.8 $\pm$  0.5 $\pm$   1.5&2.466e-02 $\pm$  0.6 $\pm$  0.3 $\pm$   1.2&2.541e-02  $\pm$  0.6 $\pm$  0.3 $\pm$   1.2&2.618e-02 $\pm$  0.6 $\pm$  0.3 $\pm$   1.2&2.606e-02 $\pm$  0.5 $\pm$  0.3 $\pm$   0.9 \\ 
  2.0 --  4.0 &5.324e-02 $\pm$  0.5 $\pm$  0.3 $\pm$   1.1&5.454e-02 $\pm$  0.5 $\pm$  0.3 $\pm$   1.2&5.139e-02 $\pm$  0.4 $\pm$  0.2 $\pm$   0.6&5.254e-02  $\pm$  0.4 $\pm$  0.2 $\pm$   0.6&5.382e-02 $\pm$  0.4 $\pm$  0.2 $\pm$   0.7&5.394e-02 $\pm$  0.3 $\pm$  0.2 $\pm$   0.6 \\ 
  4.0 --  6.0 &5.345e-02 $\pm$  0.5 $\pm$  0.3 $\pm$   0.6&5.408e-02 $\pm$  0.5 $\pm$  0.3 $\pm$   0.6&5.307e-02 $\pm$  0.4 $\pm$  0.2 $\pm$   0.3&5.366e-02  $\pm$  0.4 $\pm$  0.2 $\pm$   0.3&5.433e-02 $\pm$  0.4 $\pm$  0.2 $\pm$   0.5&5.425e-02 $\pm$  0.3 $\pm$  0.2 $\pm$   0.4 \\ 
  6.0 --  8.0 &4.640e-02 $\pm$  0.6 $\pm$  0.3 $\pm$   0.9&4.646e-02 $\pm$  0.6 $\pm$  0.3 $\pm$   0.9&4.678e-02 $\pm$  0.4 $\pm$  0.3 $\pm$   0.5&4.674e-02  $\pm$  0.4 $\pm$  0.3 $\pm$   0.5&4.679e-02 $\pm$  0.4 $\pm$  0.3 $\pm$   0.9&4.667e-02 $\pm$  0.3 $\pm$  0.2 $\pm$   0.6 \\ 
  8.0 -- 10.0 &3.960e-02 $\pm$  0.6 $\pm$  0.3 $\pm$   0.8&3.931e-02 $\pm$  0.6 $\pm$  0.3 $\pm$   0.9&3.978e-02 $\pm$  0.5 $\pm$  0.2 $\pm$   0.4&3.950e-02  $\pm$  0.5 $\pm$  0.2 $\pm$   0.4&3.922e-02 $\pm$  0.5 $\pm$  0.2 $\pm$   0.5&3.925e-02 $\pm$  0.4 $\pm$  0.2 $\pm$   0.4 \\
\bottomrule 
\end{tabular}
}
\label{tab:CombZPtY5U2}
\end{table*}

\begin{table*}
\centering
\caption{The values of $(1/\sigma)\,d\sigma/d\Zpt$ in each bin of \Zpt\ for the electron and muon channels separately (for various generator-level definitions) and for the Born-level combination in the kinematic region $ 66 \textrm{ GeV} \leq m_{\ell\ell} < 116 \textrm{ GeV},\ 2.0 \leq |y_{\ell\ell}| < 2.4$. The associated statistical and systematic (both uncorrelated and correlated between bins) are provided in percentage form.}
\resizebox{\textwidth}{!}{
\centering 
\begin{tabular}{ c  cccccc } \toprule  
 Bin & \multicolumn{6}{c}{$(1/\sigma)\,d\sigma/d\Zpt$ $\pm$ Statistical [\%] $\pm$ Uncorrelated systematic [\%] $\pm$ Correlated systematic [\%] } \\ 
 \cmidrule(r){2-7} 
  & 	\multicolumn{2}{c}{Electron channel} & \multicolumn{3}{c}{Muon channel} & Combination \\ 
 \cmidrule(r){2-3} \cmidrule(r){4-6} \cmidrule(r){7-7}
  & Dressed & Born & Bare & Dressed & Born & Born \\ 
 0.0 --  2.0&2.582e-02 $\pm$  1.3 $\pm$  0.6 $\pm$   2.3&2.668e-02 $\pm$  1.3 $\pm$  0.6 $\pm$   2.3&2.483e-02 $\pm$  1.1 $\pm$  0.5 $\pm$   2.0&2.563e-02  $\pm$  1.1 $\pm$  0.5 $\pm$   2.0&2.646e-02 $\pm$  1.1 $\pm$  0.5 $\pm$   2.0&2.676e-02 $\pm$  0.8 $\pm$  0.4 $\pm$   1.5 \\ 
  2.0 --  4.0 &5.224e-02 $\pm$  0.8 $\pm$  0.5 $\pm$   1.5&5.354e-02 $\pm$  0.8 $\pm$  0.5 $\pm$   1.5&5.173e-02 $\pm$  0.7 $\pm$  0.3 $\pm$   1.1&5.296e-02  $\pm$  0.7 $\pm$  0.3 $\pm$   1.1&5.432e-02 $\pm$  0.7 $\pm$  0.3 $\pm$   1.2&5.429e-02 $\pm$  0.5 $\pm$  0.3 $\pm$   0.9 \\ 
  4.0 --  6.0 &5.276e-02 $\pm$  0.8 $\pm$  0.5 $\pm$   1.0&5.345e-02 $\pm$  0.8 $\pm$  0.5 $\pm$   1.0&5.203e-02 $\pm$  0.7 $\pm$  0.4 $\pm$   0.8&5.274e-02  $\pm$  0.7 $\pm$  0.4 $\pm$   0.8&5.337e-02 $\pm$  0.7 $\pm$  0.4 $\pm$   0.8&5.338e-02 $\pm$  0.6 $\pm$  0.3 $\pm$   0.6 \\ 
  6.0 --  8.0 &4.545e-02 $\pm$  0.9 $\pm$  0.5 $\pm$   0.9&4.558e-02 $\pm$  0.9 $\pm$  0.5 $\pm$   1.0&4.548e-02 $\pm$  0.8 $\pm$  0.4 $\pm$   1.0&4.556e-02  $\pm$  0.8 $\pm$  0.4 $\pm$   1.0&4.575e-02 $\pm$  0.8 $\pm$  0.4 $\pm$   1.0&4.567e-02 $\pm$  0.6 $\pm$  0.3 $\pm$   0.7 \\ 
  8.0 -- 10.0 &3.928e-02 $\pm$  1.0 $\pm$  0.5 $\pm$   0.9&3.904e-02 $\pm$  1.0 $\pm$  0.5 $\pm$   0.9&3.903e-02 $\pm$  0.8 $\pm$  0.4 $\pm$   0.7&3.872e-02  $\pm$  0.8 $\pm$  0.4 $\pm$   0.7&3.850e-02 $\pm$  0.8 $\pm$  0.4 $\pm$   0.8&3.867e-02 $\pm$  0.6 $\pm$  0.3 $\pm$   0.6 \\
\bottomrule 
\end{tabular}
}
\label{tab:CombZPtY6U2}
\end{table*}

\begin{table*}
\centering
\caption{The values of $(1/\sigma)\,d\sigma/d\Zpt$ in each bin of \Zpt\ for the electron and muon channels separately (for various generator-level definitions) and for the Born-level combination in the kinematic region $ 12 \textrm{ GeV} \leq m_{\ell\ell} <  20 \textrm{ GeV},\ 0.0 \leq |y_{\ell\ell}| < 2.4$. The associated statistical and systematic (both uncorrelated and correlated between bins) are provided in percentage form.}
\resizebox{\textwidth}{!}{
\centering 
\begin{tabular}{ c  cccccc } \toprule  
 Bin & \multicolumn{6}{c}{$(1/\sigma)\,d\sigma/d\Zpt$ $\pm$ Statistical [\%] $\pm$ Uncorrelated systematic [\%] $\pm$ Correlated systematic [\%] } \\ 
 \cmidrule(r){2-7} 
  & 	\multicolumn{2}{c}{Electron channel} & \multicolumn{3}{c}{Muon channel} & Combination \\ 
 \cmidrule(r){2-3} \cmidrule(r){4-6} \cmidrule(r){7-7}
  & Dressed & Born & Bare & Dressed & Born & Born \\ 
 45.0 -- 55.0 &2.720e-02 $\pm$  1.9 $\pm$  1.4 $\pm$   2.3&2.719e-02 $\pm$  1.9 $\pm$  1.4 $\pm$   2.3&2.820e-02 $\pm$  1.3 $\pm$  0.3 $\pm$   0.7&2.815e-02  $\pm$  1.3 $\pm$  0.3 $\pm$   0.7&2.807e-02 $\pm$  1.3 $\pm$  0.3 $\pm$   1.3&2.792e-02 $\pm$  1.1 $\pm$  0.4 $\pm$   1.1 \\ 
 55.0 -- 65.0 &2.137e-02 $\pm$  2.3 $\pm$  1.6 $\pm$   1.6&2.140e-02 $\pm$  2.3 $\pm$  1.6 $\pm$   1.8&2.166e-02 $\pm$  1.6 $\pm$  0.5 $\pm$   0.5&2.156e-02  $\pm$  1.6 $\pm$  0.5 $\pm$   0.5&2.158e-02 $\pm$  1.6 $\pm$  0.5 $\pm$   0.7&2.150e-02 $\pm$  1.3 $\pm$  0.5 $\pm$   0.8 \\ 
 65.0 -- 75.0 &1.526e-02 $\pm$  2.6 $\pm$  2.7 $\pm$   2.9&1.526e-02 $\pm$  2.6 $\pm$  2.7 $\pm$   3.0&1.534e-02 $\pm$  1.9 $\pm$  0.5 $\pm$   0.6&1.532e-02  $\pm$  1.9 $\pm$  0.5 $\pm$   0.6&1.538e-02 $\pm$  1.9 $\pm$  0.5 $\pm$   1.3&1.534e-02 $\pm$  1.6 $\pm$  0.7 $\pm$   1.2 \\ 
 75.0 -- 85.0 &1.014e-02 $\pm$  3.2 $\pm$  1.4 $\pm$   1.9&1.014e-02 $\pm$  3.2 $\pm$  1.4 $\pm$   2.1&1.038e-02 $\pm$  2.3 $\pm$  0.7 $\pm$   1.1&1.042e-02  $\pm$  2.3 $\pm$  0.7 $\pm$   1.1&1.044e-02 $\pm$  2.3 $\pm$  0.7 $\pm$   1.1&1.030e-02 $\pm$  1.9 $\pm$  0.7 $\pm$   1.1 \\ 
 85.0 -- 105.0 &6.223e-03 $\pm$  2.5 $\pm$  1.1 $\pm$   2.0&6.239e-03 $\pm$  2.5 $\pm$  1.1 $\pm$   2.1&5.958e-03 $\pm$  2.1 $\pm$  0.6 $\pm$   1.1&5.974e-03  $\pm$  2.1 $\pm$  0.6 $\pm$   1.1&5.974e-03 $\pm$  2.1 $\pm$  0.6 $\pm$   1.4&6.042e-03 $\pm$  1.6 $\pm$  0.6 $\pm$   1.3 \\
\bottomrule 
\end{tabular}
}
\label{tab:CombZPtLowM1U2}
\end{table*}

\begin{table*}
\centering
\caption{The values of $(1/\sigma)\,d\sigma/d\Zpt$ in each bin of \Zpt\ for the electron and muon channels separately (for various generator-level definitions) and for the Born-level combination in the kinematic region $ 20 \textrm{ GeV} \leq m_{\ell\ell} <  30 \textrm{ GeV},\ 0.0 \leq |y_{\ell\ell}| < 2.4$. The associated statistical and systematic (both uncorrelated and correlated between bins) are provided in percentage form.}
\resizebox{\textwidth}{!}{
\centering 
\begin{tabular}{ c  cccccc } \toprule  
 Bin & \multicolumn{6}{c}{$(1/\sigma)\,d\sigma/d\Zpt$ $\pm$ Statistical [\%] $\pm$ Uncorrelated systematic [\%] $\pm$ Correlated systematic [\%] } \\ 
 \cmidrule(r){2-7} 
  & 	\multicolumn{2}{c}{Electron channel} & \multicolumn{3}{c}{Muon channel} & Combination \\ 
 \cmidrule(r){2-3} \cmidrule(r){4-6} \cmidrule(r){7-7}
  & Dressed & Born & Bare & Dressed & Born & Born \\ 
 45.0 -- 55.0 &2.771e-02 $\pm$  2.3 $\pm$  0.9 $\pm$   2.2&2.745e-02 $\pm$  2.3 $\pm$  0.9 $\pm$   2.4&2.775e-02 $\pm$  1.6 $\pm$  0.6 $\pm$   1.1&2.766e-02  $\pm$  1.6 $\pm$  0.6 $\pm$   1.1&2.750e-02 $\pm$  1.6 $\pm$  0.6 $\pm$   1.4&2.751e-02 $\pm$  1.3 $\pm$  0.5 $\pm$   1.3 \\ 
 55.0 -- 65.0 &2.081e-02 $\pm$  2.7 $\pm$  1.4 $\pm$   1.8&2.085e-02 $\pm$  2.7 $\pm$  1.4 $\pm$   1.8&2.122e-02 $\pm$  1.9 $\pm$  0.7 $\pm$   0.9&2.114e-02  $\pm$  1.9 $\pm$  0.7 $\pm$   0.9&2.134e-02 $\pm$  1.9 $\pm$  0.7 $\pm$   1.3&2.122e-02 $\pm$  1.6 $\pm$  0.6 $\pm$   1.1 \\ 
 65.0 -- 75.0 &1.441e-02 $\pm$  3.1 $\pm$  1.3 $\pm$   2.3&1.447e-02 $\pm$  3.1 $\pm$  1.3 $\pm$   2.3&1.475e-02 $\pm$  2.4 $\pm$  1.0 $\pm$   1.5&1.480e-02  $\pm$  2.4 $\pm$  1.0 $\pm$   1.5&1.479e-02 $\pm$  2.4 $\pm$  1.0 $\pm$   1.6&1.468e-02 $\pm$  1.9 $\pm$  0.8 $\pm$   1.3 \\ 
 75.0 -- 85.0 &1.048e-02 $\pm$  3.7 $\pm$  1.7 $\pm$   3.1&1.052e-02 $\pm$  3.7 $\pm$  1.7 $\pm$   3.1&1.012e-02 $\pm$  2.9 $\pm$  1.3 $\pm$   2.0&1.010e-02  $\pm$  2.9 $\pm$  1.3 $\pm$   2.0&1.014e-02 $\pm$  2.9 $\pm$  1.3 $\pm$   2.0&1.027e-02 $\pm$  2.3 $\pm$  1.0 $\pm$   1.7 \\ 
 85.0 -- 105.0 &6.007e-03 $\pm$  3.0 $\pm$  1.4 $\pm$   3.4&6.029e-03 $\pm$  3.0 $\pm$  1.4 $\pm$   3.4&6.245e-03 $\pm$  2.5 $\pm$  1.2 $\pm$   1.9&6.240e-03  $\pm$  2.5 $\pm$  1.2 $\pm$   1.9&6.245e-03 $\pm$  2.5 $\pm$  1.2 $\pm$   2.0&6.140e-03 $\pm$  1.9 $\pm$  0.9 $\pm$   1.9 \\
\bottomrule 
\end{tabular}
}
\label{tab:CombZPtLowM2U2}
\end{table*}

\begin{table*}
\centering
\caption{The values of $(1/\sigma)\,d\sigma/d\Zpt$ in each bin of \Zpt\ for the electron and muon channels separately (for various generator-level definitions) and for the Born-level combination in the kinematic region $ 30 \textrm{ GeV} \leq m_{\ell\ell} <  46 \textrm{ GeV},\ 0.0 \leq |y_{\ell\ell}| < 2.4$. The associated statistical and systematic (both uncorrelated and correlated between bins) are provided in percentage form.}
\resizebox{\textwidth}{!}{
\centering 
\begin{tabular}{ c  cccccc } \toprule  
 Bin & \multicolumn{6}{c}{$(1/\sigma)\,d\sigma/d\Zpt$ $\pm$ Statistical [\%] $\pm$ Uncorrelated systematic [\%] $\pm$ Correlated systematic [\%] } \\ 
 \cmidrule(r){2-7} 
  & 	\multicolumn{2}{c}{Electron channel} & \multicolumn{3}{c}{Muon channel} & Combination \\ 
 \cmidrule(r){2-3} \cmidrule(r){4-6} \cmidrule(r){7-7}
  & Dressed & Born & Bare & Dressed & Born & Born \\ 
 45.0 -- 55.0 &2.853e-02 $\pm$  2.2 $\pm$  0.6 $\pm$   1.7&2.821e-02 $\pm$  2.2 $\pm$  0.6 $\pm$   2.0&2.974e-02 $\pm$  1.6 $\pm$  0.5 $\pm$   1.1&2.977e-02  $\pm$  1.6 $\pm$  0.5 $\pm$   1.1&2.953e-02 $\pm$  1.6 $\pm$  0.5 $\pm$   1.3&2.909e-02 $\pm$  1.3 $\pm$  0.4 $\pm$   1.2 \\ 
 55.0 -- 65.0 &2.153e-02 $\pm$  2.6 $\pm$  1.4 $\pm$   1.4&2.165e-02 $\pm$  2.7 $\pm$  1.4 $\pm$   1.6&2.118e-02 $\pm$  2.0 $\pm$  0.8 $\pm$   1.0&2.119e-02  $\pm$  2.0 $\pm$  0.8 $\pm$   1.0&2.132e-02 $\pm$  2.0 $\pm$  0.8 $\pm$   1.3&2.146e-02 $\pm$  1.6 $\pm$  0.7 $\pm$   1.1 \\ 
 65.0 -- 75.0 &1.517e-02 $\pm$  3.3 $\pm$  1.1 $\pm$   1.6&1.529e-02 $\pm$  3.3 $\pm$  1.1 $\pm$   1.7&1.412e-02 $\pm$  2.6 $\pm$  1.1 $\pm$   1.3&1.405e-02  $\pm$  2.6 $\pm$  1.1 $\pm$   1.3&1.410e-02 $\pm$  2.6 $\pm$  1.1 $\pm$   1.3&1.452e-02 $\pm$  2.1 $\pm$  0.8 $\pm$   1.1 \\ 
 75.0 -- 85.0 &9.428e-03 $\pm$  4.2 $\pm$  1.5 $\pm$   2.3&9.462e-03 $\pm$  4.2 $\pm$  1.5 $\pm$   2.3&1.001e-02 $\pm$  3.2 $\pm$  1.2 $\pm$   1.6&9.895e-03  $\pm$  3.2 $\pm$  1.2 $\pm$   1.6&9.917e-03 $\pm$  3.2 $\pm$  1.2 $\pm$   2.8&9.763e-03 $\pm$  2.6 $\pm$  0.9 $\pm$   2.0 \\ 
 85.0 -- 105.0 &5.793e-03 $\pm$  3.4 $\pm$  1.4 $\pm$   2.5&5.807e-03 $\pm$  3.4 $\pm$  1.4 $\pm$   2.6&5.772e-03 $\pm$  2.9 $\pm$  1.2 $\pm$   1.8&5.799e-03  $\pm$  2.9 $\pm$  1.2 $\pm$   1.8&5.808e-03 $\pm$  2.9 $\pm$  1.2 $\pm$   2.0&5.793e-03 $\pm$  2.2 $\pm$  0.9 $\pm$   1.7 \\
\bottomrule 
\end{tabular}
}
\label{tab:CombZPtLowM3U2}
\end{table*}

 \begin{table*}
\centering
\caption{The values of $(1/\sigma)\,d\sigma/d\Zpt$ in each bin of \Zpt\ for the electron and muon channels separately (for various generator-level definitions) and for the Born-level combination in the kinematic region $ 46 \textrm{ GeV} \leq m_{\ell\ell} <  66 \textrm{ GeV},\ 0.0 \leq |y_{\ell\ell}| < 2.4$. The associated statistical and systematic (both uncorrelated and correlated between bins) are provided in percentage form.}
\resizebox{\textwidth}{!}{
\centering 
\begin{tabular}{ c  cccccc } \toprule  
 Bin & \multicolumn{6}{c}{$(1/\sigma)\,d\sigma/d\Zpt$ $\pm$ Statistical [\%] $\pm$ Uncorrelated systematic [\%] $\pm$ Correlated systematic [\%] } \\ 
 \cmidrule(r){2-7} 
  & 	\multicolumn{2}{c}{Electron channel} & \multicolumn{3}{c}{Muon channel} & Combination \\ 
 \cmidrule(r){2-3} \cmidrule(r){4-6} \cmidrule(r){7-7}
  & Dressed & Born & Bare & Dressed & Born & Born \\ 
 0.0 --  2.0&3.886e-02 $\pm$  1.4 $\pm$  0.7 $\pm$   4.5&4.361e-02 $\pm$  1.4 $\pm$  0.7 $\pm$   5.2&3.680e-02 $\pm$  1.0 $\pm$  0.3 $\pm$   4.4&3.927e-02  $\pm$  1.0 $\pm$  0.3 $\pm$   4.4&4.420e-02 $\pm$  1.0 $\pm$  0.3 $\pm$   4.8&4.347e-02 $\pm$  0.8 $\pm$  0.3 $\pm$   4.4 \\ 
  2.0 --  4.0 &7.415e-02 $\pm$  1.0 $\pm$  0.5 $\pm$   1.3&8.265e-02 $\pm$  1.0 $\pm$  0.5 $\pm$   3.1&6.865e-02 $\pm$  0.7 $\pm$  0.3 $\pm$   0.8&7.302e-02  $\pm$  0.7 $\pm$  0.3 $\pm$   0.8&8.110e-02 $\pm$  0.7 $\pm$  0.3 $\pm$   2.1&8.148e-02 $\pm$  0.6 $\pm$  0.3 $\pm$   1.2 \\ 
  4.0 --  6.0 &6.674e-02 $\pm$  1.0 $\pm$  0.5 $\pm$   1.1&7.307e-02 $\pm$  1.0 $\pm$  0.5 $\pm$   2.7&6.451e-02 $\pm$  0.7 $\pm$  0.3 $\pm$   0.7&6.764e-02  $\pm$  0.7 $\pm$  0.3 $\pm$   0.7&7.414e-02 $\pm$  0.7 $\pm$  0.3 $\pm$   1.8&7.390e-02 $\pm$  0.6 $\pm$  0.2 $\pm$   1.1 \\ 
  6.0 --  8.0 &5.417e-02 $\pm$  1.1 $\pm$  0.5 $\pm$   0.9&5.822e-02 $\pm$  1.1 $\pm$  0.5 $\pm$   2.0&5.296e-02 $\pm$  0.8 $\pm$  0.3 $\pm$   0.8&5.484e-02  $\pm$  0.8 $\pm$  0.3 $\pm$   0.8&5.885e-02 $\pm$  0.8 $\pm$  0.3 $\pm$   1.5&5.877e-02 $\pm$  0.6 $\pm$  0.2 $\pm$   0.9 \\ 
  8.0 -- 10.0 &4.119e-02 $\pm$  1.3 $\pm$  0.5 $\pm$   1.1&4.293e-02 $\pm$  1.3 $\pm$  0.5 $\pm$   1.6&4.097e-02 $\pm$  0.9 $\pm$  0.4 $\pm$   0.8&4.163e-02  $\pm$  0.9 $\pm$  0.4 $\pm$   0.8&4.361e-02 $\pm$  0.9 $\pm$  0.4 $\pm$   1.3&4.355e-02 $\pm$  0.8 $\pm$  0.3 $\pm$   0.9 \\
\bottomrule 
\end{tabular}
}
\label{tab:CombZPtM1U2}
\end{table*}

\begin{table*}
\centering
\caption{The values of $(1/\sigma)\,d\sigma/d\Zpt$ in each bin of \Zpt\ for the electron and muon channels separately (for various generator-level definitions) and for the Born-level combination in the kinematic region $ 66 \textrm{ GeV} \leq m_{\ell\ell} < 116 \textrm{ GeV},\ 0.0 \leq |y_{\ell\ell}| < 2.4$. The associated statistical and systematic (both uncorrelated and correlated between bins) are provided in percentage form.}
\resizebox{\textwidth}{!}{
\centering 
\begin{tabular}{ c  cccccc } \toprule  
 Bin & \multicolumn{6}{c}{$(1/\sigma)\,d\sigma/d\Zpt$ $\pm$ Statistical [\%] $\pm$ Uncorrelated systematic [\%] $\pm$ Correlated systematic [\%] } \\ 
 \cmidrule(r){2-7} 
  & 	\multicolumn{2}{c}{Electron channel} & \multicolumn{3}{c}{Muon channel} & Combination \\ 
 \cmidrule(r){2-3} \cmidrule(r){4-6} \cmidrule(r){7-7}
  & Dressed & Born & Bare & Dressed & Born & Born \\ 
 0.0 --  2.0&2.608e-02 $\pm$  0.3 $\pm$  0.1 $\pm$   0.9&2.677e-02 $\pm$  0.3 $\pm$  0.1 $\pm$   0.9&2.568e-02 $\pm$  0.2 $\pm$  0.1 $\pm$   0.9&2.626e-02  $\pm$  0.2 $\pm$  0.1 $\pm$   0.9&2.696e-02 $\pm$  0.2 $\pm$  0.1 $\pm$   0.9&2.669e-02 $\pm$  0.2 $\pm$  0.1 $\pm$   0.5 \\ 
  2.0 --  4.0 &5.441e-02 $\pm$  0.2 $\pm$  0.1 $\pm$   0.5&5.553e-02 $\pm$  0.2 $\pm$  0.1 $\pm$   0.6&5.337e-02 $\pm$  0.1 $\pm$  0.1 $\pm$   0.3&5.428e-02  $\pm$  0.1 $\pm$  0.1 $\pm$   0.3&5.541e-02 $\pm$  0.2 $\pm$  0.1 $\pm$   0.4&5.545e-02 $\pm$  0.1 $\pm$  0.1 $\pm$   0.2 \\ 
  4.0 --  6.0 &5.505e-02 $\pm$  0.2 $\pm$  0.1 $\pm$   0.3&5.560e-02 $\pm$  0.2 $\pm$  0.1 $\pm$   0.3&5.483e-02 $\pm$  0.1 $\pm$  0.1 $\pm$   0.4&5.525e-02  $\pm$  0.1 $\pm$  0.1 $\pm$   0.4&5.579e-02 $\pm$  0.1 $\pm$  0.1 $\pm$   0.5&5.590e-02 $\pm$  0.1 $\pm$  0.1 $\pm$   0.3 \\ 
  6.0 --  8.0 &4.831e-02 $\pm$  0.2 $\pm$  0.1 $\pm$   0.4&4.831e-02 $\pm$  0.2 $\pm$  0.1 $\pm$   0.5&4.856e-02 $\pm$  0.2 $\pm$  0.1 $\pm$   0.2&4.851e-02  $\pm$  0.2 $\pm$  0.1 $\pm$   0.2&4.851e-02 $\pm$  0.2 $\pm$  0.1 $\pm$   0.6&4.847e-02 $\pm$  0.1 $\pm$  0.1 $\pm$   0.3 \\ 
  8.0 -- 10.0 &4.084e-02 $\pm$  0.2 $\pm$  0.1 $\pm$   0.4&4.052e-02 $\pm$  0.2 $\pm$  0.1 $\pm$   0.5&4.115e-02 $\pm$  0.2 $\pm$  0.1 $\pm$   0.2&4.085e-02  $\pm$  0.2 $\pm$  0.1 $\pm$   0.2&4.055e-02 $\pm$  0.2 $\pm$  0.1 $\pm$   0.6&4.047e-02 $\pm$  0.1 $\pm$  0.1 $\pm$   0.3 \\
\bottomrule 
\end{tabular}
}
\label{tab:CombZPt}
\end{table*}

\begin{table*}
\centering
\caption{The values of $(1/\sigma)\,d\sigma/d\Zpt$ in each bin of \Zpt\ for the electron and muon channels separately (for various generator-level definitions) and for the Born-level combination in the kinematic region $ 116 \textrm{ GeV} \leq m_{\ell\ell} < 150 \textrm{ GeV},\ 0.0 \leq |y_{\ell\ell}| < 2.4$. The associated statistical and systematic (both uncorrelated and correlated between bins) are provided in percentage form.}
\resizebox{\textwidth}{!}{
\centering 
\begin{tabular}{ c  cccccc } \toprule  
 Bin & \multicolumn{6}{c}{$(1/\sigma)\,d\sigma/d\Zpt$ $\pm$ Statistical [\%] $\pm$ Uncorrelated systematic [\%] $\pm$ Correlated systematic [\%] } \\ 
 \cmidrule(r){2-7} 
  & 	\multicolumn{2}{c}{Electron channel} & \multicolumn{3}{c}{Muon channel} & Combination \\ 
 \cmidrule(r){2-3} \cmidrule(r){4-6} \cmidrule(r){7-7}
  & Dressed & Born & Bare & Dressed & Born & Born \\ 
 0.0 --  2.0&2.181e-02 $\pm$  3.0 $\pm$  1.3 $\pm$   4.4&2.228e-02 $\pm$  3.0 $\pm$  1.3 $\pm$   4.4&1.967e-02 $\pm$  2.7 $\pm$  1.4 $\pm$   5.0&2.023e-02  $\pm$  2.7 $\pm$  1.4 $\pm$   5.0&2.071e-02 $\pm$  2.7 $\pm$  1.4 $\pm$   5.1&2.130e-02 $\pm$  2.0 $\pm$  1.0 $\pm$   4.3 \\ 
  2.0 --  4.0 &4.492e-02 $\pm$  1.9 $\pm$  0.9 $\pm$   1.8&4.572e-02 $\pm$  1.9 $\pm$  0.9 $\pm$   1.8&4.413e-02 $\pm$  1.7 $\pm$  0.8 $\pm$   1.2&4.488e-02  $\pm$  1.7 $\pm$  0.8 $\pm$   1.2&4.597e-02 $\pm$  1.7 $\pm$  0.8 $\pm$   1.2&4.563e-02 $\pm$  1.2 $\pm$  0.6 $\pm$   1.1 \\ 
  4.0 --  6.0 &4.604e-02 $\pm$  1.9 $\pm$  0.9 $\pm$   1.4&4.659e-02 $\pm$  1.9 $\pm$  0.9 $\pm$   1.4&4.842e-02 $\pm$  1.6 $\pm$  0.9 $\pm$   1.1&4.882e-02  $\pm$  1.6 $\pm$  0.9 $\pm$   1.1&4.949e-02 $\pm$  1.6 $\pm$  0.9 $\pm$   1.5&4.826e-02 $\pm$  1.2 $\pm$  0.6 $\pm$   1.0 \\ 
  6.0 --  8.0 &4.296e-02 $\pm$  1.9 $\pm$  0.9 $\pm$   1.3&4.332e-02 $\pm$  1.9 $\pm$  0.9 $\pm$   1.3&4.296e-02 $\pm$  1.6 $\pm$  0.8 $\pm$   1.4&4.326e-02  $\pm$  1.6 $\pm$  0.8 $\pm$   1.4&4.336e-02 $\pm$  1.6 $\pm$  0.8 $\pm$   1.4&4.306e-02 $\pm$  1.2 $\pm$  0.6 $\pm$   1.0 \\ 
  8.0 -- 10.0 &3.850e-02 $\pm$  2.1 $\pm$  1.0 $\pm$   1.4&3.849e-02 $\pm$  2.1 $\pm$  1.0 $\pm$   1.5&3.792e-02 $\pm$  1.9 $\pm$  0.8 $\pm$   1.6&3.804e-02  $\pm$  1.9 $\pm$  0.8 $\pm$   1.6&3.818e-02 $\pm$  1.9 $\pm$  0.8 $\pm$   1.6&3.838e-02 $\pm$  1.4 $\pm$  0.6 $\pm$   1.1 \\
\bottomrule 
\end{tabular}
}
\label{tab:CombZPtM3U2}
\end{table*}

\clearpage
\newpage
\begin{flushleft}
{\Large The ATLAS Collaboration}

\bigskip

G.~Aad$^\textrm{\scriptsize 85}$,
B.~Abbott$^\textrm{\scriptsize 113}$,
J.~Abdallah$^\textrm{\scriptsize 151}$,
O.~Abdinov$^\textrm{\scriptsize 11}$,
R.~Aben$^\textrm{\scriptsize 107}$,
M.~Abolins$^\textrm{\scriptsize 90}$,
O.S.~AbouZeid$^\textrm{\scriptsize 158}$,
H.~Abramowicz$^\textrm{\scriptsize 153}$,
H.~Abreu$^\textrm{\scriptsize 152}$,
R.~Abreu$^\textrm{\scriptsize 116}$
% ...
\bigskip
\\
% ...
$^{11}$ Institute of Physics, Azerbaijan Academy of Sciences, Baku, Azerbaijan\\
% ...
$^{85}$ CPPM, Aix-Marseille Universit{\'e} and CNRS/IN2P3, Marseille, France\\
% ...
$^{90}$ Department of Physics and Astronomy, Michigan State University, East Lansing MI, United States of America\\
% ...
$^{107}$ Nikhef National Institute for Subatomic Physics and University of Amsterdam, Amsterdam, Netherlands\\
% ...
$^{113}$ Homer L. Dodge Department of Physics and Astronomy, University of Oklahoma, Norman OK, United States of America\\
% ...
$^{116}$ Center for High Energy Physics, University of Oregon, Eugene OR, United States of America\\
% ...
$^{151}$ Institute of Physics, Academia Sinica, Taipei, Taiwan\\
$^{152}$ Department of Physics, Technion: Israel Institute of Technology, Haifa, Israel\\
$^{153}$ Raymond and Beverly Sackler School of Physics and Astronomy, Tel Aviv University, Tel Aviv, Israel\\
% ...
$^{158}$ Department of Physics, University of Toronto, Toronto ON, Canada\\
% ...
$^{*}$ Deceased
\end{flushleft}



\end{document}